\documentclass[12pt]{article}
\usepackage{amsmath,amssymb,amsfonts,amsthm,bbm}
\usepackage{epic,eepic,epsfig,longtable}
\usepackage{multirow,verbatim}
\usepackage{array}
\usepackage{graphicx}
\usepackage{mleftright}
\usepackage{xparse}
\usepackage{epsfig}
\usepackage{setspace}
\usepackage{CJK}
\usepackage{color}

\usepackage{lipsum}
\usepackage{mathtools}
\usepackage[authoryear,round]{natbib}
\usepackage{lineno,hyperref}
\usepackage{diagbox}

\usepackage{afterpage}
\usepackage{pdflscape}
\usepackage{stackengine}

\usepackage[normalem]{ulem}

\def\b#1{\textcolor{blue}{#1}}

\newcommand{\stkout}[1]{\ifmmode\text{\sout{\ensuremath{#1}}}\else\sout{#1}\fi}

\makeatletter
\def\singlespace{\def\baselinestretch{1}\@normalsize}

\numberwithin{equation}{section}

\renewcommand{\hat}{\widehat}

\renewcommand{\hat}{\widehat}

\newcommand{\bfm}[1]{\ensuremath{\mathbf{#1}}}

\def\bc{\bfm c}

   \def\bY{\bfm Y}

\newcommand{\bfsym}[1]{\ensuremath{\boldsymbol{#1}}}

 \def\bvartheta{\bfsym \vartheta}
              \def\bSigma{\bfsym \Sigma}

 \def\bchi{\bfsym {\chi}}

 \def\b1{\bfm 1}

\def\bvartheta{\bfsym{\vartheta}}	
\DeclareMathOperator{\argmax}{argmax}
\DeclareMathOperator{\argmin}{argmin}

\DeclareMathOperator{\cov}{cov}

\DeclareMathOperator{\var}{var}

\DeclarePairedDelimiter\norm{\lVert}{\rVert}

\def\var{\mbox{var}}

\def\today{\ifcase\month\or
  January\or February\or March\or April\or May\or June\or
  July\or August\or September\or October\or November\or December\fi
  \space\number\day, \number\year}

\newdimen\biblioindent    \biblioindent=30pt

 at 8truept

\newcommand{\beq}{\begin{equation}}
  \newcommand{\eeq}{\end{equation}}
\newcommand{\beqn}{\begin{eqnarray}}
  \newcommand{\eeqn}{\end{eqnarray}}
\newcommand{\beqnn}{\begin{eqnarray*}}
  \newcommand{\eeqnn}{\end{eqnarray*}}

\allowdisplaybreaks
\usepackage{verbatim}
\renewcommand{\baselinestretch}{1.66}
\def\tilde{\widetilde}

\def\[{\left [}  \def\]{\right ]} \def\({\left (}  \def\){\right )}
 \def\endpf{$\blacksquare$}
\def\hat{\widehat}

 \def \1 {\mathbf{1}}

\newtheorem{thm}{Theorem}

\newtheorem{lemma}{Lemma}

\theoremstyle{remark}
\newtheorem{remark}{Remark}
\theoremstyle{proposition}
\newtheorem{proposition}{Proposition}
\newtheorem{definition}{Definition}
\newtheorem{assumption}{Assumption}

\graphicspath{{figures/}{./images}}

\title{Robust Realized Integrated Beta Estimator with Application to Dynamic Analysis of Integrated Beta}

\author{ Minseog Oh$^a$, Donggyu Kim$^{b,}\footnote{Corresponding author.}$,  and Yazhen Wang$^c$ \\
{\footnotesize $^a$College of Business, Korea Advanced Institute of Science and Technology (KAIST), Seoul, South Korea}\\
{\footnotesize $^b$Department of Economics, University of California, Riverside, CA, USA}\\
{\footnotesize $^c$Department of Statistics, University of Wisconsin-Madison, WI, USA}}

\begin{document}
\maketitle
\footnotetext{\textit{E-mail address:} minsoh@kaist.ac.kr (M. Oh), donggyu.kim@ucr.edu (D. Kim), yzwang@stat.wisc.edu (Y. Wang).}
\begin{spacing}{1.45}

\begin{abstract}

In this paper, we develop a robust non-parametric realized integrated beta estimator using high-frequency financial data contaminated by microstructure noise, which is robust to the stylized features, such as the time-varying beta and the price-dependent and autocorrelated microstructure noise.
With this robust realized integrated beta estimator, we investigate dynamic structures of integrated betas and find a persistent autoregressive structure.
To model this dynamic structure, we utilize the autoregressive--moving-average (ARMA) model for daily integrated market betas.
We call this the dynamic realized beta (DR Beta).
Then, we propose a quasi-likelihood procedure for estimating the parameters of the ARMA model with the robust realized integrated beta estimator as the proxy.  
We establish asymptotic theorems for the proposed estimator and conduct a simulation study to check the performance of finite samples of the estimator.
The proposed DR Beta model with the robust realized beta estimator is also illustrated by using data from the E-mini S\&P 500 index futures and the top 50 large trading volume stocks from the S\&P 500 and an application to constructing market-neutral portfolios.

\end{abstract}

\noindent \textbf{Key words and phrases: } high-frequency financial data,   pre-averaging estimation, quasi-maximum likelihood estimation, time-varying beta.

\noindent \textbf{JEL classification:}  C14, C22, C58

\section{Introduction}\label{intro}
Market beta is a statistical measure of assets’ sensitivity to the overall market.
This measure plays a central role as the systemic risk measurement in financial applications such as asset pricing, risk management, and portfolio allocation \citep{fama2004capital,perold2004capital}.  
Thus, the characteristic of the market beta is a primary concern in empirical finance.
Especially, several empirical studies reported that market betas vary over time \citep{bos1984empirical, breen1989economic, hansen1987role, keim1986predicting}.
To account for the time-varying property, low- and high-frequency finance modeling approaches have been independently adopted.
In the low-frequency financial modeling approach, we often employ discrete-time series regression models in either a non-parametric or parametric framework based on low-frequency data such as daily, weekly, and monthly return data.
For example, \citet{fama1973risk} used a rolling window regression approach with the ordinary least square (OLS) method, 
and \citet{black1992uk} employed the state-space model by using the Kalman filter method.
In addition, to account for market beta dynamics, several studies proposed autoregressive time series models, such as generalized autoregressive conditional heteroskedasticity (GARCH) model-type structures \citep{engle2016dynamic, gonzalez1996time, koutmos1994time, ng1991tests}.
In contrast,  \citet{bollerslev2016roughing} showed that incorporating high-frequency financial data offers more benefits while capturing beta dynamics.
Specifically, intraday data provide accurate estimations with sufficient data even within a short time period.
To exploit this property, several non-parametric market beta estimators based on high-frequency data under continuous-time series regression models have been developed.
For example, \citet{ barndorff2004econometric} employed the OLS method by calculating a ratio of the integrated covariance between assets and systematic factors to the integrated variation of systematic factors.
See also  \citet{andersen2006realized, li2017adaptive, mykland2006anova, reiss2015nonparametric}.
\citet{mykland2009inference} further computed the market beta as the aggregation of market betas estimated over local blocks. 
\citet{ait2020high} proposed an integrated beta approach, using spot market betas in the absence of market microstructure noise,
and \citet{andersen2021recalcitrant} investigated intraday variation of spot market betas.
\citet{jacod2013quarticity} introduced the non-parametric inference for nonlinear volatility functionals of general multivariate It\^o semimartingales in a high-frequency, but without the presence of noise.
Recently, \citet{chen2018inference} extended this non-parametric inference to contexts with the presence of microstructure noise.
They do not allow for any dependent structure of the microstructure noise on the true latent price, nor do they account for its autocorrelation.
However, several studies indicated that the microstructure noise is not only dependent on the true latent price but also exhibits autocorrelation \citep{hautsch2013preaveraging, jacod2019estimating, li2020dependent, li2022remedi, li2023robust}.
Thus, to measure the market beta accurately, we need to develop a robust realized beta estimation procedure.

In this paper, to accommodate the stylized features, such as the time-varying beta and the price-dependent and autocorrelated microstructure noise, we develop a robust realized integrated beta ($RIB$) estimator for integrated betas with high-frequency data contaminated by price-dependent and autocorrelated microstructure noise.
For example, to handle the time-varying spot beta process and the price-dependent and autocorrelated microstructure noise, we estimate spot volatilities using the robust pre-averaging method \citep{jacod2019estimating}.
Then,  we can calculate the spot betas using spot volatility estimators.
However, due to the microstructure noise, they have asymptotically diverged bias with a convergence rate of $m^{-1/4}$, which is known as the optimal with the presence of microstructure noise.
To overcome this problem, we introduce a bias adjustment scheme and integrate the bias-adjusted spot beta estimators to obtain the realized integrated beta estimator.  
We show its asymptotic properties and obtain the convergence rate $m^{-1/4}$.
To the best of our knowledge, the proposed $RIB$ is the first integrated beta estimator, which is robust to the financial features, such as the time-varying beta and the price-dependent and autocorrelated microstructure noise.
Since the proposed $RIB$ estimation procedure provides an accurate and robust market beta estimator, it may help us study the dynamic structures of integrated market betas.

With the $RIB$ estimator, we find that the realized betas have persistent autoregressive (AR) structures (see Figure \ref{Figure-1} in Section \ref{sec-3}).
This result coincides with the previous literature.
The literature on beta dynamics predominantly employs two approaches; modeling conditional covariance \citep{engle2016dynamic,gonzalez1996time,hansen2014realized,koutmos1994time,ng1991tests} and directly modeling conditional beta.
\citet{adrian2009learning,ang2007capm,blume1971assessment} employed AR(1) structure to analyze beta dynamics based on low-frequency data, such as monthly or quarterly stock returns.
\citet{andersen2006realized,becker2021memory,hollstein2016estimating} employed the class of ARFIMA structure on monthly, quarterly, and semiannual beta, estimated from 30-minute and daily returns.
In this paper, we model the daily integrated betas using the ARMA($p,q$) model to capture the persistent AR structure and call this dynamic realized beta (DR Beta).
To estimate the parameters of the ARMA model, we suggest a quasi-maximum likelihood estimation procedure with the robust non-parametric $RIB$ estimator. 
For example, we use $RIB$ as the proxy for the corresponding conditional expected integrated beta and employ the well-known least square loss function.
It is crucial to use a consistent estimator when working with the ARMA model, as measurement errors can significantly jeopardize estimation and prediction accuracy \citep{koreisha1999impact}.
Since incorporating ultra-high-frequency data contaminated by microstructure noise is essential for obtaining consistent estimators of daily integrated beta, the analysis of these betas presents different aspects of asymptotic behavior compared to previous literature that uses at least monthly beta.
To address these points, we establish asymptotic theorems for the proposed estimation procedure and further discuss how to conduct hypothesis tests. 

The rest of the paper is organized as follows.
In Section \ref{sec-2}, we propose statistical inference procedures for the integrated beta.
In Section \ref{sec-3}, we suggest the DR Beta model and examine the parameter estimation procedure with their asymptotic theorems.
In Section \ref{sec-4}, we provide a simulation study to check the finite sample performance for the proposed estimators.
In Section \ref{sec-5}, we carry out an empirical study with the E-mini S\&P 500 index futures and 50 individual stocks to investigate the advantage of the proposed model. 
In Section \ref{sec-6}, we conclude.
The proofs and supplementary materials are collected in the online Appendix.

\section{Robust realized integrated beta estimator} \label{sec-2}

 \subsection{Model setup}
We first fix some notations that we will use.
Let $\mathbb{R}_+=[0,\infty)$ and $\mathbb{N}$ be the set of all positive integers.
Let $A_{ij}$ denote the $(i,j)$th element of a matrix $A$, $A^\top$ denote its transpose matrix, and $\det(A)$ denote the determinant of $A$.
We use the superscripts $c$ and $d$ for the continuous and jump processes, respectively.

We consider the following diffusion regression model, as originally introduced in \citet{mykland2006anova} (see also \citet{li2017adaptive,li2016generalized,reiss2015nonparametric}):
\begin{equation}\label{Equation-2.1}
    dX_{2,t} = \beta_{t-}^c dX_{1,t}^c +  \beta^d_{t-} \Delta X^d_{1,t} +dV_t ,
\end{equation}
where $X_2$ is a dependent process, $X_1$ is a covariate process, and $V$ is a residual process. 
Further, $X^c_{1,t}$ denotes the continuous part of the covariate process, and $\Delta X^d_{1,t}$ is its jump at time $t$.  
Then, $\beta^c_t$ and $\beta^d_t$ are the time-varying factor loadings with respect to the continuous and jump parts of $X_1$, respectively.
We assume that $X_{1,t}$ and $V_t$ admit the following Grigelionis decomposition of the forms:
\begin{eqnarray*}\label{Equation-2.2:Grigelionis}
  &&X_{1,t}  =   X_{1,0} + \int_{0}^{t} \mu_{1,s}ds + \int_{0}^{t}  \sigma_{s}dB_{s} +  \mathfrak{d}_{1}  \b1_{\left\lbrace \left|\mathfrak{d}_{1}\right| \leq 1  \right\rbrace} \ast (\mathfrak{p} - \mathfrak{q})_{t} +  \mathfrak{d}_{1}  \b1_{\left\lbrace \left|\mathfrak{d}_{1}\right| > 1  \right\rbrace} \ast \mathfrak{p}_{t}, \\
  &&V_t  =  V_{0} + \int_{0}^{t}  \mu_{2,s}ds + \int_{0}^{t}  q _{s}dW_{s}  +  \mathfrak{d}_{2}  \b1_{\left\lbrace \left|\mathfrak{d}_{2}\right| \leq 1  \right\rbrace} \ast (\mathfrak{p} - \mathfrak{q})_{t} +  \mathfrak{d}_{2}  \b1_{\left\lbrace \left|\mathfrak{d}_{2}\right| > 1  \right\rbrace} \ast \mathfrak{p}_{t},
\end{eqnarray*}
where $\mu_{1,t}$ and $\mu_{2,t}$ are c\`adl\`ag, progressively measurable, and locally bounded drifts, $\sigma_t$ and $q_t$ are adapted c\`adl\`ag processes,  $B_t$ and $W_t$ are independent standard Brownian motions, $\mathfrak{p}$ is a Poisson random measure on $\mathbb{R}^{+} \times E$, with the compensator $\mathfrak{q}(dt,dx) = dt \otimes \lambda (dx)$ and the Polish space $(E, \mathcal{E})$, $\lambda$ is a $\sigma$-finite measure, $\mathfrak{d}_{1}$ and $\mathfrak{d}_{2}$ are predictable functions on $\Omega \times \mathbb{R}^{+} \times E$.
All random quantities above are defined on a fixed filtered probability space $(\Omega, \mathcal{F}, (\mathcal{F}_{t})_{t\geq 0}, \mathbb{P})$.
Furthermore, $\sigma^2_t$ stays away from 0.

For the proposed time series regression model in \eqref{Equation-2.1}, we assume that time-varying market betas follow a stochastic process defined on $(\Omega, \mathcal{F}, P)$ as follows:
\begin{equation} \label{Equation-3.beta}
	d \beta^c_t = \mu_{\beta,t}dt+\sigma_{\beta, t}dB_{\beta, t},
\end{equation} 
where $\mu_{\beta,t}$ is progressively measurable and locally bounded drift, $\sigma_{\beta,t}$ is c\`adl\`ag,  $B_{\beta, t}$ is a standard Brownian motion with $dB_{\beta, t}dW_t=0$ and $dB_{\beta, t} dB_t=\rho_{\beta,t} dt$ a.s.
To measure the daily market beta, we use the following integrated beta ($I\beta$):
\begin{equation} \label{Equation-2.4}
    I\beta_i=\int_{i-1}^{i} \beta^c_t dt,    \quad i \in \mathbb{N}.
\end{equation} 
In this paper, the parameter of interest is the daily integrated beta.
When the beta process is constant over time with no price jumps, the integrated beta returns to the usual market beta of the capital asset pricing model (CAPM).
That is, the diffusion regression model includes the traditional discrete-time CAPM regression.
\begin{remark}\label{remark-1}
In this paper, we separate the continuous and jump parts and mainly consider the continuous part. 
We also investigate market betas corresponding to the jump part in the empirical study, which is calculated based on the jump beta estimation method suggested by \citet{li2017robust}.
However, unlike the beta for the continuous part,  the beta for the jump part does not have significant time series structures (see Figure \ref{Figure-A}). 
Thus, we focus on the beta process for the continuous part. 
\end{remark}

For the high-frequency observations, one of the stylized features is that the transaction prices are polluted by the market microstructure noise due to the discreteness of the price, bid-ask spread bounce, and adverse selection effects, such as clearing costs \citep{ait2008high}.
To reflect this, we assume that the observed log prices have the additive microstructure noise as follows: 
\begin{equation} \label{Equation-3.1}
    Y_{1,i}^{m}=X_{1,t_i}+\epsilon_{1,i}^{m} \quad \text{and} \quad    Y_{2,i}^{m}=X_{2,t_i}+\epsilon_{2,i}^{m} ,  
\end{equation} 
where $\epsilon_{1,i}^{m}$ and $\epsilon_{2,i}^{m}$ are the noise.
Empirical studies reveal that the microstructure noise is dependent on the true price \citep{ait2011ultra,hansen2006realized,ubukata2009estimation} and has positive autocorrelation \citep{jacod2017statistical,li2022remedi}.
To capture this, we allow microstructure noise to have a dependence on the true latent price, diurnal features, and polynomial decaying autocorrelation.
Before describing our assumption about microstructure noise, we state the $\rho$-mixing property of a stationary random vector $\bfsym\chi = (\bfsym\chi_i)_{i \in \mathbb{Z}} = ((\chi_{1,i},\chi_{2,i})^\top)_{i \in \mathbb{Z}}$.
\begin{definition}\label{rho-mixing}
  For a stationary process $\bfsym\chi$, let $\mathcal{G}_j = \sigma(\bfsym\chi_i : i \leq j)$ and $\mathcal{G}^j = \sigma(\bfsym\chi_i : i \geq j)$ be the pre- and post-$\sigma$-fields at time $j$.
  A stationary process $\bfsym\chi$ is $v$-polynomially $\rho$-mixing if for some $C > 0$, $\rho_k(\bfsym\chi) \leq C/k^v$ for all $k \geq 1$, where
  \begin{eqnarray*}
       \rho_k(\bfsym\chi)  & =& \sup \big\{ |\mathbb{E}(UV)| :  U \text{ and } V \text{ are random variables, measurable  with respect to}  \\
      && \mathcal{G}_0 \text{ and } \mathcal{G}^k, \text{ respectively} , \quad \mathbb{E}(U) = \mathbb{E}(V) = 0,    \mathbb{E}(U^2) \leq 1,  \mathbb{E}(V^2) \leq 1 \big\}.
  \end{eqnarray*}
\end{definition}

\begin{assumption}\label{assumption-noise}
  The noise $(\bfsym {\epsilon}_{i}^m)_{i \in \mathbb{Z}} = ((\epsilon_{1,i}^{m}, \epsilon_{2,i}^{m})^\top)_{i \in \mathbb{Z}}$ is realized as
  \begin{equation}\label{noise-form}
      \epsilon_{1,i}^{m} = \vartheta_{1,t_i} \chi_{1,i} \quad \text{and} \quad \epsilon_{2,i}^{m} = \vartheta_{2,t_i} \chi_{2,i} ,
  \end{equation}
  where $\vartheta_1$ and $\vartheta_2$ are non-negative It\^{o} semimartingales with locally bounded drift and c\`adl\`ag diffusion terms.
  Furthermore, $(\bfsym\chi_i)_{i \in \mathbb{Z}}$ is a stationary process, independent of the $\sigma$-field $\mathcal{F}_{\infty} = \bigvee_{t>0} \mathcal{F}_{t}$ and $v$-polynomially $\rho$-mixing for some $v \geq 5 $.
  $\chi_{1,i}$ and $\chi_{2,i}$ are mean 0 and variance 1 with finite moments of all orders.
\end{assumption}
\begin{remark}
  Assumption \ref{assumption-noise} implies that there exists a constant $C$ such that $\left|r_{ab}(i)\right| \leq \frac{C}{\left( \left|i\right| +1 \right)^v }$ for all $i \in \mathbb{Z}$ and $a,b \in \left\lbrace 1,2 \right\rbrace$, where $r_{ab}(i) = \mathbb{E}[\chi_{a,0} \chi_{b,i}]$.
  Thus, $R_{ab} = \sum_{i\in \mathbb{Z}} r_{ab}(i)$ is well defined.
\end{remark}

\subsection{Robust realized integrated beta estimator} \label{sec-3.1}

When it comes to estimating the integrated beta based on the observed high-frequency financial data, there are a couple of obstacles. 
One is the microstructure noise, and the other is the intraday dynamics of the spot beta process. 
In this section, we discuss how to overcome these issues for the general stochastic beta process in \eqref{Equation-3.beta}.

For simplicity, we temporarily assume that the distance between adjacent observations is equal to $\Delta_{m}={1}/{m}$, where $m$ is the number of high-frequency observations. 
We denote the high-frequency observed time points $t_l={l}/{m}$ for $l=1,  \ldots , m$. 
This equally spaced observation time assumption can be easily extended to irregular observation time points. 
We discuss this later.
To manage the intraday dynamics--that is, the time-varying spot beta process--we can use the following relationship:
$$
  \frac{d} {dt}[X^c_{2,t}, X^c_{1,t}] = \beta_t^c \frac{d} {dt} [X^c_{1,t}, X^c_{1,t}] ,
$$ 
where $X^c_{2,t}$ is the continuous part of the individual asset log price process, and   $[\cdot, \cdot]$ denotes the quadratic covariation.
If $\beta_{t}^{c}$ is identifiable--that is, the spot volatility of $X_1^c$ is nonzero--, we can obtain the spot beta by comparing the spot volatility of $X_1^c$ and the spot covolatility between $X_1^c$ and $X_2^c$  as follows:
\begin{equation}\label{eq-beta}
 \beta_t^c  = \frac{ \frac{d} {dt} [X^c_{2,t}, X^c_{1,t}]}{  \frac{d} {dt}[X^c_{1,t}, X^c_{1,t}]}.
\end{equation}
This is similar to the result of the usual regression coefficient, which is the covariance of the dependent and covariate variables over the variance of the covariate variable. 
The difference is that the spot beta is defined by the spot volatility and covolatility. 
Thus, it can represent the linear relationship at time $t$ between the dependent and covariate processes. 
If the spot volatility and covolatility are constant over time, the spot beta is the same as the usual regression coefficient.
By integrating the spot beta process, we finally obtain the integrated beta. 
Therefore, as long as the spot volatility estimators perform well, we can estimate the integrated beta.

To estimate spot volatilities, we employ the estimation method developed for estimating integrated volatility with microstructure noise \citep{ait2010high,barndorff2008designing,barndorff2011multivariate,christensen2010pre,fan2018robust,jacod2009microstructure,jacod2019estimating, shin2023adaptive, xiu2010quasi, zhang2006efficient, zhang2005tale, zhang2016jump}.
In order to handle the autocorrelation structure of the microstructure noise, we employ the pre-averaging method in \citet{jacod2019estimating} as follows.
We choose a sequence of integers, $k_m$, such that ${k_m}= C_k \Delta_m^{-1/2}$ for some positive constant $C_k$.
We select a weight function $g(\cdot)$ on $[0,1]$ satisfying that $g (\cdot)$
is continuous, piecewise continuously differentiable with a piecewise Lipschitz derivative ${g}'$ with $g(0)=g(1)=0$ and  $\int_{0}^{1}g^2(s)ds>0$. 
Let
\begin{eqnarray*}
    &&\phi_0(s)=\int ^1_s g(u)g(u-s)du,  \;  \psi_0=\phi_0(0),  \phi_1(s)=\int ^1_s g'(u)g'(u-s)du, \; \psi_1=\phi_1(0), \cr
    &&\Phi_{00}=\int^1_0 \phi_0^2(s)ds, \; \Phi_{01}=\int^1_0 \phi_0(s) \phi_1(s)ds, \; \Phi_{11}=\int^1_0 \phi_1^2(s)ds.
\end{eqnarray*}
We also choose a sequence of integers, $l_m$, such that $l_m=C_l \Delta_m ^{-\varsigma}$ for some positive constant $C_l$ and $\varsigma \in [\frac{1}{8}, \frac{1}{5} ]$.
Then, for $l =1, \ldots, m$, $d = 1, \ldots, l_m$, and any processes $P$ and $P'$, we define
\begin{eqnarray*}
  && g_{d}^{m} =  g \left( \frac{d}{k_m}  \right) , \quad P_{l}^{m} = P_{t_l }, \quad
  \tilde{P}_{l}^{m}=\sum_{j=1}^{k_m -1}g_{j}^{m} \left( P_{l+j}^{m} - P_{l+j-1}^{m} \right)  , \quad \bar{P}_{l}^{m}  = \frac{1}{l_m} \sum_{i=0}^{l_m-1} P_{l+i}^{m}, \\
  && \mathcal{E}_{PP',l}^{m,d}  = \left( P_{l}^{m} - \bar{P}_{l+2l_m}^{m}   \right) \left( P_{l+d}^{'m} - \bar{P}_{l+4l_m}^{'m} \right), \quad \mathcal{E}_{PP',l}^{m,-d}  = \left( P_{l}^{'m} - \bar{P}_{l+2l_m}^{'m} \right) \left( P_{l+|d|}^{m} - \bar{P}_{l+4l_m}^{m}   \right) .
\end{eqnarray*}
The spot covariance matrix $\bSigma$ of $(X_1^c, X_2^c)^\top$ at time $t_l$ is estimated with
\begin{equation*} \label{Equation-3.3}
    \hat{\bSigma}_{l}^{m}  = \hat{\bSigma}_{t_l} =
    \begin{pmatrix}
      v(Y_1,Y_1,u_{1,m},u_{1,m},t_l) & v(Y_1,Y_2,u_{1,m},u_{2,m},t_l) \\
      v(Y_2,Y_1,u_{2,m},u_{1,m},t_l) & v(Y_2,Y_2,u_{2,m},u_{2,m},t_l)
    \end{pmatrix},
  \end{equation*}
  where
  \begin{eqnarray*}
    && v(P,P',a, a', t_l) = 
    \frac{1}{(b_m - 2 k_m) \Delta_m k_m \psi_0} 
    \Bigg\lbrace  \sum_{i=0}^{b_m - 2 k_m -1} \tilde{P}_{l+i}^{m}  \tilde{P}_{l+i}^{'m}  \mathbf{1}_{\{|\tilde{P}_{l+i}^{m}|\leq a, \, |\tilde{P}_{l+i}^{'m}|\leq a' \}} \\
    && \qquad \qquad \qquad \qquad \qquad \qquad \quad - \frac{1}{k_m}  \sum_{i=0}^{b_m - 6 l_m}  \sum_{d=-k_m'}^{k_m'} \phi_{d}^{m}  \mathcal{E}_{PP',l+i}^{m,d}  \Bigg\rbrace,\\
    && \phi_{d}^{m} = k_m \sum_{i\in\mathbb{Z}} \left( g_{i+1}^{m} - g_{i}^{m} \right) \left( g_{i-d+1}^{m} - g_{i-d}^{m} \right)  , %
  \end{eqnarray*}
$b_m = C_b \Delta_m^{-\kappa}$, $k'_m = C_{k'} \Delta_m ^{-\tau}$ for some positive constant $C_b$ and $C_{k'}$, tuning parameters  %
$\kappa \in ( \frac{2}{3}  , \frac{3}{4}  )$ and $\tau \in  (\frac{1}{4v-4}  , \frac{1}{8}   ]$, 
$u_{1,m}$ and $u_{2,m}$ are the thresholds chosen as $u_{1,m} = a_{1} (k_m \Delta_m )^{\varpi_1}$ and $u_{2,m} = a_{2} (k_m \Delta_m )^{\varpi_1}$ for some $a_{1}, a_{2} > 0$, and $\varpi_1 \in (\frac{[v]-1}{2[v]-r}, \frac{2[v]-5}{4[v]-8} )$ for $r$ defined in Assumption \ref{assumption-formal}(c).
Under some mild conditions, we can show the consistency of the spot volatility estimator (see Theorem \ref{Theorem-1} and \citet{figueroa2022kernel}).
Using the plug-in method, we can estimate the spot beta with the above spot volatility estimators. 
However, due to the microstructure noise, the functional form of the spot volatility estimators has a bias term.
This fact prevents obtaining the asymptotic distribution with the convergence rate of $m^{-1/4}$ when estimating the integrated beta by the simple integration of the biased spot beta estimators. 
To overcome this, we introduce a bias adjustment scheme and construct a realized integrated beta ($RIB$) estimator as follows:
\begin{equation} \label{Equation-3.4}
RIB_{1}=\hat{I \beta}_{1}=b_m\Delta_m \sum ^{\lfloor \frac{1}{b_m\Delta_m} \rfloor -1 }_{i=0} 
\left( \hat{\beta}_{i b_m }^{m}  - \hat{B}^{m}_{ib_m} \right), \quad \hat{\beta}^m_{i}=\frac{\hat{\bSigma}_{12,i}^{m}}{\hat{\bSigma}_{11,i}^{m,*}}, \quad  \hat{\bSigma}_{11,i}^{m,*} = \max( \hat{\bSigma}_{11,i}^{m}, \delta_m),
\end{equation}
where $\delta_m$ is a sequence of positive real numbers converging to zero and 
$\hat{B}^{m}_{ib_m}$ is a de-biasing term of the form
\begin{align}\label{estimators-for-debiasing}
    & \hat{B}^{m}_{i} = \hat{B}_{t_{i}} =   \frac{4}{\psi_0^{2} {C_k}^3 b_m\Delta_m^{1/2}} \left[ \left( \frac{{C_k}^2 \Phi_{01}}{\hat{\bSigma}_{11,i}^{m,*}} + \frac{\Phi_{11} \hat{\bvartheta}_{11,i}^{m}}{\left( \hat{\bSigma}_{11,i}^{m,*} \right)^2 }   \right) \left( \hat{\bvartheta}_{11,i}^{m} \frac{\hat{\bSigma}_{12,i}^{m}}{\hat{\bSigma}_{11,i}^{m,*}} - \hat{\bvartheta}_{12,i}^{m}  \right)  \right]   ,\\
    & \hat{\bvartheta}^{m}_{11,i} = (b_m-6l_m)^{-1} \sum_{j=i}^{i + b_m - 6 l_m}  \sum_{d=-k_m'}^{k_m'} \mathcal{E}_{Y_1 Y_1,j}^{m,d}, \quad
    \hat{\bvartheta}^{m}_{12,i} = (b_m-6l_m)^{-1} \sum_{j=i}^{i + b_m - 6 l_m}  \sum_{d=-k_m'}^{k_m'} \mathcal{E}_{Y_1 Y_2,j}^{m,d}, \nonumber\\
    & \hat{\bvartheta}^{m}_{22,i} = (b_m-6l_m)^{-1} \sum_{j=i}^{i + b_m - 6 l_m}  \sum_{d=-k_m'}^{k_m'} \mathcal{E}_{Y_2 Y_2,j}^{m,d}. \nonumber
\end{align}
We utilize $\hat{\bSigma}_{11,l}^{m,*}$ instead of $\hat{\bSigma}_{11,l}^{m}$ for estimating spot beta in order to prevent the denominator of $\hat{\beta}^m_{i}$ from being a non-positive value.
Thanks to the de-biasing term $\hat{B}^{m}_{i b_m }$, the average form in \eqref{Equation-3.4} can achieve the optimal convergence rate $m^{-1/4}$. 

\begin{remark}
  The $RIB$ estimator is developed along the lines of the estimators in \citet{chen2018inference,jacod2013quarticity}, which proposed estimators of integrated volatility functionals.
  Specifically, \citet{jacod2013quarticity} considered the estimator in the absence of microstructure noise, and \citet{chen2018inference} addressed the case of i.i.d. microstructure noise presence using the traditional pre-averaging scheme.
  The traditional pre-averaging scheme utilizes the property that microstructure noise dominates the high-frequency returns, $Y_{1,i+1}^{m} - Y_{1,i}^{m} \approx \epsilon_{1,i+1}^{m} - \epsilon_{1,i}^{m}$ as $m \rightarrow \infty$, to remove the effect of microstructure noise.
  Specifically, to remove the effect of microstructure noise, \citet{chen2018inference} utilized the summation of squared high-frequency returns with proper normalizations, since $(Y_{1,i+1}^{m} - Y_{1,i}^{m})^{2} \approx (\epsilon_{1,i+1}^{m})^2 + (\epsilon_{1,i}^{m})^2 - 2 \epsilon_{1,i+1}^{m} \epsilon_{1,i}^{m}$ as $m \rightarrow \infty$ and the part related to the cross-product of the noise, $- 2 \epsilon_{1,i+1}^{m} \epsilon_{1,i}^{m}$, becomes asymptotically negligible when the noise is i.i.d.
  However, in the presence of autocorrelated microstructure noise, this approach faces challenges since the part related to the cross-product of the noise cannot correctly remove the autocorrelated noise effect.
  The $RIB$ estimator, adapting the approach presented in \citet{jacod2019estimating}, handles the dependent structure of microstructure noise by directly utilizing a proxy for the microstructure noise, $Y_{1,i}^{m} - \bar{Y}_{1,i+2l_m}^{m} \approx \epsilon_{1,i}^{m}$ as $m \rightarrow \infty$ \citep{jacod2017statistical}.
  Specifically, the autocorrelated microstructure noise effect can be estimated by the summation of $(Y_{1,i}^{m} - \bar{Y}_{1,i+2l_m}^{m})(Y_{1,i}^{m} - \bar{Y}_{1,i+d+4l_m}^{m}) \approx \epsilon_{1,i}^{m}\epsilon_{1,i+d}^{m}$ over the high-frequency observation index $i$ and the lag index $d$ with proper normalizations.
  Therefore, the $RIB$ estimator is a consistent estimator of the integrated beta, whereas the others in \citet{chen2018inference,jacod2013quarticity} are not consistent estimators in the presence of autocorrelated microstructure noise.
  The consistency of estimators plays a crucial role in time series analysis, as it contributes to capturing the time series dynamics.
\end{remark}
\begin{remark}
  To better understand the robustness of the autocorrelation of microstructure noise, it would be helpful to compare the $RIB$ estimator with the estimator of \citet{chen2018inference} in the presence of i.i.d. noise.
  When the microstructure noise is i.i.d., both estimators are consistent estimators of the integrated beta.
  However, the denoising term of $RIB$ is more complex than that of \citet{chen2018inference} even in the case of $k'_m=0$ due to the estimation step for the proxy of microstructure noise.
  This complexity in the denoising term may lead to worse finite sample performance for i.i.d. noise cases due to the estimation variance, although it achieves the same asymptotic convergence rate.
  That is, the $RIB$ estimator is robust to autocorrelated microstructure noise at the expense of estimating the proxy of the microstructure noise.
\end{remark}

To investigate the asymptotic behavior of the $RIB$ estimator, we need the following technical conditions.
\begin{assumption}\label{assumption-formal}
~
\begin{enumerate}
\item[(a)]  The processes $\sigma_{t}$ and $\sigma_{t}^{-1}$ are locally bounded.

\item[(b)]  We have
\begin{eqnarray*}
  && \sigma^2_{t}= \sigma^2_0 + \int_{0}^{t}  \tilde{\mu}_{1,s}ds + \int_{0}^{t}  \tilde{\sigma}_{s}d\tilde{B}_{s} + \tilde{\mathfrak{d}}_{1} \b1_{\left\lbrace \left|\tilde{\mathfrak{d}}_{1}  \right| \leq 1  \right\rbrace} \ast (\mathfrak{p} - \mathfrak{q})_{t} + \tilde{\mathfrak{d}}_{1} \b1_{\left\lbrace \left|\tilde{\mathfrak{d}}_{1}  \right| > 1  \right\rbrace} \ast \mathfrak{p}_{t} \;  \text{ and } \\
  && q^2_{t}= q^{2}_{0} + \int_{0}^{t}  \tilde{\mu}_{2,s}ds + \int_{0}^{t}  \tilde{q}_{s} d\tilde{W}_{s} + \tilde{\mathfrak{d}}_{2} \b1_{\left\lbrace \left|\tilde{\mathfrak{d}}_{2}  \right| \leq 1  \right\rbrace} \ast (\mathfrak{p} - \mathfrak{q})_{t} + \tilde{\mathfrak{d}}_{2} \b1_{\left\lbrace \left|\tilde{\mathfrak{d}}_{2}  \right| > 1  \right\rbrace} \ast \mathfrak{p}_{t},
\end{eqnarray*} 
where $\tilde{\mu}_{1,t}$ and  $\tilde{\mu}_{2,t}$ are progressively measurable and locally bounded drifts; $\tilde{\sigma}_{t}$ and   $\tilde{q}_{t}$ are adapted c\`adl\`ag processes.
 The standard Brownian motions $\tilde{B}_{t}$ and $\tilde{W}_{t}$ satisfy almost surely
\begin{eqnarray*}
    && d\tilde{B}_{t}dB_{\beta,t}=0, \quad d\tilde{B}_{t}dW_{t}=0, \quad d\tilde{B}_{t}d\tilde{W}_{t}=0, \quad d\tilde{B}_{t}dB_{t}=\tilde{\rho}_{1,t}dt, \\
    && d\tilde{W}_{t}dB_{\beta, t}=0, \quad d\tilde{W}_{t}dB_{t}=0, \quad d\tilde{W}_{t}d\tilde{B}_{t}=0, \quad d\tilde{W}_{t}dW_{t}=\tilde{\rho}_{2,t}dt,
\end{eqnarray*}
where $\tilde{\rho}_{1,t}$ and $\tilde{\rho}_{2,t}$ are bounded.
The stochastic processes $\tilde{\mu}_{1,t}, \tilde{\mu}_{2,t}, \tilde{\sigma}_{t}$, $\tilde{q}_{t}$, $\tilde{\rho}_{1,t}$, and $\tilde{\rho}_{2,t}$  are defined on $(\Omega,\mathcal{F}, P)$.

\item[(c)]
For some $r \in [0,\frac{2[v]-8}{2[v]-5} )$, there are a sequence $T_{k}$ of stopping times increasing to $\infty$, a sequence of deterministic nonnegative $\lambda$-integrable functions $\mathcal{J}_{k}$ on $\mathbb{R}^{2}$ such that $\left|\mathfrak{d}_{i}(\omega,t,z)\right|^{r} \land 1 \leq \mathcal{J}_{k}(z)$, $\left|\beta_{t-}^{d} \mathfrak{d}_{1}(\omega,t,z)\right|^{r} \land 1 \leq \mathcal{J}_{k}(z)$ and $\left|\tilde{\mathfrak{d}}_{i}(\omega,t,z)\right|^{2} \land 1 \leq \mathcal{J}_{k}(z) $ for $i \in \{1,2\}$ and all $(\omega,t,z)$ with $t \leq T_{k}(\omega)$.

\item[(d)]
If $P_{t}$ is one of the processes $\tilde{\sigma}_{t}$, $\tilde{q}_{t}$, $\tilde{\rho}_{1,t}$, or $\rho_{\beta,t}$, then it satisfies the property (P-2), where
\begin{enumerate}
  \item [(P-k)] There exist $C > 0$, such that $\mathbb{E}\left[ \sup_{u \in [t,t+s]} (P_{u} - P_{t})^{k} | \mathcal{F}_{t} \right] \leq C s$ a.s.  for any $t,s \geq 0$.
\end{enumerate}
\end{enumerate}
\end{assumption} 
\begin{remark}
  The locally bounded condition of $\sigma_{t}^{-1}$ in Assumption \ref{assumption-formal}(a) is required to identify the beta from the processes.
  Assumption \ref{assumption-formal}(c) is required to bound the degree of activity of jumps \citep{ait2009estimating}.
  The parameter $r$ should be less than $\frac{2[v]-8}{2[v]-5}$, whereas \citet{jacod2019estimating} requires $r < \frac{2[v]-4}{2[v]-3} $.
  This is because the $RIB$ estimator requires the second-moment condition for the jump-truncation error, whereas \citet{jacod2019estimating} only requires convergence in probability for the jump-truncation error.
  If $\bchi$ is $\rho$-mixing with exponential decay rate, we only require $r < 1$.
  Assumption \ref{assumption-formal}(d) holds for any It\^{o} semimartingale process with bounded drift, diffusion, and jump terms.
\end{remark}

The following theorem establishes the convergence rate and asymptotic distributions for the proposed $RIB$ estimator.
\begin{thm} \label{Theorem-1}
    Under Assumptions \ref{assumption-noise} and \ref{assumption-formal}, we have
    \begin{equation*}
        m ^{1/4}(RIB_{1}-I \beta_1) \rightarrow \int_{0}^{1} \mathcal{R}_s d\tilde{Z}_s \quad \mathcal{F}_{\infty}\text{-stably as }  m \rightarrow \infty ,
    \end{equation*}
    where $\tilde{Z}$ is a standard Brownian motion independent of $\mathcal{F}$, $\mathcal{R}_s$ is the square root of
    \begin{equation*}
      \mathcal{R}_s = \frac{2}{\psi_0^2} \left( \Phi_{00}  \frac{{C_k} q_s^2}{\sigma_s^2} + \Phi_{01} \frac{A_{1,s}}{C_k} + \Phi_{11} \frac{A_{2,s}}{{C_k}^3}   \right) 
      ,
    \end{equation*}
    and
    \begin{eqnarray*}
      A_{1,s} &=&  \frac{\vartheta_{1,s}^2 R_{11} q_s^2 - 2 \beta_{s}^{c} \vartheta_{1,s}\vartheta_{2,s} R_{12} + \vartheta_{2,s}^2 R_{22}}{\sigma_s^2} + \vartheta_{1,s}^2 R_{11} (\beta_{s}^{c})^2 , \\
      A_{2,s} &=&  \frac{\vartheta_{1,s}^{2}}{\sigma_{s}^{4}} \left( 2(\beta_{s}^{c})^2 \vartheta_{1,s}^{2} R_{11}^{2} - 4 \beta_{s}^{c} \vartheta_{1,s} \vartheta_{2,s} R_{11} R_{12} + \vartheta_{2}^{2} (R_{11}R_{22} + R_{12}^2) \right) 
      .
    \end{eqnarray*}
\end{thm}

Theorem \ref{Theorem-1} shows that the convergence rate of the $RIB$ estimator is $m^{-1/4}$, which is known as the optimal rate with the presence of microstructure noise and establishes its asymptotic normality.
To extend the estimator over all periods, we set $m_i$ to be the total number of high-frequency observations for the $i$th day and rewrite $m = \sum_{i=1}^n m_i /n$. 
We further let the $t_{i,j}$'s be the high-frequency observed time points for the $i$th day, such that $i-1=t_{i,0}<t_{i,1}< \cdots <t_{i,m_i}=i$, where   $\left| t_{i,j}-t_{i,j-1}\right| =O\left( m^{-1}\right)$ for all $i, j$. 
Then,  we can construct the $RIB$ estimator as follows:
\begin{equation}  \label{Equation-3.5}
  RIB_{i}=\frac{1}{m_i} \sum_{l=0}^{m_i - b_{m_i}} 
  \left[ \hat{\beta}_{i,t_{i,l}} - \hat{B}_{i,t_{i,l}} \right]  \quad \text{and} \quad  \hat{\beta}_{i,t_{i,l}}=\frac{\hat{\bSigma}_{21, t_{i,l}}}{\max(\hat{\bSigma}_{22,t_{i,l}},\delta_{m_i})}.
\end{equation}
Moreover, we can show Theorem \ref{Theorem-1} for each $RIB_i$ under the usual assumption in the asynchronous high-frequency data analysis (see Assumption \ref{Assumption-2}(e)). 
We utilize these well-performing $RIB$ estimators to analyze the dynamic structures of integrated betas in the following section.

To utilize the asymptotic distribution result, we need to construct a consistent asymptotic variance estimator.
In the following proposition, we propose the asymptotic variance estimator and show its consistency.

\begin{proposition}\label{prop-AsympVar}
  Under Assumptions \ref{assumption-noise} and \ref{assumption-formal}, $\hat{S}_m =  b_m \Delta_m  \sum_{i=0}^{\left[ \frac{1}{b_m\Delta_m} \right]  -1 } \hat{\mathcal{R}}^{2,m}_{i b_m} $ is a consistent asymptotic variance estimator of the $RIB$ estimator, where
  \begin{eqnarray*}
    \hat{\mathcal{R}}^{2,m}_{i} &=& \frac{2C_k}{\psi_{0}^{2}} \Biggl[ \Phi_{00}  \left( \frac{\hat{\bSigma}_{22,i}^{m}}{\hat{\bSigma}_{11,i}^{m,*}} - \frac{(\hat{\bSigma}_{12,i}^{m})^2}{(\hat{\bSigma}_{11,i}^{m,*})^2}  \right)     + \frac{\Phi_{01}}{C_k^{2}}  \left( \frac{\hat{\bvartheta}_{22,i}^{m}}{\hat{\bSigma}_{11,i}^{m,*}} -  \frac{2\hat{\bSigma}_{12,i}^{m} \hat{\bvartheta}_{12,i}^{m}  }{ ( \hat{\bSigma}_{11,i}^{m,*}  )^2  }  + \frac{\hat{\bSigma}_{22,i}^{m} \hat{\bvartheta}_{11,i}^{m} }{( \hat{\bSigma}_{11,i}^{m,*}  )^2 }   \right) \\
    &&  + \frac{\Phi_{11}}{C_k^3}  \left( \frac{2  (\hat{\bSigma}_{12,i}^{m} \hat{\bvartheta}_{11,i}^{m} )^2    }{(\hat{\bSigma}_{11,i}^{m,*} )^4} + \frac{\hat{\bvartheta}_{11,i}^{m} \hat{\bvartheta}_{12,i}^{m} }{(\hat{\bSigma}_{11,i}^{m,*} )^2} - 4 \frac{\hat{\bSigma}_{12,i}^{m} \hat{\bvartheta}_{11,i}^{m} \hat{\bvartheta}_{12,i}^{m}  }{(\hat{\bSigma}_{11,i}^{m,*} )^3}  + \frac{(\hat{\bvartheta}_{11,i}^{m} )^2}{(\hat{\bSigma}_{11,i}^{m,*} )^2}  \right)     \Biggl]
    ,
  \end{eqnarray*}
  where $\hat{\bvartheta}_{11,i}^{m}$, $\hat{\bvartheta}_{12,i}^{m}$, and $\hat{\bvartheta}_{12,i}^{m}$ are defined in \eqref{estimators-for-debiasing}.
  That is, we have $\hat{S}_m \xrightarrow[]{p} \int_{0}^{1} \mathcal{R}_{s}^2 ds$.
\end{proposition}

\section{Dynamic analysis of  integrated betas} \label{sec-3}

\subsection{Dynamic realized beta models}
In this section, we conduct a dynamic analysis of integrated betas.
\begin{figure}[t] 
\centering
\includegraphics[width = 1\textwidth]{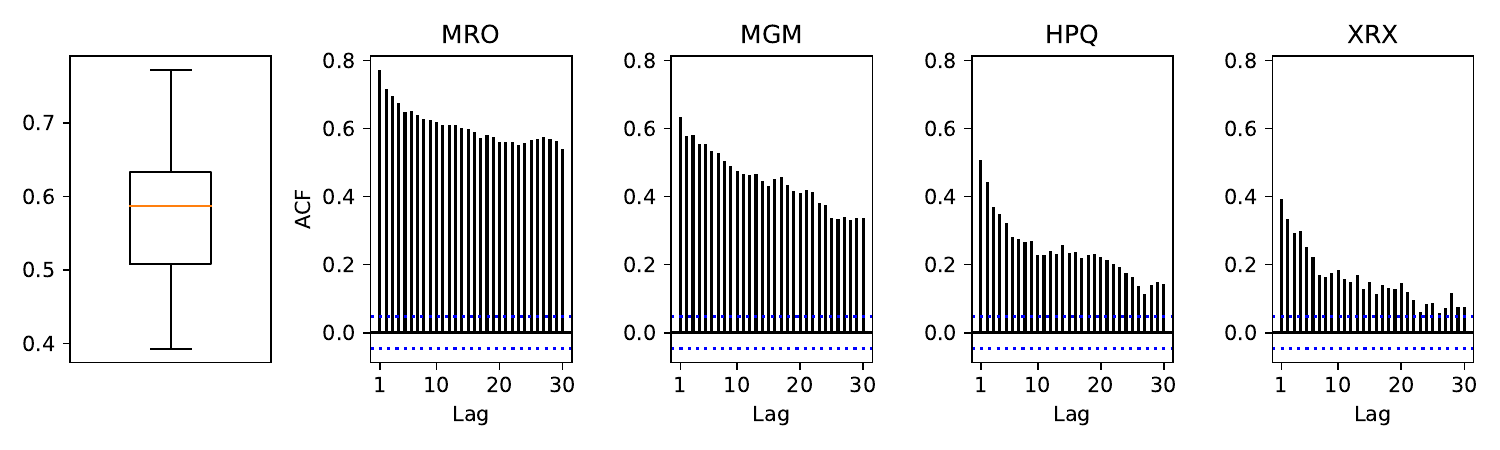}
\caption{The box plot (left) of the first-order autocorrelations of the daily realized integrated betas for the top 50 large trading volume assets among the S\&P 500 from January 1, 2010, to December 31, 2016, and the ACF plots for the largest, 75th, 25th, and smallest first-order autocorrelation among the 50 assets excluding outliers.}
\label{Figure-1}
\end{figure} 
To check the low-frequency time series structure of high-frequency-based market betas, we draw autocorrelation function (ACF) plots for daily realized integrated betas for the top 50 large trading volume stocks in Figure \ref{Figure-1}.
Figure \ref{Figure-1} shows that the realized beta has a persistent autoregressive structure. 
To account for this beta dynamics, we consider the ARMA($p,q$) structure on integrated betas as follows:
\begin{equation}\label{eq:ARMApq-beta}
  I\beta_{n} = \omega + \sum_{i=1}^{p}  \alpha_{i} I\beta_{n-i} + \sum_{i=1}^{q} \gamma_{i}  D_{n-i} + D_{n}
  ,
\end{equation}
where $D_n = I\beta_{n} - h_n$ is martingale difference and $h_n$ is $\mathcal{F}_{n-1}$-adapted.
We call this dynamic realized beta (DR Beta) model.
The model \eqref{eq:ARMApq-beta} is equivalent to the following model:
\begin{eqnarray}\label{eq:DRBeta}
  && I\beta_{n} = h_n(\theta) + D_n, \cr
  && h_n(\theta) = \omega + \sum_{i=1}^{q} \gamma^{g}_{i} h_{n-i}(\theta) + \sum_{i=1}^{p \lor q} \alpha^{g}_{i} I\beta_{n-i}
  ,
\end{eqnarray}
where $\gamma_{i}^{g} = -\gamma_{i}$, $\alpha_{i}^{g} = \alpha_{i} \mathbf{1}_{\{i \leq p\}} + \gamma_{i} \mathbf{1}_{\{i \leq q\}}$, and $\theta = (\omega, \gamma^{g}_{1}, \ldots, \gamma^{g}_{q}, \alpha^{g}_{1}, \ldots, \alpha^{g}_{p \lor q})$ is model parameter.
Since \eqref{eq:DRBeta} is more practicable than \eqref{eq:ARMApq-beta}, we focus on estimating parameters of \eqref{eq:DRBeta}.

\subsection{Parametric estimation for the DR Beta model} \label{sec-3.2}

\subsubsection{Estimation procedure based on high-frequency data and a low-frequency structure}

According to the strong autoregressive structure of the $RIB_i$'s in Figure \ref{Figure-1}, we now assume that integrated market betas follow the DR Beta model defined in \eqref{eq:DRBeta}.
To estimate the true parameters $\theta_0=(\omega_0, \gamma^{g}_{0,1}, \ldots, \gamma^{g}_{0,q}, \alpha^g_{0,1},\ldots, \alpha^g_{0,p\vee q})$, we consider the well-known  ordinary least squares (OLS) estimation, which compares the conditional expectations of integrated betas and its non-parametric estimators $RIB_{i}$'s as follows:
\begin{equation}  \label{Equation-3.6}
  L_{n,m}(\theta)=-\frac{1}{n}\sum^n_{i=1} \left\{ RIB_i-h_i(\theta) \right\}^2.
\end{equation}
The difference between the $h_{i}(\theta)$ and the $RIB_{i}$ can be decomposed into the martingale difference and the estimation error.
The estimation error is asymptotically negligible.
Furthermore, with some technical assumptions, the martingale difference terms have a negligible effect on the estimation result.
Thus, the $RIB$ estimator can be utilized as the proxy of $h_{i}(\theta)$.
To harness the quasi-likelihood function above, we first need to evaluate the conditional expectation term $h_i(\theta)$.
Unfortunately, the true integrated betas are not observable.
Thus, we adopt their non-parametric estimators $RIB_{i}$ to evaluate $h_i(\theta)$ as follows:
\begin{eqnarray*}
  \hat{h}_n(\theta) = \omega + \sum_{i=1}^{q}  \gamma^{g}_{i} \hat{h}_{n-i}(\theta) + \sum_{j=1}^{p \lor q} \alpha_{j}^{g} RIB_{n-j}
  .
\end{eqnarray*}
Then,  we define the quasi-likelihood function as follows:
\begin{equation}  \label{Equation-3.7}
    \hat{L}_{n,m}(\theta)=-\frac{1}{n}\sum^n_{i=1} \left\{ RIB_i-\hat{h}_i(\theta) \right\}^2,
\end{equation}
and estimate the model parameters by maximizing  the quasi-likelihood function $\hat{L}_{n,m}(\theta)$ as follows:
 \begin{equation*}
    \hat{\theta} =\argmax_{\theta \in \Theta} \hat{L} _{n,m}(\theta),
\end{equation*}
where  $\Theta$ is the parameter space of $\theta$.

To estimate $\hat{h}_{i}(\theta)$, we need initial values $\hat{h}_{0}(\theta), \ldots, \hat{h}_{-q+1}(\theta)$, and $RIB_{0}, \ldots, RIB_{-p\lor q +1}$, which we cannot obtain from given information whereas it is required to get $\hat{h}_{1}(\theta), \ldots, \hat{h}_{p\lor q}(\theta)$.
Meanwhile, similar to Lemma 1 in \citet{kim2016unified}, we can show that the dependence of $h_{i}(\theta)$ on initial values decays with the order $n^{-1}$.
Thus, we can utilize, for example, $\frac{1}{n} \sum_{i=1}^{n} RIB_{i}$ as initial values.

\subsubsection{Asymptotic theory}

In this subsection, we establish asymptotic theorems for the proposed estimator $\hat{\theta}$.
We first define some notations. 
Define $\left\| A\right\| _{\max}=\smash{\displaystyle\max_{1 \leq i \leq k, 1 \leq j \leq k'}}  \left| A_{ij}\right|$ for a $k \times k'$ matrix $A$.
Let $C>0$ be generic constants whose values are free of $n$ and $m$ and may change from occurrence to occurrence.

To explore the asymptotic behaviors of $\hat{\theta}$, the following technical conditions are required.

\begin{assumption} \label{Assumption-2} 
~
\begin{enumerate}
\item [(a)] Let
\begin{eqnarray*}
   \Theta &=& \{ (\omega, \gamma_{1}^{g}, \ldots, \gamma_{q}^{g}, \alpha_{1}^{g}, \ldots, \alpha_{p \lor q}^{g}): \omega_l<\omega<\omega_u, \, \gamma_l^{g}< \gamma_{1}^{g} , \ldots, \gamma_{q}^{g}  <\gamma_u^{g}, \\
   && \alpha^g_l < \alpha_{1}^g, \ldots, \alpha_{p \lor q}^{g} <\alpha^g_u,  \sum_{i=1}^{q} \left|\gamma_{i}^{g} \right| < 1,  \sum_{i=1}^{p \lor q} \left | \mathbf{1}_{\left\lbrace i \leq q \right\rbrace} \gamma_{i}^{g} + \mathbf{1}_{\left\lbrace i \leq p \lor q \right\rbrace} \alpha_{i}^{g} \right |  <1 \},
\end{eqnarray*}
where $\omega_l, \omega_u, \alpha^g_l, \alpha^g_u, \gamma^g_l,\gamma^g_u$ are known constants such that $\alpha^g_l, \alpha^g_u, \gamma^g_l$ and $\gamma^g_u$ are in  $(-1,1)$.

\item [(b)] For all  $\theta \in \Theta$, $|\varphi_{\theta} (x) | = 0  \Rightarrow |x| > 1 $,  where
 $
\varphi _{\theta} (x)= 1-  \sum_{i=1}^{p \lor q}  (\mathbf{1}_{\left\lbrace i \leq q \right\rbrace} \gamma_{i}^{g} + \mathbf{1}_{\left\lbrace i \leq p \lor q \right\rbrace} \alpha_{i}^{g} ) x^i.
$

\item [(c)]    $\Upsilon_{ \theta_0} (x)$ does not have no common root with $\varphi _{\theta_0} (x)$, where 
 $
\Upsilon_{ \theta} (x)=  \sum_{i=1}^{1 }  \gamma_i^{g}  x^i.
$

\item [(d)] $D_{i}$ is stationary ergodic process satisfying $\smash{\displaystyle\sup_{i \leq n }}  \mathbb{E} \left[ D_{i}^{2} | \mathcal{F}_{i-1} \right]  \leq C$.

\item [(e)] There exist some fixed constants $C_1$, $C_2$ such that $C_1 m \leq m_i \leq C_2 m$, and \\
$\smash{\displaystyle\sup_{1\leq j \leq m_i}} \left| t_{i,j}-t_{i,j-1}\right| =O\left( m^{-1}\right)$ and $n^2 m^{-1} \rightarrow 0$ as $m,n \rightarrow \infty$.

\item [(f)] We have $\smash{\displaystyle\sup_{i \leq n }} \mathbb{E} \left[ \left| RIB_i-I\beta_i \right|^{2} \right]  \leq C m^{-1/2}$.

\end{enumerate}
\end{assumption}

\begin{remark} 
Assumption \ref{Assumption-2}(a)--(d) are usually imposed when analyzing asymptotic properties of the ARMA-type models. 
For example, Assumption \ref{Assumption-2}(b) implies the stationarity of $I\beta_i$ and $h_i(\theta)$, and Assumption \ref{Assumption-2}(c) is required to identify the parameter space.
Finally, Assumption \ref{Assumption-2}(f) is required to handle the estimation errors of the unobserved integrated betas. 
Under some bounded moment conditions on the random quantities, we can show that Assumption \ref{Assumption-2}(f) holds.
We elaborate on conditions of the long-span asymptotic behavior of the $RIB$ estimator in  Assumption \ref{assumption-bounded-moment}.
\end{remark}

The following theorems provide the asymptotic results including the convergence rate and asymptotic normality for the proposed parameters $\hat{\theta}$. 
\begin{thm} \label{Theorem-2}
Under Assumption \ref{Assumption-2} (except for $n^2 m^{-1} \rightarrow 0$), we have
\begin{equation*}
    \| \hat{\theta} -\theta_{ 0} \|_{max}=O_p( m^{-1/4} +n^{-1/2}).
\end{equation*}
\end{thm}

\begin{thm} \label{Theorem-3}
Under Assumption \ref{Assumption-2}, we have, as $m,n\rightarrow \infty$, 
\begin{equation*}
    \sqrt{n}(\hat{\theta} -\theta_0) \overset{d}{\rightarrow} N(0,V),
\end{equation*} where
\begin{equation} \label{eq-V}
  V = \mathbb{E}\left[ D_{1}^{2} \right]   \left ( \mathbb{E} \left[  \left. 
 {\dfrac{\partial h_1(\theta)}{\partial \theta} \dfrac{\partial h_1(\theta)}{\partial \theta^\top} }\right|_{\theta=\theta_0}  \right]\right )^{-1}.
\end{equation}
\end{thm}

\begin{remark}
    Theorem \ref{Theorem-2} shows that the quasi-maximum likelihood estimator $\hat{\theta}$ has the converge rate $m^{-1/4} +n^{-1/2}$. 
    The first term $m^{-1/4}$ comes from estimating the integrated beta, which is known as the optimal convergence rate with the presence of market microstructure noise.
    The second term $n^{-1/2}$ is the typical parametric convergence rate based on the low-frequency observations. 
    Theorem \ref{Theorem-3} establishes the asymptotic normality of $\hat{\theta}$.
\end{remark}

In the asymptotic analysis of low-frequency dynamics, the sample size, $n$, is allowed to go to infinity.
Thus, we need the long-span asymptotic behavior of the $RIB$ estimator, such as  Assumption \ref{Assumption-2}(f). 
However, this condition is not satisfied under the locally bounded condition such as Assumption \ref{assumption-formal}. 
To coincide with the asymptotic results of the proposed estimation procedures, we investigate the long-span asymptotic behavior of $RIB$ as follows.

\begin{assumption}\label{assumption-bounded-moment}
  ~
  \begin{enumerate}
    \item [(a)] We have bounded 64th moment of $\sigma, \mu_{1}, \mathfrak{d}_{1}, \beta^{c}, \beta^{d}, \mu_{\beta}$, and $\sigma_{\beta}$ and bounded 32nd moment of $\sigma^{-1},\tilde{\mu}_{1}, \tilde{\sigma}, \tilde{\mathfrak{d}}_{1}, \vartheta_{1}, \vartheta_{2}, \mu_{2}, q$, and $\mathfrak{d}_{2}$.
    \item [(b)] The process $\mu_{1}$ satisfies (P-64) and the processes $\mu_{2}, \vartheta_{1}$, and $\vartheta_{2}$ satisfy (P-32) in Assumption \ref{assumption-formal}(d).
    \item [(c)] For some $r \in [0,\frac{2[v]-8}{2[v]-5} )$, there are deterministic nonnegative $\lambda$-integrable functions $\mathcal{J}$ on $\mathbb{R}^{2}$ such that
    \begin{eqnarray*}
      && \mathbb{E}[ |\mathfrak{d}_{1}(t,z)|^{r} \land 1 ] \lor \mathbb{E}[ |\mathfrak{d}_{2}(t,z)|^{r} \land 1 ] \lor \mathbb{E}[ |\beta_{t-}^{d}\mathfrak{d}_{1}(t,z)|^{r} \land 1 ] \leq \mathcal{J}(z) , \cr
      && \mathbb{E}[ |\tilde{\mathfrak{d}}_{1}(t,z)|^{2} \land 1 ] \lor \mathbb{E}[ |\tilde{\mathfrak{d}}_{2}(t,z)|^{2} \land 1 ]  \leq \mathcal{J}(z) , \cr
      && \mathbb{E}[ |{\mathfrak{d}}_{1}(t,z)|^{64} ] \lor \mathbb{E}[ |{\mathfrak{d}}_{2}(t,z)|^{32} ] \lor \mathbb{E}[ |\tilde{\mathfrak{d}}_{2}(t,z)|^{32} ] \lor \mathbb{E}[ |\tilde{\mathfrak{d}}_{1}(t,z)|^{32} ]  \leq \mathcal{J}(z)
      .
    \end{eqnarray*}
  \end{enumerate}
\end{assumption}

\begin{remark}
To establish the convergence in the second mean in Theorem \ref{thm:BoundedMoment}, we need moment conditions on the spot error terms.
  For example, we consider the squared error of the de-biasing term, $(\hat{B}_{ib_m}^{m} - B_{ib_m}^{m})^2$, from which the highest order error terms comes, where $B_{ib_m}^{m}$ is defined in the online Appendix equation \eqref{eq:def-B}.
  Technically, after applying Talyor's theorem, $(\hat{B}_{ib_m}^{m} - B_{ib_m}^{m})^2$ become an octic function of the errors of spot variations with denominator, since $\hat{B}_{ib_m}^{m}$ is a cubic function of estimated spot variations with denominator.
 Thus, we need 16th-moment conditions (32nd-moment conditions) on spot variation terms (drift and diffusion terms).
  Further, $\beta^{c}$ and $\beta^{d}$ are random processes which are multiplied by $dX_{1}^{c}$ and $dX_{1}^{d}$.
Therefore, we need 64th moment conditions for some random quantities.
If we assume that $\sigma_{t}^{-1}$ is bounded, then we can reduce the 64th and 32nd-moment conditions in Assumptions \ref{assumption-bounded-moment}(a), (b), and (c) by half, since we need one less H\"older's inequality.
 On the other hand, unlike the case of asset price processes, it is economically sensible to consider the random quantities in Assumption \ref{assumption-bounded-moment} as mean-reverting processes.
When a mean-reverting process follows a generalized Ornstein-Uhlenbeck process with Brownian motion, the high-order moment condition, such as Assumption \ref{assumption-bounded-moment}, can be satisfied.
Thus, it is not restrictive. 
\end{remark}

The following theorem establishes the long-span asymptotic behavior for the proposed $RIB$ estimator.
\begin{thm}\label{thm:BoundedMoment}
  Under Assumptions \ref{assumption-noise}, \ref{assumption-formal}(b) and (d), and \ref{assumption-bounded-moment} with $v \geq 7$, we have for $\delta_m = C_{\delta} \Delta_m ^{\frac{1-\kappa}{16}}$ defined in \eqref{Equation-3.4},
  \begin{equation*}
    \sup_{i \in \mathbb{N}}\mathbb{E}\left[ \left( RIB_{i} - I\beta_{i} \right)^2  \right] \leq C m^{-1/2}
    .
  \end{equation*}
\end{thm}

Theorem \ref{thm:BoundedMoment} shows that under some moment conditions, Assumption \ref{Assumption-2}(f) is satisfied. 
That is, the condition in Assumption \ref{Assumption-2}(f) can be replaced by Assumption \ref{assumption-bounded-moment}.

\subsubsection{Hypothesis tests}\label{sec-3.3}

In financial practices, we are interested in model validity and making statistical inferences, such as hypothesis tests. 
To do this, we can harness the asymptotic normality result in Theorem \ref{Theorem-3} as follows:
 \begin{equation*}
    T_n = \sqrt{n} \hat{V}^{-1/2}(\hat{\theta} -\theta_0)  \overset{d}{\rightarrow} N(0,\mathbf{I}),
\end{equation*}
where $\hat{V}$ is a consistent estimator of the asymptotic variance $V$ defined in \eqref{eq-V}, and $\mathbf{I}$ is a $A \times A$  identical matrix, where $A = p \lor q + q + 1$. 
Then,  with the test statistics $T_n$, we can conduct hypothesis tests based on the standard normal distribution.
To evaluate the statistics $T_n$, we use the following asymptotic variance estimator,
\begin{eqnarray} \label{Equation-3.8}
  \hat{V} &=& \frac{1}{n} \sum^{n}_{i=1} \left \{ RIB_i-\hat{h}_i (\hat{\theta}) \right \}^2 \left( \frac{1}{n} \sum_{j=1}^{n} \frac{\partial \hat{h}_{j}(\hat{\theta})}{\partial \theta} \frac{\partial \hat{h}_{j}(\hat{\theta})^{\top}}{\partial \theta} \right)^{-1} 
  .
\end{eqnarray}
Its consistency can be derived similarly to the proof of Theorem \ref{Theorem-3}.

\section{A simulation study} \label{sec-4}

We conducted simulations to check the finite sample performance of the proposed statistical inference procedures.  
For simplicity, we chose $p=q=1$ for the DR Beta model.
In the online Appendix, we provide a high-frequency data-generating process of the beta diffusion process, whose integrated beta follows the DR Beta model.
Using this data-generating process, we generated the beta processes $\beta^c_{t_{i,j}}$ and the jump-diffusion processes $X_{1,t_{i,j}}$ and $X_{2,t_{i,j}}$ for $ t_{i,j}=i-1+j/m$, $i=1,2, \ldots, n, \, j=1,2, \ldots, m$ as follows:
\begin{eqnarray}\label{eq:simulation-process}
  && dX_{2,t}=\beta^c _{t} (\theta ) dX^c_{1,t}+\beta^d_t J_{1,t} d\Lambda_{1,t}+dV_t,  \cr  &&dX_{1,t}=\sigma_tdB_t+J_{1,t}d\Lambda_{1,t},  \quad   dV_t=q_tdW_t+J_{2,t}d\Lambda_{2,t}, \cr 
  && d \beta_{t}^{c}(\theta) =   \Bigg \{  2  \left( t - [t] \right)  \left( \omega_1  + \gamma_{1} \beta_{[t]}^{c} (\theta)      \right)  - \(  \omega_ 2 +  \beta_{[t]}^{c}(\theta) \)  - \nu  (Z_t - Z_{[t]}) + \alpha_1 \beta_t^c (\theta)  \Bigg \}  dt  \cr
  && \qquad\qquad + \nu \left( [t] + 1 - t \right)  dZ_t, \cr
  &&dZ_t dB_t = dZ_t dW_t= dB_t dW_t=0,
\end{eqnarray}
where $(\omega_1, \omega_2, \gamma_1, \alpha_1, \nu)=(-1.0, -1.5, 0.6, 0.2, 0.9)$, $q_t = 0.008$, and $B_t$, $Z_t$, and $W_t$ are standard Brownian motions.
The parameters $\omega_{1}$ and $\omega_{2}$ control the deterministic quadratic time-trend of the spot beta process, thereby determining its mean level.
The persistent feature of the beta process is governed by the parameters $\gamma_{1}$ and $\alpha_{1}$, with $\alpha_{1}$ playing a key role in regulating the intraday level autoregressive characteristic of the spot beta process.
The parameter $\nu$ controls the degree of intraday variation in the beta process.
More detailed explanations of the process and its properties can be found in the online Appendix.
With the chosen diffusion process parameters, the parameter of the DR Beta model becomes $\theta_0=(\omega^g_0, \gamma_0,\alpha^g_0)=(0.84, 0.20,$ $0.50)$. 
We generated the individual asset log price process $X_{1,t}$ based on the realized GARCH-It\^o model \citep{song2021volatility} as follows:
\begin{eqnarray*}
 d\sigma^2_t &= &\left \{2 \tilde{\gamma}(t-\lceil t-1\rceil)(\tilde{\omega}_1+\sigma^2_{\lceil  t-1\rceil})-(\tilde{\omega}_2+\sigma^2_{\lceil t-1\rceil})+\tilde{\alpha}\sigma^2_t -\tilde{\nu} \tilde{Z}_t^2 \right \} dt \cr
  &&  +\tilde{\beta}  J_{1,t}^2 d\Lambda_{1,t} +2\tilde{\nu}(\lceil t-1\rceil+1-t)\tilde{Z}_t d\tilde{B}_t,
\end{eqnarray*}
where a standard Brownian motion $\tilde{B}_t$  satisfies $d\tilde{B}_t dW_t=d\tilde{B}_t dU_t=0, d\tilde{B}_t dB_t=\tilde{\rho}dt$,       $(\tilde{\omega}_1,\tilde{\omega}_2,\tilde{\gamma},\tilde{\alpha}, \tilde{\beta},\tilde{\nu},\tilde{\rho})=(6.04\times 10^{-5}, 9.00\times 10^{-6}, 0.35, 0.4, 0.1,1 \times 10^{-5}, -0.5)$, and $\tilde{Z}_t=  \tilde{B}_t - \tilde{B}_{\lceil t-1\rceil}$.
The initial values for the simulation data were chosen to be $\beta^c_0 (\theta) = \mathbb{E}\left[ \beta^c_1 (\theta) \right] = 2.72$, $\sigma_0^2= \mathbb{E}\left[ \sigma_1^2 \right] = 7.55\times 10^{-5}$,  $X_{1,0}=16$, and $X_{2,0}=10$. 
For the jump part, we consider the finite activity jumps.
Specifically, $\Lambda_{1,t}$ is a standard Poisson process with the intensities $\lambda_{1,t}=5$ and $\lambda_{2,t}=1$, and  the jump sizes $J_{1,t}$  and $J_{2,t}$ were generated as follows:
\begin{eqnarray*}
 J_{1,t}^2= \max(4 \times 10^{-5} + M_{1,t}, \quad 4 \times 10^{-6})  \quad  \text{and} \quad J^{2}_{2,t}   = \max(8 \times 10^{-6} + M_{2,t}, \quad 8 \times 10^{-7}) ,
\end{eqnarray*}
where $M_{1,t}$ and $M_{2,t}$ follow   $N(0,(5.5 \times 10^{-6})^2)$ and $N(0,(1 \times 10^{-6})^2)$, respectively.
For each $J_{1,t}$ and $J_{2,t}$, we further assigned a positive (negative) sign with probability $0.5$ to make a positive (negative) jump.
Finally, $\beta^d_t$ was chosen to be   $2.4$, and we generated Brownian motions using the Euler scheme.

The noisy high-frequency data $Y_{1,t_{i,j}}$ and  $Y_{2,t_{i,j}}$ were generated from the model \eqref{Equation-3.1}, where the true log price processes  $X_{1,{t_{i,j}}}$ and $X_{2,{t_{i,j}}}$ were generated from \eqref{eq:simulation-process}, and the microstructure noise $\epsilon_{1,t_{i,j}}$  and  $\epsilon_{2,t_{i,j}}$ follow \eqref{noise-form}, where $\vartheta_{1,t}$, $\vartheta_{2,t}$, and $\bchi_{i}$ follow Ornstein--Uhlenbeck-type processes with an U-shaped pattern and the AR$(1)$ process with Gaussian innovations as follows:
\begin{align*}
& d\vartheta_{1,t} = 10(\mu_{\vartheta_1,t} - \vartheta_{1,t})dt + s_1 dB_t, \quad d\vartheta_{2,t} = 10(\mu_{\vartheta_2,t} - \vartheta_{2,t})dt + 0.6 s_2 dB_t + 0.8 s_2 dW_t, \\
& \mu_{\vartheta_1,t} = s_1 (1 + 0.1\cos(2\pi t)), \quad  \mu_{\vartheta_2,t} = s_2 (1 + 0.1\cos(2\pi t)), \\
& s_1 = 2.234 \times 10^{-4} ,\quad s_2 = 5.464 \times 10^{-3}, \\
& \bchi_i = \begin{pmatrix} 0.5 & 0.1 \cr 0.1 & 0.5 \end{pmatrix} \bchi_{i-1} + e_{i}, \quad e_{i} \sim_{i.i.d.} N\left[\begin{pmatrix} 0 \cr 0 \end{pmatrix},\begin{pmatrix} 0.815 & -0.652 \cr -0.652 & 0.815 \end{pmatrix}\right]
.
\end{align*}
In this specification of the noise, the noise-to-signal ratio in the returns is predominantly determined by the parameters $s_1$ and $s_2$.
Additionally, the cross-autocovariance structure of the noise is influenced by the VAR coefficients and the covariance of their innovations.
This simulation setting satisfies Assumptions 3(a)--(f), and specifically, Assumption 3(f) can be verified by confirming that it aligns with Assumption 4.
We repeated the simulation process $1000$ times.
We normalized one second to $1/23400$ so that the unit time contains $6.5$ hours.
For each simulation process, we generated high-frequency data with $m=23400$ for $500$ consecutive days and used the subsampled log prices of the last $n=125,250,500$ days with high-frequency observations $m=7800, 11700, 23400$ per day.

\begin{figure}[!ht] 
  \centering
  \includegraphics [width = 0.7\textwidth]{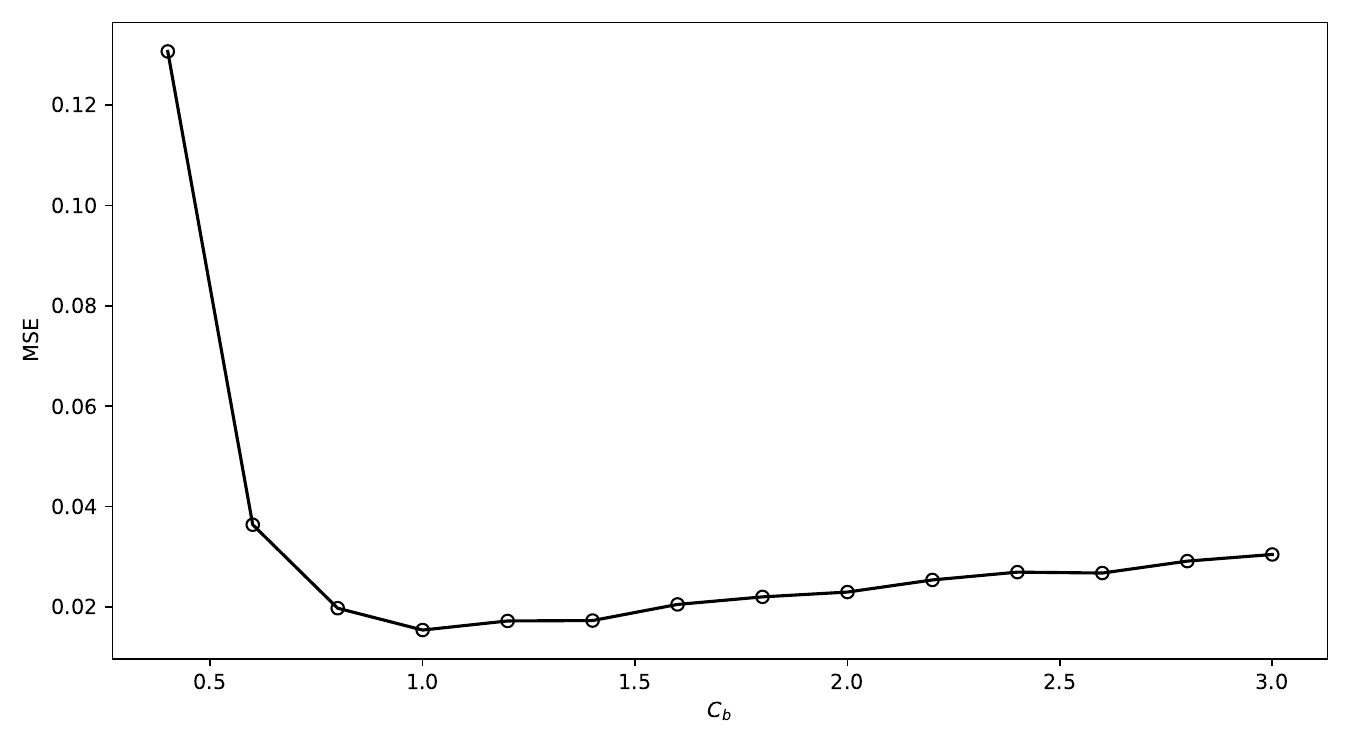}
  \caption{The MSEs of $RIB$ estimator with $m= 23400$ against varying $C_b$.}
  \label{Figure-sensitivity}
\end{figure}
For the $RIB$ estimator, we used the usual triangular weight function $g(x)=\{ x \wedge (1-x)\}$, and set $k_m =  [\Delta_m ^{-0.5}]$ and $\varpi_1=0.47$ as recommended by \citet{christensen2010pre} and \citet{ait2016increased}, respectively.
For each estimation of the daily integrated beta, we chose $l_m$ using the heuristic criterion presented in Section 5.1.2 of \citet{jacod2017statistical}, where the distance between two sequences is measured as the sum of their squared differences.
To determine $k'_m$, we utilized the test for autocovariance of noise as presented in Corollary 3.5 in \citet{jacod2017statistical}.
Details can be found in the online Appendix \ref{sec-tuning-parameter}.
In addition, for the truncation, we chose $a_1$ and $a_2$ as four times the sample standard deviation of the pre-averaged prices $k_m^{-1/2} \tilde{Y}_{1,t_{d,k}}^{m}$ and $k_m^{-1/2} \tilde{Y}_{2,t_{d,k}}^{m}$, respectively.
We then needed to determine $C_b$.
To do this, we checked the effect of the choice of $C_b$ of the $RIB$ estimator.
Figure \ref{Figure-sensitivity} depicts the estimated mean squared errors (MSE) of the $RIB$ estimator with $m=23400$ against varying $C_b$ from $0.4$ to $3.0$, where $\kappa=0.67$ and the integrated beta $I\beta_i$ is calculated as the Riemann sum of the true beta values for each trading days.
From Figure \ref{Figure-sensitivity}, we find that for $C_b < 1$, the MSEs decrease as $C_b$ increases, and for $C_b \geq 1$, the MSEs slightly increase as $C_b$ increases.
This may be because the window size for the spot betas should be large enough to estimate spot betas, but too large a window size hinders the capture of the intraday dynamics of the beta processes.
From this analysis, we set $C_b=1$.

We first checked the performance of the non-parametric integrated beta estimator, $RIB$,  proposed in Section \ref{sec-3.1}.
For comparison, we employed other integrated beta estimators proposed by \citet{chen2018inference} and \citet{christensen2010pre}.
\citet{chen2018inference} proposed the estimator for volatility functionals and the integrated beta (CHEN) is a specific example.
\citet{christensen2010pre} calculated the integrated beta as a ratio of the integrated covariance between assets and systematic factors to the integrated variation of systematic factors.
The proposed estimator utilizes a pre-averaged realized covariance estimator that is robust to i.i.d. microstructure noise but is not to autocorrelated noise and price jump.
On the other hand, \citet{jacod2019estimating} proposed a robust pre-averaged integrated volatility estimator that is robust to price-dependent and autocorrelated microstructure noise and price jump.
We employed the integrated beta estimator (PRVB), which adopts the robust pre-averaging integrated volatility estimator of \citet{jacod2019estimating} as the input of the integrated beta estimator in \citet{christensen2010pre}.
The details of estimators can be found in the online Appendix \ref{sec-detail-estimator}.
We note that  PRVB is a consistent estimator of the ratio of the integrated covariance between assets and systematic factors to the integrated variation of systematic factors.
That is, while  PRVB is a consistent estimator of the integrated beta when the intraday beta or market volatility is constant over time, the PRVB is not a consistent estimator of the integrated beta in general.
On the other hand, CHEN is designed for estimating time-varying beta, but it does not consider the autocorrelated microstructure noise.
\begin{figure}[t] 
\centering
\includegraphics [height=0.4\textwidth,width = 0.4\textwidth]{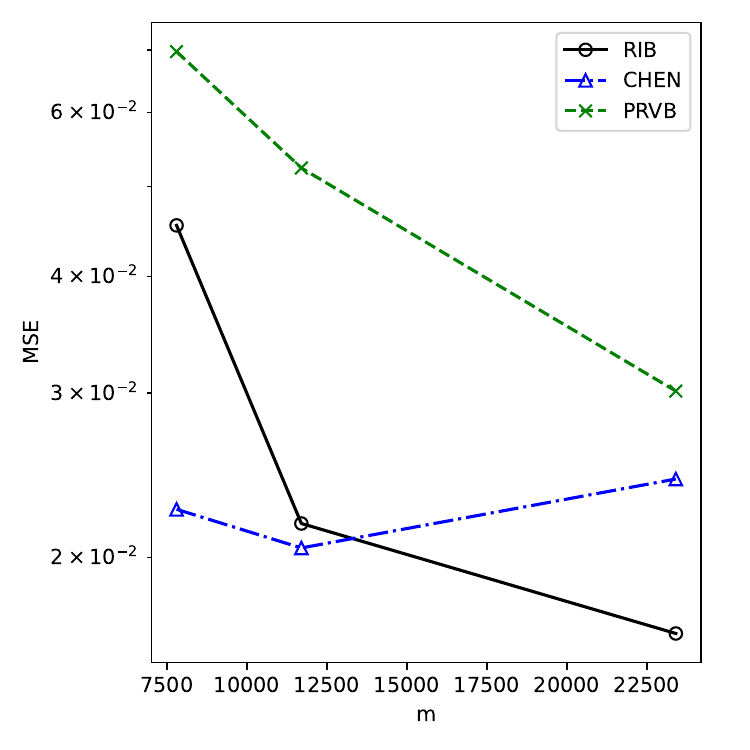}
\caption{The MSEs  of  $RIB$, CHEN, and  PRVB  for  $m=7800, 11700, 23400$.
The average value of the true integrated beta was 2.802.}
\label{Figure-nonparam}
\end{figure}
Figure \ref{Figure-nonparam} shows the MSEs of the non-parametric integrated beta estimators,  $RIB$, CHEN, and PRVB, for   $m=7800, 11700, 23400$.
We note that the average value of the true integrated beta was 2.802.
Figure \ref{Figure-nonparam} shows that the MSEs of $RIB$ and PRVB decrease as the number of high-frequency observations increases, whereas the MSEs of CHEN do not.
This is because the $RIB$ and PRVB estimators can account for the autocorrelation structure of the microstructure noise, whereas CHEN fails to handle it. 
Further, $RIB$ and CHEN perform better than PRVB since PRVB fails to deal with the time-varying beta.
The magnitude of the difference in performance between the $RIB$ and PRVB estimators may depend on how volatile the intraday beta and market volatility processes are.
When comparing the performances of the $RIB$ and CHEN, the $RIB$ estimator shows better performance for $m=23400$, while CHEN does for $m=7800, 11700$.
It may be because the effect of the autocorrelated microstructure noise increases as $m$ increases, while the estimation variance of the denoise term of $RIB$, which decreases as $m$ increases, is larger than that of CHEN.
These results support the theoretical results derived in Section \ref{sec-3.1}.

\begin{figure}[t] 
  \centering
  \includegraphics [width = 0.9\textwidth]{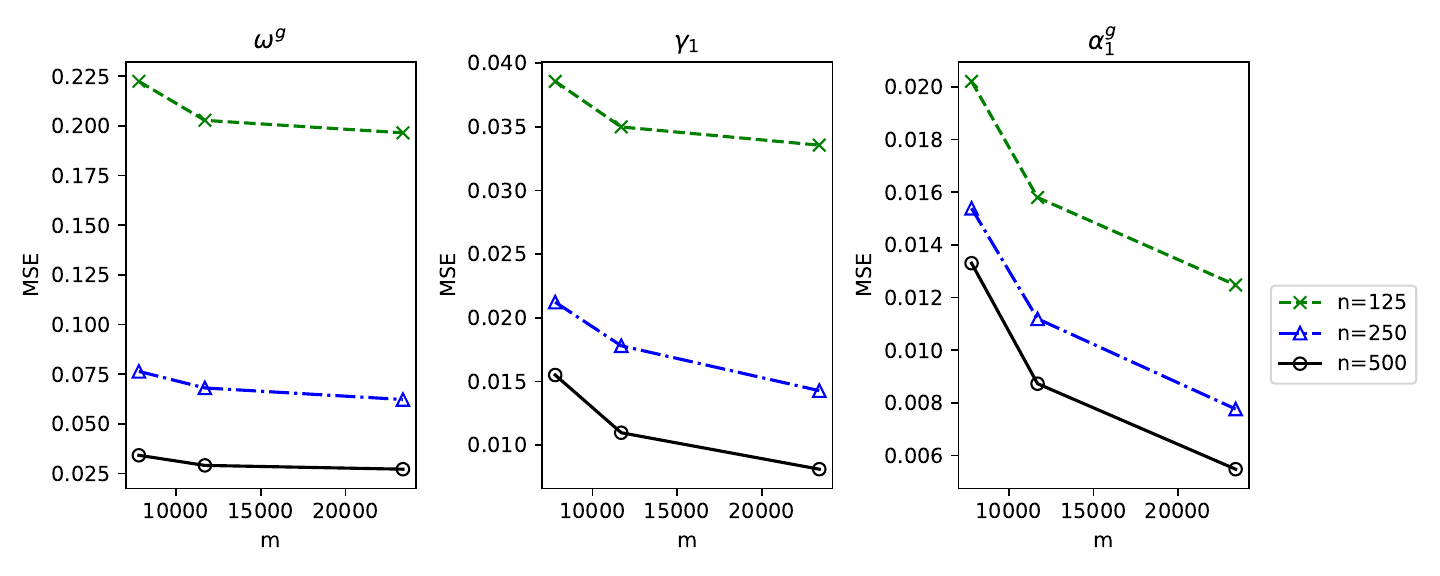}
  \caption{The MSEs of the least squared estimates with $m=2340, \,4680, \, 23400$ and $\, \, $ $n=100, \, 250, \, 500$.}
  \label{Figure-parMSE}
\end{figure}

Next, we checked the finite sample performances of the proposed DR Beta model.
We first estimated the model parameters using the proposed quasi-maximum likelihood estimation in Section \ref{sec-3.2}  for $n=100, \, 250, \, 500$ and $m=7800, 11700, 23400$.
To estimate $\hat{h}_{i}(\theta)$, we set initial values $\hat{h}_{0}(\theta) = RIB_{0} = \frac{1}{n} \sum_{i=1}^{n} RIB_{i}$.
Figure \ref{Figure-parMSE} draws the MSEs of the least squared estimates $\hat{\theta}$'s for the model parameter $\theta_0$. 
From Figure \ref{Figure-parMSE}, we find that the MSEs decrease as $n$ or $m$ increases.
These results match the theoretical findings in Section \ref{sec-3.2}.

\begin{figure}[t] 
\centering
\includegraphics[width = 0.85\textwidth]{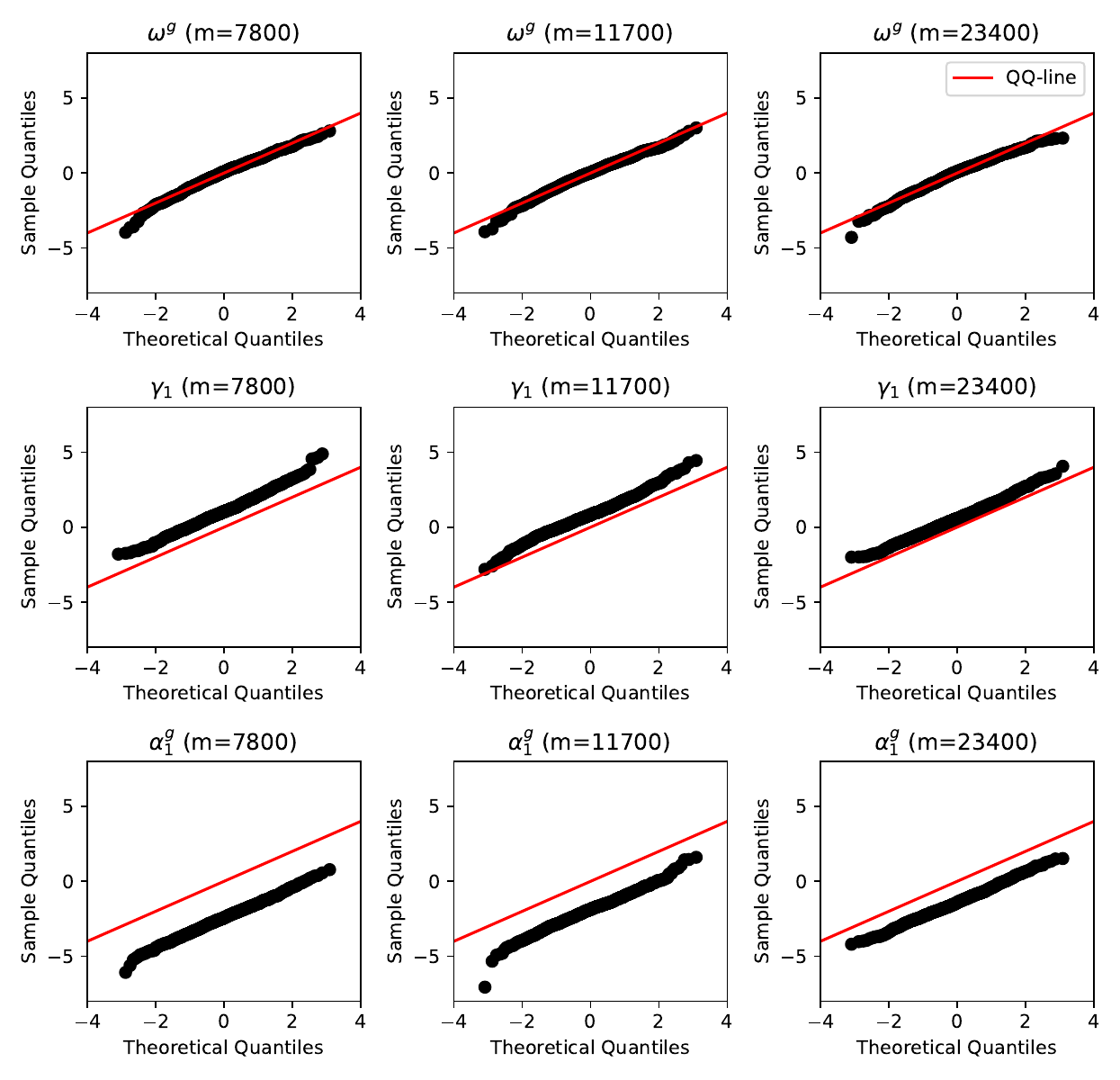}
 \caption{The standard normal (original) quantile-quantile plots of the $Z$-statistics estimates of $\omega^g$, $\gamma_1$, and $\alpha^g_1$ for $n=500$, $m=7800, 11700, 23400$.}
 \label{QQplot}
\end{figure}
 To check the asymptotic normality of the model parameters $(\omega^g, \gamma_1, \alpha^g_1)$, we calculated the $Z$-statistics  proposed in Section \ref{sec-3.3}.
Figure \ref{QQplot} shows standard normal quantile-quantile plots of the $Z$-statistics estimates of $\omega^g_1$, $\gamma_1$, and $\alpha^g_1$  for $n=500$ and $m=7800, 11700, 23400$. 
From Figure \ref{QQplot}, we find that the $Z$-statistics  close to the standard normal distribution as $m$ increases--that is,  the non-parametric integrated beta estimator $RIB$ closes to the true integrated beta $I\beta$.  
This result agrees with the theoretical findings in Section \ref{sec-3}.
Thus, based on the proposed $Z$-statistics, we can conduct hypothesis tests for the model parameters using the standard normal distribution.

The DR Beta model is an ARMA model for the integrated beta, utilizing the $RIB$, which is a consistent estimator of the integrated beta.
One of the advantages of employing this consistent estimator to predict future market betas lies in its ability to effectively capture the low-frequency autoregressive dynamic structure, which helps improve the predictability of future beta values.
Thus, we examined the out-of-sample performance of estimating the one-day-ahead conditional expected integrated beta $h_{n+1}(\theta_0)$ to check the predictability of the DR Beta model. 
We compared the DR Beta with three parametric models that employ high-frequency data and two parametric models that use low-frequency data.
For the parametric model with high-frequency data,  we considered the ARMA$(1,1)$ models, which utilize CHEN (ARMAC) or PRVB (ARMAP) as daily realized beta, and Realized Beta GARCH (RBG) model \citep{hansen2014realized}, which is a multivariate GARCH model utilizing realized measures of volatility and correlation.
For the input covariance matrix of the RBG model, we used realized covariance, the sum of squared log-returns, with 5-min, 1-min, and 30-sec data ($m=78, 390,780$, respectively) to reduce the impact of the microstructure noise.
We also used the robust pre-averaging realized covariance \citep{jacod2019estimating} as the input of the RBG model (PRBG).
Details of the RBG model can be found in \citet{hansen2014realized}.
For the parametric models with low-frequency data, we used the dynamic conditional beta (DCB) model framework proposed by \citet{engle2016dynamic}.
Specifically, the beta prediction can be established by comparing the conditional covariance between assets and systematic factors to the conditional variance of systematic factors.
The details of the procedure can be found in the online Appendix \ref{sec-detail-estimator}.
We employed the BEKK(1,1) and DCC(1,1) models as the conditional covariance matrix models, as suggested by \citet{engle1995multivariate} and \citet{bali2010intertemporal}, respectively.
We call the beta estimators with BEKK(1,1) and DCC(1,1) BEKK and DCC, respectively. 
For each model, we calculated the mean squared forecast errors (MSFEs) with the one-day-ahead forecasted beta across 1000 repeated simulations as follows:
\begin{equation*}
    \frac{1}{1000}\sum_{i=1}^{1000} \left (\hat{Beta}_{n+1, i} -h_{ n+1,i}(\theta_0) \right )^2,
\end{equation*}
where $h_{n+1,i} (\theta_0)$ is the true conditional expectation of the $(n+1)$th integrated beta and $\hat{Beta}_{n+1, i}$ denotes one of the forecasted beta obtained using a parametric model such as DR Beta, ARMAC, ARMAP, PRBG, RBG, DCC, and BEKK models at the $i$th sample-path given the available information at time $n$.
We note that the target of the benchmarks, except for ARMAC, is the ratio of the integrated covariance between assets and systematic factors to the integrated variation of systematic factors.
Therefore, the MSFEs of the benchmarks additionally include the error from the discrepancy between the true integrated beta and the true ratio of the integrated covariance between assets and systematic factors to the integrated variation of systematic factors.
\begin{figure}[t] 
\centering
\includegraphics[width = 0.9\textwidth]{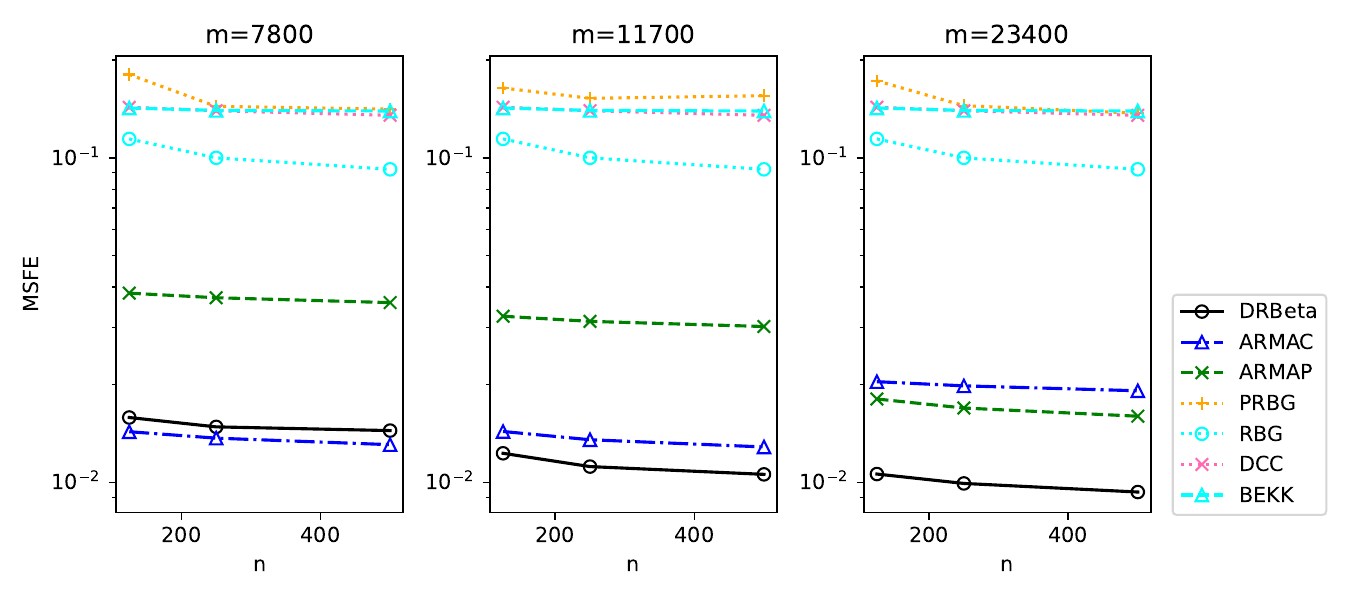} 
\caption{ The MSFEs of DR Beta, ARMAC, ARMAP, PRBG, RBG, DCC, and BEKK with $n=100, 250, 500$ and $m=7800, 11700, 23400$.
The average value of $h_{501,i}(\theta_0)$ was 2.813.}
\label{Figure-pred}
\end{figure} 
In Figure \ref{Figure-pred}, the MSFEs of DR Beta, ARMAC, ARMAP, PRBG, RBG, DCC, and BEKK are plotted for $n=100, \, 250, \, 500$ and $m=7800, 11700, 23400$.
The average value of the true conditional expectation of $501$st integrated beta was 2.813.
For the RBG model, we plotted only the MSFEs with $m=780$, which is the lowest MSFEs among $m=78,390,780$.
Figure \ref{Figure-pred} shows the MSFEs of the DR Beta and ARMAP decrease as $n$ or $m$ increases, but other estimators do not have any strong pattern. 
This may be because the other benchmarks cannot account for the autocorrelated microstructure noise well.
When comparing the DR Beta and ARMAP models, the DR Beta consistently outperforms the ARMAP.
This is because the target variable of the PRVB estimator differs from the integrated beta under the time-varying spot beta and market volatility processes, thereby resulting in a less effective capture of integrated beta dynamics by the ARMAP model.
Meanwhile, the high-frequency-based ARMA models show better performance than other competitors.
When comparing the high-frequency-based ARMA models, the ARMAC and the DR Beta models show the best performance for $m=7800$ and $m=11700,23400$, respectively, even though the CHEN has lower MSEs than the $RIB$ for $m=11700$.
This may be because CHEN cannot account for the autocorrelation structure of the microstructure noise, which may cause some bias in the integrated beta estimation.
From this result, we can conclude that estimating the ratio of integrated covariance to integrated variance cannot be a good proxy of integrated beta and the robust non-parametric integrated beta estimator helps account for the market beta dynamics.

We end this section by remarking that the proposed $RIB$ estimator is not only a consistent estimator of the integrated beta under autocorrelated microstructure noise but also consistent even in the absence of autocorrelation in microstructure noise.
To assess the finite sample performance of the proposed estimator when the microstructure noise has zero autocorrelation structure, we conducted an additional analysis under a setting of zero autocorrelation in the microstructure noise and fixed all other parameters.
The full methodology and results of this analysis are presented in the online Appendix \ref{sec-additional-simulation}.

\section{Empirical analysis} \label{sec-5}

In this section, we apply the proposed DR Beta model to real high-frequency trading data.
We obtained high-frequency data for the top 50 large trading volume stocks among the S\&P 500  from the TAQ database in the Wharton Research Data Services (WRDS) system from January 1, 2010, to December 31, 2016, $1762$ trading days in total. 
We used the E-mini S\&P 500 index futures as the market portfolio, which was obtained from Refinitiv Tick History.
We used 1-sec log-returns, which were subsampled by the previous tick \citep{zhang2011estimating} scheme.
High-frequency data were available between the open and close of the market, so the number of high-frequency observations for a full trading day is $m = 23400$.

\begin{figure}[t] 
\centering
\includegraphics[width = 1\textwidth]{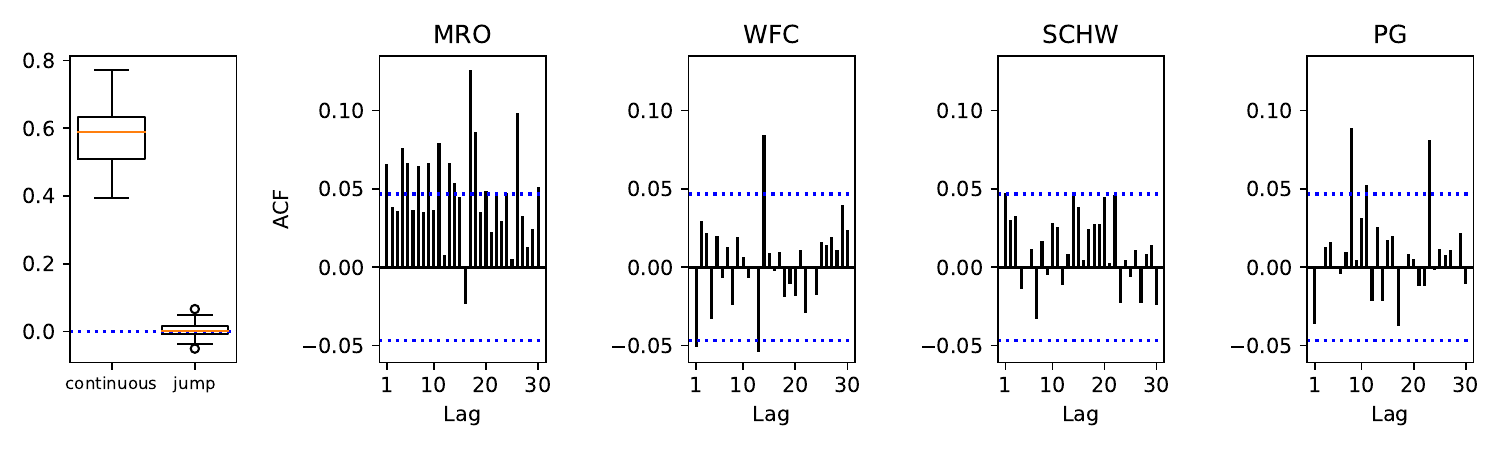} 
\caption{The box plot (left) of the first-order autocorrelations of $RIB$ (continuous) and jump beta (jump) from January 1, 2010, to December 31, 2016, and the ACF plots for the top four first-order autocorrelation stocks.}
\label{Figure-A}
\end{figure} 

\begin{table}[t]
  \caption{ Estimated parameters from the DR Beta model. 
  The numbers in parentheses indicate the $p$-value, multiplied by 10, from the hypothesis tests.}\label{table-2}
  \centering
  \scalebox{0.9}{

    \begin{tabular}{lrrrllrrr}
      \hline
      \multicolumn{1}{c}{Stock} & \multicolumn{1}{c}{$\omega$} & \multicolumn{1}{c}{$\gamma_1^g$} & \multicolumn{1}{c}{$\alpha_1^g$} & \multicolumn{1}{c}{} & \multicolumn{1}{c}{Stock} & \multicolumn{1}{c}{$\omega$} & \multicolumn{1}{c}{$\gamma_1^g$} & \multicolumn{1}{c}{$\alpha_1^g$} \\ \cline{1-4} \cline{6-9} 
      AAPL                 & 0.126 (0.0)                       & 0.539 (0.0)                      & 0.325 (0.0)                      &                      & JPM                       & 0.076 (0.0)                  & 0.608 (0.0)                      & 0.314 (0.0)                         \\
      AIG                  & 0.063 (0.0)                       & 0.623 (0.0)                      & 0.308 (0.0)                      &                     & KEY                        & 0.111 (0.0)                  & 0.542 (0.0)                      & 0.354 (0.0)                        \\
      AMAT                 & 0.190 (0.0)                       & 0.491 (0.0)                      & 0.323 (0.0)                      &                      & KO                        & 0.030 (0.0)                  & 0.653 (0.0)                      & 0.288 (0.0)                        \\
      AMD                  & 0.011 (1.0)                       & 0.826 (0.0)                      & 0.163 (0.0)                      &                     & MGM                        & 0.045 (0.0)                  & 0.715 (0.0)                      & 0.248 (0.0)                        \\
      ATVI                 & 0.088 (0.0)                       & 0.618 (0.0)                      & 0.283 (0.0)                      &                      & MRK                       & 0.056 (0.0)                  & 0.622 (0.0)                      & 0.299 (0.0)                         \\
      BAC                  & 0.083 (0.0)                       & 0.608 (0.0)                      & 0.315 (0.0)                      &                     & MRO                        & 0.042 (0.0)                  & 0.618 (0.0)                      & 0.352 (0.0)                        \\
      BMY                  & 0.070 (0.0)                       & 0.604 (0.0)                      & 0.310 (0.0)                      &                     & MS                         & 0.117 (0.0)                  & 0.547 (0.0)                      & 0.359 (0.0)                       \\
      BSX                  & 0.121 (0.0)                       & 0.540 (0.0)                      & 0.330 (0.0)                      &                     & MSFT                       & 0.072 (0.0)                  & 0.619 (0.0)                      & 0.301 (0.0)                         \\
      CSCO                 & 0.112 (0.0)                       & 0.541 (0.0)                      & 0.326 (0.0)                      &                      & MU                        & 0.091 (0.0)                  & 0.683 (0.0)                      & 0.254 (0.0)                        \\
      CSX                  & 0.073 (0.0)                       & 0.607 (0.0)                      & 0.314 (0.0)                      &                     & NEM                        & 0.025 (0.0)                  & 0.588 (0.0)                      & 0.339 (0.0)                        \\
      DAL                  & 0.096 (0.0)                       & 0.604 (0.0)                      & 0.304 (0.0)                      &                     & NFLX                       & 0.047 (0.0)                  & 0.720 (0.0)                      & 0.246 (0.0)                         \\
      DIS                  & 0.071 (0.0)                       & 0.619 (0.0)                      & 0.293 (0.0)                      &                     & NVDA                       & 0.176 (0.0)                  & 0.456 (0.0)                      & 0.374 (0.0)                         \\
      DOW                  & 0.051 (0.0)                       & 0.698 (0.0)                      & 0.249 (0.0)                      &                     & ORCL                       & 0.111 (0.0)                  & 0.540 (0.0)                      & 0.326 (0.0)                         \\
      EBAY                 & 0.106 (0.0)                       & 0.554 (0.0)                      & 0.327 (0.0)                      &                      & PFE                       & 0.050 (0.0)                  & 0.650 (0.0)                      & 0.281 (0.0)                         \\
      F                    & 0.071 (0.0)                       & 0.605 (0.0)                      & 0.311 (0.0)                      &                   & PG                           & 0.030 (0.0)                  & 0.653 (0.0)                      & 0.288 (0.0)                     \\
      FCX                  & 0.031 (0.1)                       & 0.675 (0.0)                      & 0.303 (0.0)                      &                     & QCOM                       & 0.077 (0.0)                  & 0.568 (0.0)                      & 0.335 (0.0)                         \\
      FITB                 & 0.066 (0.0)                       & 0.614 (0.0)                      & 0.323 (0.0)                      &                      & RF                        & 0.070 (0.0)                  & 0.652 (0.0)                      & 0.286 (0.0)                        \\
      GE                   & 0.129 (0.0)                       & 0.489 (0.0)                      & 0.332 (0.0)                      &                    & SCHW                        & 0.112 (0.0)                  & 0.554 (0.0)                      & 0.355 (0.0)                        \\
      GILD                 & 0.070 (0.0)                       & 0.607 (0.0)                      & 0.327 (0.0)                      &                      & T                         & 0.026 (0.0)                  & 0.706 (0.0)                      & 0.240 (0.0)                       \\
      GLW                  & 0.142 (0.0)                       & 0.564 (0.0)                      & 0.287 (0.0)                      &                     & VZ                         & 0.027 (0.0)                  & 0.707 (0.0)                      & 0.241 (0.0)                       \\
      HAL                  & 0.067 (0.0)                       & 0.644 (0.0)                      & 0.296 (0.0)                      &                     & WFC                        & 0.111 (0.0)                  & 0.540 (0.0)                      & 0.326 (0.0)                        \\
      HBAN                 & 0.047 (0.0)                       & 0.674 (0.0)                      & 0.278 (0.0)                      &                      & WMB                       & 0.026 (0.1)                  & 0.699 (0.0)                      & 0.278 (0.0)                         \\
      HPQ                  & 0.178 (0.0)                       & 0.526 (0.0)                      & 0.301 (0.0)                      &                     & WMT                        & 0.038 (0.0)                  & 0.625 (0.0)                      & 0.298 (0.0)                        \\
      HST                  & 0.077 (0.0)                       & 0.618 (0.0)                      & 0.295 (0.0)                      &                     & XOM                        & 0.063 (0.0)                  & 0.550 (0.0)                      & 0.368 (0.0)                        \\
      INTC                 & 0.108 (0.0)                       & 0.552 (0.0)                      & 0.324 (0.0)                      &                      & XRX                       & 0.138 (0.0)                  & 0.568 (0.0)                      & 0.283 (0.0)                         \\ \hline
      \end{tabular}

  }
\end{table}

To examine the goodness of fit, we conducted in-sample validation.
We draw autocorrelation plots for $RIB$ and jump beta in Figures \ref{Figure-1} and \ref{Figure-A}, where the jump beta is estimated by the method suggested by \citet{li2017robust}.
As we discussed in Section \ref{sec-3}, the integrated beta for the continuous part has a strong autocorrelation structure, but the beta for the jumps does not.
Thus, it is reasonable to focus on modeling the beta for the continuous part. 
To conduct the validation of the DR Beta model, we first selected the $(p,q) \in \left\lbrace (p,q): 0 \leq p, q \leq 5 \right\rbrace$ of the DR Beta for each stock by BIC, and we estimated the parameters of the DR Beta using the sample over the last 1000 trading days.
Then, we conducted the hypothesis tests proposed in Section \ref{sec-3.3}.
Table \ref{table-2} reports the parameter estimates of the DR Beta model with the selected $(p,q)$ for each stock and their $p$-values.
The BIC values were minimized when $(p,q)=(1,1)$ for all stocks.
All coefficients are significant at a significance level of 0.05, except for the case of the AMD.
On the other hand, the higher $\gamma_{1}^{g} + \alpha_{1}^{g}$ implies more persistent integrated beta process.
From Table 1, we find that all stocks have the $\gamma_{1}^{g} + \alpha_{1}^{g}$ greater than 0.8, and for 36 stocks out of 50, $\gamma_{1}^{g} + \alpha_{1}^{g}$ is greater than 0.9.
Thus, we can conclude that the proposed DR Beta model is statistically valid and may capture the persistent autoregressive structure.

To check the economic benefits of predicting future market beta, we analyzed the out-of-sample performance of the market-neutral portfolios.
We considered the close-to-close log-returns of market-neutral portfolios constructed by holding a share of stock, simultaneously taking a short position in E-mini S\&P 500 index futures contracts.
The amount of the futures contracts, namely the hedging ratio, was calibrated using the one-day-ahead forecasted beta.
While the daily integrated beta defined in \eqref{Equation-2.4} is an effective measure to capture the time-series dynamics of market beta, it cannot be directly used as the hedging ratio due to the price jumps and overnight returns in the market returns.
Particularly, when the jump beta or the overnight beta differs from the integrated beta, using the predicted integrated beta as the hedging ratio may not adequately minimize the portfolio's exposure to market variations due to the jump and overnight parts.
To reflect this in constructing the market-neutral portfolio, the hedging ratio should be a weighted average of the predicted integrated betas corresponding to the continuous, jump, and overnight parts.
However, there is an obstacle to obtaining the weighted average since the beta for the jump part does not have a significant time series structure.
Additionally, a complex structure may lead to serious estimation errors.
Thus, we assume that the jump beta and the overnight beta can be forecasted by a linear form of one-day-ahead forecasted betas. 
That is, the hedging ratio for a stock $A$ was calibrated  as follows: 
\begin{eqnarray*}
  (\hat{a}, \hat{b})  = \argmin_{a,b} \sum_{i=1}^{n}  (R_{A,i}- (a +b \tilde{h}_{A,i} )R_{M,i})^2
  ,
\end{eqnarray*}
where $(\tilde{h}_{A,i})_{i=1,\ldots,n}$ are in-sample fitted betas, and $R_{A,i}$ and $R_{M,i}$ are the $i$th close-to-close log-returns of a stock $A$ and the market portfolio, respectively.
That is, the forecasted hedging ratio is $\hat{a} + \hat{b} \hat{Beta}_{A,n+1}$, where the $\hat{Beta}_{A,i}$'s are one-day-ahead forecasted betas for a stock $A$, obtained using one of the DR Beta, ARMAC, ARMAP, PRBG, RBG, DCC, and BEKK, and utilizing 500 in-sample observations.
We note that this procedure is unbiased toward any specific market beta prediction method, although it might not accurately capture the jump and overnight beta dynamics.
It would be interesting to develop a robust and simple model that can simultaneously capture the dynamics of the jump, overnight, and continuous integrated betas.
We leave this for a future study. 
To evaluate the effectiveness of the hedging, we calculated the absolute correlation, the hedging effectiveness \citep{ederington1979hedging}, and the ex-post portfolio beta between the hedged portfolio of stock $A$ and the market portfolio as follows:
\begin{align*}
  & \text{Absolute correlation} = \left|\frac{\cov(R_{A}^{H}, R_{M})}{\sqrt{\var(R_{A}^{H}) \var(R_{M})}}\right| , \cr
  & \text{Hedging effectiveness} = 1 - \frac{\var(R_{A}^{H})}{\var(R_{A})}  ,\quad \text{and} \quad \text{Ex-post beta} = \frac{\cov(R_{A}^{H},R_M)}{\var(R_M)} 
  ,
\end{align*}
where $R_{A,i}^{H} = R_{A,i} - (\hat{a} + \hat{b} \hat{Beta}_{A,i})  R_{M,i}$ denotes the $i$th day out-of-sample log-return for the hedged portfolio of stock $A$.
We then averaged these evaluation measures across the 50 stocks.

\begin{table}[h]
\caption{The mean absolute correlation, hedging effectiveness, and ex-post beta between the hedged portfolios and the market portfolio, where the hedged portfolios are constructed based on the predicted beta using the OLS regression beta, DR Beta, ARMAC, ARMAP, PRBG, RBG, DCC, and BEKK.
Unhedged indicates the unhedged single-stock portfolio.
}\label{table-application}
\centering
\scalebox{0.9}{
\setlength\tabcolsep{2.5pt}
\begin{tabular}{lrrrrrrrrr}
  \hline
\multicolumn{1}{r}{Measure $\backslash$ Model} & \multicolumn{1}{r}{Unhedged} & \multicolumn{1}{r}{OLS} & \multicolumn{1}{r}{DR Beta} & \multicolumn{1}{r}{ARMAC} & \multicolumn{1}{r}{ARMAP} & \multicolumn{1}{r}{PRBG} & \multicolumn{1}{r}{RBG} & \multicolumn{1}{r}{DCC} & \multicolumn{1}{r}{BEKK}  \\ \cline{1-10} 
Absolute Correlation  & 0.569 & 0.031 & 0.020 & 0.021 & 0.023 & 0.064 & 0.060 & 0.033 & 0.029 \\
Hedging Effectiveness & 0.000 & 0.334 & 0.338 & 0.337 & 0.339 & 0.330 & 0.331 & 0.333 & 0.330 \\
Ex-post beta          & 1.154 & 0.057 & 0.034 & 0.035 & 0.038 & 0.124 & 0.115 & 0.058 & 0.054 \\
\hline
\end{tabular}
}
\end{table}
Table \ref{table-application} reports the mean of absolute correlation, hedging effectiveness, and ex-post beta for the unhedged single-stock portfolio (Unhedged), hedged portfolio using the one-day-ahead forecasted beta from the regression beta (OLS), DR Beta, ARMAC, ARMAP, PRBG, RBG, DCC, and BEKK models.
For the forecasted beta using OLS, we employed the beta derived from OLS regression on daily close-to-close log-returns, using  500 in-sample observations.
From Table \ref{table-application}, we find that the ARMA models incorporating high-frequency-based non-parametric estimators as inputs show the best performance in hedging the market factor.
While the ARMA models provided comparable performances, incorporating the $RIB$ estimator led to an improvement in the absolute correlation and ex-post beta measures.
In the case of hedging effectiveness, the ARMAP is slightly better than the DR Beta model.

\section{Conclusion}\label{sec-6}
This paper investigates integrated market betas based on high-frequency financial data.
We first develop a robust non-parametric integrated beta estimation procedure, $RIB$, which can handle the price-dependent and autocorrelated microstructure noise and time-varying beta.
Then, we establish its asymptotic properties. 
With this robust non-parametric $RIB$ estimator, we find the time-series structure of the integrated betas. 
To account for this beta dynamics, we propose the DR Beta model.
To estimate the model parameters, we propose a quasi-likelihood estimation procedure and establish its asymptotic theorems.
From the empirical study, we demonstrate that using the proposed DR Beta model with the robust realized integrated beta estimator to predict future integrated beta helps construct market-neutral portfolios.

 \subsection*{Acknowledgment}
 The authors thank the co-Editor Professor Torben Andersen,  and anonymous associate editor and two referees for their careful reading of this paper and valuable comments. 
 The research of Yazhen Wang was supported in part by NSF grant DMS-1913149.

\bibliography{references}

\clearpage

\appendix
\setcounter{figure}{0}
\renewcommand{\thefigure}{A\arabic{figure}}
\section*{Appendix.} \label{Appendix}

\section{Choosing $l_m$ and $k'_m$ in practice}\label{sec-tuning-parameter}
To choose $l_m$ for calculating market volatility, we follow the heuristic criterion presented in Section 5.1.2 of \citet{jacod2017statistical}.
On the other hand, to choose $l_m$ for obtaining the covariance estimation, we follow the bivariate version of the heuristic criterion presented in Section 5.1.2 of \citet{jacod2017statistical}.
Specifically, we utilize
\begin{equation*}
  \hat{l}_m = \argmin_{1 \leq l \leq 2 [\Delta_m ^{-\frac{1}{5}}]} \sum_{j=1}^{2 [\Delta_m ^{-\frac{1}{8} }]} \left( \hat{R}_{12,l}(0) + \hat{R}_{21,l}(0) - \hat{R}_{12,l}(j) - \hat{R}_{21,l}(j) - \hat{\Delta R(j)}^{\text{adj}} \right)^2 
  ,
\end{equation*}
where $\bar{P}_{i}^{m,l} = l^{-1} \sum_{k=i}^{i+l-1} P^{m}_{k}$ for any process $P$, $\hat{R}_{12,l}(j) = \sum_{i=0}^{N-4l} (Y^{m}_{1,i} - \bar{Y}_{1,i+2l}^{m,l}) (Y^{m}_{2,i+j} - \bar{Y}_{2,i+4l}^{m,l})$, $\hat{R}_{21,l}(j) = \sum_{i=0}^{N-4l} (Y^{m}_{2,i} - \bar{Y}_{2,i+2l}^{m,l}) (Y^{m}_{1,i+j} - \bar{Y}_{1,i+4l}^{m,l})$,
and $\hat{\Delta R(j)}^{\text{adj}} = m^{-1} \sum_{i=0}^{m-j} (Y_{1,i+j}^{m}-Y_{1,i}^{m}) (Y_{2,i+j}^{m}-Y_{2,i}^{m}) $.
To choose $k'_m$, we utilized the test for (cross) autocovariance of noise as presented in Corollary 3.5 in \citet{jacod2017statistical}.
When considering $k'_m$ for cross autocovariance, we can consider two parameters, $\hat{k}'_{Y_1 Y_2,m}$ and $\hat{k}'_{Y_2 Y_1,m}$, which are related to $\mathbb{E}\left[ \chi_{1,0} \chi_{2,j} \right]$ and $\mathbb{E}\left[ \chi_{2,0} \chi_{1,j} \right]$, respectively, where $j$ is a positive integer.
Specifically, we chose 
\begin{equation*}
  \hat{k}'_{PQ,m} = \max \left\lbrace k \in \mathbb{N} : \forall j \leq k , |T_{PQ,j} | > 1.96    \right\rbrace
  ,
\end{equation*}
where $P$ and $Q$ are one of the processes $Y_1$ and $Y_2$, 
\begin{align*}
  & T_{PQ,j} = \frac{U_{PQ}(j)}{\sqrt{{S}_{pq}(j)}} ,\quad
  U_{PQ}(j) = \sum_{i=0}^{m-4\hat{l}_{m}} (P^{m}_{i} - \bar{P}^{m,\hat{l}_{m}}_{i+\hat{l}_{m}})(Q^{m}_{i+j} - \bar{Q}^{m,\hat{l}_{m}}_{i+3\hat{l}_{m}}) ,\cr
  & S_{PQ}(j) = U(0,j,0,j) + \sum_{k=1}^{j} \left( U_{PQ}(0,j,m,m+j) + U_{PQ}(m,m+j,0,j) \right) - (2j+1) U_{PQ}^4(j) ,\cr
  & U_{PQ}(j_1,j_2,j_3,j_4) = \sum_{i=0}^{m-8\hat{l}_m} (P^{m}_{i+j_1} - \bar{P}^{m,\hat{l}_{m}}_{i+\hat{l}_{m}})(Q^{m}_{i+j_2} - \bar{Q}^{m,\hat{l}_{m}}_{i+3\hat{l}_{m}}) (P^{m}_{i+j_3} - \bar{P}^{m,\hat{l}_{m}}_{i+5\hat{l}_{m}})(Q^{m}_{i+j_4} - \bar{Q}^{m,\hat{l}_{m}}_{i+7\hat{l}_{m}}) ,\cr
  & U_{PQ}^{4}(j) = \sum_{i=0}^{m-9\hat{l}_m} (P^{m}_{i} - \bar{P}^{m,\hat{l}_{m}}_{i+\hat{l}_{m}})(Q^{m}_{i+j} - \bar{Q}^{m,\hat{l}_{m}}_{i+3\hat{l}_{m}})  (P^{m}_{i+5\hat{l}_m} - \bar{P}^{m,\hat{l}_{m}}_{i+6\hat{l}_{m}})(Q^{m}_{i+j+5\hat{l}_m} - \bar{Q}^{m,\hat{l}_{m}}_{i+8\hat{l}_{m}}) 
  .
\end{align*}

\section{High-frequency data-generating diffusion process}\label{sec-HFDGP}

The DR Beta model in Section \ref{sec-3} can capture the low-frequency dynamics using the high-frequency-based measure, the integrated beta, which is essentially developed based on the continuous diffusion model as described in Section \ref{sec-2}.
Thus, there is a gap between the  DR Beta model in \eqref{eq:DRBeta} and the continuous diffusion model in Section \ref{sec-2}.
Also, we need a high-frequency data-generating process to investigate the $RIB$ estimator in terms of the dynamic analysis through a simulation study.
Therefore, we provide a high-frequency data-generating example diffusion process for market beta processes whose integrated betas follow the DR Beta model.
This data-generating process will serve to bridge the gap between low- and high-frequency models and will also provide a rigorous mathematical background for the DR Beta model.

We introduce a spot beta process whose integrated betas satisfy the DR Beta model, using the framework of the unified GARCH-It\^o-type models \citep{kim2016unified, song2021volatility}.
Similar to the work by \citet{kallsen1998option}, who interpolate the ARCH model as a piecewise continuous process, the unified GARCH-It\^{o} model is a continuous-time diffusion process with a continuous-time volatility process embedding a GARCH volatility.
We extend the unified GARCH-It\^o-type model from the ARMA(1,1) structure to the ARMA($p,q$) structure as follows:
\begin{definition}  \label{Definition-1}
For the proposed time-series regression model in \eqref{Equation-2.1}, a beta process $\beta_t^c(\theta)$,  $t \in \mathbb{R}_+$, follows the DR Beta diffusion process if it satisfies:
\begin{eqnarray} \label{EXTENDpq}
  \beta_{t}^{c}(\theta) &=& \beta_{[t]}^{c}(\theta) + \left( t - [t] \right) ^2 \left( \omega_1  + \sum_{i=1}^{q} \gamma_{i} \beta_{[t]+1-i}^{c} (\theta)  + \sum_{i=2}^{p} \alpha_{i} \int_{[t]-i+1}^{[t]-i+2} \beta_{s}^{c}(\theta) ds    \right)    \nonumber\\
  && - \left( t - [t] \right)    \(  \omega_ 2 +  \beta_{[t]}^{c}(\theta) \)  + \alpha_{1} \int_{[t]}^{t} \beta_s^c (\theta) ds  + \nu \left( [t] + 1 - t \right) \int_{[t]}^{t} dZ_t ,
\end{eqnarray}
where $[t]$ denotes the integer part of $t$ and $Z_t$ is a standard Brownian motion with $dZ_t dB_t=\rho dt$ and $dZ_t dW_t=0$  a.s.
$\beta_{0}^{c} (\theta), \ldots, \beta_{-q+1}^{c} (\theta)  $ and $ \int_{-1}^{0} \beta_{s}^{c}(\theta) ds, \ldots, \int_{1-p}^{2-p} \beta_{s}^{c}(\theta) ds $ are initial values.
We denote the model parameter by $\theta = \left( \omega_1, \omega_2, \gamma_1, \ldots, \gamma_{q} , \alpha_{1}, \ldots, \alpha_{p}, \nu \right) $.
\end{definition}

The DR Beta diffusion process is continuous at all times $t \in \mathbb{R}^+$ and has a quadratic shape pattern within the intraday.
For example, $\omega_{1}$ and $\omega_{2}$ govern the deterministic quadratic time-trend of the spot beta process.
For the non-deterministic part of the quadratic interpolation, the interpolation gives more weight to the persistent terms (related to the square term) and reduces the weight of the past information (related to the linear term).
The persistent feature of the beta process is determined through the parameters $\gamma_{1},\ldots,\gamma_{q}$ and $\alpha_{1}, \ldots, \alpha_{p}$.
By choosing appropriate parameters, high (low) initial betas form a downward (upward) convex shape with respect to time $t$.
This intraday structure can accommodate the intraday spot beta dynamics found in \citet{andersen2021recalcitrant}.
Moreover, parameter $\alpha_{1}$ plays a key role in controlling the intraday level autoregressive characteristic of the spot beta process.
We introduce $Z_t$ to account for the random fluctuations of the spot beta process.
On the other hand, the spot beta process can be considered as a generalized Ornstein-Uhlenbeck process whose existence and uniqueness have been proven in \citet{jacod1979calcul}.
Specifically, we have
\begin{equation*}
  d\beta^{c}_{t} = dH_{t} + \beta^{c}_{t-} dS_{t}
  ,
\end{equation*}
where
\begin{align*}
  & dS_{t} = \alpha_{1}dt \quad \text{and} \quad \cr
  & dH_{t} = \Bigg (2  \left( t - [t] \right)  \left( \omega_1  + \sum_{i=1}^{q} \gamma_{i} \beta_{[t]+1-i}^{c} (\theta)  + \sum_{i=2}^{p} \alpha_{i} \int_{[t]-i+1}^{[t]-i+2} \beta_{s}^{c}(\theta) ds    \right) - (\omega_{2} + \beta^{c}_{[t]}(\theta)) \cr
  & \qquad\qquad - \nu (Z_{t} - Z_{[t]}) \Bigg) dt + \nu ([t] + 1 - t) dZ_{t} 
  ,
\end{align*}
with the initial values $\beta_{0}^{c} (\theta), \ldots, \beta_{-q+1}^{c} (\theta)  $ and $ \int_{-1}^{0} \beta_{s}^{c}(\theta) ds, \ldots, \int_{1-p}^{2-p} \beta_{s}^{c}(\theta) ds $.
We note that, to define the general ARMA($p,q$) model, for $p\geq2$ or $q\geq2$, we need to define the additional initial values, such as $\beta_{-1}^{c} (\theta), \ldots, \beta_{-q+1}^{c} (\theta)  $ and $ \int_{-1}^{0} \beta_{s}^{c}(\theta) ds, \ldots, \int_{1-p}^{2-p} \beta_{s}^{c}(\theta) ds $. 
Finally, when the process is restricted to low-frequency time points, the spot beta adopts the following realized ARMA$(p,q)$ model-type structure: %
\begin{equation}\label{spot-int}
  \beta^c_n(\theta) = \omega + \sum_{i=1}^{q} \gamma_{i} \beta^c_{n-i} (\theta) + \sum_{j=1}^{p} \alpha_{j} \int_{n-j}^{n-j+1} \beta_{s}^{c} (\theta) ds \quad \text{ for any } n \in \mathbb{N} ,
\end{equation}
where $\omega= \omega_1- \omega_2$.
The spot beta process is along the lines of the unified GARCH-It\^o type processes  \citep{kim2016unified, song2021volatility}.
That is, the DR Beta diffusion process is developed to explain the low-frequency beta dynamics, which we find in the empirical study using the proposed robust non-parametric realized beta estimator, and fill the mathematical gap between the low-frequency discrete-time series and continuous-time series regression models.
Unlike the unified GARCH-It\^o type processes  \citep{kim2016unified, song2021volatility}, we develop the ARMA$(p,q)$ model-type structure to capture a more general dynamic structure.

The following proposition presents properties of the integrated betas for the DR Beta diffusion process, which show the existence of a diffusion process satisfying the DR Beta model.
\begin{proposition}\label{Proposition-1}
  For $\sum_{i=1}^{q} \left|\gamma_{i} \right| < 1$, $\sum_{i=1}^{p \lor q} \left | \mathbf{1}_{\left\lbrace i \leq q \right\rbrace} \gamma_{i} + \mathbf{1}_{\left\lbrace i \leq p \lor q \right\rbrace} \alpha_{i}^{g} \right |   <1$, and $n \in \mathbb{N}$,  integrated betas for the DR Beta diffusion process in Definition \ref{Definition-1} have the following properties:
  \begin{enumerate}
    \item[(a)] We have
    \begin{equation}\label{HandD}
      I\beta_n(\theta)=\int ^{n}_{n-1}\beta_t^c(\theta)dt=h_n(\theta)+D_n\quad a.s.
      ,
    \end{equation}
    where
    \begin{align}
      & h_n(\theta) = \omega^g + \sum_{i=1}^{q}  \gamma_{i} h_{n-i}(\theta) + \sum_{i=1}^{p \lor q} \alpha_{i}^{g} I\beta_{n-i} , \label{GARCH-pq} \\
      & \omega^g = \left( \varrho_{1} - \varrho_{2} + 2 \varrho_{3} \right) \omega + (2 \varrho_{3} - \varrho_{2}) \left( 1 - \sum_{i=1}^{q} \gamma_{i} \right)  \omega_{2} , \nonumber \\
      & \alpha_{i}^{g} = \mathbf{1}_{\left\lbrace i \leq q \right\rbrace} 2 \varrho_{3} \gamma_{i} \alpha_{1} + \mathbf{1}_{\left\lbrace i \leq p \right\rbrace} (\varrho_{1} - \varrho_{2}) \alpha_{i} + \mathbf{1}_{\left\lbrace i \leq p-1 \right\rbrace} 2 \varrho_{3} \alpha_{i+1} , \nonumber \\
      & \varrho_1 = \alpha_{1}^{-1} \left( e^{\alpha_{1}} - 1 \right) , \quad \varrho_{2} = \alpha_{1}^{-2} \left( e^{\alpha_{1}} - 1 - \alpha_{1} \right) , \quad \varrho_{3} = \alpha_{1}^{-3} \left( e^{\alpha_{1}} - 1 - \alpha_{1} - \frac{\alpha_{1}^{2}}{2}  \right) , \nonumber
    \end{align}
    and
    \begin{equation*}
      D_n = \nu \int_{n-1}^{n} \left[ (n-t) \alpha_{1}^{-1} e^{\alpha_{1} (n-t)} - \alpha_{1}^{-2} e^{\alpha_{1}(n-t)} + \alpha_{1}^{-2} \right] dZ_t
    \end{equation*}
    is a martingale difference.

    \item[(b)] $\beta^c_n$ and $  I\beta_{n}$ have a finite moment for any given order, and we have 
    \begin{eqnarray*}
      &&\mathbb{E}[h_n(\theta)]=\frac{\omega^g}{1-\sum_{i=1}^{q}  \gamma_{i} -  \sum_{j=1}^{p \lor q} \alpha^g_{j}}, \\
      &&\mathbb{E}[\beta^c_n]=\frac{\omega \left( 1-\sum_{i=1}^{q}  \gamma_{i} -  \sum_{j=1}^{p \lor q} \alpha^g_{j} \right) +  \omega^g \sum_{i=1}^{p} \alpha_{i}}{\left(1-\sum_{i=1}^{q}  \gamma_{i} -  \sum_{j=1}^{p \lor q} \alpha^g_{j} \right)(1- \sum_{i=1}^{q} \gamma_{i})}, \quad \text{and} \\
      &&\mathbb{E}\left[ I\beta_{n}(\theta) | \mathcal{F}_{n-1} \right] = h_n(\theta) \quad \text{a.s.} 
    \end{eqnarray*}

    \item[(c)] We have
    \begin{equation*}
      \sup_{i\in \mathbb{N}} \mathbb{E}\left[ D_{i}^{2} | \mathcal{F}_{i-1} \right] \leq C
      .
    \end{equation*}
    
  \end{enumerate}

\end{proposition}

Proposition \ref{Proposition-1}(a) indicates that the integrated betas $I\beta_n$'s can be decomposed into the conditional expectation $h_n(\theta)$ and the martingale difference $D_n$, where $h_n(\theta)$ is adapted to the filtration $\mathcal{F}_{n-1}$. %
Further, the conditional expectation $h_n(\theta)$ and the integrated beta $I\beta_n$ have the relationship \eqref{GARCH-pq}, which is the same form as \eqref{eq:DRBeta}.
That is, if spot betas follow the DR Beta diffusion, then its integrated betas follow the DR Beta model.
Proposition \ref{Proposition-1}(a), typically assumed in the asymptotic analysis of ARMA models, is inherently satisfied under the DR Beta diffusion process.
We end this section by remarking that the DR Beta diffusion process is not the only solution of the DR Beta model \eqref{eq:DRBeta}.
For example, at each integer point, \eqref{spot-int} is satisfied, and between integer points, we can interpolate.
In this paper, we adopt the quadratic interpolation.
Alternatively, we can use linear interpolation or higher interpolation.
On the other hand, we can also use a step function form for the spot beta over each low-frequency period--that is, it does not need to be continuous.

\section{Detailed descriptions of benchmark estimators}\label{sec-detail-estimator}
In Sections \ref{sec-4} and \ref{sec-5}, we employed the benchmark estimators, CHEN and PRVB, for comparison purposes.
The integrated beta (CHEN) in \citet{chen2018inference} was estimated as follows:
\begin{equation*}
  \hat{I\beta}^{CHEN} = b_m \Delta_m  \sum_{i=0}^{[\frac{1}{b_m \Delta_m }] -1} \left[ \hat{\beta}_{ib_m}^{m,C} - \hat{B}_{ib_m}^{m,C} \right], \quad \hat{\beta}_{ib_m}^{m,C} = \frac{\hat{\bSigma}_{12,i}^{m,C}}{\hat{\bSigma}_{11,i}^{m,C,*}}, \quad \hat{\bSigma}_{11,i}^{m,C,*} = \max(\hat{\bSigma}_{11,i}^{m,C},\delta_m)
  ,
\end{equation*}
where
\begin{eqnarray}\label{def-chen}
  && \hat{\bSigma}_{i}^{m,C} = \frac{1}{(b_m - k_m) \Delta_m k_m \psi_0} \sum_{l=0}^{b_m - k_m + 1}  \left(  \tilde{\bold{Y}}_{i+l}^{m} \tilde{\bold{Y}}_{i+l}^{m\top} \mathbf{1}_{\left\lbrace \norm{\tilde{\bold{Y}}_{i+l}^{m}} \leq u_m \right\rbrace} - \hat{\bold{Y}}_{i+l}^{m} \right), \nonumber\\
  && \hat{\bold{Y}}_{i}^{m} = \frac{1}{2} \sum_{l=1}^{k_m} (g_{l}^{m} - g_{l-1}^{m})^2 (\bold{Y}_{i+l}^{m} - \bold{Y}_{i+l-1}^{m}) (\bold{Y}_{i+l}^{m} - \bold{Y}_{i+l-1}^{m})^{\top}, \nonumber\\
  && \hat{B}^{m, C}_{i b_m } =   \frac{4}{\psi_0^{2} {C_k}^3 b_m\Delta_m^{1/2}} \left[ \left( \frac{{C_k}^2 \Phi_{01}}{\hat{\bSigma}_{11,ib_m}^{m,C,*}} + \frac{\Phi_{11} \hat{\bvartheta}_{11,ib_m}^{m,C}}{\left( \hat{\bSigma}_{11,ib_m}^{m,C,*} \right)^2 }   \right) \left( \hat{\bvartheta}_{11,ib_m}^{m,C} \frac{\hat{\bSigma}_{12,ib_m}^{m,C}}{\hat{\bSigma}_{11,ib_m}^{m,C,*}} - \hat{\bvartheta}_{12,ib_m}^{m,C}  \right)  \right], \nonumber\\
  && \hat{\bvartheta}_{ib_m}^{m,C} = \frac{1}{2k_m}  \sum_{l=1}^{k_m} (\bold{Y}_{i+l}^{m} - \bold{Y}_{i+l-1}^{m}) (\bold{Y}_{i+l}^{m} - \bold{Y}_{i+l-1}^{m})^{\top}, \quad \bold{Y} = (Y_1, Y_2)^{\top} \nonumber
  ,
\end{eqnarray}
and $b_m$, $k_m$, and the truncation parameters are the same as that of the $RIB$ estimator.
On the other hand, adopting the robust pre-averaging integrated volatility estimator of \citet{jacod2019estimating} as the input of the beta estimator in \citet{christensen2010pre}, the integrated beta (PRVB) can be estimated as follows:
\begin{align*}
  & \hat{I\beta}^{PRVB} = \frac{\hat{\bSigma}_{12}^{m,P}}{\hat{\bSigma}_{11}^{m,P}}, \cr
  & \hat{\bSigma}^{m,P} = \frac{1}{k_m \psi_0} \Bigg[  \sum_{l=0}^{m - k_m + 1}  \tilde{\bold{Y}}_{l}^{m} \tilde{\bold{Y}}_{l}^{m\top} \mathbf{1}_{\left\lbrace \left| \tilde{\bold{Y}}_{l,1}^{m} \right| \leq u_{1,m}, \left| \tilde{\bold{Y}}_{l,2}^{m} \right| \leq u_{2,m} \right\rbrace} \cr
  & \qquad \qquad \qquad \qquad - \frac{1}{k_m}  \sum_{d=-k'_m}^{k'_m} \phi_{d}^{m} \sum_{l=0}^{m-5l_m} (\bY_{l}^{m} - \bar{\bY}_{i+2l_m}^{m})^{\top} (\bY_{l+d}^{m} - \bar{\bY}_{i+4l_m}^{m})   \Bigg]  ,
\end{align*}
where $k_m$, $k'_m$, $l_m$, and the truncation parameters are the same as that of the $RIB$ estimator.

Adopting the dynamic conditional beta (DCB) model framework proposed by \citet{engle2016dynamic}, the beta prediction can be established as follows:
\begin{eqnarray*}
    && \hat{Beta}_{n+1}=\left(\mathbf{H}_{n+1,11}(\hat{\theta})\right)^{-1}\mathbf{H}_{n+1,12}(\hat{\theta}), \cr 
    &&\hat{\theta}=\argmax_{\theta} \left(-\frac{1}{2} \sum_{i=1}^{n}\log |\det(\mathbf{H}_{i}(\theta))| -\frac{1}{2} \sum_{i=1}^{n}(\mathbf{y}_{i}-\bar{\mathbf{y}})^\top \left(\mathbf{H}_{i}(\theta)\right)^{-1}(\mathbf{y}_{i}-\bar{\mathbf{y}}) \right ), 
\end{eqnarray*}
where $\mathbf{y}_{i} =(Y_{1,i}-Y_{1,i-1}, \; Y_{2,i}-Y_{2,i-1})^{\top}$, $\bar{\mathbf{y}}=\frac{1}{n}\sum^{n}_{j=1}\mathbf{y}_j$, and $\mathbf{H}_i(\theta) $ denotes a conditional covariance matrix of $\mathbf{y}_{i}$.

\section{Additional simulation analyses}\label{sec-additional-simulation}

The proposed $RIB$ estimator is not only a consistent estimator of the integrated beta under autocorrelated microstructure noise but also consistent even in the absence of autocorrelation in microstructure noise.
To assess the finite sample performance in the absence of autocorrelation in microstructure noise, we conducted an additional simulation analysis.
We used the same simulation setting used in Section \ref{sec-4}, except for the process $$\bchi_{i} \sim_{i.i.d.} N\left[\begin{pmatrix} 0 \cr 0 \end{pmatrix},\begin{pmatrix} 1 & 0 \cr 0 & 1 \end{pmatrix}\right].$$
That is, we consider the scenario where the microstructure noise exhibits zero autocorrelations.

\begin{figure}[h] 
\centering
\includegraphics [height=0.5\textwidth,width = 0.55\textwidth]{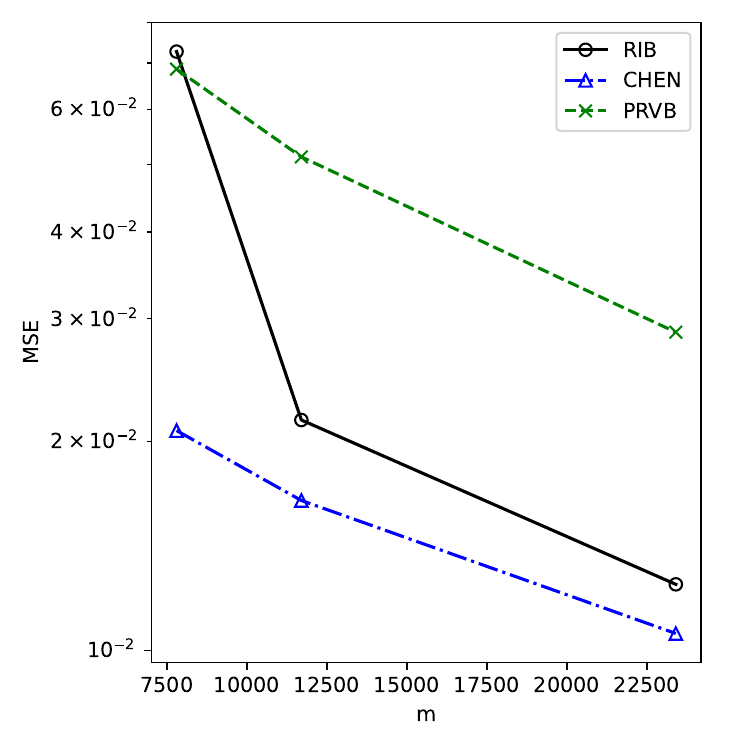}
\caption{The MSEs  of  $RIB$, CHEN, and  PRVB  for  $m=7800, 11700, 23400$ in the absence of autocorrelation in microstructure noise. 
The average value of the true integrated beta was 2.802.}
\label{Figure-nonparam-absence}
\end{figure}

Figure \ref{Figure-nonparam-absence} shows the MSEs of the non-parametric integrated beta estimators, $RIB$, CHEN, and PRVB, for $m=7800, 11700, 23400$.
We note that the average value of the true integrated beta was 2.802.
From Figure \ref{Figure-nonparam-absence}, we find that the CHEN estimator shows the best performance.
This is because the CHEN estimator is a consistent estimator of integrated beta and has a simple structure since it does not consider the autocorrelation in microstructure noise.
On the other hand, as the number of high-frequency observations increases, the MSEs of the $RIB$ and CHEN estimators become comparable.
This may be because the disadvantage associated with the complexity of the $RIB$ estimator due to accounting for the autocorrelated structure of noise diminishes as the number of observations increases.

\begin{figure}[t] 
\centering
\includegraphics[width = 0.9\textwidth]{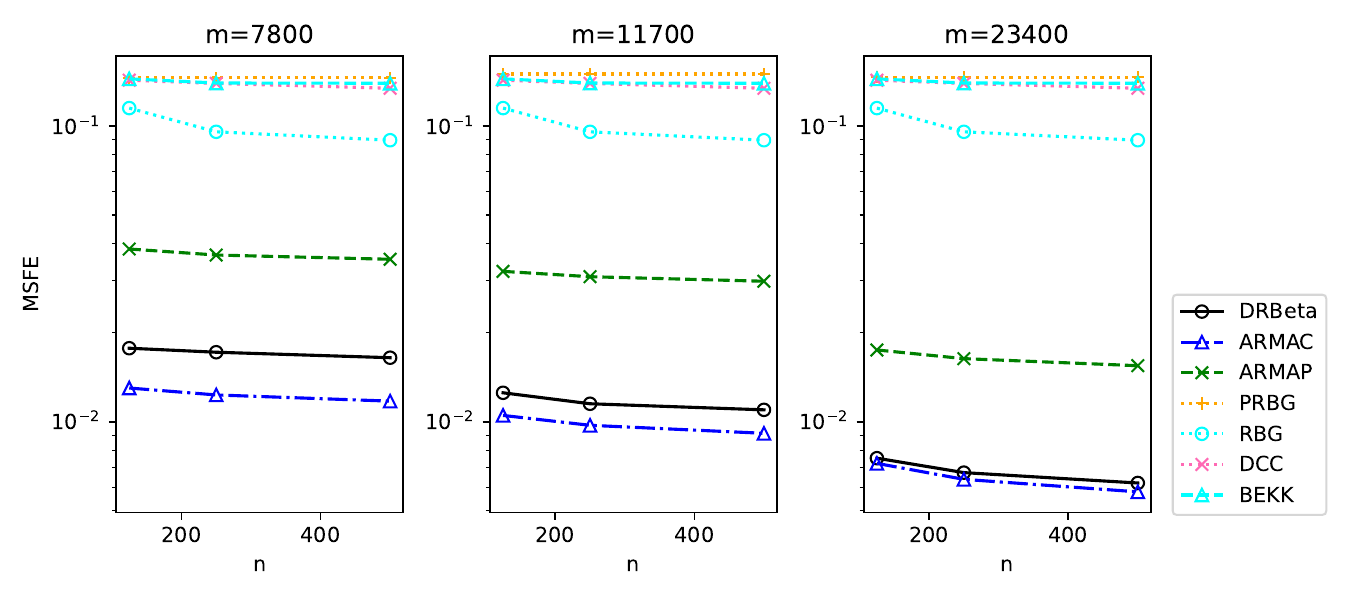} 
\caption{ The MSFEs of DR Beta, ARMAC, ARMAP, PRBG, RBG, DCC, and BEKK with $n=100, 250, 500$ and $m=7800, 11700, 23400$ in the absence of autocorrelation in microstructure noise.}
\label{Figure-pred-absence}
\end{figure} 
Figure \ref{Figure-pred-absence} draws the MSFEs of DR Beta, ARMAC, ARMAP, PRBG, RBG, DCC, and BEKK for $n=100, \, 250, \, 500$ and $m=7800, 11700, 23400$.
Similar to the result shown in Figure \ref{Figure-nonparam-absence}, we find that the MSFEs of the DR Beta model get close to that of the ARMAC model as the number of high-frequency observations increases.
From this additional analysis, we can conclude that the proposed $RIB$ estimator demonstrates satisfactory finite sample performance, even in the absence of autocorrelation in microstructure noise.

\section{Additional empirical analyses}\label{sec-additional-empric}

\afterpage{
\begin{landscape}
  \begin{table}[ptbh]
  \caption{The mean of $RIB$ estimates in the out-of-sample period and the MAPEs for the DR Beta, ARMAC, ARMAP, PRBG, RBG, DCC, and BEKK for each stock.
  }\label{table-3}
  \centering
  \scalebox{0.85}{
  \setlength\tabcolsep{2.5pt}
  \begin{tabular}{lrrrrrrrrllrrrrrrrr}
    \hline
  \multicolumn{1}{c}{Stock} & \multicolumn{1}{c}{$RIB$}    & \multicolumn{1}{c}{DRBeta} & \multicolumn{1}{c}{ARMAC} & \multicolumn{1}{c}{ARMAP} & \multicolumn{1}{c}{PRBG} & \multicolumn{1}{c}{RBG} & \multicolumn{1}{c}{DCC} & \multicolumn{1}{c}{BEKK} & \multicolumn{1}{c}{} & \multicolumn{1}{c}{Stock} & \multicolumn{1}{c}{$RIB$}    & \multicolumn{1}{c}{DRBeta} & \multicolumn{1}{c}{ARMAC} & \multicolumn{1}{c}{ARMAP} & \multicolumn{1}{c}{PRBG} & \multicolumn{1}{c}{RBG} & \multicolumn{1}{c}{DCC} & \multicolumn{1}{c}{BEKK} \\ \cline{1-9} \cline{11-19} 
  AAPL                      & 0.985                        & 0.238                      & 0.254                     & 0.259                     & 0.300                    & 0.279                    & 0.324                  & 0.326                   &                      & JPM                        & 1.030                        & 0.170                     & 0.203                    & 0.211                    & 0.293                   & 0.274                  & 0.325                  & 0.314                   \\
  AIG                       & 0.989                        & 0.183                      & 0.206                     & 0.225                     & 0.348                    & 0.326                    & 0.346                  & 0.407                   &                      & KEY                        & 1.064                        & 0.226                     & 0.252                    & 0.276                    & 0.334                   & 0.366                  & 0.406                  & 0.363                   \\
  AMAT                      & 1.028                        & 0.215                      & 0.235                     & 0.262                     & 0.388                    & 0.397                    & 0.443                  & 0.365                   &                      & KO                         & 0.499                        & 0.117                     & 0.127                    & 0.134                    & 0.168                   & 0.167                  & 0.214                  & 0.186                   \\
  AMD                       & 0.950                        & 0.371                      & 0.382                     & 0.444                     & 0.779                    & 0.892                    & 0.936                  & 0.953                   &                      & MGM                        & 1.295                        & 0.280                     & 0.302                    & 0.347                    & 0.468                   & 0.434                  & 0.583                  & 0.590                   \\
  ATVI                      & 0.847                        & 0.208                      & 0.225                     & 0.242                     & 0.313                    & 0.331                    & 0.359                  & 0.293                   &                      & MRK                        & 0.674                        & 0.142                     & 0.162                    & 0.159                    & 0.341                   & 0.185                  & 0.241                  & 0.218                   \\
  BAC                       & 1.181                        & 0.240                      & 0.261                     & 0.303                     & 0.431                    & 0.382                    & 0.465                  & 0.490                   &                      & MRO                        & 1.400                        & 0.295                     & 0.325                    & 0.349                    & 0.404                   & 0.412                  & 0.562                  & 0.595                   \\
  BMY                       & 0.740                        & 0.163                      & 0.185                     & 0.184                     & 0.248                    & 0.237                    & 0.312                  & 0.271                   &                      & MS                         & 1.350                        & 0.228                     & 0.273                    & 0.286                    & 0.369                   & 0.365                  & 0.424                  & 0.456                   \\
  BSX                       & 0.873                        & 0.219                      & 0.240                     & 0.265                     & 0.432                    & 0.430                    & 0.486                  & 0.396                   &                      & MSFT                       & 0.889                        & 0.168                     & 0.191                    & 0.194                    & 0.260                   & 0.269                  & 0.293                  & 0.223                   \\
  CSCO                      & 0.871                        & 0.169                      & 0.193                     & 0.197                     & 0.238                    & 0.264                    & 0.273                  & 0.250                   &                      & MU                         & 1.444                        & 0.329                     & 0.363                    & 0.379                    & 0.531                   & 0.525                  & 0.653                  & 0.593                   \\
  CSX                       & 0.943                        & 0.180                      & 0.205                     & 0.224                     & 0.302                    & 0.282                    & 0.404                  & 0.385                   &                      & NEM                        & 0.466                        & 0.277                     & 0.279                    & 0.286                    & 0.407                   & 0.452                  & 0.535                  & 0.514                   \\
  DAL                       & 1.016                        & 0.263                      & 0.280                     & 0.305                     & 0.626                    & 0.422                    & 0.511                  & 0.431                   &                      & NFLX                       & 1.334                        & 0.315                     & 0.347                    & 0.356                    & 0.437                   & 0.406                  & 0.575                  & 0.578                   \\
  DIS                       & 0.810                        & 0.140                      & 0.162                     & 0.172                     & 0.230                    & 0.222                    & 0.287                  & 0.260                   &                      & NVDA                       & 1.104                        & 0.237                     & 0.256                    & 0.283                    & 0.376                   & 0.374                  & 0.496                  & 0.460                   \\
  DOW                       & 0.995                        & 0.186                      & 0.206                     & 0.232                     & 0.343                    & 0.311                    & 0.405                  & 0.478                   &                      & ORCL                       & 0.868                        & 0.160                     & 0.179                    & 0.195                    & 0.264                   & 0.240                  & 0.303                  & 0.293                   \\
  EBAY                      & 0.936                        & 0.185                      & 0.207                     & 0.212                     & 0.300                    & 0.279                    & 0.360                  & 0.307                   &                      & PFE                        & 0.700                        & 0.149                     & 0.166                    & 0.170                    & 0.188                   & 0.204                  & 0.235                  & 0.199                   \\
  F                         & 0.903                        & 0.206                      & 0.212                     & 0.262                     & 0.377                    & 0.388                    & 0.483                  & 0.438                   &                      & PG                         & 0.483                        & 0.119                     & 0.131                    & 0.128                    & 0.178                   & 0.162                  & 0.204                  & 0.187                   \\
  FCX                       & 1.398                        & 0.315                      & 0.343                     & 0.368                     & 0.562                    & 0.565                    & 0.708                  & 0.768                   &                      & QCOM                       & 0.845                        & 0.157                     & 0.181                    & 0.179                    & 0.250                   & 0.238                  & 0.349                  & 0.273                   \\
  FITB                      & 1.046                        & 0.199                      & 0.229                     & 0.254                     & 0.290                    & 0.299                    & 0.346                  & 0.346                   &                      & RF                         & 1.156                        & 0.250                     & 0.270                    & 0.322                    & 0.456                   & 0.477                  & 0.495                  & 0.517                   \\
  GE                        & 0.769                        & 0.151                      & 0.164                     & 0.190                     & 0.241                    & 0.264                    & 0.321                  & 0.298                   &                      & SCHW                       & 1.210                        & 0.221                     & 0.258                    & 0.276                    & 0.375                   & 0.350                  & 0.418                  & 0.347                   \\
  GILD                      & 0.996                        & 0.210                      & 0.242                     & 0.247                     & 0.505                    & 0.304                    & 0.371                  & 0.410                   &                      & T                          & 0.498                        & 0.120                     & 0.128                    & 0.136                    & 0.187                   & 0.213                  & 0.246                  & 0.201                   \\
  GLW                       & 0.958                        & 0.197                      & 0.212                     & 0.249                     & 0.330                    & 0.354                    & 0.430                  & 0.371                   &                      & VZ                         & 0.520                        & 0.129                     & 0.138                    & 0.144                    & 0.162                   & 0.168                  & 0.215                  & 0.182                   \\
  HAL                       & 1.195                        & 0.238                      & 0.264                     & 0.281                     & 0.365                    & 0.375                    & 0.451                  & 0.485                   &                      & WFC                        & 0.876                        & 0.151                     & 0.173                    & 0.188                    & 0.276                   & 0.259                  & 0.326                  & 0.337                   \\
  HBAN                      & 0.957                        & 0.218                      & 0.236                     & 0.286                     & 0.471                    & 0.509                    & 0.475                  & 0.483                   &                      & WMB                        & 1.073                        & 0.257                     & 0.277                    & 0.301                    & 0.351                   & 0.336                  & 0.553                  & 0.568                   \\
  HPQ                       & 1.050                        & 0.226                      & 0.252                     & 0.262                     & 0.304                    & 0.278                    & 0.375                  & 0.330                   &                      & WMT                        & 0.483                        & 0.113                     & 0.126                    & 0.127                    & 0.176                   & 0.159                  & 0.224                  & 0.166                   \\
  HST                       & 0.923                        & 0.198                      & 0.211                     & 0.245                     & 0.404                    & 0.411                    & 0.530                  & 0.514                   &                      & XOM                        & 0.778                        & 0.136                     & 0.159                    & 0.163                    & 0.191                   & 0.190                  & 0.226                  & 0.217                   \\
  INTC                      & 0.895                        & 0.163                      & 0.186                     & 0.197                     & 0.228                    & 0.257                    & 0.297                  & 0.246                   &                      & XRX                        & 0.919                        & 0.235                     & 0.242                    & 0.289                    & 0.509                   & 0.537                  & 0.566                  & 0.447                   \\
  \hline
  \end{tabular}
  }
  \end{table}
\end{landscape}
}

To evaluate the out-of-sample performance of predicting future integrated beta, we computed the mean absolute prediction error (MAPE)  as follows: 
\begin{equation*}
 \frac{1}{n-500}\sum^n_{i=501}\left|\hat{Beta}_{i}-RIB_i \right|,
\end{equation*}
where $\hat{Beta}_{i}$ denotes the one-day-ahead forecasted beta from parametric models such as DR Beta, ARMAC, ARMAP, PRBG, RBG, DCC, and BEKK, as defined in Section \ref{sec-4}, using 500 in-sample observations.
Unlike in a simulation study where the true integrated beta is known, it is impossible to obtain the true integrated beta in the empirical study.
Therefore, we need to use the proxy of the true integrated beta when calculating MAPE.
Since, to the best of our knowledge,  the proposed $RIB$ estimator is the only consistent estimator of integrated beta that accounts for the existing empirical feature of price observations, the autocorrelated microstructure noise \citep{jacod2017statistical,li2022remedi}, we employed the $RIB$ estimator as the proxy of the true integrated beta.
It is worth noting that  using the $RIB$ estimator as a proxy for the target integrated beta in out-of-sample performance evaluations could potentially introduce a bias favoring the proposed DR Beta model.
For each stock, we used the selected $(p,q)$ order for the DR Beta, ARMAC, and ARMAP models. %
In the case of ARMAC, we also checked their performance with input integrated betas estimated by CHEN with data subsampled at 1, 5, 10, 30, and 60-second frequencies to deal with the autocorrelated microstructure noise.
Then, we reported the best performance results among the different frequencies.
For RBG, we used realized covariance, the sum of squared log-returns, with 5-min, 1-min, and 30-sec data ($m=78, 390,780$, respectively) to handle the microstructure noise, and reported the best results among them. 
The in-sample period is 500 days, and we estimated the models using the rolling window scheme. 
Table \ref{table-3} reports the mean of $RIB$ estimates in the out-of-sample period and the MAPEs for DR Beta, ARMAC, ARMAP, PRBG, RBG, DCC, and BEKK for 50 stocks.
From Table \ref{table-3}, we find that the models using high-frequency information show better performance than the models using only low-frequency information.
Further, the ARMA-type models utilizing realized betas usually perform better than the RBG and PRBG models.
When comparing the ARMA-type models using realized betas, MAPEs for the proposed DR Beta or ARMAC have the smallest values for every stock,
and DR Beta always shows the lowest MAPE among the benchmarks.
It may be because the proposed DR Beta and ARMAC model can account for the time-varying beta by incorporating high-frequency data.
These results indicate that accommodating the time-varying beta feature helps account for the beta dynamics, and the DR Beta holds advantages in predicting future integrated beta by utilizing the autoregressive structure with consistent $RIB$ estimates.

We evaluated how well the proposed methodologies capture the autoregressive structure.
Adopting the idea of the Durbin-Watson test, we took into account regression residuals between the non-parametric and out-of-sample predicted values using DR Beta, ARMAC, ARMAP, PRBG, RBG, DCC, and BEKK. 
Specifically, for each model, we fitted the following linear regression model:
\begin{equation*}
    RIB_i = a+b \times  \hat{Beta} _{i} + e_i , 
\end{equation*}
where the $\hat{Beta}_i$'s are one-day-ahead forecasted betas obtained using one of the DR Beta, ARMAC, ARMAP, PRBG, RBG, DCC, and BEKK.
Then, we calculated the regression residuals for each model and checked their autocorrelations.

\begin{figure}[t] 
\centering
\includegraphics[width =.98\textwidth]{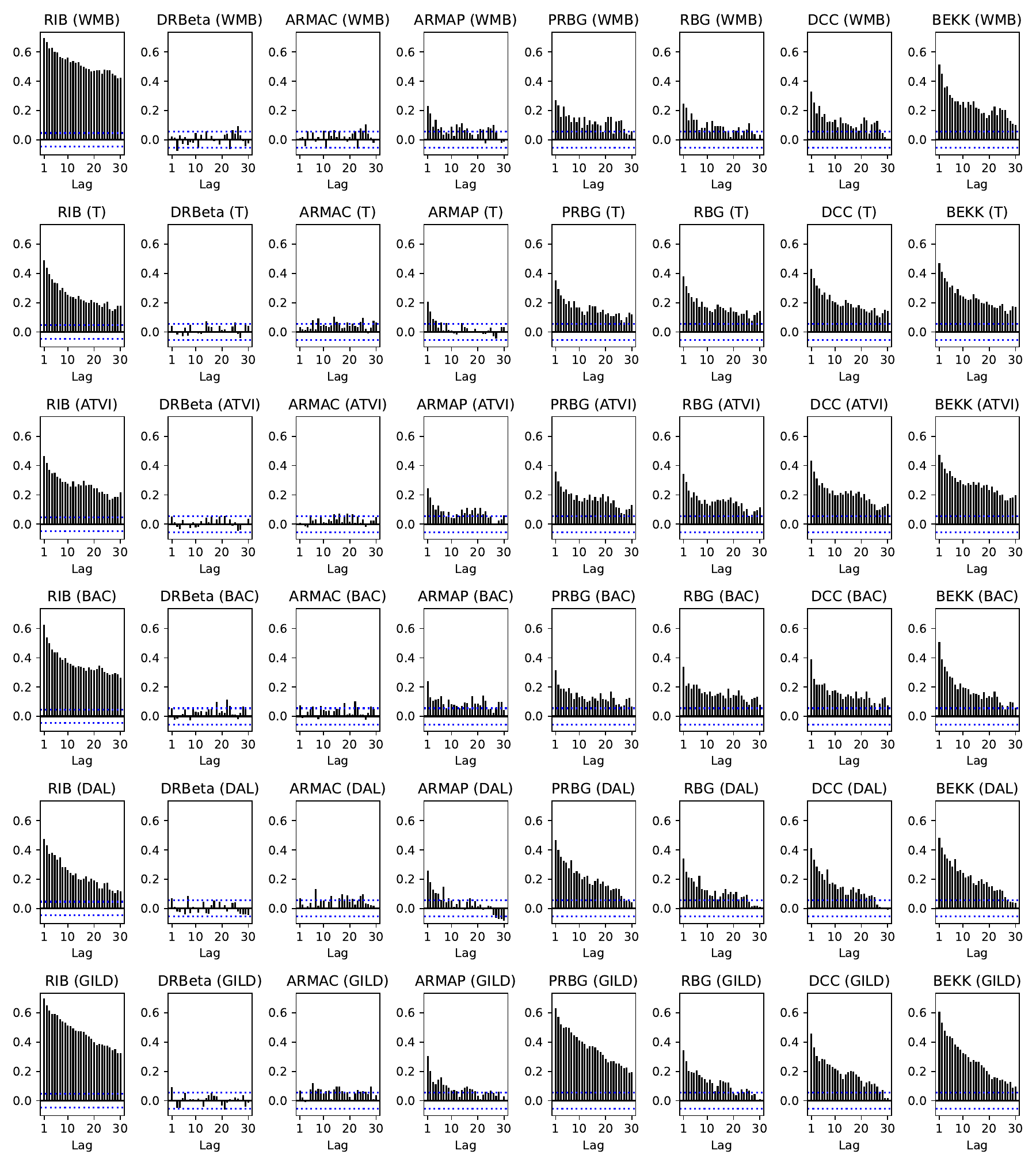} 
\caption{ACF plots for the non-parametric integrated beta, $RIB$, and the regression residuals between $RIB$ and the predicted integrated beta from DR Beta, ARMAC, ARMAP, PRBG, RBG, DCC, and BEKK for six stocks, which have the smallest, $20$th, $40$th, $60$th, $80$th, and largest first-order autocorrelations among the 50 assets in order. }
\label{Figure-7}
\end{figure}

\begin{figure}[t] 
\centering
\includegraphics[width = 0.9\textwidth]{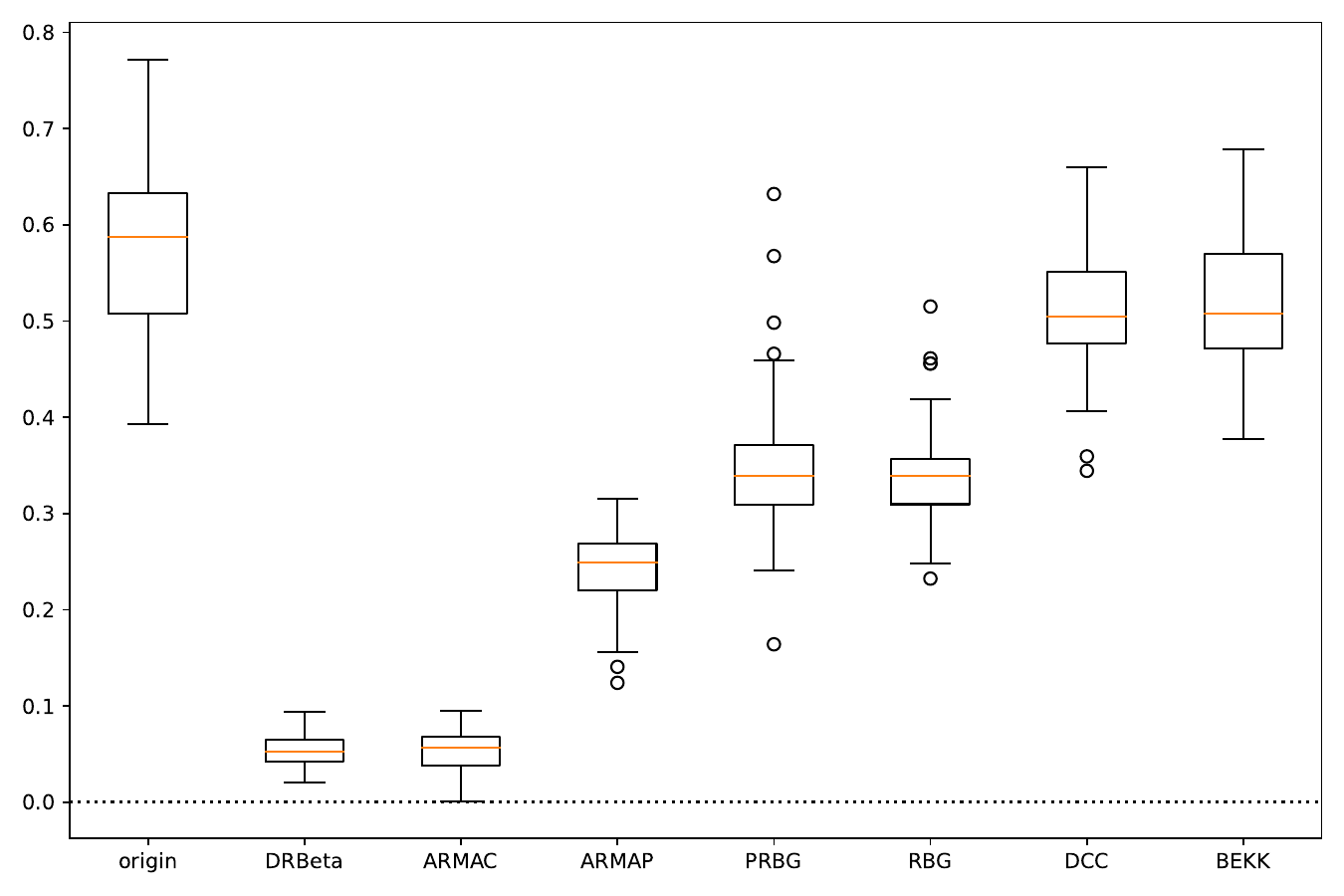} 
\caption{The box plots of the first-order autocorrelations for the regression residuals between the non-parametric integrated beta and the predicted integrated beta from DR Beta, ARMAC, ARMAP, PRBG, RBG, DCC, and BEKK. 
The origin is the first-order autocorrelations of the non-parametric integrated beta.}
\label{Figure-6}
\end{figure}

\afterpage{
\begin{landscape}
  \begin{table}[t]
  \caption{The first-order autocorrelations for the regression residuals between the non-parametric integrated beta estimates, $RIB$, and the predicted integrated beta from DR Beta, ARMAC, ARMAP, PRBG, RBG, DCC, and BEKK. 
  }\label{table-4}
  \centering
  \scalebox{0.85}{
  \setlength\tabcolsep{2.5pt}
  \begin{tabular}{lrrrrrrrrllrrrrrrrr}
    \hline
  \multicolumn{1}{c}{Stock} & \multicolumn{1}{c}{$RIB$} & \multicolumn{1}{c}{DRBeta} & \multicolumn{1}{c}{ARMAC} & \multicolumn{1}{c}{ARMAP} & \multicolumn{1}{c}{PRBG} & \multicolumn{1}{c}{RBG} & \multicolumn{1}{c}{DCC} & \multicolumn{1}{c}{BEKK} & \multicolumn{1}{c}{}  & \multicolumn{1}{c}{Stock}  & \multicolumn{1}{c}{$RIB$} & \multicolumn{1}{c}{DRBeta} & \multicolumn{1}{c}{ARMAC} & \multicolumn{1}{c}{ARMAP} & \multicolumn{1}{c}{PRBG} & \multicolumn{1}{c}{RBG} & \multicolumn{1}{c}{DCC} & \multicolumn{1}{c}{BEKK} \\ \cline{1-9} \cline{11-19} 
  AAPL                      & 0.583                     & 0.048                     & 0.095                    & 0.173                    & 0.259                   & 0.292                  & 0.512                  & 0.571                   &                              & JPM                        & 0.548                     & 0.036                     & 0.045                    & 0.262                    & 0.319                   & 0.319                  & 0.497                  & 0.513                   \\
  AIG                       & 0.620                     & 0.056                     & 0.054                    & 0.217                    & 0.241                   & 0.232                  & 0.485                  & 0.472                   &                              & KEY                        & 0.554                     & 0.062                     & 0.079                    & 0.264                    & 0.335                   & 0.412                  & 0.552                  & 0.556                   \\
  AMAT                      & 0.415                     & 0.047                     & 0.059                    & 0.267                    & 0.347                   & 0.344                  & 0.406                  & 0.415                   &                              & KO                         & 0.598                     & 0.050                     & 0.061                    & 0.206                    & 0.309                   & 0.348                  & 0.598                  & 0.598                   \\
  AMD                       & 0.578                     & 0.052                     & 0.070                    & 0.124                    & 0.331                   & 0.456                  & 0.517                  & 0.571                   &                              & MGM                        & 0.584                     & 0.086                     & 0.065                    & 0.276                    & 0.386                   & 0.297                  & 0.583                  & 0.570                   \\
  ATVI                      & 0.476                     & 0.051                     & 0.007                    & 0.245                    & 0.361                   & 0.345                  & 0.473                  & 0.428                   &                              & MRK                        & 0.574                     & 0.049                     & 0.024                    & 0.252                    & 0.568                   & 0.299                  & 0.537                  & 0.513                   \\
  BAC                       & 0.644                     & 0.057                     & 0.075                    & 0.240                    & 0.315                   & 0.339                  & 0.508                  & 0.615                   &                              & MRO                        & 0.777                     & 0.080                     & 0.095                    & 0.266                    & 0.283                   & 0.339                  & 0.600                  & 0.578                   \\
  BMY                       & 0.610                     & 0.038                     & 0.001                    & 0.213                    & 0.317                   & 0.305                  & 0.569                  & 0.519                   &                              & MS                         & 0.628                     & 0.085                     & 0.076                    & 0.280                    & 0.331                   & 0.330                  & 0.507                  & 0.623                   \\
  BSX                       & 0.498                     & 0.063                     & 0.068                    & 0.237                    & 0.459                   & 0.461                  & 0.476                  & 0.487                   &                              & MSFT                       & 0.474                     & 0.038                     & 0.039                    & 0.259                    & 0.327                   & 0.320                  & 0.473                  & 0.428                   \\
  CSCO                      & 0.466                     & 0.061                     & 0.079                    & 0.255                    & 0.340                   & 0.392                  & 0.466                  & 0.457                   &                              & MU                         & 0.484                     & 0.054                     & 0.061                    & 0.244                    & 0.312                   & 0.294                  & 0.469                  & 0.484                   \\
  CSX                       & 0.535                     & 0.080                     & 0.064                    & 0.290                    & 0.341                   & 0.347                  & 0.530                  & 0.525                   &                              & NEM                        & 0.689                     & 0.057                     & 0.062                    & 0.141                    & 0.301                   & 0.398                  & 0.660                  & 0.679                   \\
  DAL                       & 0.485                     & 0.070                     & 0.068                    & 0.257                    & 0.466                   & 0.341                  & 0.483                  & 0.471                   &                              & NFLX                       & 0.544                     & 0.032                     & 0.036                    & 0.213                    & 0.327                   & 0.321                  & 0.542                  & 0.543                   \\
  DIS                       & 0.472                     & 0.040                     & 0.021                    & 0.272                    & 0.345                   & 0.337                  & 0.470                  & 0.465                   &                              & NVDA                       & 0.544                     & 0.077                     & 0.084                    & 0.257                    & 0.301                   & 0.293                  & 0.503                  & 0.493                   \\
  DOW                       & 0.558                     & 0.040                     & 0.015                    & 0.228                    & 0.268                   & 0.280                  & 0.478                  & 0.495                   &                              & ORCL                       & 0.503                     & 0.046                     & 0.028                    & 0.239                    & 0.351                   & 0.274                  & 0.478                  & 0.481                   \\
  EBAY                      & 0.514                     & 0.080                     & 0.032                    & 0.228                    & 0.278                   & 0.312                  & 0.500                  & 0.493                   &                              & PFE                        & 0.470                     & 0.033                     & 0.034                    & 0.156                    & 0.270                   & 0.305                  & 0.457                  & 0.455                   \\
  F                         & 0.499                     & 0.052                     & 0.037                    & 0.258                    & 0.354                   & 0.354                  & 0.487                  & 0.461                   &                              & PG                         & 0.615                     & 0.043                     & 0.061                    & 0.196                    & 0.378                   & 0.322                  & 0.597                  & 0.609                   \\
  FCX                       & 0.753                     & 0.026                     & 0.049                    & 0.220                    & 0.164                   & 0.256                  & 0.494                  & 0.549                   &                              & QCOM                       & 0.590                     & 0.042                     & 0.042                    & 0.270                    & 0.306                   & 0.310                  & 0.570                  & 0.563                   \\
  FITB                      & 0.603                     & 0.053                     & 0.080                    & 0.289                    & 0.375                   & 0.418                  & 0.559                  & 0.593                   &                              & RF                         & 0.571                     & 0.064                     & 0.092                    & 0.270                    & 0.448                   & 0.456                  & 0.556                  & 0.570                   \\
  GE                        & 0.474                     & 0.056                     & 0.037                    & 0.264                    & 0.334                   & 0.350                  & 0.475                  & 0.462                   &                              & SCHW                       & 0.508                     & 0.040                     & 0.045                    & 0.269                    & 0.371                   & 0.315                  & 0.508                  & 0.503                   \\
  GILD                      & 0.661                     & 0.094                     & 0.071                    & 0.306                    & 0.632                   & 0.343                  & 0.610                  & 0.643                   &                              & T                          & 0.470                     & 0.040                     & 0.034                    & 0.207                    & 0.354                   & 0.379                  & 0.470                  & 0.470                   \\
  GLW                       & 0.391                     & 0.034                     & 0.047                    & 0.222                    & 0.338                   & 0.345                  & 0.359                  & 0.383                   &                              & VZ                         & 0.480                     & 0.077                     & 0.069                    & 0.224                    & 0.301                   & 0.330                  & 0.480                  & 0.479                   \\
  HAL                       & 0.616                     & 0.066                     & 0.066                    & 0.269                    & 0.346                   & 0.338                  & 0.587                  & 0.593                   &                              & WFC                        & 0.579                     & 0.076                     & 0.081                    & 0.285                    & 0.376                   & 0.358                  & 0.515                  & 0.524                   \\
  HBAN                      & 0.547                     & 0.047                     & 0.049                    & 0.246                    & 0.498                   & 0.515                  & 0.546                  & 0.516                   &                              & WMB                        & 0.715                     & 0.021                     & 0.010                    & 0.230                    & 0.272                   & 0.248                  & 0.516                  & 0.477                   \\
  HPQ                       & 0.454                     & 0.077                     & 0.052                    & 0.270                    & 0.316                   & 0.309                  & 0.447                  & 0.445                   &                              & WMT                        & 0.514                     & 0.051                     & 0.061                    & 0.198                    & 0.379                   & 0.353                  & 0.514                  & 0.513                   \\
  HST                       & 0.517                     & 0.065                     & 0.053                    & 0.245                    & 0.342                   & 0.378                  & 0.486                  & 0.501                   &                              & XOM                        & 0.603                     & 0.064                     & 0.068                    & 0.315                    & 0.379                   & 0.348                  & 0.598                  & 0.603                   \\
  INTC                      & 0.499                     & 0.051                     & 0.055                    & 0.280                    & 0.371                   & 0.416                  & 0.499                  & 0.498                   &                              & XRX                        & 0.379                     & 0.061                     & 0.046                    & 0.196                    & 0.357                   & 0.364                  & 0.344                  & 0.377                   \\
  \hline
  \end{tabular}
  }
  \end{table}
\end{landscape}
}

Figure \ref{Figure-7} shows the ACF plots for $RIB$ and the models' regression residuals for six stocks, which have the smallest, $20$th, $40$th, $60$th, $80$th, and the largest first-order autocorrelations among the 50 stocks.
Figure \ref{Figure-6} depicts the box plot of the first-order autocorrelations of the regression residuals for each model, and Table \ref{table-4} reports their numerical values. 
For ARMAC, only the result of the case with the lowest first-order autocorrelations of the regression residuals among the different sample frequencies is reported for each stock.
From Table \ref{table-4} and Figures  \ref{Figure-7} and \ref{Figure-6}, we find that the proposed DR Beta and ARMAC models have much smaller autocorrelations for most of the stocks, but the other models still yield significantly non-zero autocorrelations for most of the stocks.
This may be because the other competitors could not appropriately estimate the integrated beta due to the time-varying beta feature.
When comparing the DR Beta and ARMAC models, the DR Beta model usually has smaller autocorrelation than the ARMAC model.
Specifically, for 28 stocks out of 50, DR Beta shows the best performance among the benchmarks.
One of the possible explanations is that the CHEN estimator, which is used in the ARMAC model as the non-parametric beta estimator, cannot handle the autocorrelation structure of the microstructure noise; thus, some autocorrelation may remain in the regression residuals.
From these numerical results, we can conjecture that incorporating the stylized features, such as the time-varying beta and the autocorrelation structure of the microstructure noise, helps account for the integrated beta dynamics. 
Thus, the proposed DR Beta model can explain the integrated beta dynamics well by incorporating the proposed robust realized integrated beta estimator.

\section{Sketch of proof}\label{sec-proof}
In this section, we provide sketches of proof for the main theorems.
We show Theorems \ref{Theorem-1}, \ref{Theorem-2}, and \ref{Theorem-3} using the ideas in \citet{chen2018inference,jacod2019estimating} and \citet{kim2016unified}, respectively.
Let $C$ be a generic constant whose values are free of  $n$ and $m$.
We denote the matrix differentiation $\partial_{jk}f(A) = \partial f(A)/ \partial A_{jk}$ for any $2 \times 2$ matrix $A$ and generic differentiable function $f$ defined on the $2 \times 2$ matrix space.
In addition, we define $\mathbf{1}_{\{statement\}}$ as follows:
\begin{equation*}
    \mathbf{1}_{\{statement\}}=\begin{cases}1,\quad \text{if the statement is true} \\ 0, \quad \text{otherwise.}\end{cases}
\end{equation*}
We use generic random variables $\varPsi_{par}^{m,w}$, depending on $m$ and parameters ``$par$'', nonnegative, $\mathcal{G}$-measurable, and satisfying $\mathbb{E}\left[ \left( \varPsi_{par}^{m,w} \right)^w  \right] \leq 1$, where $\mathcal{G}=\mathcal{G}_{\infty}=\mathcal{G}^{\infty}$.
Similarly, we use generic generic random variables $\varPsi_{par}^{m}$, depending on $m$ and parameters ``$par$'', nonnegative, $\mathcal{G}$-measurable, but satisfying $\mathbb{E}\left[ \left( \varPsi_{par}^{m} \right)^w  \right] \leq C_w$ for any $w > 0$.
We also use $O_u(x)$ for a random quantity smaller than $Cx$ for some constant $C$.

\subsection{Proof of Theorem \ref{Theorem-1}}\label{pf-thm1}
Note that the spot covariance matrix $\bSigma_t$ of $(X_1^c, X_2^c)^\top$ can be written as 
\begin{equation*}
    \bSigma_t
    =\begin{pmatrix} \sigma^2_t & \beta^c_t \sigma^2_t \\ \beta^c_t \sigma^2_t & (\beta^c_t)^2 \sigma^2_t+q_t^2 \end{pmatrix},
\end{equation*}
for all $t \in \mathbb{R}^+$.
Moreover, similar to \eqref{Equation-3.3}, we can construct an estimator for $\bSigma(t)$ as follows:
\begin{eqnarray*}
    \hat{\bSigma}^c(t)
= \begin{pmatrix} v^{c}(Y_1,Y_1,t) & v^{c}(Y_1,Y_2,t)  \\
  v^{c}(Y_2,Y_1,t) & v^{c}(Y_2,Y_2,t)  \end{pmatrix} ,
\end{eqnarray*}
where
\begin{equation*}
  v^{c}(P,P', t_l) = 
  \frac{1}{(b_m - 2 k_m) \Delta_m k_m \psi_0} 
  \Bigg\lbrace  \sum_{i=0}^{b_m - 2 k_m -1} \tilde{P}_{l+i}^{c,m}  \tilde{P}_{l+i}^{'c,m}   \\
  - \frac{1}{k_m}  \sum_{i=0}^{b_m - 6 l_m}  \sum_{d=-k_m'}^{k_m'} \phi_{d}^{m}  \mathcal{E}_{PP',l+i}^{m,d}  \Bigg\rbrace,
\end{equation*}
and continuous processes $Y_1^c$ and $Y_2^c$ satisfy $Y_{1,t}^{c,m}=X^c_{1,t}+\epsilon_{1,i}^{m}$ and  $Y_{2,t}^{c,m}=X^c_{2,t}+\epsilon_{2,i}^{m}$.
Note that $Y_{1,t}=Y^c_{1,t}+X^d_{1,t}$, $Y_{2,t}=Y^c_{2,t}+X^d_{2,t}$. 

Define $\Xi (v, \mu) : \mathbb{R}^{2\times 2} \times \mathbb{R}^{2\times 2} \rightarrow \mathbb{R}^{2\times 2}$ such that for any $x,y \in \left\lbrace 1,2 \right\rbrace$,
\begin{eqnarray*}
  \Xi (v, \mu)_{x,y} &=&   \frac{2}{\psi_0^2 {C_k}^3} \big[ {C_k}^4 \Phi_{00} \left( v_{11} v_{xy} + v_{1x} v_{1y} \right)  + {C_k}^2 \Phi_{01} \left( v_{11} \mu_{xy} + v_{1x} \mu_{1y} + v_{1y} \mu_{1x} + v_{xy} \mu_{11}  \right) \\
  &&  + \Phi_{11} \left( \mu_{11} \mu_{xy} + \mu_{1x} \mu_{1y} \right)   \big]
  ,
\end{eqnarray*}
and for any matrix $A^{m} \in \mathbb{R}^{2\times 2}$,
\begin{equation*}
  A^{m,*} =
  \begin{pmatrix}
    \max (A_{11}^{m}, \delta_m) & A_{12}^{m} \\
    A_{21}^{m} & A_{22}^{m}
  \end{pmatrix}.
\end{equation*}
Then, we obtain
\begin{equation*}
  \hat{B}^{m}_{ib_m} = \frac{1}{2 b_m \Delta_m ^{1/2}} \sum_{x,y=1}^{2} \Xi (\hat{\bSigma}_{ib_m}^{m,*} , \hat{\bvartheta}_{i b_m }^{m} )_{x,y} \partial^2_{1x,1y}f(\hat{\bSigma}_{ib_m}^{m,*})  ,
\end{equation*}
where $f(\bc) = (c_{11})^{-1} c_{12}$.
Furthermore, let
\begin{eqnarray}\label{eq:def-B}
  && \hat{B}^{c,m}_{ib_m} = \frac{1}{2 b_m \Delta_m ^{1/2}} \sum_{x,y=1}^{2}  \Xi (\hat{\bSigma}_{ib_m}^{c,m,*} , \hat{\bvartheta}_{i b_m }^{m} )_{x,y} \partial^2_{1x,1y}f(\hat{\bSigma}_{ib_m}^{c,m,*}) , \cr
  && {B}^{c,m}_{ib_m} = \frac{1}{2 b_m \Delta_m ^{1/2}} \sum_{x,y=1}^{2}  \Xi ({\bSigma}_{ib_m}^{m} , {\bvartheta}_{i b_m }^{m} )_{x,y} \partial^2_{1x,1y}f({\bSigma}_{ib_m}^{m}) ,
\end{eqnarray}
and $\bvartheta_{xy,i}^{m}  = \vartheta_{x,i}^{m} \vartheta_{y,i}^{m} R_{x y}  $ for any $x, y \in \left\lbrace 1,2 \right\rbrace$.

Lemma 4.4.9 in \citet{jacod2012discretization} indicates that if the asymptotic result, such as convergence in probability or stable convergence in law, is satisfied under the boundedness condition, it is also satisfied under the local boundedness condition.
Thus, without loss of generality, we assume that the drift, spot volatility, and its inverse processes are bounded in the following proofs.

\subsubsection{Properties of spot volatility: Continuous part}

We first show some properties of spot volatility estimator $\hat{\bSigma}^{c,m}_{xy,i}$ that can be proved similarly to the one in \citet{jacod2009microstructure, jacod2019estimating}.
We introduce some notations to follow the ``big blocks and small blocks"-technique \citep{jacod2009microstructure}.
For $p,i \in \mathbb{N}$ and $x,y \in \left\lbrace 1,2 \right\rbrace$, we define
\begin{eqnarray*}\label{block-notation}
  &&C_{xy,t} =  \int _{0}^{t} \bSigma_{xy,s} ds, \qquad \breve{C}_{xy,i}^{m} = \sum_{l=1}^{k_m-1} (g _{l}^{m})^2 \left( C_{xy,i+l}^{m} - C_{xy,i+l-1}^{m} \right) , \\
  &&\tilde{\Gamma}^{m}_{xy,i} = \vartheta^{m}_{x,i} \vartheta^{m}_{y,i} \sum_{l_1,l_2 = 0}^{k_m-1} r_{xy}(l_1,l_2) h_{l_1}^m h_{l_2}^m ,\qquad %
  r_{xy}(l_1,l_2) = \mathbb{E}\left[ \chi_{x,l_1} \chi_{y,l_2} \right], \qquad h_{i}^{m} = g_{i+1}^{m} - g_{i}^{m}, \cr
  &&\zeta_{xy,i}^{m} = \tilde{Y}_{x,i}^{c} \tilde{Y}_{y,i}^{c} - \breve{C}_{xy,i}^{m} - \tilde{\Gamma}^{m}_{xy,i}, \qquad
\zeta(p)_{xy,i}^{m} = \sum_{l=i}^{i+pk_m-1} \zeta _{xy,l}^{m}. 
\end{eqnarray*}
The estimation error of spot volatility, $\hat{\bSigma}_{xy,i}^{c,m} - \bSigma_{xy,i}^{m}$ can be decomposed as follows:
\begin{equation}\label{SpotErrorDecomp}
  e_{xy,i}^{m} = \hat{\bSigma}_{xy,i}^{c,m} - \bSigma_{xy,i}^{m} = M(p)_{xy,i}^{m} + M'(p)_{xy,i}^{m} + \xi_{xy,i}^{m,1} + \xi_{xy,i}^{m,2},
\end{equation}
where
\begin{align}\label{details}
  &L(m,p) = \left[ \frac{b_m - 2 k_m}{(p+2)k_m} \right] , \quad \mathcal{K}_{i}^{m} = \mathcal{F}_{i}^{m} \otimes \mathcal{G}_{i-k_m}, \cr
  & \mathcal{H}(p)_{j}^{m,i} = \mathcal{K}_{i+j(p+2)k_m}^{m} ,\quad \mathcal{H}'(p)_{j}^{m,i} = \mathcal{K}_{i+j(p+2)k_m+p}^{m},  \cr
  &\eta(p)_{xy,j}^{m,i} =  \frac{1}{(b_m - 2 k_m)\Delta_m k_m \psi_0} \zeta(p)_{xy,i+j(p+2)k_m}^{m}, \quad \bar{\eta}(p)_{xy,j}^{m,i} = \mathbb{E}\left[ \eta(p)_{xy,j}^{m,i} | \mathcal{H}(p)_{j}^{m,i} \right], \cr
  &\eta'(p)_{xy,j}^{m,i} =  \frac{1}{(b_m - 2 k_m)\Delta_m k_m \psi_0} \zeta(2)_{xy,i+j(p+2)k_m+p}^{m}, \quad \bar{\eta}'(p)_{xy,j}^{m,i} = \mathbb{E}\left[ \eta'(p)_{xy,j}^{m,i} | \mathcal{H}'(p)_{j}^{m,i} \right], \cr
  &\hat{\eta}(p)_{xy,j}^{m,i} = \eta(p)_{xy,j}^{m,i} - \bar{\eta}(p)_{xy,j}^{m,i}, \quad M(p)_{xy,i}^{m} =   \sum_{j=0}^{L(m,p)-1} {\eta}(p)_{xy,j}^{m,i} ,\cr%
  &\hat{\eta}'(p)_{xy,j}^{m,i} = \eta'(p)_{xy,j}^{m,i} - \bar{\eta}'(p)_{xy,j}^{m,i}, \quad M'(p)_{xy,i}^{m} =   \sum_{j=0}^{L(m,p)-1} {\eta}'(p)_{xy,j}^{m,i} , \cr
  &\xi_{xy,i}^{m,1} = \frac{1}{(b_m - 2 k_m) \Delta_m k_m \psi_0} \sum_{l=0}^{b_m- 2 k_m - 1}  \breve{C}_{xy,i+l}^{m} -\bSigma_{xy,i}^{m}, \cr
  &\xi_{xy,i}^{m,2} = \frac{1}{(b_m - 2 k_m) \Delta_m k_m \psi_0}  \sum_{l=1}^{b_m - 2 k_m} \tilde{\Gamma}_{xy,i+l}^{m} - \frac{1}{(b_m - 2 k_m) \Delta_m k_m^2 \psi_0}  \sum_{d=-k_m'}^{k_m'} \phi_{d}^{m}  U_{m,i}^{Y_x Y_y} (|d|), \cr
  & U_{m,i}^{Y_x Y_y} (|d|) = \sum_{l=0}^{b_m - 6 l_m} \mathcal{E}_{Y_x Y_y,i+l}^{m,d}
  .
\end{align}
In the decomposition of \eqref{SpotErrorDecomp}, the leading term is $M(p)_{xy,i}^{m} + M'(p)_{xy,i}^{m}$ and the others are residual terms.

We first prove that the terms $\xi_{xy,i}^{m,1}$ and $\xi_{xy,i}^{m,2}$ in \eqref{SpotErrorDecomp} are negligible once multiplied by the rate $\Delta_m ^{-1/4}$.
\begin{lemma}\label{negligible-xi}
  Under Assumption 1, we have for any $x,y \in \left\lbrace 1,2 \right\rbrace$
  \begin{enumerate}
    \item[(a)] $\left| \mathbb{E} \left[ \xi_{xy,i}^{m,1} | \mathcal{F}_{i}^{m} \right] \right| \leq C  b_m \Delta_m$ and $ \mathbb{E}\[ | \xi_{xy,i}^{m,1} |^q  | \mathcal{F}_{i}^{m} \]  \leq C_q (b_m \Delta_m)^{(q/2) \wedge 1} \text{ a.s.}$ for any $q \in \mathbb{N}$;
    \item[(b)] $\left| \mathbb{E}\[ \xi_{xy,i}^{m,2} | \mathcal{K}_{i}^{m} \]  \right| \leq C  (k_m'^{-(v-1)} + \varPsi_{i}^{m,4}  k'_m l_m ^{-(v+\frac{1}{2} )}) $ and \\
    $ \mathbb{E}[ \left| \xi_{xy,i}^{m,2} \right|^{w} | \mathcal{K}_{i}^{m} ]  \leq C_{w,\varepsilon} ({k'_m}^{-w(v-1)} + \varPsi_{i}^{m,2} \Delta_m ^{\frac{(\kappa-\varsigma-2\tau)w}{2}  \land (\varsigma v - \tau w - \varepsilon)})\text{ a.s.}$ for any $w\in \mathbb{N}$ and $\varepsilon > 0$.
  \end{enumerate}
\end{lemma}

Now, we provide estimates on various moments of the variables $\zeta(p)_{1x,i}^{m}$, $M(p)_{1x,i}^{m}$, $M'(p)_{1x,i}^{m}$, and $e_{1x,i}^{m}$ in below lemmas.
\begin{lemma}\label{lemma:zeta}
  Under Assumption 1, we have for any $x,y \in \left\lbrace 1, 2 \right\rbrace$, $p \in \mathbb{N} \cap [2,b_m /k_m -2]$, $w \in \mathbb{N}$, and $\varepsilon > 0$,
  \begin{eqnarray}\label{zeta-moment}
    && \mathbb{E}\left[ \left| \zeta(p)_{1x,i}^{m} \right|^w | \mathcal{K}_{i}^{m}  \right] \leq C_{w} \varPsi_{i}^{m,2}  p^w \Delta_m ^{\frac{v-w}{2} \land 0} ,\cr
    && \left|\mathbb{E}\left[ \zeta(p)_{1x,i}^{m} | \mathcal{K}_{i}^{m} \right]\right|^{w} \leq C_{w,\varepsilon} \varPsi_{i}^{m,2} p^{w} \Delta_m ^{\frac{(v-w -\varepsilon) \land w}{2} } ,\cr
    && \left| \mathbb{E}\left[  \zeta(p)_{1x,i}^{m} \zeta(p)_{1y,i}^{m}    - \varXi(p)_{x,y,i}^{m} \Big| \mathcal{K}_{i}^{m}  \right] \right| \leq C p^2 \varPsi_{i}^{m,2} \Delta_m^{1/4} \text{ a.s.}
    ,
  \end{eqnarray}
  where $\varXi(p)_{x,y,i}^{m} = 2 (p\Phi_{00}-\bar{\Phi}_{00})  (\bSigma_{11,i}^{m} \bSigma_{xy,i}^{m} + \bSigma_{1x,i}^{m} \bSigma_{1y,i}^{m} )      k_m^4 \Delta_m^2  + 2 (p\Phi_{01}-\bar{\Phi}_{01}) (\bSigma_{11,i}^{m} \bvartheta_{xy,i}^{m} + \bSigma_{1x,i}^{c,m} \bvartheta_{1y,i}^{m} + \bSigma_{1y,i}^{m} \bvartheta_{1x,i}^{m} + \bSigma_{xy,i}^{m} \bvartheta_{11,i}^{m}  ) k_m^2 \Delta_m  + 2 \left( p \Phi_{11} - \bar{\Phi}_{11} \right) (\bvartheta_{11,i}^{m} \bvartheta_{xy,i}^{m} + \bvartheta_{1x,i}^{m} \bvartheta_{1y,i}^{m})$.
  Furthermore, if $p_1, p_2 \in \mathbb{N} \cap [2,b_m /k_m -2]$ and $i_1 + (p_1 +2) k_m \leq i_2$, we have
  \begin{eqnarray}\label{zeta-p1p2}
    && \left|\mathbb{E}\left[ \zeta(p_1)_{1x,i_1}^{m} \zeta(p_2)_{1y,i_2}^{m} | \mathcal{K}_{i_1}^{m} \right]\right|  \leq C \varPsi_{i}^{m,2} p_{1}p_{2} \Delta_m ^{1/2}  \text{ a.s.} %
  \end{eqnarray}
\end{lemma}

\begin{lemma}\label{lemma:M}
  Under Assumption 1, we have for any $x \in \left\lbrace 1, 2 \right\rbrace$, $w \in \mathbb{N}$, $p \in \mathbb{N} \cap [2,b_m /k_m -2]$, and $\varepsilon > 0$,
  \begin{align}\label{M-moment}
    & \left|\mathbb{E}\left[ M(p)_{1x,i}^{m} | \mathcal{K}_{i}^{m}  \right]\right|^{w} \leq C_{w,\varepsilon} \varPsi_{i}^{m,2}  \Delta_m ^{\frac{(v-w-\varepsilon) \land w}{2} }, \quad \left|\mathbb{E}\left[ M'(p)_{1x,i}^{m} | \mathcal{K}_{i}^{m}  \right]\right|^{w} \leq C_{w,\varepsilon} \varPsi_{i}^{m,2} p^{-w}  \Delta_m ^{\frac{(v-w-\varepsilon) \land w}{2} }, \cr
    & \mathbb{E}\left[ \left|M(p)_{11,i}^{m}\right|^{w}  | \mathcal{K}_{i}^{m}  \right] \leq C_{w,\varepsilon} \varPsi_{i}^{m,2} (p^{w/2} b_m ^{-w/2} \Delta_m ^{(\frac{v}{2} - \frac{3}{4} w) \land -\frac{w}{4} } + \Delta_m ^{\frac{(v-w-\varepsilon) \land w}{2} }) , \cr
    & \mathbb{E}\left[ \left|M(p)_{11,i}^{m}\right|^{w}    \right] \leq C_{w,\varepsilon} (p^{w/2} b_m ^{-w/2} \Delta_m ^{(\frac{v}{2}  - \frac{3}{4} w) \land -\frac{w}{4} } + \Delta_m ^{ (v - \frac{w}{2} - \varepsilon) \land \frac{w}{2}  }) , \cr
    & \mathbb{E}\left[ \left|M'(p)_{11,i}^{m}\right|^{w}  | \mathcal{K}_{i}^{m}  \right] \leq C_{w,\varepsilon} \varPsi_{i}^{m,2} (p^{-w/2} b_m ^{-w/2} \Delta_m ^{(\frac{v}{2} - \frac{3}{4} w) \land -\frac{w}{4} } + \Delta_m ^{\frac{(v-w - \varepsilon) \land w}{2} })  ,\cr
    & \mathbb{E}\left[ \left|M'(p)_{11,i}^{m}\right|^{w}    \right] \leq C_{w,\varepsilon} (p^{-w/2} b_m ^{-w/2} \Delta_m ^{(\frac{v}{2}  - \frac{3}{4} w) \land -\frac{w}{4} } + \Delta_m ^{ (v - \frac{w}{2} - \varepsilon) \land \frac{w}{2}  })\quad \text{ a.s. }
  \end{align}
\end{lemma}

\begin{lemma}\label{lemma:e}
  Under Assumption 1, we have for any $x,y \in \left\lbrace 1, 2 \right\rbrace$ and $\varepsilon > 0$, we have
  \begin{align}\label{e-moment}
    & \left| \mathbb{E}\left[ e_{1x,i}^{m} | \mathcal{K}_{i}^{m}   \right] \right| \leq C \varPsi_{i}^{m,4}  b_m \Delta_m , \cr
    & \mathbb{E}\left[ \left| e_{1x,i}^{m} \right|^{w}  \right] \leq
    \begin{cases}
      C_{w} (b_m \Delta_m + b_m ^{-w/2} \Delta_m ^{-w/4}), & \text{ if } w \leq v , \\
      C_{w,\varepsilon} (b_m \Delta_m + b_m ^{-w/2} \Delta_m ^{\frac{v}{2}  - \frac{3}{4} w} + \Delta_m ^{(v-\frac{w}{2} - \varepsilon )}), & \text{ if } w > v , \\
    \end{cases}\cr
    & \mathbb{E}\left[ \left| e_{1x,i}^{m} \right|^{w} | \mathcal{K}_{i}^{m} \right] \leq C_{w,\varepsilon} \varPsi_{i}^{m,2} ( b_m \Delta_m  +  b_m ^{-w/2} \Delta_m ^{(\frac{v}{2} - \frac{3}{4} w) \land -\frac{w}{4} } + \Delta_m ^{\frac{(v-w-\varepsilon) \land w}{2} }) \quad \text{a.s.}
  \end{align}
  Furthermore, there exists $\varepsilon > 0 $ such that
  \begin{equation}\label{e-debiased}
    \left| \mathbb{E}\left[   e_{1x,i}^{m} e_{1y,i}^{m}    - b_m ^{-1} \Delta_{m}^{-1/2}  \Xi(\bSigma_{i}^{m}, \bvartheta_{i}^{m}  )_{x,y} \Big| \mathcal{K}_{i}^{m}  \right] \right| \leq C \varPsi_{i}^{m,2}  \Delta_m ^{\frac{1}{4} + \varepsilon} \text{ a.s.}
  \end{equation}
\end{lemma}

\subsubsection{Properties of spot volatility: Jump part}
In this subsection, we estimate moments of jump-related terms that come from $\hat{\bSigma}_{xy,i}^{m} - \hat{\bSigma}_{xy,i}^{c,m}$.
\begin{lemma}\label{lemma:jump-preavg-element}
  Under Assumption \ref{assumption-formal}, we have for any $x,y \in \{1,2\}$ and $1 \leq z < [v]$
  \begin{equation*}
    \mathbb{E}\left[ \left|\tilde{Y}_{x,i}^{m}\tilde{Y}_{y,i}^{m} \b1_{\lbrace |\tilde{Y}_{x,i}^{m}| \leq u_{x,m} , |\tilde{Y}_{x,i}^{m}| \leq u_{y,m} \rbrace} - \tilde{Y}_{x,i}^{c,m}\tilde{Y}_{y,i}^{c,m} \right|^{z} \big| \mathcal{K}_{i}^{m}  \right] \leq C \varPsi_{i}^{m,2}  \Delta_m ^{[v](\frac{1}{2} - \varpi_{1}) + \varpi_1 z}
  \end{equation*}
  and
  \begin{equation*}
    \mathbb{E}\left[ \left|\hat{\bSigma}_{xy,i}^{m} - \hat{\bSigma}_{xy,i}^{c,m}\right|^{z} \big| \mathcal{K}_{i}^{m} \right] \leq C \varPsi_{i}^{m,2}  \Delta_m ^{([v] -z)(\frac{1}{2} - \varpi_1)}
    .
  \end{equation*}
\end{lemma}

\begin{lemma}\label{spot-noise-estimate}
  Under Assumption \ref{assumption-formal}, for any $w \in \mathbb{N}$ and $x,y \in \left\lbrace 1,2 \right\rbrace$, we have almost surely
  \begin{equation*}
    \mathbb{E}\left[ \left( \hat{\bvartheta}_{11,i}^{m} - (\vartheta_{1,i}^{m})^2 R_{11} \right)^{w} \big| \mathcal{K}_{i}^{m} \right] \leq C_{w} \varPsi_{i}^{m,2}  ((k'_m)^{w} l_m^{w/2} b_m ^{-w/2} + (k'_m)^{-(v-1)w} + b_m \Delta_m )
    .
  \end{equation*}  
\end{lemma}

\subsubsection{A key decomposition}
To prove Theorem \ref{Theorem-1}, we decompose the estimation error of $RIB$.
Let $N_m = \left\lfloor 1/b_m \Delta_m  \right\rfloor$, $\beta_{i b_m }^{c,m}  = \bSigma^{c,m}_{12,i b_m } / \bSigma^{c,m}_{11,i b_m }$,  $\hat{\beta}^{c,m}_{i} = \hat{\bSigma}_{12,ib_m}^{c,m} / \hat{\bSigma}_{11,ib_m}^{c,m,*} $, $\hat{\bSigma}_{11,ib_m}^{c,m,*} = \max (\hat{\bSigma}_{11,ib_m}^{c,m},\delta_m)$, $f(\bc) = (c_{11})^{-1} c_{12}$, $e_{11,i}^{m,*} = \hat{\bSigma}_{11,i}^{c,m,*} - \bSigma_{11,i}^{c,m} $, and $e_{12,i}^{m,*} = e_{12,i}^{m} $.
Simple algebra shows that
\begin{equation*}
  \Delta_m ^{-1/4} \left( RIB - \int_{0}^{1} \beta_{t}^{c} dt \right) = \mathcal{D}_{m,1} + \mathcal{D}_{m,2} + \mathcal{D}_{m,3} + \mathcal{D}_{m,4} + \mathcal{D}_{m,5} 
  ,
\end{equation*}
where
\begin{eqnarray}\label{key-decomp-thm1}
  \mathcal{D}_{m,1} &=&  b_m \Delta_m ^{3/4} \sum_{i=0}^{N_m-1} \left[ \hat{\beta}_{i b_m } - \hat{\beta}_{i b_m }^{c,m} - \left( \hat{B}^{m}_{i} - \hat{B}^{c,m}_{i}  \right)   \right] ,\cr
  \mathcal{D}_{m,2} &=& b_m \Delta_m ^{3/4} \sum_{i=0}^{N_m-1} \left[ \hat{\beta}_{i b_m }^{c,m} - \beta_{i b_m }^{c,m} - \sum_{x=1}^{2}  \partial_{1x}f(\bSigma_{ib_m}^{c,m}) e_{1x,i b_m}^{m,*}     - \hat{B}^{c,m}_{i} \right] ,\cr
  \mathcal{D}_{m,3} &=& \Delta_m ^{-1/4} \left[ \sum_{i=0}^{N_m - 1} \int_{i b_m \Delta_m }^{(i+1) b_m \Delta_m } \beta_{i b_m }^{c,m} - \beta_{s }^{c} ds + \int_{N_m b_m \Delta_m }^{1}  \beta_{s }^{c}  ds \right] ,\cr
  \mathcal{D}_{m,4}(p) &=& b_m \Delta_m ^{3/4} \sum_{i=0}^{N_m-1} \sum_{x=1}^{2}  \partial_{1x}f(\bSigma_{ib_m}^{c,m}) \left[  M'(p)_{1x,i b_m }^{m} + \xi_{1x,i b_m }^{m,1} + \xi_{1x,i b_m }^{m,2} \right] ,\cr
  \mathcal{D}_{m,5}(p) &=& b_m \Delta_m ^{3/4} \sum_{i=0}^{N_m-1} \sum_{x=1}^{2}  \partial_{1x}f(\bSigma_{ib_m}^{c,m})  M(p)_{1x,i b_m }^{m} .
\end{eqnarray}
We note that Lemmas \ref{lemma:e} and \ref{lemma:jump-preavg-element} also hold in view of $\hat{\bSigma}^{*}$ and $\hat{\bSigma}^{c,*}$ due to the fact that
\begin{equation*}
  \left|\hat{\bSigma}^{m,*}_{1x,i} - \hat{\bSigma}^{c,m,*}_{1x,i}\right| \leq \left|\hat{\bSigma}^{m}_{1x,i} - \hat{\bSigma}^{c,m}_{1x,i}\right| \quad \text{and} \quad \left|\hat{\bSigma}^{c,m,*}_{1x,i} - \bSigma^{c,m}_{1x,i}\right| \leq \left|\hat{\bSigma}^{c,m}_{1x,i} - \bSigma^{c,m}_{1x,i}\right|
\end{equation*}
for sufficiently large $m$ and $x \in \left\lbrace 1,2 \right\rbrace$.
We can show the following lemmas using Lemmas \ref{lemma:e}, \ref{lemma:jump-preavg-element}, and \ref{spot-noise-estimate}.
\begin{lemma}\label{lemma-D123}
  As $m\rightarrow\infty$, we have $\mathcal{D}_{m,1} \xrightarrow[]{p} 0$, $\mathcal{D}_{m,2} \xrightarrow[]{p} 0$, and $\mathcal{D}_{m,3} \xrightarrow[]{p} 0$.
\end{lemma}

\begin{lemma}\label{lemma-D4}
  As $m\rightarrow\infty$, we have $\mathbb{E}\left[ \left|{\mathcal{D}}_{m,4}(p)\right|  \right] \leq C/\sqrt{p} $.
\end{lemma}

\begin{lemma}\label{lemma-D5}
  For any fixed $p\geq 2$, the sequence $\mathcal{D}_{m,5}(p)$ of processes converges $\mathcal{F}_{\infty}$-stably in law to the process
  \begin{equation*}
    Z(p) = \int_{0}^{1} \mathcal{R}(p)_s d\tilde{Z}_s
    ,
  \end{equation*}
  where $\tilde{Z}$ is a standard Brownian motion independent of $\mathcal{F}$, $\mathcal{R}(p)_s$ is the square root of
  \begin{eqnarray*}
    && \mathcal{R}(p)_s^2 = \frac{2}{\psi_0^2} \left( \frac{p \Phi_{00} - \bar{\Phi}_{00}}{p+2} \frac{{C_k} q_s^2}{\sigma_s^2} +   \frac{p \Phi_{01} - \bar{\Phi}_{01}}{p+2} \frac{A_{1,s}}{{C_k}}  + \frac{p \Phi_{11} - \bar{\Phi}_{11}}{p+2} \frac{A_{2,s}}{{C_k}^3}    \right) ,
  \end{eqnarray*}
  and
  \begin{eqnarray*}
    A_{1,s} &=&  \frac{\vartheta_{1,s}^2 R_{11} q_s^2 - 2 \beta_{s}^{c} \vartheta_{1,s}\vartheta_{2,s} R_{12} + \vartheta_{2,s}^2 R_{22}}{\sigma_s^2} + \vartheta_{1,s}^2 R_{11} (\beta_{s}^{c})^2 , \\
    A_{2,s} &=&  \frac{\vartheta_{1,s}^{2}}{\sigma_{s}^{4}} \left( 2(\beta_{s}^{c})^2 \vartheta_{1,s}^{2} R_{11}^{2} - 4 \beta_{s}^{c} \vartheta_{1,s} \vartheta_{2,s} R_{11} R_{12} + \vartheta_{2}^{2} (R_{11}R_{22} + R_{12}^2) \right) 
    .
  \end{eqnarray*}
\end{lemma}
\textbf{Proof of Theorem \ref{Theorem-1}.}
By Lemmas \ref{lemma-D123} and \ref{lemma-D4}, we have
\begin{equation*}
  \lim_{p\rightarrow\infty} \limsup_{m \rightarrow \infty} \mathbb{P}\left(\left|\Delta_m ^{-1/4} \left( RIB - \int_{0}^{1} \beta_{t}^{c} dt \right) - \mathcal{D}_{m,5}(p) \right| > \varepsilon \right) = 0
  ,
\end{equation*}
for all $\varepsilon > 0$.
In addition, with a fixed sample path of Brownian motion $\tilde{Z}$ independent of $\mathcal{F}$ we have $\mathcal{R}(p)_s(\omega)^2 \rightarrow \mathcal{R}_s^2$ for all $s$ and $\omega$, and $\mathcal{R}(p)_s^2 \leq C$.
Thus, we have $Z(p) \xrightarrow[]{p} \int_{0}^{1} \mathcal{R}_s d\tilde{Z}_{s}$.
Finally, Lemma \ref{lemma-D5} concludes that
\begin{equation*}
  m ^{1/4}(RIB_{1}-I \beta_1) \rightarrow \int_{0}^{1} \mathcal{R}_s d\tilde{Z}_s \quad \mathcal{F}_{\infty}\text{-stably as }  m \rightarrow \infty
  .
\end{equation*}
$\blacksquare$

\subsection{Proof of Theorem \ref{Theorem-2}}

For simplicity, we denote derivatives of any given function $f$ at $x_0$ by
\begin{equation*} 
    \frac{\partial f(x_0)}{\partial x}= \left. \dfrac{\partial f\left( x\right) }{\partial x} \right| _{x=x_{0}} ,
\end{equation*}
and define
\begin{eqnarray*}
    &&\hat{L}_{n,m}(\theta)=-\frac{1}{n}\sum^n_{i=1} \{ RIB_i-\hat{h}_i(\theta) \}^2\quad \text{and} \quad \hat{s}_{n,m} (\theta)=\dfrac{\partial \hat{L}_{n,m} (\theta)}{\partial \theta}; \cr
    &&\hat{L}_{n} (\theta)=-\frac{1}{n}\sum^n_{i=1} 
    \left\{ I\beta_i-h_i(\theta) \right \}^2  \quad  \text{and} \quad \hat{s}_{n} (\theta)=\dfrac{\partial \hat{L}_{n} (\theta)}{\partial \theta}; \cr
    &&L_{n} (\theta)=-\frac{1}{n}\sum^n_{i=1} \left[ 
    \left\{ h_i(\theta_0)-h_i(\theta) \right \}^2+D^2_i  
     \right]\quad  \text{and} \quad s_{n} (\theta)=\dfrac{\partial L_{n} (\theta)}{\partial \theta}.
\end{eqnarray*}
Since the dependence of $h_i(\theta)$ on the initial value decays with the order $n^{-1}$, without loss of the generality, we suppose that $h_1(\theta_0)$ is given during the rest of the proofs.
We first establish the below lemmas.
\begin{lemma}\label{Lemma-6}
\begin{enumerate}
Under the assumption of Theorem \ref{Theorem-2}, we have 
\item [(a)] $\sup_{i \in \mathbb{N}} \mathbb{E}[\left|I\beta_i \right|] < \infty$ and $ \sup_{i \in \mathbb{N}}\mathbb{E}[ \sup_{\theta \in \Theta} | h_i(\theta) |] < \infty$ a.s. 
 
\item [(b)]for any $j,k,l \in \{ 1, 2, \ldots, q+ p\vee q + 1 \}$,  
\begin{eqnarray*}
    && \sup_{i \in \mathbb{N}} \mathbb{E}\left[ \sup_{\theta \in \Theta} \left| \frac{\partial {{h}_i(\theta)}}{\partial {\theta_{ j}}} \right| \right]  \leq C, \quad 
    \sup_{i \in \mathbb{N}} \mathbb{E}\left[ \sup_{\theta \in \Theta} \left| \frac{\partial^2 {{h}_i(\theta)}}{\partial {\theta_{j}} {\partial {\theta_{ k}}}}\right|  \right] \leq C, \quad \text{and} \cr
    && \sup_{i \in \mathbb{N}} \mathbb{E}\left[ \sup_{\theta \in \Theta} \left| 
    \frac{\partial^3 {{h}_i(\theta)}}{\partial {\theta_{ j}} {\partial {\theta_{ k}}}{\partial {\theta_{ l}}}} \right|  \right] \leq C, 
\end{eqnarray*}
where $=(\theta_{1},\theta_{ 2}, \ldots, \theta_{q + p \vee q +1})=(\omega, \gamma_1^g, \ldots, \gamma_q^g, \alpha_1^g, \ldots,  \alpha_{p\vee q} ^g)$.
\end{enumerate}
\end{lemma}

\begin{lemma} \label{Lemma-7}
Under  the assumption of Theorem \ref{Theorem-2}, we have
\begin{align}
    & \sup_{\theta \in \Theta} \left| \hat{L}_{n,m} (\theta) - \hat{L}_{n} (\theta) \right|= O_{p}\left( m^{-1/4}  \right), \label{Equation-A.29} \\ 
    & \sup_{\theta \in \Theta} \left| \hat{L}_{n} (\theta) - L_{n} (\theta) \right| = o_{p}(1), \label{Equation-A.30}\\
    & \sup_{\theta \in \Theta} \left| \hat{L}_{n,m} (\theta) - L_{n} (\theta)  \right| =  O_{p}\left( m^{-1/4}  \right) + o_{p}(1). \label{Equation-A.31}
\end{align}
\end{lemma}

\begin{proposition} \label{Proposition-2}
Under Assumptions \ref{Assumption-2}  (except for $n^2 m^{-1} \rightarrow 0$), there is a unique maximizer of $L_n (\theta)$ and as $m,n \rightarrow \infty$, $\hat{\theta} \rightarrow \theta_0$ in probability.
\end{proposition}

\textbf{Proof of Theorem \ref{Theorem-2}.}
By the mean value theorem and Taylor expansion, there exists $\theta^{\ast }$ between $\hat{\theta}$ and $\theta_0$ such that
\begin{equation*}
    \hat{s}_{n,m}(\theta_0)-\hat{s}_{n,m}(\hat{\theta})=\hat{s}_{n,m}(\theta_0)=-\triangledown \hat{s}_{n,m}(\theta^{\ast })(\hat{\theta}-\theta_0).
\end{equation*}
Similar to the proofs of Proposition \ref{Proposition-2}, we can show 
\begin{equation*} 
    -\triangledown \hat{s}_{n,m}(\theta^{\ast }) \overset{p}{\rightarrow} -\triangledown s_{n}(\theta_0).
\end{equation*}
Then,  from the concavity of $\hat{L}_{n,m}(\theta)$, the convergence rate of $\left\|\hat{\theta}-\theta_0 \right \|_{\max}$ is the same as that of $ s_{n}(\theta_0)$.
Thus, it is enough to show 
\begin{equation} \label{Equation-A.33}
    \hat{s}_{n,m} (\theta_0)=O_p(m^{-1/4}  )+O_p(n^{-1/2}).
\end{equation}
Similar to the proof of Lemma \ref{Lemma-7}, we can show that
\begin{eqnarray} \label{Equation-A.34}
    \hat{s}_{n,m} (\theta_0)
    &=&s_{n} (\theta_0)+ \frac{2}{n}\sum_{i=1}^{n} \dfrac{\partial h_i(\theta_0)}{\partial \theta} D_i+O_p(m^{-1/4} ) \cr
    &=&\frac{2}{n}\sum_{i=1}^{n} \dfrac{\partial h_i(\theta_0)}{\partial \theta}  D_i+O_p(m^{-1/4} ).
\end{eqnarray}
Since $D_i$ is a martingale difference and $h_i(\theta_0)$ is $\mathcal{F}_{i-1}$-adaptive, we have 
\begin{equation*}
    \frac{2}{n}\sum_{i=1}^{n} \dfrac{\partial h_i(\theta_0)}{\partial \theta}  D_i=O_p(n^{-1/2}),
\end{equation*}
which shows \eqref{Equation-A.33} with \eqref{Equation-A.34}.
\endpf 
\\

\subsection{Proof of Theorem \ref{Theorem-3}}
\textbf{Proof of Theorem \ref{Theorem-3}.}
By the mean value theorem and Taylor expansion, we obtain, for some $\theta^{\ast }$ between $\theta_0$ and $\hat{\theta} $,
\begin{eqnarray*}
    -\triangledown \hat{s} _{n,m}(\theta^{\ast })(\hat{\theta} -\theta_0)
    &&= \hat{s} _{n}(\theta_0)+\left\{ \hat{s} _{n,m}(\theta_0)- \hat{s} _{n}(\theta_0)\right\} \cr
    &&= \frac{2}{n} \sum ^n_{i=1} \dfrac{\partial h_i(\theta_0)}{\partial \theta}  D_i+O_p(m^{-1/4}),
\end{eqnarray*}
where the last equality is due to \eqref{Equation-A.34}. 
By Assumption \ref{Assumption-2}(b), we can show that $I\beta_i$ and $h_i(\theta)$ can be represented by MA($\infty$) with $D_i$'s. 
Thus,  $I\beta_i$'s and $h_i(\theta)$'s are strictly stationary. 
By the ergodic theorem and the result in the proof of Theorem \ref{Theorem-2}, we have 
\begin{equation*} 
    -\triangledown \hat{s} _{n,m}(\theta^{\ast })  
    \overset{p} \rightarrow 
    2 \mathbb{E} \left[ \left. {\dfrac{\partial h_1(\theta)}{\partial \theta} \dfrac{\partial h_1(\theta)}{\partial \theta^\top} }  \right|_ {\theta=\theta_0} \right],
\end{equation*}
where its asymptotic covariance matrix is positive definite.
The ergodic theorem also provides that
\begin{equation*}
    \sqrt{n}\hat{s}_i (\theta_0)=\sqrt{n}\frac{2}{n} \sum ^n_{i=1} \left\{ \dfrac{\partial h_i(\theta_0)}{\partial \theta}  D_i \right\}
    \overset{d} \rightarrow  N(0,S),
\end{equation*}
where
\begin{eqnarray*}
  &&S = 4\mathbb{E}\left[ D_{1}^{2} \right] \mathbb{E} \left[ 
 \left. {\dfrac{\partial h_1(\theta)}{\partial \theta} \dfrac{\partial h_1(\theta)}{\partial \theta^\top} }\right|_{\theta=\theta_0} \right]. \cr
\end{eqnarray*}
Therefore, by Slutsky's theorem, we obtain
\begin{equation*}
    \sqrt{n}(\hat{\theta} -\theta_0) \overset{d} \rightarrow  N(0,V),
\end{equation*}
where
\begin{eqnarray*}
  V
  &=&\left(2 \mathbb{E} \left[ \left. {\dfrac{\partial h_1(\theta)}{\partial \theta} \dfrac{\partial h_1(\theta)}{\partial \theta^\top} }  \right|_ {\theta=\theta_0} \right]\right)^{-1}
  S
  \left(2 \mathbb{E} \left[ \left. {\dfrac{\partial h_1(\theta)}{\partial \theta} \dfrac{\partial h_1(\theta)}{\partial \theta^\top} }  \right|_ {\theta=\theta_0} \right]\right)^{-1} \\
  &=& \mathbb{E}\left[ D_{1}^{2} \right]   \left ( \mathbb{E} \left[  \left. 
  {\dfrac{\partial h_1(\theta)}{\partial \theta} \dfrac{\partial h_1(\theta)}{\partial \theta^\top} }\right|_{\theta=\theta_0}  \right]\right )^{-1}.
\end{eqnarray*}
\endpf

\subsection{Proof of Theorem \ref{thm:BoundedMoment}}
The proof of Theorem \ref{thm:BoundedMoment} is almost the same as that of Theorem \ref{Theorem-1}.
The main difference is that the localization procedure cannot be applied in Theorem \ref{thm:BoundedMoment}. %
Under the moment conditions in Assumption \ref{assumption-bounded-moment}, however, using Lemmas \ref{lemma:BoundE2BoundCE} and \ref{lemma:moment-itoproperty} which we will state below, we can prove Lemmas \ref{lemma:e}, \ref{lemma:jump-preavg-element}, and \ref{spot-noise-estimate} with additional term $\varPhi_{i}^{m,4}$ for the bounds, where $\varPhi_{i}^{m,w}$ is generic $\mathcal{F}_{i}^{m}$-measurable random variable satisfying $\mathbb{E}\left[ (\varPhi_{i}^{m,w} )^{w} \right] \leq 1$.
\begin{lemma}\label{lemma:BoundE2BoundCE}
  Suppose that a process $P_{t}$ is defined on a fixed filtered probability space $(\Omega,\mathcal{F},$ $(\mathcal{F}_{t})_{t\geq 0},$ $\mathbb{P} )$, and for some $k,w \in [1,\infty)$ and $C>0$, $P_{t}$ satisfies
  \begin{equation*}
    \mathbb{E}\left[ |P_t|^{kw} \right] \leq C \quad \text{ for any } t \geq 0
    .
  \end{equation*}
  Then, we have $C' > 0$ such that
  \begin{eqnarray*}
    \mathbb{E}\left[ |P_{t+u}|^{w} | \mathcal{F}_{t} \right] \leq C' \varPhi_{t}^{k}  \quad \text{for any } t,u \geq 0
    .
  \end{eqnarray*}
\end{lemma}
\begin{lemma}\label{lemma:moment-itoproperty}
  Suppose that a process $X$ is It\^{o} semimartingale with Grigelionis form, where $\mu, \sigma$, and $\mathfrak{d}$ are drift, diffusion, and jump terms, respectively, and for some $k,w \geq 1$, there exists $C > 0$ and a deterministic nonnegative $\lambda$-integrable function $\mathfrak{J}$ on $E$ such that
  \begin{eqnarray*}
    \mathbb{E}\left[ |\mu_{t}|^{kw} \right] < C, \quad \mathbb{E}\left[ |\sigma_{t}|^{kw} \right] < C, \quad \mathbb{E}\left[ |\mathfrak{d}(t, z)|^{kw} \right] \leq \mathfrak{J}(z)
    ,
  \end{eqnarray*}
  for any $t \in \mathbb{R}^{+}$ and $z \in E$.
  Then, we have for $s \leq 1$,
  \begin{eqnarray}\label{eq:moment-itoproperty}
    \mathbb{E}\left[ \sup_{u \in [0,s]} |X_{t+u} - X_{t}|^{w} \Big| \mathcal{F}_{t} \right] \leq C \varPhi_{t}^{k}  s \quad \text{and} \quad \left| \mathbb{E}\left[  X_{t+u} - X_{t} | \mathcal{F}_{t} \right] \right| \leq C \varPhi_{t}^{kw}  s
    \quad \text{a.s.}
  \end{eqnarray}
\end{lemma}

We then need to check that the lemmas hold for $\hat{\bSigma}^{*}$ instead of $\hat{\bSigma}$.
Due to the fact that $|\hat{\bSigma}_{11,i}^{m,*} - \hat{\bSigma}_{11,i}^{c,m,*}| \leq |\hat{\bSigma}_{11,i}^{m} - \hat{\bSigma}_{11,i}^{c,m}|$, Lemma \ref{lemma:jump-preavg-element} hold in view of $\hat{\bSigma}^{*}$ instead of $\hat{\bSigma}$.
Further, unlike the proof of Theorem \ref{Theorem-1} that we can apply the localization procedure, we observe that $|\hat{\bSigma}_{11,i}^{c,m,*} - {\bSigma}_{11,i}^{c,m,*}| \leq |\hat{\bSigma}_{11,i}^{c,m} - {\bSigma}_{11,i}^{c,m}|$ if $\bSigma_{11,i}^{c,m} > \delta_{m}$, and $|\hat{\bSigma}_{11,i}^{c,m,*} - {\bSigma}_{11,i}^{c,m,*}| \leq \delta_m$, if $\bSigma_{11,i}^{c,m} \leq \delta_{m}$ with probability
\begin{eqnarray*}
  \mathbb{P}\left( \bSigma_{11,i}^{c,m} \leq \delta_{m}  \right) &=& \mathbb{P}\left( (\bSigma_{11,i}^{c,m})^{-16} \geq \delta_{m}^{-16}  \right) \cr
  &\leq& \delta_m^{16} \mathbb{E}\left[ \left|\sigma_{\Delta_m i}^{-1}\right|^{32}  \right] \cr
  &\leq& C \delta_m^{16}
  ,
\end{eqnarray*}
where the first inequality is due to Markov's inequality.
Thus, we have for any $w \in \mathbb{N}$,
\begin{eqnarray*}
  && \mathbb{E}\left[ \left|e_{11,i}^{m,*}\right|^{w} | \mathcal{K}_{i}^{m}  \right] \cr
  &=& \mathbb{E}\left[ \left|e_{11,i}^{m,*}\right|^{w} , \bSigma_{11,i}^{c,m} \leq \delta_m | \mathcal{K}_{i}^{m} \right] \mathbb{P}\left( \bSigma_{11,i}^{c,m} \leq \delta_{m}  \right) + \mathbb{E}\left[ \left|e_{11,i}^{m,*}\right|^{w} , \bSigma_{11,i}^{c,m} > \delta_m  | \mathcal{K}_{i}^{m} \right] \mathbb{P}\left( \bSigma_{11,i}^{c,m} > \delta_{m}  \right) \cr
  &\leq& C \delta_m^{16+w} + \mathbb{E}\left[ \left|e_{11,i}^{m}\right|^{w} , \bSigma_{11,i}^{c,m} > \delta_m | \mathcal{K}_{i}^{m}  \right] \mathbb{P}\left( \bSigma_{11,i}^{c,m} > \delta_{m}  \right) \cr
  &\leq& C \delta_m^{16+w} + \mathbb{E}\left[ \left|e_{11,i}^{m}\right|^{w} | \mathcal{K}_{i}^{m} \right] \cr
  &\leq& C \mathbb{E}\left[ \left|e_{11,i}^{m}\right|^{w} | \mathcal{K}_{i}^{m} \right] \quad \text{a.s.}
\end{eqnarray*}
Therefore, Lemma \ref{lemma:e} also holds in view of $\hat{\bSigma}^{c,*}$ instead of $\hat{\bSigma}^{c}$.
Similar to the proof of Theorem \ref{Theorem-1}, we can show the rest of the proof by decomposing the estimation error of $RIB$.
Details can be found in Appendix \ref{key-decomp-longspan}.
$\blacksquare$

\section{Details of proofs}
\subsection{Proofs of Lemmas in Theorem \ref{Theorem-1}}\label{pf-thm1-lemmas}
Below three lemmas help prove the lemmas in Theorem \ref{Theorem-1}.
\begin{lemma}\label{smooth-process}
  If $P_{1,t},\ldots,P_{k,t}$, $k \in \mathbb{N}$ are bounded stochastic processes defined on a filtered probability space $(\Omega, \mathcal{F}, \left\lbrace \mathcal{F}_{t}, t \in [0,\infty) \right\rbrace, P)$, satisfying the property (P-2) defined in Assumption \ref{assumption-formal}(d),
  then $Q_{t} = \prod_{i=1}^{k} P_{i,t}$ also satisfy (P-2).
\end{lemma}  
\textbf{Proof of Lemma \ref{smooth-process}.}
If $k=2$, $(P)$ is satisfied for $Q_{t}$, since
\begin{eqnarray*}
  \mathbb{E}\left[ \left( Q_{t+s} - Q_{t} \right)^2  | \mathcal{F}_{t} \right] &\leq& C \mathbb{E}\left[P_{1,t+s}^{2} \left( P_{2,t+s} - P_{2,t}   \right)^2  | \mathcal{F}_{t} \right] + C \mathbb{E}\left[P_{2,t}^{2} \left( P_{1,t+s} - P_{1,t}   \right)^2  | \mathcal{F}_{t} \right] \\
  &\leq& C \left(  \mathbb{E}\left[ \left( P_{2,t+s} - P_{2,t}   \right)^2 | \mathcal{F}_{t} \right] +   \mathbb{E}\left[ \left( P_{1,t+s} - P_{1,t}   \right)^2 | \mathcal{F}_{t} \right] \right) \\
  &\leq& Cs \quad \text{a.s.}
  ,
\end{eqnarray*}
where the first inequality is due to Jensen's inequality, and the second and third inequalities are due to the fact that the process $P_1$ and $P_2$ are bounded and satisfy $(P)$, respectively.
Further, $Q_{t}$ is bounded.
We can prove the cases $k>2$ by using the mathematical induction method.
$\blacksquare$

\begin{lemma}\label{gmeasurable-bound}
  Let $\xi_{i}^{m} $ be random variables, measurable with respect to $\mathcal{G}^{i}$.
  We have
  \begin{enumerate}
    \item [(a)] $\left| \mathbb{E}\left[ \xi_{i}^{m} | \mathcal{K}_{i}^{m}  \right] \right|  \leq \varPsi_{i}^{m} \text{ a.s.}$, if $\mathbb{E}\left[ \left|\xi_{i}^{m}\right| ^{w} \right] \leq C_{w}$ for any $w \in \mathbb{N}$;
    \item [(b)] $\mathbb{E}\left[ \left|\mathbb{E}\left[ \xi_{i}^{m} | \mathcal{G}_{i-j}^{m}  \right]\right|^{w} \right]  \leq \frac{C_{w,\varepsilon}}{j^{2v-\varepsilon}}$ for any $w \in \mathbb{N}$ and $\varepsilon > 0$, if $\xi _{i}^{m} $ is centered with finite moment of all orders;
    \item [(c)] if $k,i', j \in \mathbb{N}$, $i \leq i' < i + j$, and $\xi_{i}^{m}$ and $\xi_{i'+j}^{m}$ are measurable with respect to $\mathcal{G}_{i'}$ and $\mathcal{G}^{i'+j}$, respectively, then we have 
    \begin{eqnarray*}
      \mathbb{E}\left[\xi_{i}^{m} \xi_{i'+j}^{m} | \mathcal{G}_{i-k}  \right] &\leq& \left| \mathbb{E}\left[\xi_{i}^{m} \right] \mathbb{E}\left[ \xi_{i'+j}^{m}   \right] \right|  +  C j^{-v} \mathbb{E}\left[ \left( \xi_{i'+j}^{m} \right)^2  \right]^{1/2} \mathbb{E}\left[ \left( \xi_{i}^{m} \right)^2  \right]^{1/2}   \\
      && + C \varPsi_{i}^{m,2} k^{-v} \mathbb{E}\left[ \left( \xi_{i'+j}^{m} \right)^4  \right]^{1/4} \mathbb{E}\left[ \left( \xi_{i}^{m} \right)^4  \right]^{1/4}\text{ a.s.}
    \end{eqnarray*}
    
  \end{enumerate}
\end{lemma}
\textbf{Proof of Lemma \ref{gmeasurable-bound}.}
Consider (a).
By (A.3) of \citet{jacod2017statistical}, we have 
\begin{eqnarray}\label{rho-mixing2}
  \mathbb{E}\left[ \left|\mathbb{E}\left[ \xi_{i}^{m} | \mathcal{K}_{i}^{m} \right]\right|^{2}   \right] &\leq& \frac{C}{k_m^{v}} \mathbb{E}\left[ \mathbb{E}\left[ \xi_{i}^{m}  | \mathcal{K}_{i}^{m}  \right]^{2} \right]^{1/2} \mathbb{E}\left[ \mathbb{E}\left[ \xi_{i}^{m} \right]^{2} \right]^{1/2} + \left|\mathbb{E}\left[ \mathbb{E}\left[ \xi_{i}^{m} | \mathcal{K}_{i}^{m}  \right] \right] \mathbb{E}\left[ \xi_{i}^{m} \right] \right| \nonumber\\
  &\leq& \frac{C}{k_m^{v}} \mathbb{E}\left[  \left(   \xi_{i}^{m} \right)^2  \right]  + \mathbb{E}\left[ \xi_{i}^{m} \right]^2 \nonumber \\
  &\leq& C_2 \text{ a.s.}
\end{eqnarray}
Similarly, we have  
\begin{eqnarray}\label{rho-mixing3}
  \mathbb{E}\left[ \left| \mathbb{E}\left[ \xi_{i}^{m} | \mathcal{K}_{i}^{m} \right]\right|^3  \right] &\leq& \frac{C}{k_m^v} \mathbb{E}\left[ \mathbb{E}\left[ \xi_{i}^{m} | \mathcal{K}_{i}^{m} \right]^4 \right]^{1/2} \mathbb{E}\left[ \left( \xi_{i}^{m} \right)^2 \right]^{1/2} + \left| \mathbb{E}\left[ \mathbb{E}\left[ \xi_{i}^{m} | \mathcal{K}_{i}^{m} \right]^2  \right] \mathbb{E}\left[ \xi_{i}^{m} \right] \right| \nonumber\\
  &\leq& \frac{C}{k_m^v} \mathbb{E}\left[ \left( \xi_{i}^{m} \right) ^4 \right]^{1/2} \mathbb{E}\left[ \left( \xi_{i}^{m} \right)^2 \right]^{1/2}  + \left| \mathbb{E}\left[ \mathbb{E}\left[ \xi_{i}^{m} | \mathcal{K}_{i}^{m} \right]^2  \right] \mathbb{E}\left[ \xi_{i}^{m} \right] \right| \nonumber\\
  &\leq& \frac{C}{k_m^v} \mathbb{E}\left[ \left( \xi_{i}^{m} \right) ^4 \right]^{1/2} \mathbb{E}\left[ \left( \xi_{i}^{m} \right)^2 \right]^{1/2}  +   \frac{C}{k_m^{v}} \mathbb{E}\left[  \left(   \xi_{i}^{m} \right)^2  \right] \left| \mathbb{E}\left[ \xi_{i}^{m} \right]  \right| + \left| \mathbb{E}\left[ \xi_{i}^{m} \right]\right|^3     \nonumber\\
  &\leq& C_3 \text{ a.s.}
\end{eqnarray}
and  
\begin{eqnarray}\label{rho-mixing4}
  \left|\mathbb{E}\left[ \mathbb{E}\left[ \xi_{i}^{m} | \mathcal{K}_{i}^{m} \right]^4 \right]\right| &\leq& \frac{C}{k_m^v} \mathbb{E}\left[ \mathbb{E}\left[ \xi_{i}^{m} | \mathcal{K}_{i}^{m} \right]^6 \right]^{1/2} \mathbb{E}\left[ \left( \xi_{i}^{m} \right)^2 \right]^{1/2} + \left| \mathbb{E}\left[ \mathbb{E}\left[ \xi_{i}^{m} | \mathcal{K}_{i}^{m} \right]^3  \right] \mathbb{E}\left[ \xi_{i}^{m} \right] \right| \nonumber \\
  &\leq& \frac{C}{k_m^v} \mathbb{E}\left[ \mathbb{E}\left[ \xi_{i}^{m} | \mathcal{K}_{i}^{m} \right]^6 \right]^{1/2} \mathbb{E}\left[ \left( \xi_{i}^{m} \right)^2 \right]^{1/2} + \frac{C}{k_m^v} \mathbb{E}\left[ \left( \xi_{i}^{m} \right) ^4 \right]^{1/2} \mathbb{E}\left[ \left( \xi_{i}^{m} \right)^2 \right]^{1/2} \left| \mathbb{E}\left[ \xi_{i}^{m} \right] \right| \nonumber\\
  && + \frac{C}{k_m^{v}} \mathbb{E}\left[  \left(   \xi_{i}^{m} \right)^2  \right] \left| \mathbb{E}\left[ \xi_{i}^{m} \right]  \right|^{2} + \left| \mathbb{E}\left[ \xi_{i}^{m} \right]\right|^4 \nonumber \\
  &\leq& \frac{C}{k_m^v} \mathbb{E}\left[ \left( \xi_{i}^{m} \right) ^6 \right]^{1/2} \mathbb{E}\left[ \left( \xi_{i}^{m} \right)^2 \right]^{1/2} + \frac{C}{k_m^v} \mathbb{E}\left[ \left( \xi_{i}^{m} \right) ^4 \right]^{1/2} \mathbb{E}\left[ \left( \xi_{i}^{m} \right)^2 \right]^{1/2} \left| \mathbb{E}\left[ \xi_{i}^{m} \right] \right| \nonumber\\
  && + \frac{C}{k_m^{v}} \mathbb{E}\left[  \left(   \xi_{i}^{m} \right)^2  \right] \left| \mathbb{E}\left[ \xi_{i}^{m} \right]  \right|^{2} + \left| \mathbb{E}\left[ \xi_{i}^{m} \right]\right|^4 \nonumber \\
  &\leq& C_4 \text{ a.s.}
\end{eqnarray}
Using the iterative relationship, we can show that 
$\mathbb{E}\left[ \left|\mathbb{E}\left[ \xi_{i}^{m} | \mathcal{K}_{i}^{m}  \right]\right| ^{w} \right] \leq C_w \text{ a.s.}$ for any $w \in \mathbb{N}$.
Due to the fact that $\xi_{i}^{m}$ is measurable with respect to $\mathcal{G}^{i}$ and independent of $\mathcal{F}_{i}^{m}$, $\left|\mathbb{E}\left[ \xi_{i}^{m} | \mathcal{K}_{i}^{m}  \right]\right| $ is $\mathcal{G}$-measurable.
That is, we can consider $\left|\mathbb{E}\left[ \xi_{i}^{m} | \mathcal{K}_{i}^{m}  \right]\right| $ as $\varPsi_i^m$, a nonnegative and $\mathcal{G}$-measurable random vaiable satisfying $\mathbb{E}\left[ (\varPsi_i^m)^{w} \right] \leq C_{w}$ for any $w>0$.
To simplify notation, we write $\left|\mathbb{E}\left[ \xi_{i}^{m} | \mathcal{K}_{i}^{m}  \right]\right|  \leq \varPsi_i^{m}$, which is exactly the same as Lemma \ref{gmeasurable-bound}(a).
We further note that we can similarly show that if $\xi_{i}^{m}$ is centered and has finite moment of all orders, then we have for any $w \in \mathbb{N}$
\begin{equation}\label{rho-mixing-centered-finite}
  \mathbb{E}\left[ \left|\mathbb{E}\left[ \xi_{i}^{m} | \mathcal{G}_{i-j} \right]\right|^{w}  \right] \leq C j^{-v}
  .
\end{equation}

Consider (b).
By (A.4) of \citet{jacod2019estimating}, we only need to consider $w \geq 3$.
Using the above iterative relationship, we can show that
\begin{equation}\label{CE-w-E-2}
  \mathbb{E}\left[ \left|\mathbb{E}\left[ \xi_{i}^{m} | \mathcal{G}_{i-j} \right]\right|^{w}  \right] \leq \frac{C_w}{j^{v}} \mathbb{E}\left[ \left|\mathbb{E}\left[ \xi_{i}^{m} | \mathcal{G}_{i-j} \right]\right|^{2(w-1)}  \right]^{1/2}
  .
\end{equation}
Then, using \eqref{CE-w-E-2} and the mathematical induction method, we can show that for any $k \in \mathbb{N} \cup \{0\}$, we have
\begin{equation}\label{CE-w-E-induction}
  \mathbb{E}\left[ \left|\mathbb{E}\left[ \xi_{i}^{m} | \mathcal{G}_{i-j} \right]\right|^{w}  \right] \leq  \frac{C_{w,k}}{j^{v(2-2^{-k} )}} \quad \text{ for any } w \in \mathbb{N} 
  .
\end{equation}
Due to the fact that for any $\varepsilon > 0$, there exists $k \in \mathbb{N}$ such that ${j^{-v(2-2^{-k} )}} < {j^{-2v+\varepsilon}} $ and \eqref{CE-w-E-induction}, we have
\begin{equation*}
  \mathbb{E}\left[ \left|\mathbb{E}\left[ \xi_{i}^{m} | \mathcal{G}_{i-j}^{m}  \right]\right|^{w} \right]  \leq \frac{C_{w,\varepsilon}}{j^{2v-\varepsilon}}
  .
\end{equation*}

Consider (c).
By (A.3) of \citet{jacod2019estimating}, we have
\begin{equation*}
  \left| \mathbb{E}\left[ \xi_{i}^{m} \xi_{i'+j}^{m} \right] \right| \leq C j^{-v} \mathbb{E}\left[ \left( \xi_{i}^{m} \right)^2  \right]^{1/2} \mathbb{E}\left[ \left( \xi_{i'+j}^{m} \right)^2  \right]^{1/2} + \left|\mathbb{E}\left[ \xi_{i}^{m} \right] \mathbb{E}\left[ \xi_{i'+j}^{m} \right] \right| 
  .
\end{equation*}
Since $\xi_{i}^{m} \xi_{i'+j}^{m}$ is $\mathcal{G}^{i}$-measurable, we have
\begin{eqnarray*}
  \mathbb{E}\left[ \left( \mathbb{E}\left[ \xi_{i}^{m} \xi_{i'+j}^{m} | \mathcal{G}_{i-k} \right] - \mathbb{E}\left[ \xi_{i}^{m} \xi_{i'+j}^{m} \right] \right)^2  \right] &=& \mathbb{E}\left[ \mathbb{E}\left[ \xi_{i}^{m} \xi_{i'+j}^{m} - \mathbb{E}\left[ \xi_{i}^{m} \xi_{i'+j}^{m} \right] | \mathcal{G}_{i-k} \right]^{2} \right] \\
  &\leq& C k^{-2v} \mathbb{E}\left[ \left( \xi_{i}^{m} \xi_{i'+j}^{m} \right)^2  \right] \\
  &\leq& C  k^{-2v} \mathbb{E}\left[ \left( \xi_{i}^{m} \right)^{4} \right]^{1/2} \mathbb{E}\left[ \left( \xi_{i'+j}^{m} \right)^4  \right]^{1/2} \text{ a.s.}
  ,
\end{eqnarray*}
where the first and second inequalities are due to (A.4) of \citet{jacod2019estimating} and H\"older's inequality, respectively.
Thus, we have
\begin{eqnarray*}
  \mathbb{E}\left[\xi_{i}^{m} \xi_{i'+j}^{m} | \mathcal{G}_{i-k}  \right] &\leq& \left| \mathbb{E}\left[ \xi_{i}^{m} \xi_{i'+j}^{m} \right] \right| + \left|\mathbb{E}\left[ \xi_{i}^{m} \xi_{i'+j}^{m} | \mathcal{G}_{i-k} \right] - \mathbb{E}\left[ \xi_{i}^{m} \xi_{i'+j}^{m} \right]\right|  \\
  &\leq&  C j^{-v} \mathbb{E}\left[ \left( \xi_{i'+j}^{m} \right)^2  \right]^{1/2} \mathbb{E}\left[ \left( \xi_{i}^{m} \right)^2  \right]^{1/2} + C \varPsi_{i,2}^{m} k^{-v} \mathbb{E}\left[ \left( \xi_{i'+j}^{m} \right)^4  \right]^{1/4} \mathbb{E}\left[ \left( \xi_{i}^{m} \right)^4  \right]^{1/4} \\
  && + \left| \mathbb{E}\left[\xi_{i}^{m} \right] \mathbb{E}\left[ \xi_{i'+j}^{m}   \right] \right| 
\quad \text{a.s.}
\end{eqnarray*}
$\blacksquare$

\begin{lemma}\label{chi-property}
  Let $\mathcal{X}_{xy,i,d} = \chi_{x,i} \chi_{y,i+d} - r_{xy}(|d|)$ for any $x,y \in \left\lbrace 1,2 \right\rbrace$ and $d\geq 0$, $\mathcal{X}_{xy,i,d} = \mathcal{X}_{yx,i,-d}$ for $d < 0$.
  Under Assumption \ref{assumption-noise}, for any $i,j,d,w \in \mathbb{N}$, we have
  \begin{enumerate}
    \item [(a)] $ \mathbb{E}[ |\bar{\chi}_{x,i}^m |^{w}  ] \leq C l_m^{-w/2} $, $ \mathbb{E}[ |\bar{\chi}_{x,i}^m |^{w} | \mathcal{K}_{i}^{m}  ] \leq C \varPsi_{i}^{m,2} l_m ^{-w/2}   $, and $\mathbb{E}\left[ \chi_{x,i} \bar{\chi}_{y,i+j}^m | \mathcal{G}_{i-k_m} \right]  \leq C j^{-v} l_m^{-1/2}$ $+ C \varPsi_{i}^{m,2} k_m^{-v} l_m ^{-1/2} $ \text{ a.s.};
    \item [(b)] $|\mathbb{E}\[\mathcal{X}_{xy,i,d} \mathcal{X}_{xy,j,d} | \mathcal{G}_{i-k_m} \]| \leq C \left( (j-i-d)^{-v} + \varPsi_{i}^{m,2}  k_m^{-v} \right)  \text{ a.s.} $ if $j-i > d$;
  \end{enumerate}
\end{lemma}

\textbf{Proof of Lemma \ref{chi-property}.}
Consider (a).
By Theorem 1.1 in \citet{shao1995maximal}, the first part of Lemma \ref{chi-property}(a) holds.
For the second part of Lemma \ref{chi-property}(a), using (A.4) in \citet{jacod2019estimating}, we can show that
\begin{eqnarray}\label{unconditional-to-conditional}
  \mathbb{E}\left[ \left( \mathbb{E}\left[ |\bar{\chi}_{x,i}^m |^{w} | \mathcal{K}_{i}^{m}  \right] - \mathbb{E}\left[ |\bar{\chi}_{x,i}^m |^{w} \right] \right)^2  \right] &=&  \mathbb{E}\left[ \mathbb{E}\left[ |\bar{\chi}_{x,i}^m |^{w} -  \mathbb{E}\left[ |\bar{\chi}_{x,i}^m |^{w}   \right] | \mathcal{K}_{i}^{m}  \right]^2  \right] \nonumber\\
  &\leq& C k_m^{-2v} \mathbb{E}\left[ |\bar{\chi}_{x,i}^m |^{2w} \right]
  \text{ a.s.}
\end{eqnarray}
By the first part of Lemma \ref{chi-property}(a) and \eqref{unconditional-to-conditional}, the second part of Lemma \ref{chi-property}(a) holds.
For the third part of the Lemma \ref{chi-property}(a), using Lemma \ref{gmeasurable-bound}(c) with the finiteness of all moments of $\chi_i$, we have
\begin{eqnarray*}
  \mathbb{E}[\chi_{x,i} \bar{\chi}_{y,i+j}^m | \mathcal{G}_{i-k_m} ] &\leq&  C j^{-v} + C \varPsi_{i}^{m,2} k_m^{-v}\text{ a.s.}
\end{eqnarray*}

Consider (b).
Using Lemma \ref{gmeasurable-bound}(c), we can show Lemma \ref{chi-property}(b).
$\blacksquare$

\textbf{Proof of Lemma \ref{negligible-xi}.}
Lemma \ref{negligible-xi}(a) is a trivial consequence of (B.8) and (B.9) in \citet{chen2018inference}, so we only need to prove (b).
We consider the case $x=y=1$.
Simple algebra shows that
$\tilde{\Gamma}^{m}_{11,i} = \frac{\left( \vartheta_{1,i}^{m} \right)^2 }{k_m} \sum_{m\in\mathbb{Z}} \phi_{d}^{m} r_{11}(m)$.
Then, we can write
\begin{equation*}
  \xi_{11,i}^{m,2} = \frac{1}{(b_m- 2 k_m)\Delta_m k_m^2 \psi_0}  \left(  V_{11,i}^{m} + V_{11,i}^{'m}    \right),
\end{equation*}
where $V_{11,i}^{m} = \mathcal{U}_{11,i}^{m} \sum_{|d|>k_m'} \phi _{d}^{m} r_{11}(|d|)$, $V_{11,i}^{'m} = \sum_{d=-k_m'}^{k_m'} \phi _{d}^{m} \left\{ r_{11}(|d|) \mathcal{U} _{11,i}^{m} - U _{m,i}^{Y_{1} Y_{1}} (|d|) \right\}$, and $\mathcal{U} _{11,i}^{m} = \sum_{l=0}^{b_m - 6l_m} (\vartheta _{1,i+l}^{n} )^2$.
Using $\rho$-mixing property and the facts that $|\mathcal{U} _{11,i}^{m}| \leq C b_m $ and $|\phi _{d}^{m}| \leq C$, we can show that
\begin{equation}\label{V11-bound}
  |V_{11,i}^{m}| \leq C b_m (k_m') ^{-(v-1)} \quad \text{a.s.}
\end{equation}
Let
\begin{eqnarray}\label{micro-autocorrelation-theta}
  T (j,l) _{i}^{m,1} &=& X _{1,i+j}^{m} - \bar{X} _{1,i+l}^{m} + \chi_{1,i+j} (\vartheta_{1,i+j}^{m} - \vartheta_{1,i}^{m}) - \frac{1}{l_m} \sum_{s=0}^{l_m -1} \chi_{1,i+l+s} (\vartheta_{1,i+l+s}^{m} - \vartheta_{1,i}^{m}) \quad \text{and} \nonumber \\
  T (j,l) _{i}^{m,2} &=& \vartheta _{1,i}^{m} (\chi_{1,i+j} - \bar{\chi} _{1,i+l}^{m}).
\end{eqnarray}
By the finiteness of all moments of $\chi_{i}$ and the fact that $X_{1}$ and $\vartheta_{1}$ are It\^{o} semimartingales, we have for any $w \geq 2$ and $l > j$,
\begin{equation}\label{theta-moment}
\mathbb{E}\left[ | T (j,l) _{i}^{m,1} |^w | \mathcal{K}_{i}^{m} \right] \leq C_w \varPsi_{i}^{m}  (l-j+ l_m ) \Delta_m, \quad \mathbb{E}\left[ | T (j,l) _{i}^{m,2} |^w  | \mathcal{K}_{i}^{m} \right] \leq C_w \varPsi_{i}^{m} \text{ a.s.}
\end{equation}
On the other hand, we have
\begin{align}
  U _{m,i}^{Y_{1}Y_{1}} (|d|) - r_{11} (|d| ) \mathcal{U} _{11,i}^{m} =& \sum_{l=0}^{b_m -6 l_m} (Y _{1,i+l}^{m} - \bar{Y} _{1,i+l+2 l_m}^{m}) (Y _{1,i+l+d}^{m} - \bar{Y} _{1,i+l+4 l_m}^{m})- r_{11}(|d|) \mathcal{U} _{11,i}^{m} \notag\\
  =& \sum_{l=0}^{b_m -6 l_m } \left( T (0,2 l_m ) _{i+l}^{m,1} + T (0, 2 l_m ) _{i+l}^{m,2} \right)  \notag \\
  & \times \left( T (d,4 l_m ) _{i+l}^{m,1} + T (d, 4 l_m ) _{i+l}^{m,2} \right) - r_{11}(|d|) \mathcal{U} _{11,i}^{m} \notag\\
  =& \mathcal{V}_{11,i}^{m,1}(d) + \mathcal{V}_{11,i}^{m,2}(d) + \mathcal{V}_{11,i}^{m,3}(d) + \mathcal{V}_{11,i}^{m,4}(d), \notag
\end{align}
where
\begin{eqnarray}\label{mathcalVs}
  \mathcal{V}_{11,i}^{m,1}(d) &=& \sum_{l=0}^{b_m -6 l_m} T (0,2 l_m ) _{i+l}^{m,1} T (d,4 l_m ) _{i+l}^{m,1}, \nonumber\\
  \mathcal{V}_{11,i}^{m,2}(d) &=&  \sum_{l=0}^{b_m -6 l_m} T (0,2 l_m ) _{i+l}^{m,1} T (d,4 l_m ) _{i+l}^{m,2}, \nonumber\\
  \mathcal{V}_{11,i}^{m,3}(d) &=& \sum_{l=0}^{b_m -6 l_m} T (0,2 l_m ) _{i+l}^{m,2} T (d,4 l_m ) _{i+l}^{m,1}, \nonumber\\ %
  \mathcal{V}_{11,i}^{m,4}(d) &=&  \sum_{l=0}^{b_m -6 l_m} T (0,2 l_m ) _{i+l}^{m,2} T (d,4 l_m ) _{i+l}^{m,2}   - r_{11}(|d|) \mathcal{U} _{11,i}^{m}.
\end{eqnarray}
By \eqref{theta-moment}, we have
\begin{eqnarray}\label{mathV1E}
  &&\left| \mathbb{E}\left[  \mathcal{V}_{11,i}^{m,1}(d)   | \mathcal{K}_{i}^{m} \right] \right| \cr
  &=& \left|  \sum_{l=0}^{b_m -6 l_m} \mathbb{E}\left[  T (0,2 l_m ) _{i+l}^{m,1} T (d,4 l_m ) _{i+l}^{m,1}   | \mathcal{K}_{i}^{m}  \right]  \right| \cr
  &\leq&   \sum_{l=0}^{b_m -6 l_m} \mathbb{E}\left[ \left( T (0,2 l_m ) _{i+l}^{m,1}\right)^{2}  | \mathcal{K}_{i}^{m}  \right]^{1/2} \mathbb{E}\left[ \left( T (d,4 l_m ) _{i+l}^{m,1} \right)^{2}  | \mathcal{K}_{i}^{m}  \right]^{1/2}\nonumber\\
  &\leq& C_w \varPsi_{i}^{m}  b_m l_m \Delta_m  ,%
\end{eqnarray}
where the second inequality is due to H\"older's inequality.
Let $\mathcal{T}(d)_{i,l}^{m,12} = T (0,2 l_m ) _{i+l}^{m,1}$ $\times T (d,4 l_m ) _{i+l}^{m,2}$ and $\mathcal{T}(d)_{i,l}^{m,21} = T (0,2 l_m ) _{i+l}^{m,2} T (d,4 l_m ) _{i+l}^{m,1}$.
By H\"older's inequality and \eqref{theta-moment}, we have almost surely
\begin{equation}\label{mathT-w}
  \mathbb{E}\left[ \left|\mathcal{T}(d)_{i,l}^{m,12}\right|^{w}   | \mathcal{K}_{i}^{m}  \right] \leq C_{w} \varPsi_{i}^{m} (l_m \Delta_m)^{1/2}  \quad \text{and} \quad \mathbb{E}\left[ \left|\mathcal{T}(d)_{i,l}^{m,21}\right|^{w}   | \mathcal{K}_{i}^{m}  \right] \leq C_{w} \varPsi_{i}^{m}  (l_m \Delta_m )^{1/2}
  .
\end{equation}
Since the process $\chi$ is independent of the $\sigma$-field $\mathcal{F}_{\infty}$, we obtain
\begin{align}\label{mathT-CE}
  & \quad \left|\mathbb{E}\left[ \mathcal{T}(d)_{i,l}^{m,21}  | \mathcal{K}_{i}^{m}  \right]\right|  \cr
  & \leq
  \left|\mathbb{E}\left[ \mathbb{E}\left[  X _{1,i+l+d}^{c,m} - \bar{X} _{1,i+l+4 l_m}^{c,m}  | \mathcal{K}_{i+l+d}^{m}  \right]  \vartheta _{1,i+l}^{m} | \mathcal{K}_{i}^{m}  \right] \mathbb{E}\left[   (\chi_{1,i+l} - \bar{\chi} _{1,i+l+2 l_m}^{m}) | \mathcal{K}_{i}^{m}  \right]\right|  \cr
  &\quad + \left|\mathbb{E}\left[ \mathbb{E}\left[ \vartheta_{1,i+l+d}^{m} - \vartheta_{1,i+l}^{m} | \mathcal{K}_{i+l}^{m}  \right] \vartheta _{1,i}^{m} | \mathcal{K}_{i}^{m}  \right] \mathbb{E}\left[ (\chi_{1,i+j} - \bar{\chi} _{1,i+l}^{m}) \chi_{1,i+l+d} | \mathcal{K}_{i}^{m} \right]\right|  \cr
  &\quad + \frac{1}{l_m} \sum_{s=0}^{l_m -1} \left| \mathbb{E}\left[ \mathbb{E}\left[ \vartheta_{1,i+l+4 l_m+s}^{m} - \vartheta_{1,i+l}^{m} | \mathcal{K}_{i+l}^{m}  \right] \vartheta _{1,i}^{m} | \mathcal{K}_{i}^{m}  \right] \mathbb{E}\left[  \chi_{1,i+l+4 l_m+s}^{w} (\chi_{1,i+j} - \bar{\chi} _{1,i+l}^{m}) | \mathcal{K}_{i}^{m}  \right]\right|  \cr
  &\leq C l_m \Delta_m \left( \left| \mathbb{E}\left[   (\chi_{1,i+l} - \bar{\chi} _{1,i+l+2 l_m}^{m}) | \mathcal{K}_{i}^{m}  \right]\right| + \left| \mathbb{E}\left[ (\chi_{1,i+j} - \bar{\chi} _{1,i+l}^{m}) \chi_{1,i+l+d} | \mathcal{K}_{i}^{m} \right]\right| \right) \cr
  &\quad + C \Delta_m  \sum_{s=0}^{l_m -1} \left|  \mathbb{E}\left[  \chi_{1,i+l+4 l_m+s}^{w} (\chi_{1,i+j} - \bar{\chi} _{1,i+l}^{m}) | \mathcal{K}_{i}^{m}  \right]\right|  \cr
  & \leq C \varPsi_{i}^{m}  l_m \Delta_m \text{ a.s.} ,
\end{align}
where the third inequality is due to Lemma \ref{gmeasurable-bound}(a).
Similarly, we can show that
\begin{equation}\label{mathcalT-CE}
  \left|\mathbb{E}\left[ \mathcal{T}(d)_{i,l}^{m,12} | \mathcal{K}_{i}^{m} \right]\right|  \leq C \varPsi_{i}^{m}   l_m \Delta_m   \text{ a.s.}
  \end{equation}
Thus, we have
\begin{equation}\label{mathV23E}
  \left| \mathbb{E}\left[ \mathcal{V}_{11,i}^{m,2}(d) | \mathcal{K}_{i}^{m} \right] \right| \leq C \varPsi_{i}^{m} b_m l_m \Delta_m \quad \text{and} \quad \left| \mathbb{E}\left[ \mathcal{V}_{11,i}^{m,3}(d) | \mathcal{K}_{i}^{m} \right] \right| \leq C \varPsi_{i}^{m} b_m l_m \Delta_m \text{ a.s.}
\end{equation}
Let $\mathcal{V}_{11,i}^{m,2}(d) = \hat{\mathcal{V}}_{11,i}^{m,2}(d) + \bar{\mathcal{V}}_{11,i}^{m,2}(d) + \tilde{\mathcal{V}}_{11,i}^{m,2}(d) $, where
\begin{eqnarray*}
  && \hat{\mathcal{V}}_{11,i}^{m,2}(d) = \sum_{l=0}^{5 l_m - 1} \sum_{j=0}^{\lfloor \frac{(b_m -6 l_m)}{5 l_m}  \rfloor  -1} \bar{\mathcal{T}}(d)_{i,5 l_mj + l}^{m,12},  \\
  && \bar{\mathcal{V}}_{11,i}^{m,2}(d) = \sum_{l=0}^{5 l_m - 1} \sum_{j=0}^{\lfloor \frac{(b_m -6 l_m)}{5 l_m} \rfloor  -1}  \mathbb{E}\left[ \mathcal{T}(d)_{i,5 l_mj + l}^{m,12} | \tilde{\mathcal{K}}_{i+5 l_mj + l}^{m,5l_m} \right], \\
  && \tilde{\mathcal{V}}_{11,i}^{m,2}(d) = \sum_{l=\lfloor \frac{(b_m -6 l_m)}{5 l_m}  \rfloor 5l_m}^{b_m - 6 l_m}   \mathcal{T}(d)_{i, l}^{m,12},  \\
  && \bar{\mathcal{T}}(d)_{i,l}^{m,12} =  \mathcal{T}(d)_{i, l}^{m,12} - \mathbb{E}\left[ \mathcal{T}(d)_{i,l}^{m,12} | \tilde{\mathcal{K}}_{i+ l}^{m,5l_m} \right], \cr
  && \tilde{\mathcal{K}}_{i}^{m,a} = \mathcal{F}_{i}^{m} \otimes {\mathcal{G}}_{i-a}
  .
\end{eqnarray*}
Similar to the proof of \eqref{mathT-CE}, we can show that for any $w > 0$,
\begin{equation}\label{mathT-CEt}
  \left|\mathbb{E}\left[ \mathcal{T}(d)_{i,l}^{m,12} | \tilde{\mathcal{K}}_{i+ l}^{m,5l_m} \right]\right|^{w} \leq C \varPsi_{i}^{m} l_m^{w} \Delta_m^{w} 
  .
\end{equation}
Further, we have for any $a \geq 1$
\begin{eqnarray}\label{mathT-CE-w-WC}
  \mathbb{E}\left[ \left| \mathbb{E}\left[ \mathcal{Y} | \mathcal{K}_{i}^{m} \right] - \mathbb{E}\left[ \mathcal{Y} \right]  \right|^{a}  \right] &\leq& C_a \mathbb{E}\left[ \mathbb{E}\left[ \mathcal{Y} | \mathcal{K}_{i}^{m} \right]^{a} + \mathbb{E}\left[ \mathcal{Y} \right]^{a} \right] \cr
  &\leq& C_a \mathbb{E}\left[ \mathbb{E}\left[ \mathcal{Y}^{a} | \mathcal{K}_{i}^{m} \right] + \mathbb{E}\left[ \mathcal{Y}^{a} \right] \right] \cr
  &\leq& C_a \mathbb{E}\left[ \mathcal{Y}^{a} \right] \cr
  &\leq& C_a  l_m^{wa} \Delta_m ^{wa}
  ,
\end{eqnarray}
where $\mathcal{Y} = |\mathbb{E}[ \mathcal{T}(d)_{i,l}^{m,12} | \tilde{\mathcal{K}}_{i+ l}^{m,5l_m} ]|^{w}$, the first and second are due to Jensen's inequality, and third and fourth inequalities are due to tower property and \eqref{mathT-CEt}, respectively.
Using \eqref{mathT-CEt} and \eqref{mathT-CE-w-WC}, we have
\begin{equation}\label{mathT-CE-w-CE}
  \mathbb{E}\left[ \left|\mathbb{E}\left[ \mathcal{T}(d)_{i,l}^{m,12}  | \tilde{\mathcal{K}}_{i+l}^{m,5l_m}  \right]\right|^{w} \Big| \mathcal{K}_{i}^{m} \right] \leq C \varPsi_{i}^{m} l_m^{w} \Delta_m ^{w} 
  ,
\end{equation}
For $\hat{\mathcal{V}}_{11,i}^{m,2}(d)$, we have for any integer $w \geq 2$,
\begin{eqnarray*}
  \mathbb{E}\left[ \left| \hat{\mathcal{V}}_{11,i}^{m,2}(d) \right|^w \big| \mathcal{K}_{i}^{m} \right] &\leq& \left( 5 l_m \right)^{w-1}   \sum_{l=0}^{5 l_m - 1} \mathbb{E}\left[ \left| \sum_{j=0}^{\lfloor \frac{(b_m -6 l_m)}{5 l_m}  \rfloor} \bar{\mathcal{T}}(d)_{i,5 l_mj + l}^{m,12} \right|^w \Bigg| \mathcal{K}_{i}^{m}  \right] \\
  &\leq& C_w \left( 5 l_m \right)^{w-1}   \sum_{l=0}^{5 l_m - 1} \mathbb{E}\left[ \left| \sum_{j=0}^{\lfloor \frac{(b_m -6 l_m)}{5 l_m}  \rfloor} \left( \bar{\mathcal{T}}(d)_{i,5 l_mj + l}^{m,12} \right)^2  \right|^{w/2} \Bigg| \mathcal{K}_{i}^{m}  \right] \\
  &\leq& C_w \left( 5 l_m \right)^{w-1} \frac{b_m ^{w/2 -1}}{\left( 5 l_m \right)^{w/2 -1} }  \sum_{l=0}^{5 l_m - 1}   \sum_{j=0}^{\lfloor \frac{(b_m -6 l_m)}{5 l_m}  \rfloor} \mathbb{E}\left[ \left| \bar{\mathcal{T}}(d)_{i,5 l_mj + l}^{m,12} \right|^w  | \mathcal{K}_{i}^{m}  \right] \\
  &\leq& C_{w} \varPsi_{i}^{m} (l_m b_m )^{w/2} (l_m \Delta_m )^{1/2}
  ,
\end{eqnarray*}
where the first, second, third, and fourth inequalities are due to Jensen's inequality, Burkholder-Davis-Gundy inequality, Jensen's inequality, and \eqref{mathT-w}, respectively.
In case of $\bar{\mathcal{V}}_{11,i}^{m,2}(d)$, we have
\begin{eqnarray*}
  \mathbb{E}\left[ \left| \bar{\mathcal{V}}_{11,i}^{m,2}(d) \right|^w | \mathcal{K}_{i}^{m}  \right] &\leq& C b_m ^{w-1} \sum_{l=0}^{b_m - 6 l_m} \mathbb{E}\left[ \left|\mathbb{E}\left[ \mathcal{T}(d)_{i,l}^{m,12} | \tilde{\mathcal{K}}_{i+l}^{m,5l_m} \right]\right| ^{w} | \mathcal{K}_{i}^{m}  \right] \\
  &\leq& C_w \varPsi_{i}^{m}  b_m ^{w} \left(  l_m \Delta_m \right)^w \text{ a.s.},
\end{eqnarray*}
where the first and second inequalities are due to Jensen's inequality and \eqref{mathT-CE-w-CE}, respectively.
For $\tilde{\mathcal{V}}_{11,i}^{m,2}(d)$, we have
\begin{eqnarray*}
  \mathbb{E}\left[ \left( \tilde{\mathcal{V}}_{11,i}^{m,2}(d) \right)^w | \mathcal{K}_{i}^{m}  \right] &\leq&  (5 l_m)^{w-1} \sum_{l=\lfloor \frac{(b_m -6 l_m)}{5 l_m}  \rfloor 5l_m}^{b_m - 6 l_m} \mathbb{E}\left[ \left( \mathcal{T}(d)_{i,5 l_mj + l}^{m,12} \right)^w | \mathcal{K}_{i}^{m}  \right] \\
  &\leq& C_{w} \varPsi_{i}^{m}  l_m^{w} (l_m \Delta_m )^{1/2}
  \text{ a.s.},
\end{eqnarray*}
where the first and second inequalities are due to Jensen's inequality and \eqref{mathT-w}, respectively.
Therefore, using Jensen's inequality, we have
\begin{eqnarray}\label{mathV-w-res}
 && \mathbb{E}\left[ \left( \mathcal{V}_{11,i}^{m,2}(d) \right)^w | \mathcal{K}_{i}^{m} \right] \cr
   &&\leq C_w \left( \mathbb{E}\left[ \left( \hat{\mathcal{V}}_{11,i}^{m,2}(d) \right)^w  |  \mathcal{K}_{i}^{m} \right] + \mathbb{E}\left[ \left( \bar{\mathcal{V}}_{11,i}^{m,2}(d) \right)^w | \mathcal{K}_{i}^{m} \right] + \mathbb{E}\left[ \left( \tilde{\mathcal{V}}_{11,i}^{m,2}(d) \right)^w | \mathcal{K}_{i}^{m} \right] \right)  \nonumber\\
  &&\leq C_{w} \varPsi_{i}^{m} (l_m b_m )^{w/2} (l_m \Delta_m )^{1/2}
  ,
\end{eqnarray}
for any $w \geq 2$.
Similarly, we can show that for any $z \in \left\lbrace 1,2,3 \right\rbrace$
\begin{eqnarray}\label{mathV23-2w}
  &&\mathbb{E}\left[ \left( \mathcal{V}_{11,i}^{m,z}(d) \right)^w | \mathcal{K}_{i}^{m}  \right] \leq C_{w,\varepsilon} \varPsi_{i}^{m} (b_m l_m)^{w/2} (l_m \Delta_m)^{1/2} \quad\text{ a.s.}
\end{eqnarray}
In case of $\mathcal{V}_{11,i}^{m,4}(d)$, we have
\begin{eqnarray*}
  \mathcal{V}_{11,i}^{m,4}(d) &=&  \sum_{l=0}^{b_m -6 l_m}  (\vartheta _{1,i+l}^{m})^2 \left\{ (\chi_{1,i+l} - \bar{\chi} _{1,i+l+2 l_m }^{m}) (\chi_{1,i+l+d} - \bar{\chi} _{1,i+l+ 4 l_m }^{m})   - r_{11}(|d|) \right\} \\
  &=& \mathcal{S} _{11,i}^{m,1}(d) + \mathcal{S} _{11,i}^{m,2} + \mathcal{S} _{11,i}^{m,3} + \mathcal{S} _{11,i}^{m,4},
\end{eqnarray*}
where
\begin{align}\label{S-sub}
  & \mathcal{S} _{11,i}^{m,1}(d) = \sum_{l=0}^{b_m -6 l_m}  (\vartheta _{1,i+l}^{m})^2 (\chi_{1,i+l} \chi_{1,i+l+d} - r_{11}(|d|)), \qquad
  \mathcal{S} _{11,i}^{m,2} = \sum_{l=0}^{b_m -6 l_m}  (\vartheta _{1,i+l}^{m})^2 \chi_{1,i+l} \bar{\chi} _{1,i+l+ 4 l_m }^{m}, \nonumber\\
  &\mathcal{S} _{11,i}^{m,3} = \sum_{l=0}^{b_m -6 l_m}  (\vartheta _{1,i+l+d}^{m})^2 \chi_{1,i+l} \bar{\chi} _{1,i+l+ 2 l_m }^{m}, \quad \text{and} \quad
  \mathcal{S} _{11,i}^{m,4} = \sum_{l=0}^{b_m -6 l_m}  (\vartheta _{1,i+l}^{m})^2 \bar{\chi} _{1,i+l+ 2 l_m }^{m} \bar{\chi} _{1,i+l+ 4 l_m }^{m}.
\end{align}
Since the process $\chi$ is independent of the $\sigma$-field of the process $\vartheta$, we have for any $\varepsilon > 0$,
\begin{eqnarray*}
  \left|\mathbb{E}\left[ \mathcal{S} _{11,i}^{m,1}(d)  | \mathcal{K}_{i}^{m}  \right]\right|^{4}  &\leq& C \left( \sum_{l=0}^{b_m -6 l_m} \mathbb{E}\left[  \chi_{1,i+l} \chi_{1,i+l+d} - r_{11}(|d|) | \mathcal{K}_{i}^{m}  \right] \right)^{4}  \cr
  &\leq& C b_m ^{3} \sum_{l=0}^{b_m - 6l_m} \left|\mathbb{E}\left[  \chi_{1,i+l} \chi_{1,i+l+d} - r_{11}(|d|) | \mathcal{K}_{i}^{m}  \right]\right| ^{4} \cr
  &\leq& C_{\varepsilon} \varPsi_{i}^{m,1}  b_m^{4} \Delta_m ^{v-\varepsilon} \text{ a.s.}
  ,
\end{eqnarray*}
where the last inequality is due to Lemma \ref{gmeasurable-bound}(b).
Using Lemma \ref{gmeasurable-bound}(b), we have for any $\varepsilon >0$,
\begin{equation*}
  \mathbb{E}\left[ \left|\mathbb{E}\left[ \chi_{1,i+l} \bar{\chi} _{1,i+l+ 4 l_m }^{m} | \mathcal{K}_{i}^{m}  \right] - \mathbb{E}\left[ \chi_{1,i+l} \bar{\chi} _{1,i+l+ 4 l_m }^{m}  \right]\right|^{4}  \right] \leq C_{\varepsilon} \Delta_m ^{v - \varepsilon}
  .
\end{equation*}
Thus, we have for any $\varepsilon > 0$,
\begin{eqnarray*}
  \left|\mathbb{E}\left[ \chi_{1,i+l} \bar{\chi} _{1,i+l+ 4 l_m }^{m} | \mathcal{K}_{i}^{m}  \right]\right| &\leq&  l_{m}^{-(v+\frac{1}{2} )} + C_{\varepsilon} \varPsi_{i}^{m,4} \Delta_m ^{\frac{v-\varepsilon}{4} } \cr
  &\leq& C_{\varepsilon} \varPsi_{i}^{m,4} l_{m}^{-(v+\frac{1}{2} )} \quad \text{a.s.}
\end{eqnarray*}
Therefore, we have
\begin{eqnarray*}
  \left|\mathbb{E}\left[ \mathcal{S} _{11,i}^{m,2}(d)  | \mathcal{K}_{i}^{m}  \right]\right|^{4} &\leq& \left( \sum_{l=0}^{b_m -6 l_m}  \mathbb{E}\left[ (\vartheta _{1,i+l}^{m})^2 \chi_{1,i+l} \bar{\chi} _{1,i+l+ 4 l_m }^{m} | \mathcal{K}_{i}^{m} \right] \right)^4 \cr
  &\leq& C b_m ^{3} \sum_{l=0}^{b_m -6 l_m} \left|\mathbb{E}\left[ \chi_{1,i+l} \bar{\chi} _{1,i+l+ 4 l_m }^{m} | \mathcal{K}_{i}^{m} \right]\right|^{4} \cr
  &\leq& C \varPsi_{i}^{m,1}  b_m ^{4} l_m^{-4v-2 } \quad \text{a.s.}
\end{eqnarray*}
Similarly, we can show that for any $z \in \left\lbrace 2,3,4 \right\rbrace$
\begin{equation*}
  \left|\mathbb{E}\left[ \mathcal{S} _{11,i}^{m,z}(d)  | \mathcal{K}_{i}^{m}  \right]\right|^{4} \leq C \varPsi_{i}^{m,1}  b_m ^{4} l_m^{-4v-2 } \quad \text{a.s.}
\end{equation*}
Thus, we have
\begin{eqnarray}\label{mathV4E}
  \left| \mathbb{E}\left[ \mathcal{V}_{11,i}^{m,4}(d) | \mathcal{K}_{i}^{m} \right] \right| &\leq& \sum _{z=1}^{4} \left| \mathbb{E}\left[ \mathcal{S} _{11,i}^{m,z}  | \mathcal{K}_{i}^{m} \right] \right| \nonumber\\
  &\leq& C \varPsi_{i}^{m,4} b_m l_m ^{-(v+\frac{1}{2} )} \text{ a.s.}
\end{eqnarray}
By \eqref{V11-bound}, \eqref{mathV1E}, \eqref{mathV23E}, \eqref{mathV4E}, we have
\begin{eqnarray*}
  \left| \mathbb{E}\left[ \xi_{11,i}^{m,2} | \mathcal{K}_{i}^{m} \right] \right| &\leq&  \frac{\left|\mathbb{E}  \left[  V_{11,i}^{m} | \mathcal{K}_{i}^{m} \right] \right| + \left| \mathbb{E}  \left[ V_{11,i}^{'m} | \mathcal{K}_{i}^{m} \right] \right|}{(b_m- 2 k_m)\Delta_m k_m^2 \psi_0}  , \\
  &\leq& C k_m'^{-(v-1)} + \frac{C \sum_{d=-k'_m}^{k'_m} \sum _{z=1}^{4} \left| \mathbb{E}\left[ \mathcal{V}_{11,i}^{m,z}(d) | \mathcal{K}_{i}^{m} \right] \right|}{(b_m- 2 k_m)\Delta_m k_m^2 \psi_0}\\
  &\leq& C (k_m'^{-(v-1)} + \varPsi_{i}^{m,4} k'_m l_m \Delta_m  ) \text{ a.s.}
\end{eqnarray*}
Similar to proof of \eqref{mathV-w-res}, using Burkholder-Davis-Gundy inequality, we have
\begin{eqnarray}\label{S1-w-jensen}
  && \mathbb{E}\left[ \left( \mathcal{S} _{11,i}^{m,1} (d) \right)^w \big| \mathcal{K}_{i}^{m}  \right] \cr
  &=&   \mathbb{E}\[ \left( \sum_{l=0}^{b_m -6 l_m}  (\vartheta _{1,i+l}^{m})^2 \mathcal{X}_{11,i+l,d}  \right)^{w} \bigg| \mathcal{K}_{i}^{m} \] \cr
  &\leq& C_w {l_m}^{w-1}\sum_{l=0}^{2{l_m}-1} \mathbb{E}\left[ \left( \sum_{j=0}^{\lfloor \frac{b_m - 6l_m}{2{l_m}} \rfloor - 1} \hat{\mathcal{X}}_{11,i+2{l_m}j+l,d} \right)^w \bigg| \mathcal{K}_{i}^{m}  \right] \cr
  &&+ C_w b_m^{w-1}  \sum_{l=0}^{b_m - 6l_m} \mathbb{E}\left[ \left(    \bar{\mathcal{X}}_{11,i+2{l_m}j+l,d} \right)^w  \big| \mathcal{K}_{i}^{m} \right]\cr
  &\leq& C_w {l_m}^{w-1}\sum_{l=0}^{2{l_m}-1} \mathbb{E}\left[ \left( \sum_{j=0}^{\lfloor \frac{b_m - 6l_m}{2{l_m}} \rfloor - 1} \hat{\mathcal{X}}_{11,i+2{l_m}j+l,d}^{2} \right)^{w/2} \bigg| \mathcal{K}_{i}^{m}  \right] \cr
  &&+ C_w b_m^{w-1}  \sum_{l=0}^{b_m - 6l_m} \mathbb{E}\left[ \left(    \bar{\mathcal{X}}_{11,i+2{l_m}j+l,d} \right)^w  \big| \mathcal{K}_{i}^{m} \right]\cr
  &\leq& C_w {l_m}^{w/2} b_m ^{w/2-1} \sum_{l=0}^{2{l_m}-1} \sum_{j=0}^{\lfloor \frac{b_m - 6l_m}{2{l_m}} \rfloor - 1} \mathbb{E}\left[ \left|  \hat{\mathcal{X}}_{11,i+2{l_m}j+l,d} \right|^{w} \big| \mathcal{K}_{i}^{m} \right] \cr
  &&+ C_w b_m^{w-1}  \sum_{l=0}^{b_m - 6l_m} \mathbb{E}\left[ \left|    \bar{\mathcal{X}}_{11,i+2{l_m}j+l,d} \right|^w  \big| \mathcal{K}_{i}^{m} \right]
  ,
\end{eqnarray}
where
\begin{eqnarray*}
  \bar{\mathcal{X}}_{11,i,d} = \mathbb{E}\left[ (\vartheta_{1,i}^{m})^{2} \mathcal{X}_{11,i,d} | \tilde{\mathcal{K}}_{i}^{m,{l_m}} \right] ,\quad  \hat{\mathcal{X}}_{11,i,d} = (\vartheta_{1,i}^{m})^{2} \mathcal{X}_{11,i,d} - \bar{\mathcal{X}}_{11,i,d} 
  ,
\end{eqnarray*}
and the first and third inequalities are due to Jensen's inequality.
Similar to proof of \eqref{mathT-CE-w-CE}, we can show that
\begin{equation}\label{X-w-CE}
  \mathbb{E}\left[ \left|\hat{\mathcal{X}}_{11,i,d}\right|^{w} \big| \mathcal{K}_{i}^{m}  \right] \leq C_w \varPsi_{i}^{m,2}
  .
\end{equation}
Further, we have for any $w \in \mathbb{N}$ and $\varepsilon >0$,
\begin{eqnarray}\label{X-CE-w-E}
  \mathbb{E}\left[ \left|\mathbb{E}\left[ \mathcal{X}_{11,i,d} | \tilde{\mathcal{K}}_{i}^{m,{l_m}} \right]\right|^{w}  \right] &\leq&  C_{w,\varepsilon} {l_m}^{-2v+2\varepsilon/\varsigma} \cr
  &\leq& C_{w,\varepsilon} \Delta_m ^{2\varsigma v - 2\varepsilon}
  ,
\end{eqnarray}
where the first inequality is due to Lemma \ref{gmeasurable-bound}(b).
Using \eqref{X-CE-w-E}, we have
\begin{eqnarray*}
  &&\mathbb{E}\left[ \left( \mathbb{E}\left[ \left|\mathbb{E}\left[ \mathcal{X}_{11,i,d} | \tilde{\mathcal{K}}_{i}^{m,{l_m}} \right]\right|^{w} \big| \mathcal{K}_{i}^{m} \right] - \mathbb{E}\left[ \left|\mathbb{E}\left[ \mathcal{X}_{11,i,d} | \tilde{\mathcal{K}}_{i}^{m,{l_m}} \right]\right|^{w}  \right] \right)^2  \right] \cr
  &\leq& \mathbb{E}\left[ \mathbb{E}\left[ \left|\mathbb{E}\left[ \mathcal{X}_{11,i,d} | \tilde{\mathcal{K}}_{i}^{m,{l_m}} \right]\right|^{2w}  \right] \right]
  ,
\end{eqnarray*}
and thus,
\begin{equation}\label{X-CE-w-CE}
  \mathbb{E}\left[ \left|\mathbb{E}\left[ \mathcal{X}_{11,i,d} | \tilde{\mathcal{K}}_{i}^{m,{l_m}} \right]\right|^{w} \big| \mathcal{K}_{i}^{m} \right] \leq C_w \varPsi_{i}^{m,2} \Delta_m ^{\varsigma v - \varepsilon} \quad \text{a.s.}
\end{equation}
Using \eqref{S1-w-jensen}, \eqref{X-w-CE}, and \eqref{X-CE-w-CE}, we have
\begin{equation}\label{S1-w}
  \mathbb{E}\left[ \left( \mathcal{S} _{11,i}^{m,1} (d) \right)^w \big| \mathcal{K}_{i}^{m}  \right] \leq C_{w} \varPsi_{i}^{m,2} (b_m ^{w/2} l_m ^{w/2} + b_m^w \Delta_m ^{\varsigma v - \varepsilon})
  ,
\end{equation}
Similar to the proof of \eqref{mathV-w-res}, using Burkholder-Davis-Gundy inequality, Jensen's inequality, and Lemma \ref{chi-property}(a), we can show that
\begin{equation}\label{mathS123-w}
  \mathbb{E}\left[ \left( \mathcal{S}_{11,i}^{m,z}(d) \right)^w | \mathcal{K}_{i}^{m} \right]  \leq C \varPsi_{i}^{m,2} b_m^{w/2}  l_m^{w/2} \quad \text{for any } w \in \mathbb{N} \text{ and } z \in \{2,3,4\}.
\end{equation}
\eqref{S1-w} and \eqref{mathS123-w} imply that
\begin{equation}\label{mathV4-w}
  \mathbb{E}\left[ \left( \mathcal{V}_{11,i}^{m,4}(d) \right)^w | \mathcal{K}_{i}^{m}  \right] \leq C_{w,\varepsilon} \varPsi_{i}^{m,2} (b_m ^{w/2} l_m^{w/2} + b_m^w \Delta_m ^{\varsigma v - \varepsilon}) \text{ for any } \varepsilon >0
  .
\end{equation}
By \eqref{mathV23-2w} and \eqref{mathV4-w}, we have for any $\varepsilon > 0$,
\begin{eqnarray*}
  \mathbb{E}\left[ \left( V_{11}^{'m} \right)^w | \mathcal{K}_{i}^{m}   \right] &\leq& (k'_m)^{w-1}  \sum_{d=-k_m'}^{k_m'} \left( \phi _{d}^{m} \right)^2   \mathbb{E}\left[ \left( r_{11}(|d|) \mathcal{U} _{11,i}^{m} - U _{m,i}^{Y_{1}^{c} Y_{1}^{c}} (|d|) \right)^w  | \mathcal{K}_{i}^{m}  \right] \\
  &\leq& C_{w} (k'_m)^{w-1} \sum_{d=-k_m'}^{k_m'} \sum_{z=1}^{4} \mathbb{E}\left[ \left( \mathcal{V}_{11,i}^{m,z}(d) \right)^w | \mathcal{K}_{i}^{m}   \right] \\
  &\leq& C_{w,\varepsilon} \varPsi_{i}^{m,2} (k'_m)^{w} (b_m ^{w/2} l_m^{w/2} + b_m^w \Delta_m ^{\varsigma v -\varepsilon}) 
  ,
\end{eqnarray*}
where the first and second inequalities are due to Jensen's inequality and the fact that $|\phi _{d}^{m} | \leq C$ for any $d \in \mathbb{Z}$.
Thus, we have for any $\varepsilon > 0$,
\begin{eqnarray*}
  \mathbb{E}[ \left| \xi_{11,i}^{m,2} \right|^{w} | \mathcal{K}_{i}^{m} ] &\leq& C_w\frac{ \mathbb{E}\left[ (V_{11,i}^{m})^{w} | \mathcal{K}_{i}^{m} \right] + \mathbb{E}\left[ (V_{11,i}^{'m})^{w} | \mathcal{K}_{i}^{m} \right]}{((b_m- 2 k_m)\Delta_m k_m^2 \psi_0)^{w}}  \cr
  &\leq& C_{w,\varepsilon} ({k'_m}^{-w(v-1)} + \varPsi_{i}^{m,2} \Delta_m ^{\frac{(\kappa-\varsigma-2\tau)w}{2}  \land (\varsigma v - \tau w - \varepsilon)}) 
  \quad \text{a.s.}
\end{eqnarray*}
Similarly, we can show the statement for the other cases of $x$ and $y$.
$\blacksquare$

\textbf{Proof of Lemma \ref{lemma:zeta}.}
Consider \eqref{zeta-moment}.
Let 
\begin{equation}\label{eq:zeta-decompose}
  \hat{X}_{1x,i}^{c,m} = \tilde{X}_{1,i}^{c,m} \tilde{X}_{x,i}^{c,m}  - \breve{C}_{1x,i}^{m} ,\quad \hat{\epsilon}_{1x,i}^{m} = \tilde{\epsilon}_{1,i}^{m} \tilde{\epsilon}_{x,i}^{m} - \tilde{\Gamma}^{m}_{1x,i} ,\quad \hat{X^{c}\epsilon}^{m}_{xy,i} = \tilde{X}_{x,i}^{c,m} \tilde{\epsilon}_{y,i}^{m}
  .
\end{equation}
Then, we have $ \zeta(p)_{1x,i}^{m} = \sum_{l=i}^{i+pk_m-1}  \zeta_{1x,l}^{m} = \sum_{l=i}^{i+pk_m-1} \hat{X}_{1x,l}^{c,m} + \hat{\epsilon}_{1x,l}^{m} + \hat{X^{c}\epsilon}^{m}_{1x,l} + \hat{X^{c}\epsilon}^{m}_{x1,l} $.
Similar to the proof of (A.19) and (A.30) of \citet{jacod2019estimating}, we have
\begin{equation}\label{decomposed-elements-of-zeta-1}
  \mathbb{E}\left[ (\hat{X}_{1x,i}^{c,m})^{w} | \mathcal{K}_{i}^{m} \right] \leq C_w \Delta_m ^{w/2} \quad \text{and} \quad \mathbb{E}\left[ (\hat{\epsilon}_{1x,i}^{m})^{w} | \mathcal{K}_{i}^{m}  \right] \leq C \varPsi_{i}^{m,2} \Delta_m ^{\frac{w \land v}{2} } \quad \text{a.s.}
\end{equation}
Similar to the proof of the second part of (A.30) of \citet{jacod2019estimating}, we have
\begin{eqnarray*}
  (\hat{X^{m}\epsilon}^{m}_{xy,i} )^{w} = (\tilde{X}_{x,i}^{c,m})^{w} (\vartheta_{y,i}^{m})^{w} \mathbb{E}\left[ (\tilde{\chi}_{y,0}^{m})^{w} \right] + (\tilde{X}_{x,i}^{c,m})^{w} \left( (\tilde{\epsilon}_{y,i}^{m} )^{w} - (\vartheta_{y,i}^{m})^{w} \mathbb{E}\left[ (\tilde{\chi}_{y,0}^{m})^{w} \right]  \right) 
  ,
\end{eqnarray*}
and thus using (A.23), (A.25), and (A.28) of \citet{jacod2019estimating}, we have
\begin{equation}\label{decomposed-elements-of-zeta-2}
  \mathbb{E}\left[ (\hat{X^c\epsilon}_{xy,i}^{m})^{w} | \mathcal{K}_{i}^{m}  \right] \leq
  \begin{cases}
    C \varPsi_{i}^{m,2} \Delta_m ^{w/2}, & \text{ if } w \leq v+2 \\
    C \varPsi_{i}^{m,2} \Delta_m ^{\frac{v+w+2}{4}}, & \text{ if } v+2 < w \leq 3v-2 \\
    C \varPsi_{i}^{m,2} \Delta_m ^{v}, & \text{ if } w > 3v-2 \\
  \end{cases}
  .
\end{equation}
Thus, we have
\begin{eqnarray*}
  \mathbb{E}\left[ \left| \zeta(p)_{1x,i}^{m} \right|^w | \mathcal{K}_{i}^{m}  \right] &\leq& C (pk_m)^{w-1} \sum_{l=i}^{i+pk_m-1} \Big( \mathbb{E}\left[ (\hat{X}_{1x,l}^{c,m})^{w} | \mathcal{K}_{i}^{m} \right] + \mathbb{E}\left[ (\hat{\epsilon}_{1x,l}^{m})^{w} | \mathcal{K}_{i}^{m} \right]  \cr
  &&\qquad\qquad\qquad + \mathbb{E}\left[ (\hat{X^{c}\epsilon}^{m}_{1x,l})^{w} | \mathcal{K}_{i}^{m} \right] + \mathbb{E}\left[ (\hat{X^{c}\epsilon}^{m}_{x1,l})^{w} | \mathcal{K}_{i}^{m}  \right] \Big)   \\
  &\leq&  C \varPsi_{i}^{m,2}  p^w \Delta_m ^{\frac{v-w}{2} \land 0}
  ,
\end{eqnarray*}
where the first inequality is due to Jensen's inequality.
Using Jensen's inequality, we have
\begin{eqnarray*}
  && \left|\mathbb{E}\left[ \zeta(p)_{11,i}^{m} | \mathcal{K}_{i}^{m} \right]\right| ^{w} \cr
  &=&  \left( \sum_{j=0}^{pk_m - 1} \mathbb{E}\left[ \hat{X}_{11,i+j}^{c,m} | \mathcal{K}_{i}^{m} \right]
  + \mathbb{E}\left[ \hat{\epsilon}_{11,i+j}^{m} | \mathcal{K}_{i}^{m} \right]
  + 2 \mathbb{E}\left[ \tilde{X}_{1,i+j}^{c,m} \tilde{\epsilon}_{1,i+j}^{m}  | \mathcal{K}_{i}^{m} \right] \right)^w \cr
  &\leq&  C (p k_m)^{w-1} \left( \sum_{j=0}^{pk_m - 1} \left|\mathbb{E}\left[ \hat{X}_{11,i+j}^{c,m} | \mathcal{K}_{i}^{m} \right]\right| ^{w}
  +  \left|\mathbb{E}\left[ \hat{\epsilon}_{11,i+j}^{m} | \mathcal{K}_{i}^{m} \right]\right| ^{w}
  +  \left|\mathbb{E}\left[ \tilde{X}_{1,i+j}^{c,m} \tilde{\epsilon}_{1,i+j}^{m} | \mathcal{K}_{i}^{m} \right]\right| ^{w} \right)
  .
\end{eqnarray*}
By (A.19) in \citet{jacod2019estimating}, we have
\begin{eqnarray*}
  \left|\mathbb{E}\left[ \hat{X}_{11,i+j}^{c,m} | \mathcal{K}_{i}^{m} \right]\right| ^{w} &\leq&  C \Delta_m
  .
\end{eqnarray*}
Using Jensen's inequality, we have
\begin{eqnarray*}
  &&\left|\mathbb{E}\left[ \tilde{X}_{1,i+j}^{c,m} \tilde{\epsilon}_{1,i+j}^{m} | \mathcal{K}_{i}^{m} \right]\right|^{w}  \cr
  &\leq& (k_m)^{2w-2} \sum_{l_1, l_2=0}^{k_m - 1} \left| \mathbb{E}\left[ h_{l_1}^{m} X_{1,i+j+l_1}^{c,m} h_{l_2}^{m} \vartheta_{1,i+j+l_2}^{m} \chi_{1,i+j+l_2} | \mathcal{K}_{i}^{m} \right]\right|^{w}  \cr
  &=& (k_m)^{2w-2} \sum_{l_1, l_2=0}^{k_m - 1}  \left|\mathbb{E}\left[ h_{l_1}^{m} X_{1,i+j+l_1}^{c,m} h_{l_2}^{m} \vartheta_{1,i+j+l_2}^{m} | \mathcal{F}_{i}^{m} \right] \mathbb{E}\left[ \chi_{1,i+j+l_2} | \mathcal{G}_{i-k_m}^{m} \right]\right|^{w}  \cr
  &\leq& C_{w,\varepsilon} \varPsi_{i}^{m,1}  \Delta_m ^{v - \varepsilon}
  ,
\end{eqnarray*}
for any $\varepsilon > 0$.
Similarly, we have
\begin{eqnarray*}
  && \left|\mathbb{E}\left[ \hat{\epsilon}_{11,i+j}^{m} | \mathcal{K}_{i}^{m} \right]\right| ^{w} \cr
  &\leq& (k_m)^{2w-2} \sum_{l_1, l_2 =0}^{k_m - 1}  \left|\mathbb{E}\left[ h_{l_1}^{m} h_{l_2}^{m} \vartheta_{1,i + j +l_1}^{m} \vartheta_{1,i + j +l_2}^{m} | \mathcal{F}_{i}^{m} \right] \mathbb{E}\left[ \chi_{1,i+j+l_1}\chi_{1,i+j+l_2} - r_{11}(l_1,l_2) | \mathcal{G}_{i-k_m}^{m} \right]\right| ^{w} \cr
  &\leq& C_{w,\varepsilon} \varPsi_{i}^{m,1} \Delta_m ^{v-\varepsilon}
  ,
\end{eqnarray*}
for any $\varepsilon > 0$.
Thus, we have
\begin{equation*}
  \left|\mathbb{E}\left[ \zeta(p)_{11,i}^{m} | \mathcal{K}_{i}^{m} \right]\right| ^{w} \leq C \varPsi_{i}^{m,1}  p^w \Delta_m ^{\frac{(2v-w-\varepsilon) \land w}{2} } \quad \text{and} \quad \left|\mathbb{E}\left[ \zeta(p)_{11,i}^{m} | \mathcal{K}_{i}^{m} \right]\right| ^{w} \leq C \varPsi_{i}^{m,2}  p^w \Delta_m ^{\frac{(v-w-\varepsilon) \land w}{2} }
  .
\end{equation*}

For the third part \eqref{zeta-moment}, we have
\begin{eqnarray*}
  && \left| \mathbb{E}\left[ \left(  \zeta(p)_{11,i}^{m} \right)^2  - \varXi\left(  p \right)_{1,1,i}^{m} | \mathcal{K}_{i}^{m}  \right] \right| \\
  && \leq \left| \mathbb{E}\left[ \left(  \zeta(p)_{11,i}^{m} \right)^2 - 4 (\bSigma_{11,i}^{c,m})^2 \rho(p,1)_{i}^{m} - 4 \bSigma_{11,i}^{c,m} (\vartheta _{1,i}^{m} )^2 \rho(p,3)_{i}^{m} - (\vartheta _{1,i}^{m}) ^4 \rho(p,2)_{i}^{m}  | \mathcal{K}_{i}^{m}   \right] \right| \\
  &&\quad + \left| \mathbb{E}\left[ 4 \left(\bSigma_{11,i}^{c,m}\right)^2 \rho(p,1)_{i}^{m}  - 4 \left( \bSigma_{11,i}^{c,m} \right)^{2}   k_m^4 \Delta_m^2\left(p\Phi_{00}-\bar{\Phi}_{00}\right) \right] \right| \\
  &&\quad + \left| \mathbb{E}\left[ 4 \bSigma_{11,i}^{c,m} \left(\vartheta _{1,i}^{m} \right)^2 \rho(p,3)_{i}^{m}  -  8 \bSigma_{11,i}^{c,m}  \left(  \vartheta_{1,i}^{m} \right)^2 R k_m^2 \Delta_m \left(p\Phi_{01}-\bar{\Phi}_{01}\right) \right] \right| \\
  &&\quad + \left| \mathbb{E}\left[ (\vartheta _{1,i}^{m}) ^4 \rho(p,2)_{i}^{m} -  4 \left( \vartheta_{1,i}^{m} \right)^{4} R^2 \left( p \Phi_{11} - \bar{\Phi}_{11} \right)  \right] \right| \\
  && \leq C p^2 \varPsi_{i,p}^{m,2} \Delta_m^{1/4}
  \text{ a.s.},
\end{eqnarray*}
where $\rho(p,1)_{i}^{m}, \rho(p,2)_{i}^{m}$, and $\rho(p,3)_{i}^{m}$ are defined in Lemma A.8 of \citet{jacod2019estimating} and the second inequality is due to Lemmas A.8--10 of \citet{jacod2019estimating} and the bounded $\bSigma _{11,i}^{c,m}$ and $\vartheta _{1,i}^{m}$.
Similarly, we can show the statement for the other cases of $x$ and $y$.

Consider \eqref{zeta-p1p2}.
Now, we only consider the case $x=y=1$, since we can similarly show the other cases.
Simple algebra shows that
\begin{eqnarray}\label{zetazeta-decomp}
  \mathbb{E}\left[ \zeta(p_1)_{11,i_1}^{m} \zeta(p_2)_{11,i_2}^{m} | \mathcal{K}_{i_1}^{m} \right] &=&  \sum_{l_1=i_1}^{i_1+p_1 k_m-1}  \sum_{l_2=i_2}^{i_2+p_2 k_m-1}  \mathbb{E}\left[ \hat{X}_{11,l_2}^{c,m} \hat{X}_{11,l_1}^{c,m} | \mathcal{K}_{i_1}^{m} \right] + \mathbb{E}\left[ \hat{\epsilon}_{11,l_2}^{m} \hat{X}_{11,l_1}^{c,m} | \mathcal{K}_{i_1}^{m} \right] \nonumber\\
  && + 2 \mathbb{E}\left[  \tilde{X}_{1,l_2}^{c,m} \tilde{\epsilon}_{1,l_2}^{m} \hat{X}_{11,l_1}^{c,m} | \mathcal{K}_{i_1}^{m} \right] + \mathbb{E}\left[ \hat{X}_{11,l_2}^{c,m} \hat{\epsilon}_{11,l_1}^{m} | \mathcal{K}_{i_1}^{m} \right] \nonumber \\
  && + \mathbb{E}\left[ \hat{\epsilon}_{11,l_2}^{m} \hat{\epsilon}_{11,l_1}^{m} | \mathcal{K}_{i_1}^{m} \right] + 2 \mathbb{E}\left[  \tilde{X}_{1,l_2}^{c,m} \tilde{\epsilon}_{1,l_2}^{m} \hat{\epsilon}_{11,l_1}^{m} | \mathcal{K}_{i_1}^{m} \right] \nonumber \\
  && + 2 \mathbb{E}\left[ \hat{X}_{11,l_2}^{c,m} \tilde{X}_{1,l_1}^{c,m} \tilde{\epsilon}_{1,l_1}^{m} | \mathcal{K}_{i_1}^{m} \right] + 2 \mathbb{E}\left[ \hat{\epsilon}_{11,l_2}^{m} \tilde{X}_{1,l_1}^{c,m} \tilde{\epsilon}_{1,l_1}^{m} | \mathcal{K}_{i_1}^{m} \right] \nonumber \\
  && + 4 \mathbb{E}\left[  \tilde{X}_{1,i_2}^{c,m} \tilde{\epsilon}_{1,l_2}^{m} \tilde{X}_{1,l_1}^{c,m} \tilde{\epsilon}_{1,l_1}^{m} | \mathcal{K}_{i_1}^{m} \right]
  \text{ a.s.}
\end{eqnarray}
For the first term of the summand on the right-hand side of \eqref{zetazeta-decomp}, we have %
\begin{eqnarray*}
  \left|\mathbb{E}\left[ \hat{X}_{11,l_2}^{c,m} \hat{X}_{11,l_1}^{c,m} | \mathcal{K}_{i_1}^{m} \right]\right|  &=& \left|\mathbb{E}\left[ \mathbb{E}\left[ \hat{X}_{11,l_2}^{c,m} | \mathcal{K}_{l_2}^{m} \right] \hat{X}_{11,l_1}^{c,m} | \mathcal{K}_{i_1}^{m} \right]\right| \\
  &\leq& C \Delta_m \mathbb{E}\left[ \left|\hat{X}_{11,l_1}^{c,m}\right|  | \mathcal{K}_{i_1}^{m} \right] \\
  &\leq& C \Delta_m ^{3/2}
  \text{ a.s.}
\end{eqnarray*}
where the equality is due to tower property and the first and second inequalities are due to Lemma A.2 of \citet{jacod2019estimating}.
For the second term of the summand on the right-hand side of \eqref{zetazeta-decomp}, we have
\begin{eqnarray*}
  \left|\mathbb{E}\left[ \hat{\epsilon}_{11,l_2}^{m} \hat{X}_{11,l_1}^{c,m} | \mathcal{K}_{i_1}^{m} \right]\right| &=&  \left| \mathbb{E}\left[  \mathbb{E}\left[ \hat{\epsilon}_{11,l_2}^{m} | \mathcal{K}_{l_2}^{m} \right] \hat{X}_{11,l_1}^{c,m} | \mathcal{K}_{i_1}^{m} \right]\right| \\
  &\leq&   \left| \mathbb{E}\left[  C \varPsi_{i}^{m,2} \Delta_m  \left|\hat{X}_{11,l_1}^{c,m}\right|  | \mathcal{K}_{i_1}^{m} \right]\right| \\
  &=&  C \varPsi_{i}^{m,2} \Delta_m   \left| \mathbb{E}\left[  \left|\hat{X}_{11,l_1}^{c,m}\right|  | \mathcal{K}_{i_1}^{m} \right]\right| \\
  &\leq& C \varPsi_{i}^{m,2} \Delta_m ^{3/2}
  \text{ a.s.},
\end{eqnarray*}
where the first and second equalities are due to tower property and the independence of $\varPsi_{i,2}^{m}$ and $\hat{X}_{11,l_1}^{c,m}$, and the first and second inequalities are due to Lemmas A.2 and A.5 of \citet{jacod2019estimating}.
Similarly, we can obtain the following inequalities:
\begin{eqnarray*}
  && \left|\mathbb{E}\left[  \tilde{X}_{1,l_2}^{c,m} \tilde{\epsilon}_{1,l_2}^{m} \hat{X}_{11,l_1}^{c,m} | \mathcal{K}_{i_1}^{m} \right]\right| \leq C \varPsi_{i}^{m,2} \Delta_m ^{3/4 + v/2}, \quad \left|\mathbb{E}\left[ \hat{X}_{11,l_2}^{c,m} \hat{\epsilon}_{11,l_1}^{m} | \mathcal{K}_{i_1}^{m} \right]\right| \leq C \varPsi_{i}^{m,2} \Delta_m ^{3/2},   \nonumber \\
  && \left|\mathbb{E}\left[ \hat{\epsilon}_{11,l_2}^{m} \hat{\epsilon}_{11,l_1}^{m} | \mathcal{K}_{i_1}^{m} \right]\right|  \leq C \varPsi_{i}^{m,2} \Delta_m ^{3/2}, \quad  \left|\mathbb{E}\left[  \tilde{X}_{1,l_2}^{c,m} \tilde{\epsilon}_{1,l_2}^{m} \hat{\epsilon}_{11,l_1}^{m} | \mathcal{K}_{i_1}^{m} \right]\right|  \leq C \varPsi_{i}^{m,2} \Delta_m ^{3/4 + v/2}, \nonumber \\
  && \left|\mathbb{E}\left[ \hat{X}_{11,l_2}^{c,m} \tilde{X}_{1,l_1}^{c,m} \tilde{\epsilon}_{1,l_1}^{m} | \mathcal{K}_{i_1}^{m} \right]\right|  \leq C \varPsi_{i}^{m,2} \Delta_m ^{3/2}, \quad  \left|\mathbb{E}\left[ \hat{\epsilon}_{11,l_2}^{m} \tilde{X}_{1,l_1}^{c,m} \tilde{\epsilon}_{1,l_1}^{m} | \mathcal{K}_{i_1}^{m} \right]\right|  \leq C \varPsi_{i}^{m,2} \Delta_m ^{3/2}, \nonumber \\
  && \left|\mathbb{E}\left[  \tilde{X}_{1,i_2}^{c,m} \tilde{\epsilon}_{1,l_2}^{m} \tilde{X}_{1,l_1}^{c,m} \tilde{\epsilon}_{1,l_1}^{m} | \mathcal{K}_{i_1}^{m} \right]\right|  \leq C \varPsi_{i}^{m,2} \Delta_m ^{3/4 + v/2} \text{ a.s.}
\end{eqnarray*}
Thus, in view of \eqref{zetazeta-decomp}, we have
\begin{equation*}
  \mathbb{E}\left[ \zeta(p_1)_{11,i_1}^{m} \zeta(p_2)_{11,i_2}^{m} | \mathcal{K}_{i_1}^{m} \right] \leq C \varPsi_{i}^{m,2} p_1 p_2 \Delta_m ^{1/2} \text{ a.s.}
\end{equation*}
$\blacksquare$

\textbf{Proof of Lemma \ref{lemma:M}.}
We only consider the case $x=1$, since we can similarly show the other case.
For simplicity, we denote $i_{l,j}^{m} = i + (p_m+2) k_m l + j$.
By the second part of Lemma \ref{lemma:zeta}, we have
\begin{eqnarray*}
  \left|\mathbb{E}\left[M(p)_{11,i}^{m} | \mathcal{K}_{i}^{m}  \right]\right|^{w}  &=&  \left|\frac{1}{(b_m - 2 k_m)\Delta_m k_m \psi_0} \sum_{l=0}^{L(m,p)-1} \mathbb{E}\left[ \mathbb{E}\left[ \zeta(p)^m_{11,i_{l,0}^{m}} | \mathcal{K}_{i_{l,0}^{m}}^{m}  \right] | \mathcal{K}_{i}^{m} \right]\right|^{w} \cr
  &\leq& C_{w} (b_m ^{-1} \Delta_m ^{-1/2})^w L(m,p)^{w-1} \sum_{l=0}^{L(m,p)-1} \left|\mathbb{E}\left[ \zeta(p)^m_{11,i_{l,0}^{m}} | \mathcal{K}_{i}^{m}  \right]\right|^{w} \cr
  &\leq& C \varPsi_{i}^{m,2} \Delta_m ^{\frac{(v-w-\varepsilon) \land w}{2} } \text{ a.s.},
\end{eqnarray*}
where the first and second inequalities are due to Jensen's inequality and the first part of Lemma \ref{lemma:zeta}, respectively.
Similarly, we can show the second part of \eqref{M-moment}.
For the third part of \eqref{M-moment}, we have
\begin{eqnarray}\label{M-w-CE-decomposed}
  && \mathbb{E}\left[ \left|M(p)_{11,i}^{m}\right|^{w}  | \mathcal{K}_{i}^{m}  \right] \cr
  &\leq& C_w \mathbb{E}\left[ \left( \sum_{j=0}^{L(m,p)-1} \hat{\eta}(p)_{xy,j}^{m,i} \right)^{w} \Bigg| \mathcal{K}_{i}^{m}  \right] + C_w \mathbb{E}\left[ \left( \sum_{j=0}^{L(m,p)-1} \bar{\eta}(p)_{xy,j}^{m,i} \right)^{w} \Bigg| \mathcal{K}_{i}^{m}  \right] \cr
  &\leq& C_w \mathbb{E}\left[ \left( \sum_{j=0}^{L(m,p)-1} (\hat{\eta}(p)_{xy,j}^{m,i})^{2} \right)^{w/2} \Bigg| \mathcal{K}_{i}^{m}  \right] + C_w \mathbb{E}\left[ \left( \sum_{j=0}^{L(m,p)-1} \bar{\eta}(p)_{xy,j}^{m,i} \right)^{w} \Bigg| \mathcal{K}_{i}^{m}  \right] \cr
  &\leq& C_w L(m,p)^{w/2-1}  \sum_{j=0}^{L(m,p)-1} \mathbb{E}\left[   |\hat{\eta}(p)_{xy,j}^{m,i}|^{w}  | \mathcal{K}_{i}^{m}  \right] \cr
  &&+ C_w L(m,p)^{w-1}  \sum_{j=0}^{L(m,p)-1} \mathbb{E}\left[ | \bar{\eta}(p)_{xy,j}^{m,i} |^{w} | \mathcal{K}_{i}^{m}  \right]
  ,
\end{eqnarray}
where the first and third inequalities are due to Jensen's inequality, and the second one is due to Burkholder-Davis-Gundy inequality.
Due to the fact that
\begin{eqnarray*}
  \mathbb{E}\left[ \left( \mathbb{E}\left[   |\bar{\eta}(p)_{xy,j}^{m,i}|^{w}  | \mathcal{K}_{i}^{m}  \right] - \mathbb{E}\left[   |\bar{\eta}(p)_{xy,j}^{m,i}|^{w}    \right] \right)^2  \right] 
  \leq \mathbb{E}\left[   |\bar{\eta}(p)_{xy,j}^{m,i}|^{2w}    \right]
  ,
\end{eqnarray*}
we have
\begin{eqnarray}\label{eta-CE-w-CE}
  && \mathbb{E}\left[   |\bar{\eta}(p)_{xy,j}^{m,i}|^{w}  | \mathcal{K}_{i}^{m}  \right] \cr
  &\leq& \mathbb{E}\left[   |\bar{\eta}(p)_{xy,j}^{m,i}|^{w}    \right] + \varPsi_{i}^{m,2} \mathbb{E}\left[   |\bar{\eta}(p)_{xy,j}^{m,i}|^{2w}    \right]^{1/2} \cr
  &\leq& C_w b_m ^{-w} \Delta_m ^{-w/2} \left( \mathbb{E}\left[\varPsi_{i}^{m,2} p^{w} \Delta_m ^{\frac{v-w-\varepsilon}{2} \land \frac{w}{2}  } \right] + \varPsi_{i}^{m,2}  \mathbb{E}\left[(\varPsi_{i}^{m,2} p^{w} \Delta_m ^{\frac{v-w-\varepsilon}{2} \land \frac{w}{2}  })^{2} \right]^{1/2} \right) \cr
  &\leq& C_w \varPsi_{i}^{m,2} p^{w}  b_m ^{-w} \Delta_m ^{(\frac{1}{2} v - w -\varepsilon) \land 0 } \quad\text{a.s.} ,%
\end{eqnarray}
where the second inequality of \eqref{eta-CE-w-CE}  is due to Lemma \ref{lemma:zeta}.
Using \eqref{M-w-CE-decomposed}, \eqref{eta-CE-w-CE}, and Lemma \ref{lemma:zeta}, we have
\begin{eqnarray*}
  &&\mathbb{E}\left[ \left|M(p)_{11,i}^{m}\right|^{w}  | \mathcal{K}_{i}^{m}  \right] \leq C_w \varPsi_{i}^{m,2} (p^{w/2} b_m ^{-w/2} \Delta_m ^{(\frac{v}{2} - \frac{3}{4} w) \land -\frac{w}{4} } + \Delta_m ^{\frac{(v-w-\varepsilon) \land w}{2} }) .
\end{eqnarray*}
Further, we have for $w \geq 2$,
\begin{eqnarray}\label{eta-CE-w-E}
  \mathbb{E}\left[   |\bar{\eta}(p)_{xy,j}^{m,i}|^{w}  \right] &=& \mathbb{E}\left[   (|\bar{\eta}(p)_{xy,j}^{m,i}|^{w/2})^{2}  \right] \cr
  &\leq& C_w b_m ^{-w} \Delta_m ^{-w/2} \mathbb{E}\left[ (\varPsi_{i}^{m,2} p^{w/2} \Delta_m ^{\frac{(2v-w - \varepsilon) \land w}{4} })^{2} \right] \cr
  &\leq& C_w p^{w} b_m ^{-w}  \Delta_m ^{(v-w-\varepsilon) \land 0}
  ,
\end{eqnarray}
where the first inequality is due to Lemma \ref{lemma:zeta}.
Thus, we have
\begin{equation*}
  \mathbb{E}\left[ \left|M(p)_{11,i}^{m}\right|^{w}    \right] \leq C_w (p^{w/2} b_m ^{-w/2} \Delta_m ^{(\frac{v}{2}  - \frac{3}{4} w) \land -\frac{w}{4} } + \Delta_m ^{ (v - \frac{w}{2} -\varepsilon) \land \frac{w}{2}  })
  .
\end{equation*}
Similarly, we can bound the last term of \eqref{M-moment}.
$\blacksquare$

\textbf{Proof of Lemma \ref{lemma:e}.}
Since \eqref{e-moment} is a trivial consequence of Lemmas \ref{negligible-xi} and \ref{lemma:M} in view of \eqref{SpotErrorDecomp}, so we consider \eqref{e-debiased}.

We can decompose $\left(e_{11,i}^{m}\right)^2$ as follows:
\begin{eqnarray}\label{e-decomp}
  \left(e_{11,i}^{m}\right)^2 &=&  \left ({M}(p_m)_{11,i}^{m} + {M}'(p_m)_{11,i}^{m} + \xi_{11,i}^{m,1} + \xi_{11,i}^{m,2}\right )^2 \nonumber\\
  &=& \left( {M}(p_m)_{11,i}^{m} \right) ^2 + 2 {M}(p_m)_{11,i}^{m} {M}'(p_m)_{11,i}^{m} + \left( {M}'(p_m)_{11,i}^{m} \right)  ^2 + \left( \xi_{11,i}^{m,1} \right) ^2 + \left( \xi_{11,i}^{m,2} \right) ^2 \nonumber\\
  && + 2 \xi_{11,i}^{m,2} \left( {M}(p_m)_{11,i}^{m} + {M}'(p_m)_{11,i}^{m}  + \xi_{11,i}^{m,1} \right) \nonumber\\
  && + 2 \xi_{11,i}^{m,1} \left( {M}(p_m)_{11,i}^{m} + {M}'(p_m)_{11,i}^{m} \right)
  ,
\end{eqnarray}
where $p_m$ is a sequence of integers that satisfies $p_m \asymp \Delta_m ^{- \iota }$ and $\iota \in (\frac{3}{2} -2\kappa, (\frac{1}{2} \kappa -\frac{1}{4}) \land (2\kappa -3 \tau - 1))$. %
Let $\zeta(p_m)_{11,i}^{m,l} = \zeta(p_m)_{11,i+l(p_m+2)k_m}^{m}$.
We have for some $\varepsilon > 0$ 
\begin{align}\label{decompM}
  \mathbb{E}\left[\left( {M}(p_m)_{11,i}^{m} \right) ^2 | \mathcal{K}_{i}^{m} \right] =& \mathbb{E}\left[ \left( \frac{1}{(b_m - 2 k_m)\Delta_m k_m \psi_0} \sum_{l=0}^{L(m,p_m)-1} \zeta(p_m)^m_{11,i+l(p_m+2)k_m} \right) ^2 | \mathcal{K}_{i}^{m}\right]  \nonumber\\
  \leq& \left( \frac{1}{(b_m - 2 k_m)\Delta_m k_m \psi_0} \right)^2 \sum_{l=0}^{L(m,p_m)-1} \mathbb{E}\left[\left( \zeta(p_m)^{m,l}_{11,i} \right) ^{2} | \mathcal{K}_{i}^{m} \right] \nonumber \\
  &  + C b_m ^{-2} \Delta_{m}^{-1}  \sum_{l \neq l'}^{L(m,p_m)-1} \mathbb{E}\left[ \zeta(p_m)^{m,l}_{11,i} \zeta(p_m)^{m,l'}_{11,i}  | \mathcal{K}_{i}^{m} \right] \nonumber\\
  \leq& \left( \frac{1}{(b_m - 2 k_m)\Delta_m k_m \psi_0} \right)^2 \sum_{l=0}^{L(m,p_m)-1} \mathbb{E}\left[\left( \zeta(p_m)^{m,l}_{11,i} \right) ^{2} |\mathcal{K}_{i}^{m} \right] \nonumber \\
  & + C \varPsi_{i}^{m,2} \Delta_m ^{1/2}   \nonumber \\
  \leq& \left( \frac{1}{(b_m - 2 k_m)\Delta_m k_m \psi_0} \right)^2 \sum_{l=0}^{L(m,p_m)-1} \mathbb{E}\left[\left( \zeta(p_m)^{m,l}_{11,i} \right) ^{2} | \mathcal{K}_{i}^{m} \right] \nonumber\\
  & + C \varPsi_{i}^{m,2} \Delta_m ^{\frac{1}{4} + \varepsilon} \text{ a.s.},
\end{align}
where the second inequality is due to Lemma \ref{lemma:zeta}.
Let $\zeta(2;p_m)^{m,l}_{11,i} = \zeta(2)^m_{11,i+l(p_m+2)k_m + p_mk_m}$.
We have
\begin{align}\label{MM'}
  & \quad \left|\mathbb{E}\left[ {M}(p_m)_{11,i}^{m} {M}'(p_m)_{11,i}^{m} | \mathcal{K}_{i}^{m}  \right]\right|  \nonumber \\
  & \leq C b_m ^{-2} \Delta_m^{-1} \left| \mathbb{E}\left[ \sum_{l=0}^{L(m,p_m)-1} \zeta(p_m)^{m,l}_{11,i} \sum_{l=0}^{L(m,p_m)-1} \zeta(2;p_m)^{m,l}_{11,i} | \mathcal{K}_{i}^{m}  \right] \right| \nonumber \\
  & \quad + C b_m ^{-2} \Delta_m^{-1} \left| \mathbb{E}\left[ \sum_{l=0}^{L(m,p_m)-1} \zeta(p_m)^{m,l}_{11,i} \sum_{l=L(m,p_m)(p_m+2)k_m+1}^{b_m-k_m} \zeta^m_{11,i+l} | \mathcal{K}_{i}^{m}  \right] \right| \nonumber\\
  & \leq C b_m ^{-2} \Delta_m^{-1} \left| \mathbb{E}\left[ \sum_{l=0}^{L(m,p_m)-1} \zeta(p_m)^{m,l}_{11,i} \zeta(2;p_m)^{m,l}_{11,i} | \mathcal{K}_{i}^{m} \right] \right| \nonumber\\
  & \quad + C b_m ^{-2} \Delta_m^{-1} \left| \mathbb{E}\left[ \sum_{l=1}^{L(m,p_m)-1} \zeta(p_m)^{m,l}_{11,i} \zeta(2;p_m)^{m,l-1}_{11,i} | \mathcal{K}_{i}^{m} \right] \right|\nonumber \\
  & \quad + C b_m ^{-2} \Delta_m^{-1} \left| \mathbb{E}\left[ \sum_{l-l' \notin \{0,1\} }^{L(m,p_m)-1} \zeta(p_m)^{m,l}_{11,i} \zeta(2;p_m)^{m,l'}_{11,i} | \mathcal{K}_{i}^{m} \right] \right| \nonumber \\
  & \quad + C b_m ^{-2} \Delta_m^{-1} \left| \mathbb{E}\left[ \sum_{l=0}^{L(m,p_m)-1} \zeta(p_m)^{m,l}_{11,i} \sum_{l=L(m,p_m)(p_m+2)k_m+1}^{b_m-k_m} \zeta^m_{11,i+l} | \mathcal{K}_{i}^{m} \right]  \right| \text{a.s.},
\end{align}
where the second inequality is due to triangular inequality.
For the first term on the right-hand side of \eqref{MM'}, we have
\begin{align*}
  & \quad \left| \mathbb{E}\left[ \sum_{l=0}^{L(m,p_m)-1} \zeta(p_m)^{m,l}_{11,i} \zeta(2;p_m)^{m,l}_{11,i} | \mathcal{K}_{i}^{m} \right] \right| \\
  & =  \left| \mathbb{E}\left[ \sum_{l=0}^{L(m,p_m)-1} \left( \zeta(p_m-2)^m_{11,i_{l,0}^{m}} + \zeta(2)^m_{11,i_{l,(p_m-2)k_m}^{m}} \right)   \zeta(2;p_m)^{m,l}_{11,i} | \mathcal{K}_{i}^{m} \right] \right| \\
  & \leq \sum_{l=0}^{L(m,p_m)-1} \left|\mathbb{E}\left[ \zeta(p_m-2)^m_{11,i_{l,0}^{m}} \zeta(2;p_m)^{m,l}_{11,i} | \mathcal{K}_{i}^{m} \right]\right|  \\
  & \quad + \mathbb{E}\left[ \left( \zeta(2)^m_{11,i_{l,(p_m-2)k_m}^{m}}  \right)^2 | \mathcal{K}_{i}^{m}  \right]^{1/2} \mathbb{E}\left[ \left( \zeta(2;p_m)^{m,l}_{11,i}  \right)^2 | \mathcal{K}_{i}^{m}  \right]^{1/2} \\
  & \leq C \varPsi_{i}^{m,2} \left( b_m \Delta_m +  p_m^{-1} b_m \Delta_m ^{1/2}  \right)  \text{ a.s.}
  ,
\end{align*}
where the first and second inequalities are due to H\"older's inequality and Lemmas \ref{lemma:zeta}, respectively.
Similarly, the second term on the right-hand side of \eqref{MM'} is bounded by
\begin{eqnarray*}
  \left| \mathbb{E}\left[ \sum_{l=0}^{L(m,p_m)-1} \zeta(p_m)^{m,l}_{11,i} \zeta(2;p_m)^{m,l-1}_{11,i} | \mathcal{K}_{i}^{m} \right] \right| \leq  C \varPsi_{i}^{m,2} \left( b_m \Delta_m + p_m^{-1} b_m \Delta_m ^{1/2} \right)  \text{ a.s.},
\end{eqnarray*}
and the third term on the right-hand side of \eqref{MM'} is bounded by
\begin{eqnarray*}
  \left| \mathbb{E}\left[ \sum_{l-l' \notin \{0,1\} }^{L(m,p_m)-1} \zeta(p_m)^{m,l}_{11,i} \zeta(2;p_m)^{m,l'}_{11,i} | \mathcal{K}_{i}^{m} \right] \right| &\leq& \sum_{l-l' \notin \{0,1\} }^{L(m,p_m)-1} \left| \mathbb{E}\left[ \zeta(p_m)^{m,l}_{11,i} \zeta(2;p_m)^{m,l'}_{11,i} |\mathcal{K}_{i}^{m} \right] \right| \\
  &\leq& C \varPsi_{i}^{m,2} p_m^{-1} b_m ^{2} \Delta_m ^{3/2} \text{ a.s.},
\end{eqnarray*}
where the second inequality is due to the Lemma \ref{lemma:zeta}.
For the fourth term on the right-hand side of \eqref{MM'}, we have
\begin{eqnarray*}
  \left| \mathbb{E}\left[ \sum_{l=0}^{L(m,p_m)-1} \zeta(p_m)^{m,l}_{11,i} \zeta(p_m)^{m}_{*} | \mathcal{K}_{i}^{m}\right] \right| &=& \left| \sum_{l=0}^{L(m,p_m)-1} \mathbb{E}\left[  \zeta(p_m)^{m,l}_{11,i} \zeta(p_m)^{m}_{*} | \mathcal{K}_{i}^{m}\right] \right| \\
  &\leq& C \varPsi_{i}^{m,2} p_m b_m \Delta_m  \text{ a.s.},
\end{eqnarray*}
where $\zeta(p_m)^{m}_{*} =  \sum_{l=L(m,p_m)(p_m+2)k_m}^{b_m-k_m} \zeta^m_{11,i+l}$.
Thus, we have for some $\varepsilon > 0$
\begin{equation}\label{EMM'}
  \left|\mathbb{E}\left[ \hat{M}(p_m)_{11,i}^{m} \hat{M}'(p_m)_{11,i}^{m} | \mathcal{K}_{i}^{m} \right]\right|   \leq C \varPsi_{i}^{m,2} \Delta_m ^{\frac{1}{4} + \varepsilon} \text{ a.s.}
\end{equation}
By Lemmas \ref{lemma:M}, \ref{negligible-xi}(a) and (b), we have for some $\varepsilon > 0$,
\begin{eqnarray}\label{square-negligible}
  && \mathbb{E}\left[ \left( M'(p_m)_{11,i}^{m} \right)^{2} | \mathcal{K}_{i}^{m} \right] \leq C \varPsi_{i}^{m,2} p_m^{-1} b_m^{-1} \Delta_m^{-1/2} \leq C \varPsi_{i}^{m,2} \Delta_m ^{\frac{1}{4} + \varepsilon}, \nonumber\\
  && \mathbb{E}\left[ \left( \xi_{11,i}^{m,1} \right)^2 | \mathcal{K}_{i}^{m} \right]  \leq C  b_m \Delta_m \leq C  \Delta_m ^{\frac{1}{4} + \varepsilon}, \nonumber\\
  && \mathbb{E}\left[ \left( \xi_{11,i}^{m,2} \right)^2 | \mathcal{K}_{i}^{m} \right]  \leq C \varPsi_{i}^{m,2} (k_m'^{-2(v-1)} + b_m ^{-1} k_m'^3) \leq C \varPsi_{i}^{m,2} \Delta_m ^{\frac{1}{4} + \varepsilon} \text{ a.s.},
\end{eqnarray}
respectively.
Similarly, we can obtain that for some $\varepsilon > 0$ 
\begin{align}\label{xi2M}
  &\quad \left|\mathbb{E}\left[ \xi_{11,i}^{m,2} \left( {M}(p_m)_{11,i}^{m} + {M}'(p_m)_{11,i}^{m} + \xi_{11,i}^{m,1} \right) | \mathcal{K}_{i}^{m}  \right]\right| \nonumber\\
  &\leq \mathbb{E}\left[ \left( \xi_{11,i}^{m,2} \right)^2 | \mathcal{K}_{i}^{m}  \right]^{1/2} \Biggl( \mathbb{E}\left[ \left( {M}(p_m)_{11,i}^{m} \right)^2 | \mathcal{K}_{i}^{m}   \right]^{1/2} + \mathbb{E}\left[ \left( {M}'(p_m)_{11,i}^{m} \right)^2 | \mathcal{K}_{i}^{m}   \right]^{1/2} \nonumber\\
  &\qquad\qquad\qquad + \mathbb{E}\left[ \left( \xi_{11,i}^{m,1} \right)^2 | \mathcal{K}_{i}^{m}   \right]^{1/2} \Biggl)  \nonumber \\
  &\leq C \varPsi_{i}^{m,2} \Delta_m ^{\frac{1}{4} +\varepsilon} \text{ a.s.}
  ,
\end{align}
where the first inequality is due to H\"older's inequality and the second inequality is due to Lemmas \ref{negligible-xi} and \ref{lemma:M}.
Now, consider $\mathbb{E}\left[ \xi_{11,i}^{m,1} {M}(p_m)_{11,i}^{m} | \mathcal{K}_{i}^{m} \right]$.
We have
\begin{eqnarray*}
  \zeta(p_m)_{11,l}^{m,i} = \zeta_1(p_m)_{11,l}^{m,i} + \zeta_2(p_m)_{11,l}^{m,i} + \zeta_3(p_m)_{11,l}^{m,i}
  ,
\end{eqnarray*}
where
\begin{equation*}
  \zeta_1(p_m)_{11,l}^{m,i} = \sum_{j=0}^{p_m k_m -1} \hat{\epsilon}_{11,i_{l,j}^{m}}, \quad \zeta_2(p_m)_{11,l}^{m,i} = 2 \sum_{j=0}^{p_m k_m -1} \tilde{X}_{1,i_{l,j}^{m}}^{c,m} \tilde{\epsilon}_{1,i_{l,j}^{m}}^{m}, \quad \zeta_3(p_m)_{11,l}^{m,i} = \sum_{j=0}^{p_m k_m -1} \hat{X}_{11,i_{j}^{m,l}}^{c,m}
  ,
\end{equation*}
and $\hat{\epsilon}_{11,i}$ and $\hat{X}_{11,i}^{c,m}$ are defined in \eqref{eq:zeta-decompose}.
Using (A.25) in \citet{jacod2019estimating}, we can show that
\begin{align}\label{xi1-zeta1}
  \left|\mathbb{E}\left[ \xi_{11,i}^{m,1} \zeta_1(p_m)_{11,l}^{m,i} | \mathcal{K}_{i}^{m} \right]\right| \leq& \sum_{j=0}^{p_m k_m -1} \left|\mathbb{E}\left[ \xi_{11,i}^{m,1} \ \hat{\epsilon}_{11,i_{l,j}^{m}} | \mathcal{K}_{i}^{m}\right]\right| \nonumber\\
  \leq& \sum_{j=0}^{p_m k_m -1} C(\varPsi_{i,j,l}^{m,2} \Delta_m ^{v/2} + \Delta_m ^{5/4}) \mathbb{E}\left[ \left( \xi_{11,i}^{m,1} \right)^2  | \mathcal{F}_{i}^{m} \right]^{1/2} \nonumber\\
  \leq& C \varPsi_{i,l}^{m,2} p_m b_m ^{1/2} \Delta_m ^{5/4} \text{ a.s.}
  ,
\end{align}
where the third inequality is due to Lemma \ref{negligible-xi}(a).
By (A.26) in \citet{jacod2019estimating}, we have
\begin{align}\label{xi1-zeta2}
  \left|\mathbb{E}\left[ \xi_{11,i}^{m,1} \zeta_2(p_m)_{11,l}^{m,i} | \mathcal{K}_{i}^{m}  \right]\right| \leq&  2 \sum_{j=0}^{p_m k_m -1} \left|\mathbb{E}\left[ \xi_{11,i}^{m,1}  \tilde{X}_{1,i_{l,j}^{m}}^{c,m} \tilde{\epsilon}_{1,i_{l,j}^{m}}^{m} | \mathcal{K}_{i}^{m} \right]\right| \nonumber\\
  \leq& C \sum_{j=0}^{p_m k_m -1} \varPsi_{i,j,l}^{m,2} \Delta_m ^{v/2} \mathbb{E}\left[ \left| \xi_{11,i}^{m,1}  \tilde{X}_{1,i_{l,j}^{m}}^{c,m}\right| | \mathcal{K}_{i}^{m} \right] \nonumber\\
  \leq& C \sum_{j=0}^{p_m k_m -1} \varPsi_{i,j,l}^{m,2} \Delta_m ^{v/2} \mathbb{E}\left[ \left| \xi_{11,i}^{m,1} \right|^2 | \mathcal{F}_{i}^{m} \right]^{1/2} \mathbb{E}\left[ \left| \tilde{X}_{1,i_{l,j}^{m}}^{c,m}\right| ^{2} | \mathcal{F}_{i}^{m} \right]^{1/2} \nonumber\\
  \leq& C \varPsi_{i,l}^{m,2} p_m b_m ^{1/2} \Delta_m ^{v/2 + 1/4} \text{ a.s.}
\end{align}
By It\^{o}'s formula, we have $\zeta_3(p_m)_{11,l}^{m,i} = 2 \left( \zeta_{3,1}(p_m)_{11,l}^{m,i} + \zeta_{3,2}(p_m)_{11,l}^{m,i} + \zeta_{3,3}(p_m)_{11,l}^{m,i} \right) $, where
\begin{eqnarray*}
  && \zeta_{3,1}(p_m)_{11,l}^{m,i} = \sum_{j=0}^{p_m k_m -1}  \hat{X}_{11,i_{l,j}^{m}}^{c,m,1}, \quad \zeta_{3,2}(p_m)_{11,l}^{m,i} = \sum_{j=0}^{p_m k_m -1}  \hat{X}_{11,i_{l,j}^{m}}^{c,m,2} ,\quad \zeta_{3,3}(p_m)_{11,l}^{m,i} = \sum_{j=0}^{p_m k_m -1}  \hat{X}_{11,i_{l,j}^{m}}^{c,m,3}, \\
  && \hat{X}_{11,i}^{c,m,1} = \int_{t_i}^{t_{i+k_m-1}} M_{1,u}^{m,i} \mu_{1,u} G_{u}^{m,i} du + \int_{t_i}^{t_{i+k_m-1}} B_{1,u}^{m,i} dB_{1,u}^{m,i} , \quad  \hat{X}_{11,i}^{c,m,2} = \int_{t_i}^{t_{i+k_m-1}} B_{1,u}^{m,i} dM_{1,u}^{m,i} , \\
  && \hat{X}_{11,i}^{c,m,3} = \int_{t_i}^{t_{i+k_m-1}} M_{1,u}^{m,i} dM_{1,u}^{m,i} , \quad  M_{1,u}^{m,i} = \int_{0}^{u} \sigma_{s} G_{s}^{m,i} dB_{s} , \quad B_{1,u}^{m,i} = \int_{0}^{u} \mu_{1,s} G_{s}^{m,i} ds, \\
  && G_{s}^{m,i} = \sum_{j=1}^{k_m - 1} g_{j}^{m} \mathbf{1}_{(t_{i+j-1},t_{i+j}]}(s)
  .
\end{eqnarray*}
By (S.1) in \citet{jacod2019estimating} and Lemma \ref{negligible-xi}(a), we have
\begin{align}\label{xi1-zeta31}
  \left|\mathbb{E}\left[ \xi_{11,i}^{m,1} \zeta_{3,1}(p_m)_{11,l}^{m,i} | \mathcal{K}_{i}^{m} \right]\right| \leq&  \sum_{j=0}^{p_m k_m -1} \left|\mathbb{E}\left[ \xi_{11,i}^{m,1}\hat{X}_{11,i_{l,j}^{m}}^{c,m,1} | \mathcal{F}_{i}^{m} \right]\right| \nonumber\\
  \leq&  \sum_{j=0}^{p_m k_m -1} \mathbb{E}\left[ \left( \xi_{11,i}^{m,1} \right)^2 | \mathcal{F}_{i}^{m}  \right]^{1/2} \mathbb{E}\left[ \left( \hat{X}_{11,i_{l,j}^{m}}^{c,m,1} \right)^2  | \mathcal{F}_{i}^{m} \right]^{1/2} \nonumber\\
  \leq& C p_m b_m ^{1/2} \Delta_m ^{3/4} \text{ a.s.} 
  ,
\end{align}
where the second inequality is due to H\"older's inequality.
Since $\mu_{1,s}$ is bounded and $G_{s}^{m,i}$ is zero for $s \notin (i \Delta_m  ,(i+k_m-1)\Delta_m ]$, we have $B_{1,u}^{m,i} = O_u(\Delta_m ^{1/2})$.
By It\^{o}'s isometry, we have
\begin{eqnarray*}
  \mathbb{E}\left[ \left( \zeta_{3,2}(p_m)_{11,l}^{m,i} \right)^2 | \mathcal{K}_{i}^{m}  \right] &\leq&  p_m k_m \sum_{j=0}^{p_m k_m -1} \mathbb{E}\left[ \left( \int_{t_{i_{j}^{m,l}}}^{t_{i_{j+k_m-1}^{m,l}}} B_{1,u}^{m,i_{j}^{m,l}} dM_{1,u}^{m,i_{j}^{m,l}} \right)^2 | \mathcal{F}_{i}^{m}  \right] \\
  &=& p_m k_m \sum_{j=0}^{p_m k_m -1} \mathbb{E}\left[  \int_{t_{i_{j}^{m,l}}}^{t_{i_{j+k_m-1}^{m,l}}} \left( B_{1,u}^{m,i_{j}^{m,l}} \sigma_{u} G_{u}^{m,i_{j}^{m,l}}  \right)^2  du  | \mathcal{F}_{i}^{m}  \right] \\
  &\leq& C p_m^{2} \Delta_m ^{1/2} \text{ a.s.}
  ,
\end{eqnarray*}
where $i_{j}^{m,l} = i + (p_m + 2) k_m l + j$ and the first and second inequalities are due to Jensen's inequality and the facts that $B_{1,u}^{m,i} = O_u(\Delta_m ^{1/2})$ and the boundedness of $\sigma_{u}$ and $G_{u}^{m,i_{l,j}^{m}}$, respectively.
Thus, we can show that
\begin{align}\label{xi1-zeta32}
  \left|\mathbb{E}\left[ \xi_{11,i}^{m,1} \zeta_{3,2}(p_m)_{11,l}^{m,i} | \mathcal{K}_{i}^{m} \right]\right| \leq& \mathbb{E}\left[ \left( \xi_{11,i}^{m,1} \right)^2 | \mathcal{F}_{i}^{m}  \right]^{1/2} \mathbb{E}\left[ \left( \zeta_{3,2}(p_m)_{11,l}^{m,i} \right)^2  | \mathcal{F}_{i}^{m} \right]^{1/2} \nonumber\\
  \leq&  C p_m b_m ^{1/2} \Delta_m ^{3/4}\text{ a.s.}
  ,
\end{align}
where the first and second inequalities are due to H\"older's inequality and Lemma \ref{negligible-xi}(a), respectively.

Simple algbra shows that
\begin{eqnarray*}
  \zeta_{3,3}(p_m)_{11,l}^{m,i} &=& \sum_{j=0}^{p_m k_m -1} \int_{t_{i_{j}^{m,l}}}^{t_{i_{j+k_m-1}^{m,l}}} M_{1,u}^{c,m,i_{j}^{m,l}} dM_{1,u}^{m,i_{j}^{m,l}} \\
  &=& \sum_{j=0}^{p_m k_m -1} \sum_{r=0}^{k_m -2}  \int_{t_{i_{j+r}^{m,l}}}^{t_{i_{j+r+1}^{m,l}}} M_{1,u}^{c,m,i_{j}^{m,l}} dM_{1,u}^{m,i_{j}^{m,l}} \\
  &=& \sum_{j=0}^{p_m k_m -1} \sum_{r=0}^{k_m -2}  \int_{t_{i_{j+r}^{m,l}}}^{t_{i_{j+r+1}^{m,l}}} \int_{t_{i_{j}^{m,l}}}^{u} \sigma_{s} G_{s}^{m,i_{j}^{m,l}} dB_{s} \sigma_{u} G_{u}^{m,i_{j}^{m,l}} dB_{u} \\
  &=& \sum_{j=0}^{(p_m + 1)k_m - 3} \sum_{r=0 \lor (j - p_m k_m + 1)}^{(k_m -2) \land j}  \int_{t_{i_{j}^{m,l}}}^{t_{i_{j+1}^{m,l}}} \int_{t_{i_{j-r}^{m,l}}}^{u} \sigma_{s} G_{s}^{m,i_{j-r}^{m,l}} dB_{s} \sigma_{u} G_{u}^{m,i_{j-r}^{m,l}} dB_{u} \\
  &=& \zeta_{3,3,1}(p_m)_{11,l}^{m,i} + \zeta_{3,3,2}(p_m)_{11,l}^{m,i}
  ,
\end{eqnarray*}
where
\begin{eqnarray*}
  \zeta_{3,3,1}(p_m)_{11,l}^{m,i} &=& \sum_{j=0}^{(p_m + 1)k_m - 3} \sum_{r=0 \lor (j - p_m k_m + 1)}^{(k_m -2) \land j}  \int_{t_{i_{j}^{m,l}}}^{t_{i_{j+1}^{m,l}}} \int_{t_{i_{j}^{m,l}}}^{u} \sigma_{11,s} G_{s}^{m,i_{j-r}^{m,l}} dB_{s} \sigma_{11,u} G_{u}^{m,i_{j-r}^{m,l}} dB_{u} ,\\
  \zeta_{3,3,2}(p_m)_{11,l}^{m,i} &=& \sum_{j=0}^{(p_m + 1)k_m - 3 } \sum_{r=0 \lor (j - p_m k_m + 1)}^{(k_m -2) \land j}  \int_{t_{i_{j}^{m,l}}}^{t_{i_{j+1}^{m,l}}}  \sigma_{u} G_{u}^{m,i_{j-r}^{m,l}} dB_{u} \int_{t_{i_{j-r}^{m,l}}}^{t_{i_{j}^{m,l}}} \sigma_{u} G_{u}^{m,i_{j-r}^{m,l}} dB_{u}
  .
\end{eqnarray*}
We have
\begin{eqnarray*}
  &&\quad \mathbb{E}\left[ \left(  \zeta_{3,3,1}(p_m)_{11,l}^{m,i} \right)^2 | \mathcal{K}_{i}^{m}  \right] \\ 
  &&= \sum_{j=0}^{(p_m + 1)k_m - 3} \mathbb{E}\left[ \left( \sum_{r=0 \lor (j - p_m k_m + 1)}^{(k_m -2) \land j}  \int_{t_{i_{j}^{m,l}}}^{t_{i_{j+1}^{m,l}}} \int_{t_{i_{j}^{m,l}}}^{u} \sigma_{s} G_{s}^{m,i_{j-r}^{m,l}} dB_{s} \sigma_{u} G_{u}^{m,i_{j-r}^{m,l}} dB_{u} \right)^2 | \mathcal{F}_{i}^{m} \right] \\
  &&\leq C k_m \sum_{j=0}^{(p_m + 1)k_m - 3}  \sum_{r=0 \lor (j - p_m k_m + 1)}^{(k_m -2) \land j} \mathbb{E}\left[ \left( \int_{t_{i_{j}^{m,l}}}^{t_{i_{j+1}^{m,l}}} \int_{t_{i_{j}^{m,l}}}^{u} \sigma_{s} G_{s}^{m,i_{j-r}^{m,l}} dB_{s} \sigma_{u} G_{u}^{m,i_{j-r}^{m,l}} dB_{u} \right)^2 | \mathcal{F}_{i}^{m} \right] \\
  &&= C k_m \sum_{j=0}^{(p_m + 1)k_m - 3}  \sum_{r=0 \lor (j - p_m k_m + 1)}^{(k_m -2) \land j} \mathbb{E}\left[  \int_{t_{i_{j}^{m,l}}}^{t_{i_{j+1}^{m,l}}} \left( \int_{t_{i_{j}^{m,l}}}^{u} \sigma_{s} G_{s}^{m,i_{j-r}^{m,l}} dB_{s} \sigma_{u} G_{u}^{m,i_{j-r}^{m,l}} \right)^2 du  | \mathcal{F}_{i}^{m} \right] \\
  &&\leq C k_m \sum_{j=0}^{(p_m + 1)k_m - 3}  \sum_{r=0 \lor (j - p_m k_m + 1)}^{(k_m -2) \land j}   \int_{t_{i_{j}^{m,l}}}^{t_{i_{j+1}^{m,l}}} \mathbb{E}\left[ \left( \int_{t_{i_{j}^{m,l}}}^{u} \sigma_{s} G_{s}^{m,i_{j-r}^{m,l}} dB_{s}  \right)^2 | \mathcal{F}_{i}^{m} \right] du  \\
  &&= C k_m \sum_{j=0}^{(p_m + 1)k_m - 3}  \sum_{r=0 \lor (j - p_m k_m + 1)}^{(k_m -2) \land j}   \int_{t_{i_{j}^{m,l}}}^{t_{i_{j+1}^{m,l}}} \mathbb{E}\left[ \int_{t_{i_{j}^{m,l}}}^{u} \left(  \sigma_{s} G_{s}^{m,i_{j-r}^{m,l}}  \right)^2  ds | \mathcal{F}_{i}^{m} \right] du  \\
  &&\leq C p_m \Delta_m ^{1/2}\text{ a.s.}
  ,
\end{eqnarray*}
where the second and third equalities are due to It\^{o}'s isometry, the first inequality is due to Jensen's inequality, and the second and third inequalities are due to the boundedness of $\sigma_{u}$ and $G_{u}^{m,i_{j-r}^{m,l}}$.
Thus, we can show that
\begin{align}\label{xi1-zeta331}
  \left|\mathbb{E}\left[ \xi_{11,i}^{m,1} \zeta_{3,3,1}(p_m)_{11,l}^{m,i} | \mathcal{K}_{i}^{m} \right]\right| \leq& \mathbb{E}\left[ \left( \xi_{11,i}^{m,1} \right)^2 | \mathcal{F}_{i}^{m}  \right]^{1/2} \mathbb{E}\left[ \left( \zeta_{3,3,1}(p_m)_{11,l}^{m,i} \right)^2  | \mathcal{F}_{i}^{m} \right]^{1/2} \nonumber\\
  \leq&  C p_m^{1/2} b_m ^{1/2} \Delta_m ^{3/4}\text{ a.s.}
  ,
\end{align}
where the first and second inequalities are due to H\"older's inequality and Lemma \ref{negligible-xi}(a), respectively.
By It\^{o}'s isometry, H\"older's inequality, and the boundedness of $\sigma_{u}$ and $G_{u}^{m,i}$, we have
\begin{eqnarray}\label{zeta332-2}
  && \mathbb{E}\left[ \left( \zeta_{3,3,2}(p_m)_{11,l}^{m,i} \right)^2 | \mathcal{K}_{i}^{m}  \right] \nonumber\\
  &&= \mathbb{E}\Biggl[ \sum_{j=0}^{(p_m + 1)k_m - 3 } \sum_{r,r' \geq 0 \lor (j - p_m k_m + 1)}^{(k_m -2) \land j}  \left( \int_{t_{i_{j}^{m,l}}}^{t_{i_{j+1}^{m,l}}}  \sigma_{u} G_{u}^{m,i_{j-r}^{m,l}} dB_{u} \right)^2  \int_{t_{i_{j-r}^{m,l}}}^{t_{i_{j}^{m,l}}} \sigma_{u} G_{u}^{m,i_{j-r}^{m,l}} dB_{u}  \nonumber\\
  &&\quad \times \int_{t_{i_{j-r'}^{m,l}}}^{t_{i_{j}^{m,l}}} \sigma_{u} G_{u}^{m,i_{j-r'}^{m,l}} dB_{u} | \mathcal{F}_{i}^{m} \Biggl] \nonumber\\
  &&=  \sum_{j=0}^{(p_m + 1)k_m - 3 } \sum_{r,r' \geq 0 \lor (j - p_m k_m + 1)}^{(k_m -2) \land j} \mathbb{E}\Biggl[ \mathbb{E}\Biggl[  \left( \int_{t_{i_{j}^{m,l}}}^{t_{i_{j+1}^{m,l}}}  \sigma_{u} G_{u}^{m,i_{j-r}^{m,l}} dB_{u} \right)^2 | \mathcal{F}_{i_{j}^{m,l}}^{m} \Biggl] \nonumber\\
  &&\quad \times \int_{t_{i_{j-r}^{m,l}}}^{t_{i_{j}^{m,l}}} \sigma_{u} G_{u}^{m,i_{j-r}^{m,l}} dB_{u} \int_{t_{i_{j-r'}^{m,l}}}^{t_{i_{j}^{m,l}}} \sigma_{u} G_{u}^{m,i_{j-r'}^{m,l}} dB_{u} | \mathcal{F}_{i}^{m} \Biggl] \nonumber\\
  &&\leq C \Delta_m \sum_{j=0}^{(p_m + 1)k_m - 3 } \sum_{r,r' \geq 0 \lor (j - p_m k_m + 1)}^{(k_m -2) \land j} \mathbb{E}\Biggl[ \int_{t_{i_{j-r}^{m,l}}}^{t_{i_{j}^{m,l}}} \sigma_{u} G_{u}^{m,i_{j-r}^{m,l}} dB_{u} \int_{t_{i_{j-r'}^{m,l}}}^{t_{i_{j}^{m,l}}} \sigma_{u} G_{u}^{m,i_{j-r'}^{m,l}} dB_{u} | \mathcal{F}_{i}^{m} \Biggl] \nonumber\\
  &&\leq C p_m \text{ a.s.}
\end{eqnarray}
Furthermore, we have
\begin{eqnarray}\label{zeta332-1}
  &&\quad \mathbb{E}\left[ \zeta_{3,3,2}(p_m)_{11,l}^{m,i} | \mathcal{K}_{i}^{m} \right] \nonumber\\
  &&=   \sum_{j=0}^{(p_m + 1)k_m - 3 } \sum_{r=0 \lor (j - p_m k_m + 1)}^{(k_m -2) \land j} \mathbb{E}\Bigg[ \mathbb{E}\left[ \int_{t_{i_{j}^{m,l}}}^{t_{i_{j+1}^{m,l}}}  \sigma_{u} G_{u}^{m,i_{j-r}^{m,l}} dW_{u} | \mathcal{K}_{i_{j}^{m,l}}^{m} \right] \cr
  &&\qquad\qquad \qquad \qquad \qquad \times \int_{t_{i_{j-r}^{m,l}}}^{t_{i_{j}^{m,l}}} \sigma_{u} G_{u}^{m,i_{j-r}^{m,l}} dW_{u} \Bigg | \mathcal{K}_{i}^{m} \Bigg] \nonumber\\
  &&= 0 \text{ a.s.}
\end{eqnarray}
On the other hand, we can rewrite $\xi_{11,i}^{m,1} = \tilde{\xi}_{11,i}^{m,0} + \tilde{\xi}_{11,i}^{m,1}$, where
\begin{eqnarray*}
  && \tilde{\xi}_{11,i}^{m,0} =  \frac{1}{(b_m - 2 k_m) \Delta_m k_m \psi_0} \sum_{l=0}^{b_m- 2 k_m -1}  \breve{C}_{11,i+l}^{m} - \frac{1}{b_m \Delta_m }  \int_{t_i}^{t_{i+b_m}} \bSigma_{11,t} dt ,\\
  && \tilde{\xi}_{11,i}^{m,1} = \frac{1}{b_m \Delta_m }  \int_{t_i}^{t_{i+b_m}} \bSigma_{11,t} dt - \bSigma_{11,i}^{m}
  .
\end{eqnarray*}
Simple algebra shows that
\begin{eqnarray}\label{tildeC-decomp}
  &&\sum_{l=0}^{b_m- 2 k_m -1}  \breve{C}_{11,i+l}^{m}\cr
   &&= \sum_{l=0}^{b_m- 2 k_m -1} \sum_{j=1}^{k_m-1} (g _{j}^{m})^2 \left( C_{11,i+l+j}^{m} - C_{11,i+l+j-1}^{m} \right) \nonumber\\
  &&= \sum_{j=1}^{k_m-1} (g _{j}^{m})^2 \sum_{l=k_m-1}^{b_m -2 k_m}  \left( C_{11,i+l}^{m} - C_{11,i+l-1}^{m}  \right) + \sum_{j=1}^{k_m -2} \sum_{l=1}^{j} (g _{j}^{m})^2 \left( C_{11,i+l}^{m} - C_{11,i+l-1}^{m} \right) \nonumber\\
  &&\quad  + \sum_{j=2}^{k_m -1} \sum_{l=b_m - 2 k_m - 1 + j}^{b_m- k_m -2} (g _{j}^{m})^2 \left( C_{11,i+l}^{m} - C_{11,i+l-1}^{m} \right)
  .
\end{eqnarray}
For the second and third terms on the right-hand side of \eqref{tildeC-decomp}, by the boundedness of $\bSigma$ and $g$, we have
\begin{eqnarray}\label{tildeC-side}
  &&\left|\sum_{j=1}^{k_m -2} \sum_{l=1}^{j} (g _{j}^{m})^2 \left( C_{11,i+l}^{m} - C_{11,i+l-1}^{m} \right)\right| \leq C \quad \text{and} \quad \nonumber\\
  &&\left|\sum_{j=2}^{k_m -1} \sum_{l=b_m -  2 k_m - 1 + j}^{b_m - k_m -2} (g _{j}^{m})^2 \left( C_{11,i+l}^{m} - C_{11,i+l-1}^{m} \right)\right| \leq C
  .
\end{eqnarray}
For the first term on the right-hand side of \eqref{tildeC-decomp}, we have
\begin{eqnarray}\label{tildeC-main}
  \sum_{j=1}^{k_m-1} (g _{j}^{m})^2 \sum_{l=k_m-1}^{b_m -2 k_m}  \left( C_{11,i+l}^{m} - C_{11,i+l-1}^{m}  \right) = \left( k_m \psi_0 + O(1)\right) \int_{t_{i+k_m-2}}^{t_{i+b_m -2 k_m}} \bSigma_{11,t} dt
  ,
\end{eqnarray}
by Riemann integration.
By \eqref{tildeC-side}, \eqref{tildeC-main}, and the boundedness of $\bSigma_{11,t}$, we have
\begin{align}\label{tilde-xi0}
  \left|\tilde{\xi}_{11,i}^{m,0}\right| \leq&   \Biggl| \frac{O(1) \int_{t_{i+k_m-2}}^{t_{i+b_m -2 k_m }} \bSigma_{11,t} dt}{(b_m - 2 k_m) \Delta_m k_m \psi_0} \Biggl| + \Biggl| \frac{2k_m \int_{t_{i+k_m-2}}^{t_{i+b_m -2 k_m}} \bSigma_{11,t} dt}{(b_m - 2 k_m) b_m  \Delta_m} \Biggl| \nonumber\\
  & + \Biggl| \frac{\int_{i}^{t_{i+k_m-2}} \bSigma_{11,t} dt}{b_m \Delta_m } \Biggl| + \Biggl| \frac{\int_{t_{i+b_m -2 k_m }}^{t_{i+b_m }} \bSigma_{11,t} dt}{b_m \Delta_m } \Biggl|  + C b_m^{-1} \Delta_m ^{-1/2} \nonumber\\
  \leq& C b_m^{-1} \Delta_m ^{-1/2}   .
\end{align}
Using It\^{o}'s lemma, we have
\begin{eqnarray*}
  \tilde{\xi}_{11,i}^{m,1} &=&  \frac{1}{b_m \Delta_m } \int_{t_{i}}^{t_{i+b_m }} \bSigma_{11,t} - \bSigma_{11,t_{i}} dt \\
  &=& - \frac{1}{b_m \Delta_m }  \int_{t_{i}}^{t_{i+b_m }} (t - t_{i+b_m } ) \tilde{\sigma}_{t} d\tilde{B}_{t} - \frac{1}{b_m \Delta_m }   \int_{t_{i}}^{t_{i+b_m }} (t - t_{i+b_m } ) \tilde{\mu}_{t} dt \\
  &&- \frac{1}{b_m \Delta_m }  \bigg( \int_{[t_i,t_{i+b_m}]\times E} (t - t_{i+b_m } )  \tilde{\mathfrak{d}}_{1}(t,z) \b1_{\left\lbrace \left|\tilde{\mathfrak{d}}_{1}(t,z)  \right| \leq 1  \right\rbrace} (\mathfrak{p} - \mathfrak{q})(dt,dz)  \\
  && + \int_{[t_i,t_{i+b_m}]\times E} (t - t_{i+b_m } )  \tilde{\mathfrak{d}}_{1}(t,z) \b1_{\left\lbrace \left|\tilde{\mathfrak{d}}_{1}(t,z)  \right| > 1  \right\rbrace} \mathfrak{p}(dt,dz) \bigg) \\
  &=& \int_{t_{i}}^{t_{i+b_m }} \frac{t_{i+b_m } - t}{b_m \Delta_m }   \tilde{\sigma}_{t} d\tilde{B}_{t}  + O_{u}(\Delta_m ^{1-\kappa})  \\
  &=&  \sum_{l=0}^{L(m,p_m)-1} \int_{t_{i_{0}^{m,l}}}^{t_{i_{0}^{m,l+1}}} \frac{t_{i+b_m } - t}{b_m \Delta_m } \tilde{\sigma}_{t} d\tilde{B}_{t} + \int_{t_{i_{0}^{m,L(m,p_m)}}}^{t_{i+b_m }} \frac{t_{i+b_m } - t}{b_m \Delta_m } \tilde{\sigma}_{t} d\tilde{B}_{t} + O_{u}(\Delta_m ^{1-\kappa}) \\
  &=&  \sum_{l=0}^{L(m,p_m)-1} \int_{t_{i_{0}^{m,l}}}^{t_{i_{0}^{m,l+1}}} \frac{t_{i+b_m } - t}{b_m \Delta_m } \tilde{\sigma}_{t} d\tilde{B}_{t} + O_u(\Delta_m ^{\kappa - 2 \iota}) + O_{u}(\Delta_m ^{1-\kappa}) 
  ,
\end{eqnarray*}
where the third and fifth equalities are due to the boundedness of $\tilde{\mathfrak{d}}_{1}$, $\tilde{\mu}$ and $\tilde{\sigma}$.
Thus, we have
\begin{align}\label{zeta332xi}
  &\mathbb{E}\left[  \zeta_{3,3,2}(p_m)_{11,l}^{m,i} \tilde{\xi}_{11,i}^{m,1} | \mathcal{K}_{i}^{m} \right]  \cr
  &\leq   \sum_{r=0}^{L(m,p_m)-1} \mathbb{E}\left[ \zeta_{3,3,2}(p_m)_{11,l}^{m,i} \int_{t_{i_{0}^{m,r}}}^{t_{i_{0}^{m,r+1}}} \frac{t_{i+b_m } - t}{b_m \Delta_m } \tilde{\sigma}_{t} d\tilde{B}_{t} | \mathcal{F}_{i}^{m} \right] \nonumber\\
  & \quad + C \Delta_m ^{1-\kappa} \mathbb{E}\left[ \left|\zeta_{3,3,2}(p_m)_{11,l}^{m,i}\right| | \mathcal{F}_{i}^{m}  \right] \nonumber\\
  \leq&  \sum_{r > l}^{L(m,p_m)-1} \mathbb{E}\left[ \mathbb{E}\left[ \int_{t_{i_{0}^{m,r}}}^{t_{i_{0}^{m,r+1}}} \frac{t_{i+b_m } - t}{b_m \Delta_m } \tilde{\sigma}_{t} d\tilde{B}_{t} | \mathcal{F}_{i}^{m} \right] \zeta_{3,3,2}(p_m)_{11,l}^{m,i}  | \mathcal{F}_{i_{0}^{m,r}}^{m} \right] \nonumber\\
  & \quad +  \sum_{r < l}^{L(m,p_m)-1} \mathbb{E}\left[ \mathbb{E}\left[ \zeta_{3,3,2}(p_m)_{11,l}^{m,i} | \mathcal{F}_{i_{0}^{m,r}}^{m} \right] \int_{t_{i_{0}^{m,r}}}^{t_{i_{0}^{m,r+1}}} \frac{t_{i+b_m } - t}{b_m \Delta_m } \tilde{\sigma}_{t} d\tilde{B}_{t}  | \mathcal{F}_{i}^{m} \right] \nonumber\\
  &\quad + \mathbb{E}\left[ \zeta_{3,3,2}(p_m)_{11,l}^{m,i} \int_{t_{i_{0}^{m,l}}}^{t_{i_{0}^{m,l+1}}} \frac{t_{i+b_m } - t}{b_m \Delta_m } \tilde{\sigma}_{t} d\tilde{B}_{t} | \mathcal{F}_{i}^{m} \right] + C p_m^{1/2} \Delta_m ^{1-\kappa} \nonumber\\
  &\leq   \mathbb{E}\left[ \zeta_{3,3,2}(p_m)_{11,l}^{m,i} \int_{t_{i_{0}^{m,l}}}^{t_{i_{0}^{m,l+1}}} \frac{t_{i+b_m } - t}{b_m \Delta_m } \tilde{\sigma}_{t} d\tilde{B}_{t} | \mathcal{F}_{i}^{m} \right] + C  \Delta_m ^{1-\kappa - \iota/2} \text{ a.s.},
\end{align}
where the second and third inequalities are due to \eqref{zeta332-2} and \eqref{zeta332-1}, respectively.
We have
\begin{align}\label{zeta332XtildeSigma}
  &\quad \mathbb{E}\left[ \zeta_{3,3,2}(p_m)_{11,l}^{m,i} \int_{t_{i_{0}^{m,l}}}^{t_{i_{0}^{m,l+1}}} \frac{t_{i+b_m } - t}{b_m \Delta_m } \tilde{\sigma}_{t} d\tilde{B}_{t} | \mathcal{F}_{i}^{m} \right] \nonumber\\
  &= \mathbb{E}\Biggl[ \sum_{j=0}^{(p_m + 1)k_m - 3 } \sum_{r=0 \lor (j - p_m k_m + 1)}^{(k_m -2) \land j}  \int_{i_{j}^{m,l}}^{t_{i_{j+1}^{m,l}}}  \sigma_{u} G_{u}^{m,i_{j-r}^{m,l}} dB_{u} \int_{t_{i_{j-r}^{m,l}}}^{t_{i_{j}^{m,l}}} \sigma_{u} G_{u}^{m,i_{j-r}^{m,l}} dB_{u} \nonumber\\
  &\qquad \times \sum_{j'=0}^{(p_m+2)k_m - 1}   \int_{t_{i_{j'}^{m,l}}}^{t_{i_{j'+1}^{m,l+1}}} \frac{t_{i+b_m } - t}{b_m \Delta_m } \tilde{\sigma}_{t} d\tilde{B}_{t} | \mathcal{F}_{i}^{m} \Biggl] \nonumber\\
  &= \mathbb{E}\Biggl[ \sum_{j=0}^{(p_m + 1)k_m - 3 } \sum_{r=0 \lor (j - p_m k_m + 1)}^{(k_m -2) \land j}  \mathbb{E}\left[ \int_{t_{i_{j}^{m,l}}}^{t_{i_{j+1}^{m,l}}}  \sigma_{u} G_{u}^{m,i_{j-r}^{m,l}} dB_{u} \int_{t_{i_{j}^{m,l}}}^{t_{i_{j+1}^{m,l+1}}} \frac{t_{i+b_m } - t}{b_m \Delta_m } \tilde{\sigma}_{t} d\tilde{B}_{t} | \mathcal{F}_{i_{j}^{m,l}}^{m} \right]  \nonumber\\
  &\qquad \times    \int_{t_{i_{j-r}^{m,l}}}^{t_{i_{j}^{m,l}}} \sigma_{u} G_{u}^{m,i_{j-r}^{m,l}} dB_{u} | \mathcal{F}_{i}^{m} \Biggl] \nonumber\\
  &= \mathbb{E}\Biggl[ \sum_{j=0}^{(p_m + 1)k_m - 3 } \sum_{r=0 \lor (j - p_m k_m + 1)}^{(k_m -2) \land j}  \mathbb{E}\left[ \int_{t_{i_{j}^{m,l}}}^{t_{i_{j+1}^{m,l}}} \frac{t_{i+b_m } - u}{b_m \Delta_m } G_{u}^{m,i_{j-r}^{m,l}} \sigma_{u}   \tilde{\sigma}_{u} \tilde{\rho}_{u} du | \mathcal{F}_{i_{j}^{m,l}}^{m} \right]  \nonumber\\
  &\qquad \times    \int_{t_{i_{j-r}^{m,l}}}^{t_{i_{j}^{m,l}}} \sigma_{u} G_{u}^{m,i_{j-r}^{m,l}} dB_{u} | \mathcal{F}_{i}^{m} \Biggl] \nonumber\\
  &= \mathbb{E}\Biggl[ \sum_{j=0}^{(p_m + 1)k_m - 3 } \sum_{r=0 \lor (j - p_m k_m + 1)}^{(k_m -2) \land j}   \int_{t_{i_{j}^{m,l}}}^{t_{i_{j+1}^{m,l}}} \mathbb{E}\left[  F_{u}^{m,i,l,j,r} - F_{t_{i_{j-r}^{m,l}}}^{m,i,l,j,r} | \mathcal{F}_{i_{j}^{m,l}}^{m} \right]  du \cr
  &\qquad\qquad\qquad\qquad\qquad\qquad\qquad\qquad\qquad\qquad \times \int_{t_{i_{j-r}^{m,l}}}^{t_{i_{j}^{m,l}}} \sigma_{u} G_{u}^{m,i_{j-r}^{m,l}} dB_{u} | \mathcal{F}_{i}^{m} \Biggl] \nonumber\\
  &\qquad + \mathbb{E}\Biggl[ \sum_{j=0}^{(p_m + 1)k_m - 3 } \sum_{r=0 \lor (j - p_m k_m + 1)}^{(k_m -2) \land j}  F_{t_{i_{j-r}^{m,l}}}^{m,i,l,j,r} \mathbb{E}\left[ \int_{t_{i_{j-r}^{m,l}}}^{t_{i_{j}^{m,l}}} \sigma_{u} G_{u}^{m,i_{j-r}^{m,l}} dB_{u} | \mathcal{F}_{i_{j}^{m,l}}^{m} \right] \biggl| \mathcal{F}_{i}^{m} \Biggl] \nonumber\\
  &=  \sum_{j=0}^{(p_m + 1)k_m - 3 } \sum_{r=0 \lor (j - p_m k_m + 1)}^{(k_m -2) \land j} \mathbb{E}\Biggl[  \int_{t_{i_{j}^{m,l}}}^{t_{i_{j+1}^{m,l}}} \mathbb{E}\left[  F_{u}^{m,i,l,j,r} - F_{t_{i_{j}^{m,l}}}^{m,i,l,j,r} | \mathcal{F}_{i_{j}^{m,l}}^{m} \right]  du \cr
  &\qquad\qquad\qquad\qquad\qquad\qquad\qquad\qquad\qquad\qquad \int_{t_{i_{j-r}^{m,l}}}^{t_{i_{j}^{m,l}}} \sigma_{u} G_{u}^{m,i_{j-r}^{m,l}} dB_{u} | \mathcal{F}_{i}^{m} \Biggl] \nonumber\\
  &\qquad +  \sum_{j=0}^{(p_m + 1)k_m - 3 } \sum_{r=0 \lor (j - p_m k_m + 1)}^{(k_m -2) \land j} \mathbb{E}\Biggl[   \Delta_m \left( F_{t_{i_{j}^{m,l}}}^{m,i,l,j,r} - F_{t_{i_{j-r}^{m,l}}}^{m,i,l,j,r} \right)    \int_{t_{i_{j-r}^{m,l}}}^{t_{i_{j}^{m,l}}} \sigma_{u} G_{u}^{m,i_{j-r}^{m,l}} dB_{u} | \mathcal{F}_{i}^{m} \Biggl] \text{ a.s.} ,\cr
  &
\end{align}
where $F_{u}^{m,i,l,j,r} = \frac{t_{i+b_m } - u}{b_m \Delta_m } G_{u}^{m,i_{j-r}^{m,l}} \sigma_{u}   \tilde{\sigma}_{u} \tilde{\rho}_{u}$ and the third equality is due to It\^{o}'s product rule.
For the summand of the first term on the right-hand side of \eqref{zeta332XtildeSigma}, we have
\begin{align}\label{zeta332XtildeSigma-first}
  &\quad \mathbb{E}\Biggl[  \int_{t_{i_{j}^{m,l}}}^{t_{i_{j+1}^{m,l}}} \mathbb{E}\left[  F_{u}^{m,i,l,j,r} - F_{t_{i_{j}^{m,l}}}^{m,i,l,j,r} | \mathcal{F}_{i_{j}^{m,l}}^{m} \right]  du  \int_{t_{i_{j-r}^{m,l}}}^{t_{i_{j}^{m,l}}} \sigma_{u} G_{u}^{m,i_{j-r}^{m,l}} dB_{u} | \mathcal{F}_{i}^{m} \Biggl] \nonumber\\
  &\leq \mathbb{E}\Biggl[  \int_{t_{i_{j}^{m,l}}}^{t_{i_{j+1}^{m,l}}} \mathbb{E}\left[ \left(  F_{u}^{m,i,l,j,r} - F_{t_{i_{j}^{m,l}}}^{m,i,l,j,r} \right)^2  | \mathcal{F}_{i_{j}^{m,l}}^{m} \right]^{1/2}  du  \int_{t_{i_{j-r}^{m,l}}}^{t_{i_{j}^{m,l}}} \sigma_{u} G_{u}^{m,i_{j-r}^{m,l}} dB_{u} | \mathcal{F}_{i}^{m} \Biggl] \nonumber\\
  &\leq C \Delta_m ^{5/4}\mathbb{E}\left[ \left( \int_{t_{i_{j-r}^{m,l}}}^{t_{i_{j}^{m,l}}} \sigma_{u} G_{u}^{m,i_{j-r}^{m,l}} dB_{u} \right)^2 | \mathcal{F}_{i}^{m}   \right]^{1/2} \nonumber\\
  &\leq C \Delta_m ^{3/2}\text{ a.s.}
  ,
\end{align}
where the first and second inequalities are due to H\"older's inequality and Lemma \ref{smooth-process}, respectively, and the third inequality is due to It\^{o}'s isometry and the boundedness of $\sigma_{u}$ and $G_{u}^{m,i}$.
For the summand of the second term on the right-hand side of \eqref{zeta332XtildeSigma}, we have
\begin{align}\label{zeta332XtildeSigma-second}
  &\quad \mathbb{E}\Biggl[   \Delta_m \left( F_{t_{i_{j}^{m,l}}}^{m,i,l,j,r} - F_{t_{i_{j-r}^{m,l}}}^{m,i,l,j,r} \right)    \int_{t_{i_{j-r}^{m,l}}}^{t_{i_{j}^{m,l}}} \sigma_{u} G_{u}^{m,i_{j-r}^{m,l}} dB_{u} | \mathcal{F}_{i}^{m} \Biggl] \nonumber\\
  &\leq \Delta_m  \mathbb{E}\left[  \left( F_{t_{i_{j}^{m,l}}}^{m,i,l,j,r} - F_{t_{i_{j-r}^{m,l}}}^{m,i,l,j,r} \right)^{2} | \mathcal{F}_{i}^{m}  \right]^{1/2} \mathbb{E}\left[  \left(  \int_{t_{i_{j-r}^{m,l}}}^{t_{i_{j}^{m,l}}} \sigma_{u} G_{u}^{m,i_{j-r}^{m,l}} dB_{u}  \right)^2   | \mathcal{F}_{i}^{m} \right]^{1/2} \nonumber\\
  &  \leq C \Delta_m ^{3/2} \text{ a.s.}
  ,
\end{align}
where the first inequality is due to H\"older's inequality, and the second inequality is due to Lemma \ref{smooth-process}, It\^{o}'s isometry, and the boundedness of $\sigma_{u}$ and $G_{u}^{m,i}$.
Then, by \eqref{xi1-zeta331}, \eqref{zeta332-2}, \eqref{tilde-xi0}, \eqref{zeta332xi}, \eqref{zeta332XtildeSigma}, \eqref{zeta332XtildeSigma-first}, and \eqref{zeta332XtildeSigma-second}, we have
\begin{align}\label{xi1-zeta33}
  &\quad \left|\mathbb{E}\left[  \zeta_{3,3}(p_m)_{11,l}^{m,i} \xi_{11,i}^{m,1} | \mathcal{K}_{i}^{m} \right]\right| \nonumber\\
  &\leq \left|\mathbb{E}\left[  \zeta_{3,3,1}(p_m)_{11,l}^{m,i} \xi_{11,i}^{m,1} | \mathcal{K}_{i}^{m} \right]\right| + \left|\mathbb{E}\left[  \zeta_{3,3,2}(p_m)_{11,l}^{m,i} \xi_{11,i}^{m,1} | \mathcal{K}_{i}^{m} \right]\right| \nonumber\\
  &\leq  C p_m^{1/2} b_m ^{1/2} \Delta_m ^{3/4} + \left|\mathbb{E}\left[  \zeta_{3,3,2}(p_m)_{11,l}^{m,i} \tilde{\xi}_{11,i}^{m,0} | \mathcal{K}_{i}^{m} \right]\right| + \left|\mathbb{E}\left[  \zeta_{3,3,2}(p_m)_{11,l}^{m,i} \tilde{\xi}_{11,i}^{m,1} | \mathcal{K}_{i}^{m} \right]\right|  \nonumber\\
  &\leq \mathbb{E}\left[ \zeta_{3,3,2}(p_m)_{11,l}^{m,i} \int_{t_{i_{0}^{m,l}}}^{t_{i_{0}^{m,l+1}}} \frac{t_{i+b_m } - t}{b_m \Delta_m } \tilde{\sigma}_{t} d\tilde{B}_{t} | \mathcal{F}_{i}^{m} \right] + C  \Delta_m ^{1-\kappa - \iota/2} \nonumber\\
  &\quad + \mathbb{E}\left[ \left(  \zeta_{3,3,2}(p_m)_{11,l}^{m,i} \right)^2 | \mathcal{F}_{i}^{m} \right]^{1/2} \mathbb{E}\left[ \left( \tilde{\xi}_{11,i}^{m,0} \right)^2  | \mathcal{K}_{i}^{m} \right]^{1/2}  \nonumber\\
  &\leq C \left( p_m \Delta_m ^{1/2} + \Delta_m ^{1-\kappa - \iota/2} + p_m^{1/2} b_m ^{-1} \Delta_m ^{-1/2} \right) \text{ a.s.}
\end{align}
Using \eqref{xi1-zeta1}, \eqref{xi1-zeta2}, \eqref{xi1-zeta31}, \eqref{xi1-zeta32}, and \eqref{xi1-zeta33}, we conclude that for some $\varepsilon > 0$
\begin{align}\label{M-xi1}
  &\quad \left|\mathbb{E}\left[ {M}(p_m)_{11,i}^{m} \xi_{11,i}^{m,1} | \mathcal{K}_{i}^{m} \right]\right| \nonumber\\
  &\leq C b_m^{-1} \Delta_m ^{-1/2} \sum_{l=0}^{L(m,p)-1} \left|\mathbb{E}\left[ \zeta_{1}(p_m)_{11,l}^{m,i} \xi_{11,i}^{m,1} | \mathcal{K}_{i}^{m} \right]\right| + \left|\mathbb{E}\left[ \zeta_{2}(p_m)_{11,l}^{m,i} \xi_{11,i}^{m,1} | \mathcal{K}_{i}^{m} \right]\right| \nonumber \\
  &\quad + \left|\mathbb{E}\left[ \zeta_{3,1}(p_m)_{11,l}^{m,i} \xi_{11,i}^{m,1} | \mathcal{K}_{i}^{m} \right]\right| + \left|\mathbb{E}\left[ \zeta_{3,2}(p_m)_{11,l}^{m,i} \xi_{11,i}^{m,1} | \mathcal{K}_{i}^{m} \right]\right|  + \left|\mathbb{E}\left[ \zeta_{3,3}(p_m)_{11,l}^{m,i} \xi_{11,i}^{m,1} | \mathcal{K}_{i}^{m} \right]\right| \nonumber \\
  &\leq C \left( \varPsi_{i}^{m,2} \Delta_m ^{3/4 - \kappa/2 } + \Delta_m ^{1/2 } + \Delta_m ^{1-\kappa} + \Delta_m ^{\kappa - 1/2 + \iota/2} \right) \nonumber\\
  &\leq C \varPsi_{i}^{m,2} \Delta_m ^{1/4 + \varepsilon} \quad \text{a.s.}
\end{align}
Similarly, we can show that
\begin{equation}\label{hatM-xi1}
  \left|\mathbb{E}\left[ {M}'(p_m)_{11,i}^{m} \xi_{11,i}^{m,1} | \mathcal{K}_{i}^{m} \right]\right| \leq C \varPsi_{i}^{m,2} \Delta_m ^{1/4 + \varepsilon} \text{ a.s.}
\end{equation}

Let $\Xi_{11,j}^{m} = \Xi(\bSigma^{m}_{j}, \bvartheta^{m}_{j})_{11}$.
Simple algebra shows that
\begin{eqnarray*}
  && \quad \left( \frac{1}{(b_m - 2 k_m)\Delta_m k_m \psi_0} \right)^2 \sum_{l=0}^{L(m,p_m)-1} \mathbb{E}\left[\left( \zeta(p_m)^{m,l}_{11,i} \right) ^{2} | \mathcal{K}_{i}^{m} \right] - b^{-1} \Delta_m ^{-1/2}  \Xi_{11,i}^{m} \\
  &&= \left( \frac{1}{(b_m - 2 k_m)\Delta_m k_m \psi_0} \right)^2 \left( \mathcal{A}(p_m)_{i}^{m,1} + \mathcal{A}(p_m)_{i}^{m,2} + \mathcal{A}(p_m)_{i}^{m,3} \right) + \mathcal{A}(p_m)_{i}^{m,4}
  ,
\end{eqnarray*}
where
\begin{eqnarray*}
  && \mathcal{A}(p_m)_{i}^{m,1} = \sum_{l=0}^{L(m,p_m)-1} \left( \mathbb{E}\left[ \left (\zeta(p_m)^{m,l}_{11,i}\right)^2 | \mathcal{K}_{i}^{m} \right] - \varXi(p_m)_{11,i+l(p_m+2)k_m}^{m}   \right) , \\
  && \mathcal{A}(p_m)_{i}^{m,2} = \sum_{l=0}^{L(m,p_m)-1} \left( \varXi(p_m)_{11,i+l(p_m+2)k_m}^{m}   -   {C_k}^3 p_m \psi_0^{2} \Xi_{11,i+l(p_m+2)k_m}^{m} \right)  ,  \\
  && \mathcal{A}(p_m)_{i}^{m,3} = {C_k}^3 p_m \psi_0^{2} \sum_{l=0}^{L(m,p_m)-1} \left(  \Xi_{11,i+l(p_m+2)k_m}^{m} - \Xi_{11,i}^{m} \right) , \\
  && \mathcal{A}(p_m)_{i}^{m,4} = \Xi_{11,i}^{m} \left[  \left( \frac{1}{(b_m - 2 k_m)\Delta_m k_m \psi_0} \right)^2 {C_k}^3 p_m \psi_0^{2} \times L(m,p_m) - b_m ^{-1} \Delta_m ^{-1/2}  \right]  .
\end{eqnarray*}
By Lemma \ref{lemma:zeta}, we have
\begin{equation*}
  \left|\mathbb{E}\left[ \mathcal{A}(p_m)_{i}^{m,1} | \mathcal{K}_{i}^{m} \right]\right|  \leq C \varPsi_{i}^{m,2} p_m b_m \Delta_m ^{3/4} \quad \text{ a.s.}
\end{equation*}
Since $\bSigma$ and $\bvartheta$ are bounded, we have
\begin{align*}
  &\left|\mathbb{E}\left[ \mathcal{A}(p_m)_{i}^{m,2} | \mathcal{K}_{i}^{m} \right]\right| \cr
  &= \left|\sum_{l=0}^{L(m,p_m)-1} \mathbb{E}\left[ \left( \varXi(p_m)_{11,i+l(p_m+2)k_m}^{m}   - {C_k}^3 p_m \psi_0^{2} \Xi_{11,i+l(p_m+2)k_m}^{m} \right) | \mathcal{K}_{i}^{m} \right] \right|  \\
  &= \left|\sum_{l=0}^{L(m,p_m)-1} \mathbb{E}\left[ - 4 \left( \bSigma_{11,i}^{m} \right)^{2}   {C_k}^4 \bar{\Phi}_{00} - 8 \bSigma_{11,i}^{m}  \left(  \vartheta_{1,i}^{m} \right)^2 R {C_k}^2 \bar{\Phi}_{01} - 4 \left( \vartheta_{1,i}^{m} \right)^{4} R^2  \bar{\Phi}_{11} | \mathcal{K}_{i}^{m} \right]\right|    \\
  &\leq C p_m^{-1} b_m k_m^{-1} \quad \text{ a.s.}
\end{align*}
Since $\bSigma$ and $\bvartheta$ are bounded It\^{o} semimartingale, we have
\begin{eqnarray*}
  \left|\mathbb{E}\left[ \mathcal{A}(p_m)_{i}^{m,3} \right]\right| &\leq& {C_k}^3 p \psi_0^{2} \sum_{l=0}^{L(m,p_m)-1} \left| \mathbb{E}\left[ \left(  \Xi_{11,i+l(p+2)k_m}^{m} - \Xi_{11,i}^{m} \right) \right] \right|   \\
  &\leq& C p_m \psi_0^{2} L(m,p_m) b_m \Delta_m \\
  &\leq& C b_m ^{2} \Delta_m ^{3/2} \quad \text{ a.s.}
\end{eqnarray*}
Simple algebra shows that
\begin{eqnarray*}
  &&\left| \mathbb{E}\left[ \mathcal{A}(p_m)_{i}^{m,4} | \mathcal{K}_{i}^{m} \right] \right| \cr
  &=&  \left| \mathbb{E}\left[ \Xi_{11,i+l(p_m+2)k_m}^{m} \left(  \frac{b_m }{(b_m - 2 k_m)^2}  \frac{p_m}{p_m+2} \Delta_m ^{-1/2}  - b_m ^{-1} \Delta_m ^{-1/2}  \right) | \mathcal{K}_{i}^{m} \right] \right| \\
  &=& \biggl| \mathbb{E} \biggl[ \Xi_{11,i+l(p_m+2)k_m}^{m} \Delta_m ^{-1/2} \\
  && \times \left\lbrace \left( \frac{b_m }{\left( b_m - 2 k_m \right)^2 } - b_m ^{-1} \right) \frac{p_m}{p_m+2} + b_m ^{-1} \left( \frac{p_m}{p_m+2} -1 \right)   \right\rbrace | \mathcal{K}_{i}^{m} \biggl] \biggl| \\
  &=& \left| \mathbb{E}\left[ \Xi_{11,i+l(p_m+2)k_m}^{m} \Delta_m ^{-1/2} \left\lbrace \frac{4 b_m k_m - 4 k_m^2}{\left( b_m - 2 k_m \right)^2 b_m  }  \frac{p_m}{p_m+2} - b_m ^{-1} \frac{2}{p_m+2}    \right\rbrace | \mathcal{K}_{i}^{m} \right] \right| \\
  &\leq& C p_m^{-1} \Delta_m ^{-1/2} b_m ^{-1} \quad  \text{ a.s.}
\end{eqnarray*}
Thus, we have
\begin{eqnarray}\label{ZminusX}
  && \left| \mathbb{E}\left[ \left( \frac{1}{(b_m - 2 k_m)\Delta_m k_m \psi_0} \right)^2 \sum_{l=0}^{L(m,p_m)-1} \mathbb{E}\left[\left( \zeta(p_m)^{m,l}_{11,i} \right) ^{2}\right] - b^{-1} \Delta_m ^{-1/2}  \Xi\left( \bSigma_{11,i}^{m}, \gamma_{11,i}^{m} \right) | \mathcal{K}_{i}^{m} \right] \right| \nonumber\\
  &&\leq C b_m ^{-2} \Delta_m ^{-1} \left( C \varPsi_{i}^{m,2} p_m  b_m\Delta_m ^{3/4} + p_m^{-1} b_m k_m^{-1} + b_m ^2  \Delta_m ^{3/2} \right) + C p_m^{-1} \Delta_m ^{-1/2} b_m ^{-1} \nonumber \\
  &&\leq C \varPsi_{i}^{m,2} p_m^{-1} b_m ^{-1} \Delta_m ^{-1/2} \nonumber \\
  &&\leq C \varPsi_{i}^{m,2} \Delta_m ^{\frac{1}{4} + \varepsilon} \quad \text{ a.s.}
  ,
\end{eqnarray}
for some $\varepsilon > 0$.
By \eqref{e-decomp}, \eqref{decompM}, \eqref{EMM'}, \eqref{square-negligible}, \eqref{xi2M}, \eqref{M-xi1}, \eqref{hatM-xi1}, and \eqref{ZminusX},  we establish
\begin{equation*}
  \left| \mathbb{E}\left[ \left(  e_{11,i}^{m}  \right)^2   - b_m ^{-1} \Delta_{m}^{-1/2}  \Xi_{11,i}^{m} | \mathcal{K}_{i}^{m} \right] \right| \leq C \varPsi_{i}^{m,2} \Delta_m ^{\frac{1}{4} + \varepsilon}    \quad \text{ for some } \varepsilon > 0 \quad \text{ a.s.}
\end{equation*}
$\blacksquare$

\textbf{Proof of Lemma \ref{lemma:jump-preavg-element}.}
Similar to proof of lemma A.6 in \citet{jacod2019estimating}, for any $x_{1}, y_{1},x_{1}, y_{2} \in \mathbb{R}$ and $0 < u' < u < u''$, we have constant $C$ only depending on the ratio $u'' / u'$ such that
\begin{eqnarray}\label{lemma:jump-elem-1}
  && \left| (x_{1} + y_{1}) (x_{2} + y_{2}) \b1_{ \lbrace |x_1 + y_1| \leq u_1 , |x_2 + y_2| \leq u_2  \rbrace } - x_1 x_2 \right| \nonumber\\
  &\leq& C \bigg[ | x_1 x_2 | \b1_{ \lbrace |x_1| > u'_1 /2, |x_2| \leq u'_2 /2 \rbrace} + | x_1 x_2 | \b1_{ \lbrace |x_2| > u'_2 /2 , |x_1| \leq u'_1 /2 \rbrace} + | x_1 x_2 | \b1_{ \lbrace |x_1| > u'_1 /2, |x_2| > u'_2 /2 \rbrace} \nonumber\\
  &&+ u''_1 |x_2| \left( \frac{|y_1|}{u''_1} \land 1  \right) + u''_2 |x_1| \left( \frac{|y_2|}{u''_2} \land 1  \right) + u''_1 u''_2 \left( \frac{|y_1|}{u''_1} \land 1  \right) \left( \frac{|y_2|}{u''_2} \land 1  \right)  \bigg]
  .
\end{eqnarray}
We apply this with
\begin{align*}
  & x_1 = \tilde{Y}_{x,i}^{c,m}, \quad x_2 = \tilde{Y}_{y,i}^{c,m}, \quad y_1 = \tilde{X}_{x,i}^{d,m}, \quad y_2 = \tilde{X}_{y,i}^{d,m}, \quad u_{1} = u_{1,m}, \quad u_{2} = u_{2,m}, \\
  & u'_{1} = a'_{x} (k_m \Delta_m )^{\varpi_{1}}, \quad u'_{2} = a'_{y} (k_m \Delta_m )^{\varpi_{1}}, \quad u''_{1} = a''_{x} (k_m \Delta_m )^{\varpi_{1}}, \quad u''_{2} = a''_{y} (k_m \Delta_m )^{\varpi_{1}}
  ,
\end{align*}
where $X_{x,t}^{d} = \mathfrak{d}_{x}  \b1_{\left\lbrace \left|\mathfrak{d}_{x}\right| \leq 1  \right\rbrace} \ast (\mathfrak{p} - \mathfrak{q})_{t} +  \mathfrak{d}_{x}  \b1_{\left\lbrace \left|\mathfrak{d}_{x}\right| > 1  \right\rbrace} \ast \mathfrak{p}_{t}$,
$a''_1 > a_1 > a'_1 > 0$, and $a''_2 > a_2 > a'_2 > 0$.
Using (A.28) of \citet{jacod2019estimating}, we have for $1 \leq z < [v]$,
\begin{eqnarray}\label{lemma:jump-elem-2}
  \mathbb{E}\left[ \left( \tilde{Y}_{x,i}^{c,m} \right)^{2z} \b1_{\lbrace |\tilde{Y}_{x,i}^{c,m}| > u'_1  \rbrace} \Big| \mathcal{K}_{i}^{m} \right]
  &\leq& \mathbb{E}\left[ \left|\tilde{Y}_{x,i}^{c,m}\right| ^{2[v]} \Big| \mathcal{K}_{i}^{m} \right]^{z/[v]} \mathbb{P}\left( |\tilde{Y}_{x,i}^{c,m}| > u'_1 \big| \mathcal{K}_{i}^{m} \right) \nonumber\\
  &\leq& \mathbb{E}\left[ \left|\tilde{Y}_{x,i}^{c,m}\right| ^{2[v]} \Big| \mathcal{K}_{i}^{m} \right]^{z/[v]} \mathbb{E}\left[ \left|\tilde{Y}_{x,i}^{c,m}\right| ^{2[v]} \Big| \mathcal{K}_{i}^{m} \right]^{\frac{[v] - z}{[v]} } / {u'}_1^{2([v]-z)} \nonumber\\
  &\leq&  C_{z} \varPsi_{i}^{m,2}  \Delta_m ^{[v](\frac{1}{2} - \varpi_{1}) + \varpi_1 z}
  ,
\end{eqnarray}
where the first and second inequalities are due to H\"older's inequality and Markov's inequality, respectively.
Using (A.18) and (A.28) of \citet{jacod2019estimating}, we have for $1 \leq z < 2[v]$ and $\varepsilon > 0$,
\begin{eqnarray}\label{lemma:jump-elem-3}
  && \mathbb{E}\left[ \left\lbrace u''_1 |\tilde{Y}_{y,i}^{c,m}| \left( \frac{|\tilde{X}_{x,i}^{d,m}|}{u''_1} \land 1  \right) \right\rbrace^{z} \bigg| \mathcal{K}_{i}^{m} \right] \cr
  &\leq&  (u''_1)^z \mathbb{E}\left[ \left| \tilde{Y}_{y,i}^{c,m} \right|^{2[v]} \Big| \mathcal{K}_{i}^{m} \right]^{\frac{z}{2[v]} } \mathbb{E}\left[ \left( \frac{|\tilde{X}_{x,i}^{d,m}|}{u''_1} \land 1  \right)^{\frac{2[v]z}{2[v]-z} } \Bigg| \mathcal{K}_{i}^{m} \right]^{\frac{2[v]-z}{2[v]} } \nonumber\\
  &\leq& C_{\varepsilon,z} \varPsi_{i}^{m,2}  \Delta_m ^{\frac{\varpi_1 z}{2}  + \frac{z}{4}  + \frac{(2[v]-z)(1-\varpi_1 r - \varepsilon)}{4[v]}  }
\end{eqnarray}
and
\begin{eqnarray}\label{lemma:jump-elem-4}
  && \mathbb{E}\left[ \left\lbrace u''_1 u''_2 \left( \frac{|\tilde{X}_{x,i}^{d,m}|}{u''_1} \land 1  \right) \left( \frac{|\tilde{X}_{y,i}^{d,m}|}{u''_2} \land 1  \right) \right\rbrace^{z} \bigg| \mathcal{K}_{i}^{m} \right] \nonumber\\
  &\leq& C \Delta_m ^{\varpi_1 z} \mathbb{E}\left[ \left( \frac{|\tilde{X}_{x,i}^{d,m}|}{u''_1} \land 1  \right)^{2z} \Bigg| \mathcal{K}_{i}^{m} \right]^{1/2} \mathbb{E}\left[ \left( \frac{|\tilde{X}_{y,i}^{d,m}|}{u''_2} \land 1  \right)^{2z} \Bigg| \mathcal{K}_{i}^{m} \right]^{1/2} \nonumber\\
  &\leq& C_{\varepsilon,z} \varPsi_{i}^{m,2} \Delta_m ^{\varpi_1 z + \frac{1 - \varpi_1 r - \varepsilon}{2}}  
  .
\end{eqnarray}
Since $\varpi_1 > \frac{[v] - 1}{2[v]-r}$, the rate of the bound of \eqref{lemma:jump-elem-3} and \eqref{lemma:jump-elem-4} are negligible as compared to that of \eqref{lemma:jump-elem-2}.
Using \eqref{lemma:jump-elem-1}, \eqref{lemma:jump-elem-2}, \eqref{lemma:jump-elem-3}, \eqref{lemma:jump-elem-4}, and H\"older's inequality, we have
\begin{eqnarray*}
  \mathbb{E}\left[ \left|\tilde{Y}_{x,i}^{m}\tilde{Y}_{y,i}^{m} \b1_{\lbrace |\tilde{Y}_{x,i}^{m}| \leq u_{x,m} , |\tilde{Y}_{x,i}^{m}| \leq u_{y,m} \rbrace} - \tilde{Y}_{x,i}^{c,m}\tilde{Y}_{y,i}^{c,m} \right|^{z} \big| \mathcal{K}_{i}^{m}  \right] \leq C_{z} \varPsi_{i}^{m,2}  \Delta_m ^{[v](\frac{1}{2} - \varpi_{1}) + \varpi_1 z}
  .
\end{eqnarray*}

Simple algebra shows that
\begin{equation*}
  \hat{\bSigma}_{xy,i}^{m} - \hat{\bSigma}_{xy,i}^{c,m} = \frac{1}{(b_m - 2k_m) \Delta_m k_m \psi_{0}} \sum_{l=0}^{b_m-2k_m-1} \left( \tilde{Y}_{x,i}\tilde{Y}_{y,i} \b1_{\lbrace |\tilde{Y}_{x,i}| \leq u_{x,m} , |\tilde{Y}_{x,i}| \leq u_{y,m} \rbrace} - \tilde{Y}_{x,i}^{c}\tilde{Y}_{y,i}^{c} \right) 
  .
\end{equation*}
Thus, by Jensen's inequality and the first part of Lemma \ref{lemma:jump-preavg-element}, we have
\begin{eqnarray*}
  &&\mathbb{E}\left[ \left|\hat{\bSigma}_{xy,i}^{m} - \hat{\bSigma}_{xy,i}^{c,m}\right|^{z} \big| \mathcal{K}_{i}^{m} \right] \cr
  &\leq&  C_{z}   b_m ^{-1} \Delta_m ^{-z/2} \sum_{l=0}^{b_m-2k_m-1} \mathbb{E}\left[  \left| \tilde{Y}_{x,i}\tilde{Y}_{y,i} \b1_{\lbrace |\tilde{Y}_{x,i}| \leq u_{x,m} , |\tilde{Y}_{x,i}| \leq u_{y,m} \rbrace} - \tilde{Y}_{x,i}^{c}\tilde{Y}_{y,i}^{c} \right|^{z} \bigg| \mathcal{K}_{i}^{m} \right] \\
  &\leq& C_{z} \varPsi_{i}^{m,2}  \Delta_m ^{([v] -z)(\frac{1}{2} - \varpi_1)}
  .
\end{eqnarray*}
We further note that, for the proof of Theorem \ref{thm:BoundedMoment}, we can similarly show that for $z \geq [v]$,
\begin{eqnarray*}
  &&\mathbb{E}\left[ \left( \tilde{Y}_{x,i}^{c,m} \right)^{2z} \b1_{\lbrace |\tilde{Y}_{x,i}^{c,m}| > u'_1  \rbrace} \Big| \mathcal{K}_{i}^{m} \right]
  \leq C_z \varPsi_{i}^{m,2} \Delta_m ^{[v]/2},\cr
  && \mathbb{E}\left[  \left| \tilde{Y}_{x,i}\tilde{Y}_{y,i} \b1_{\lbrace |\tilde{Y}_{x,i}| \leq u_{x,m} , |\tilde{Y}_{x,i}| \leq u_{y,m} \rbrace} - \tilde{Y}_{x,i}^{c}\tilde{Y}_{y,i}^{c} \right|^{z} \bigg| \mathcal{K}_{i}^{m} \right] \leq C_z \varPsi_{i}^{m,2} \Delta_m ^{[v] / 2} ,\cr
  && \mathbb{E}\left[ \left|\hat{\bSigma}_{xy,i}^{m} - \hat{\bSigma}_{xy,i}^{c,m}\right|^{z} \big| \mathcal{K}_{i}^{m} \right] \leq C_z \varPsi_{i}^{m,2} \Delta_m ^{\frac{[v] - z}{2} }
  .
\end{eqnarray*}
$\blacksquare$

\textbf{Proof of Lemma \ref{spot-noise-estimate}.}
First, we consider $x=1$, $y=1$.
Using the notations in \eqref{micro-autocorrelation-theta} and \eqref{mathcalVs}, we have
\begin{eqnarray}\label{spot-noise-error-decomp}
  &&\hat{\bvartheta}_{11,i}^{m} - (\vartheta_{1,i}^{m})^2 R_{11} \cr
  &=& (b_m - 6l_m)^{-1} \sum_{d=-k'_m}^{k'_m}  \sum_{l=0}^{b_m -6 l_m} (Y _{1,i+l}^{m} - \bar{Y} _{1,i+l+2 l_m}^{m}) (Y _{1,i+l+d}^{m} - \bar{Y} _{1,i+l+4 l_m}^{m}) - (\vartheta_{1,i}^{m})^2 R_{11} \cr
  &=& (b_m - 6l_m)^{-1} \sum_{d=-k'_m}^{k'_m}  \sum_{w=1}^{3} \mathcal{V}_{11,i}^{m,w}(|d|) \cr
  &&+ (b_m - 6l_m)^{-1} \sum_{l=0}^{b_m -6 l_m} \sum_{d=-k'_m}^{k'_m} T (0,2 l_m ) _{i+l}^{m,2} T (|d|,4 l_m ) _{i+l}^{m,2} - (\vartheta_{1,i}^{m})^2 R_{11}
  .
\end{eqnarray}
Using \eqref{mathV23-2w} and Jensen's inequality, we have
\begin{eqnarray*}
  \mathbb{E}\left[ \left|(b_m - 6l_m)^{-1} \sum_{d=-k'_m}^{k'_m}  \sum_{z=1}^{3} \mathcal{V}_{11,i}^{m,z}(|d|)\right|^{w} \bigg| \mathcal{K}_{i}^{m}  \right] \leq C_{w} \varPsi_{i}^{m,2}  (k'_m)^{w} (b_m^{-1} l_m)^{w/2} (l_m \Delta_m)^{1/2}
  .
\end{eqnarray*}
On the other hand, we have
\begin{eqnarray*}
  \sum_{d=-k'_m}^{k'_m} T (0,2 l_m ) _{i+l}^{m,2} T (|d|,4 l_m ) _{i+l}^{m,2} &=& (\vartheta _{1,i+l}^{m})^{2} \sum_{d=-k'_m}^{k'_m}  (\chi_{1,i+l} - \bar{\chi} _{1,i+l+2l_m}^{m})  (\chi_{1,i+l+|d|} - \bar{\chi} _{1,i+l+4l_m}^{m}) \\
  &=& \bar{\mathcal{S}}_{11,i,l}^{m,1} + \bar{\mathcal{S}}_{11,i,l}^{m,2} + \bar{\mathcal{S}}_{11,i,l}^{m,3} + \bar{\mathcal{S}}_{11,i,l}^{m,4}
  ,
\end{eqnarray*}
where
\begin{align*}
  & \bar{\mathcal{S}} _{11,i,l}^{m,1}      = (\vartheta _{1,i+l}^{m})^{2} \sum_{d=-k'_m}^{k'_m}   \chi_{1,i+l} \chi_{1,i+l+|d|}, \qquad
    \bar{\mathcal{S}} _{11,i,l}^{m,2}      = (\vartheta _{1,i+l}^{m})^{2} \sum_{d=-k'_m}^{k'_m}   \chi_{1,i+l} \bar{\chi} _{1,i+l+ 4 l_m }^{m}, \nonumber\\
  & \bar{\mathcal{S}} _{11,i,l}^{m,3}      = (\vartheta _{1,i+l}^{m})^{2} \sum_{d=-k'_m}^{k'_m}   \chi_{1,i+l+|d|} \bar{\chi} _{1,i+l+ 2 l_m }^{m}, \quad \text{and} \quad
    \bar{\mathcal{S}} _{11,i,l}^{m,4}      = (\vartheta _{1,i+l}^{m})^{2} \sum_{d=-k'_m}^{k'_m}   \bar{\chi} _{1,i+l+ 2 l_m }^{m} \bar{\chi} _{1,i+l+ 4 l_m }^{m}.
\end{align*}
Thus, we can rewrite the second term on the right-hand side of \eqref{spot-noise-error-decomp} as follows:
\begin{equation}\label{S-sub-bar-rewrite}
  (b_m - 6l_m)^{-1} \left[ \sum_{l=0}^{b_m -6 l_m} \sum_{z=2}^{4} \bar{\mathcal{S}} _{11,i,l}^{m,z}  + \sum_{l=0}^{b_m - 6l_m} \left( \bar{\mathcal{S}} _{11,i,l}^{m,1} -  (\vartheta_{1,i}^{m})^2 R_{11} \right) + (\vartheta_{1,i}^{m})^2 R_{11}  \right] 
  .
\end{equation}
Using Lemma \ref{chi-property}(a) and the mathematical induction method, we have for any $w,k \in \mathbb{N}$,
\begin{eqnarray}\label{chibarchi-CE-w-E-k}
  && \mathbb{E}\left[ \left|\mathbb{E}\left[ \chi_{1,i+l} \bar{\chi}_{1,i+l+4l_m}^{m} | \tilde{\mathcal{K}}_{i}^{m,5l_m} \right]\right|^{w}  \right] \cr
  &\leq&  C_{w,k} l_m ^{-\left( 2-2^{-(k+1)} \right) v} \mathbb{E}\left[ \left|\mathbb{E}\left[ \chi_{1,i+l} \bar{\chi}_{1,i+l+4l_m}^{m} | \tilde{\mathcal{K}}_{i}^{m,5l_m} \right]\right|^{2^{k}w-2^{k+1}+2}  \right]^{2^{-k} } \mathbb{E}\left[ (\chi_{1,i+l} \bar{\chi}_{1,i+l+4l_m}^{m})^{2} \right]^{ 1-2^{-k}} \cr
  &&+ C_{w,k}  l_m ^{-(v+\frac{1}{2} )w}.
\end{eqnarray}
Thus, we have for any $\varepsilon > 0$,
\begin{equation*}
  \left|\mathbb{E}\left[ \chi_{1,i+l} \bar{\chi}_{1,i+l+4l_m}^{m} | \tilde{\mathcal{K}}_{i+l}^{m,5l_m} \right]\right|^{w} \leq C_{w,\varepsilon} \varPsi_{i}^{m,2} ( l_m ^{(-2v-w/2+\varepsilon)} + l_m ^{ -(v+\frac{1}{2})w})
  .
\end{equation*}
Similar to \eqref{eta-CE-w-CE} and \eqref{eta-CE-w-E}, we have for any $\varepsilon > 0$ and $w \geq 2$,
\begin{eqnarray*}
 &&\mathbb{E}\left[  \left|\mathbb{E}\left[ \bar{\mathcal{S}} _{11,i,l}^{m,2} | \tilde{\mathcal{K}}_{i+l}^{m,5l_m} \right]\right| ^{w} | \mathcal{K}_{i}^{m} \right] \leq C_{w,\varepsilon} \varPsi_{i}^{m,2} (k'_m)^{w} ( l_m ^{(-2v-w/2+\varepsilon)} + l_m ^{ -(v+\frac{1}{2})w}) ,\cr
 &&\mathbb{E}\left[  \left|\mathbb{E}\left[ \bar{\mathcal{S}} _{11,i,l}^{m,2} | \tilde{\mathcal{K}}_{i+l}^{m,5l_m} \right]\right| ^{w}  \right] \leq C_{w,\varepsilon} \varPsi_{i}^{m,2} (k'_m)^{w} ( l_m ^{(-4v-w/2+\varepsilon)} + l_m ^{ -(v+\frac{1}{2})w}) .
\end{eqnarray*}
Using Lemma \ref{chi-property}(a) and (b), we have
\begin{equation*}
  \mathbb{E}\left[ (\bar{\mathcal{S}} _{11,i,l}^{m,2})^{w} | \mathcal{K}_{i}^{m} \right] \leq C_{w} \varPsi_{i}^{m,2} (k'_m)^{w} l_m^{-w/2}
  .
\end{equation*}
Thus, similar to proof of \eqref{mathV-w-res}, using Burkholder-Davis-Gundy inequality and Lemma \ref{chi-property}(a), we can show that for any $w \in \mathbb{N}$, $z \in \left\lbrace 2,3,4 \right\rbrace$, and $\varepsilon > 0$,
\begin{equation*}
  \mathbb{E}\left[ \left( (b_m - 6l_m)^{-1} \sum_{l=0}^{b_m -6 l_m} \bar{\mathcal{S}} _{11,i,l}^{m,z} \right)^{w} \bigg| \mathcal{K}_{i}^{m} \right] \leq C_{w} \varPsi_{i}^{m,2}  (k'_m)^{w} (b_m ^{-w/2} + l_m^{-2v-w/2+\varepsilon})
  .
\end{equation*}
Simple algebra shows that
\begin{eqnarray*}
  &&\sum_{l=0}^{b_m - 6l_m} \left( \bar{\mathcal{S}} _{11,i,l}^{m,1} -  (\vartheta_{1,i}^{m})^2 R_{11} \right) \\
  &=&  \sum_{l=0}^{b_m - 6l_m} (\vartheta _{1,i+l}^{m})^{2} \left( \sum_{d=-k'_m}^{k'_m} \chi_{1,i+l}\chi_{1,i+l+|d|} - R_{11}  \right)  \\
  &&+ R_{11} \sum_{l=0}^{b_m - 6l_m}  \left( (\vartheta _{1,i+l}^{m})^{2} - (\vartheta _{1,i}^{m})^{2} \right) \\
  &=&  \sum_{l=0}^{b_m - 6l_m} (\vartheta _{1,i+l}^{m})^{2} \left( \sum_{d=-k'_m}^{k'_m} \left( \chi_{1,i+l}\chi_{1,i+l+|d|} - r_{11}(|d|) \right)  + 2 \sum_{d=k'_m + 1}^{\infty} r_{11}(d)   \right)  \\
  &&+ R_{11} \sum_{l=0}^{b_m - 6l_m}  \left( (\vartheta _{1,i+l}^{m})^{2} - (\vartheta _{1,i}^{m})^{2} \right) \\
  &=&    \sum_{d=-k'_m}^{k'_m} \sum_{l=0}^{b_m - 6l_m} (\vartheta _{1,i+l}^{m})^{2} \left( \chi_{1,i+l}\chi_{1,i+l+|d|} - r_{11}(|d|) \right)  + 2 \sum_{l=0}^{b_m - 6l_m} (\vartheta _{1,i+l}^{m})^{2}  \sum_{d=k'_m + 1}^{\infty} r_{11}(d)     \\
  &&+ R_{11} \sum_{l=0}^{b_m - 6l_m}  \left( (\vartheta _{1,i+l}^{m})^{2} - (\vartheta _{1,i}^{m})^{2} \right) 
  .
\end{eqnarray*}
Since $\chi_{1,i+l}\chi_{1,i+l+|d|} - r_{11}(|d|)$ is centered and has finite moments of all orders, using Lemma \ref{gmeasurable-bound}(b), we can show that for any $w \in \mathbb{N}$ and $\varepsilon > 0$,
\begin{eqnarray*}
  && \mathbb{E}\left[ \left|\mathbb{E}\left[ \chi_{1,i+l}\chi_{1,i+l+|d|} - r_{11}(|d|) | \tilde{\mathcal{K}}_{i}^{m,l_m} \right]\right|^{w}  \right] \leq C_{w,\varepsilon} l_m^{-2v+\varepsilon} ,\cr
  && \mathbb{E}\left[ \left|\mathbb{E}\left[ \chi_{1,i+l}\chi_{1,i+l+|d|} - r_{11}(|d|) | \tilde{\mathcal{K}}_{i}^{m,l_m} \right]\right|^{w} | \mathcal{K}_{i}^{m} \right] \leq C_{w,\varepsilon} l_m^{-v+\varepsilon}
  .
\end{eqnarray*}
Similar to proof of \eqref{mathV-w-res}, using Burkholder-Davis-Gundy inequality
\begin{eqnarray*}
  && \mathbb{E}\left[ \left( \sum_{d=-k'_m}^{k'_m} \sum_{l=0}^{b_m - 6l_m} (\vartheta _{1,i+l}^{m})^{2} \left( \chi_{1,i+l}\chi_{1,i+l+|d|} - r_{11}(|d|) \right) \right)^{w} \Bigg| \mathcal{K}_{i}^{m} \right] \\
  &\leq& C_{w} (k'_m)^{w-1}  \sum_{d=-k'_m}^{k'_m} \mathbb{E}\left[ \left( \sum_{l=0}^{b_m - 6l_m} (\vartheta _{1,i+l}^{m})^{2} \left( \chi_{1,i+l}\chi_{1,i+l+|d|} - r_{11}(|d|) \right) \right)^{w} \Big| \mathcal{K}_{i}^{m} \right] \\
  &\leq& C_{w,\varepsilon} \varPsi_{i}^{m,2} ( (k'_m)^{w} l_m^{w/2} b_m ^{w/2} + b_m^{w} l_m^{-v+\varepsilon})
  ,
\end{eqnarray*}
where the first inequality is due to Jensen's inequality.
Due to the fact that $\left|\sum_{d=k'_m+1}^{\infty} r_{11}(d)\right|$ $\leq C (k'_m)^{-v+1}$ and the boundedness of $\bvartheta$, we have
\begin{equation*}
  \mathbb{E}\left[ \left( \sum_{l=0}^{b_m - 6l_m} (\vartheta _{1,i+l}^{m})^{2}  \sum_{d=k'_m + 1}^{\infty} r_{11}(d) \right) ^{w} \bigg| \mathcal{K}_{i}^{m} \right] \leq C_{w} (k'_m)^{-(v-1)w} b_m ^{w}
  .
\end{equation*}
Since $\vartheta_{1}$ is It\^{o} semimartingale, we have
\begin{equation*}
  \mathbb{E}\left[ \left( R_{11} \sum_{l=0}^{b_m - 6l_m}  \left( (\vartheta _{1,i+l}^{m})^{2} - (\vartheta _{1,i}^{m})^{2} \right)  \right)^{w} \bigg| \mathcal{K}_{i}^{m} \right] \leq C_{w} b_m ^{w+1} \Delta_m 
  .
\end{equation*}
Thus, we have for any $w \in \mathbb{N}$
\begin{eqnarray*}
  &&\mathbb{E}\left[ \left( (b_m -6l_m)^{-1}\sum_{l=0}^{b_m - 6l_m} \left( \bar{\mathcal{S}} _{11,i,l}^{m,1} -  (\vartheta_{1,i}^{m})^2 R_{11} \right)   \right)^{w} \big| \mathcal{K}_{i}^{m}  \right] \cr
  &\leq&  C_{w} \varPsi_{i}^{m,2}  ((k'_m)^{w} l_m^{w/2} b_m ^{-w/2} + (k'_m)^{-(v-1)w} + b_m \Delta_m + l_m^{-v + \varepsilon} )
  .
\end{eqnarray*}
Due to the boundedness of $\bvartheta$, we have
\begin{eqnarray*}
  \mathbb{E}\left[ \left( (b_m -6l_m)^{-1} (\vartheta_{1,i}^{m})^{2} R_{11} \right)^{w} | \mathcal{K}_{i}^{m} \right] \leq C b_m ^{-w}
  .
\end{eqnarray*}
Thus, we have for any $w \in \mathbb{N}$
\begin{align*}
  & \mathbb{E}\left[ \left( (b_m - 6l_m)^{-1} \left[ \sum_{l=0}^{b_m -6 l_m} \sum_{z=2}^{4} \bar{\mathcal{S}} _{11,i,l}^{m,z}  + \sum_{l=0}^{b_m - 6l_m} \left( \bar{\mathcal{S}} _{11,i,l}^{m,1} -  (\vartheta_{1,i}^{m})^2 R_{11} \right) + (\vartheta_{1,i}^{m})^2 R_{11}  \right]  \right) ^{w} \bigg| \mathcal{K}_{i}^{m} \right] \\
  \leq& \quad C_{w} \varPsi_{i}^{m,2}  ((k'_m)^{w} l_m^{w/2} b_m ^{-w/2} + (k'_m)^{-(v-1)w} + b_m \Delta_m )
\end{align*}
and
\begin{eqnarray*}
  \mathbb{E}\left[ \left( \hat{\bvartheta}_{11,i}^{m} - (\vartheta_{1,i}^{m})^2 R_{11} \right)^{w} \big| \mathcal{K}_{i}^{m} \right] \leq C_{w} \varPsi_{i}^{m,2}  ((k'_m)^{w} l_m^{w/2} b_m ^{-w/2} + (k'_m)^{-(v-1)w} + b_m \Delta_m )
  .
\end{eqnarray*}
$\blacksquare$

\textbf{Proof of Lemma \ref{lemma-D123}.}
Consider $\mathcal{D}_{m,1}$.
By Taylor's theorem, we have
\begin{eqnarray*}
  \left|f(\hat{\bSigma}_{ib_m}^{m,*}) - f(\hat{\bSigma}_{ib_m}^{c,m,*})\right|  &\leq&  C \left( \left|\hat{\bSigma}_{12,ib_m}^{m,*}\right| + \left|\hat{\bSigma}_{12,ib_m}^{c,m,*}\right| \right)  \left|\hat{\bSigma}_{11,ib_m}^{m,*} - \hat{\bSigma}_{11,ib_m}^{c,m,*}\right| \\
  && + C  \left|\hat{\bSigma}_{12,ib_m}^{m,*} - \hat{\bSigma}_{12,ib_m}^{c,m,*}\right| \\
  &\leq& C \left( 1 + \left|\hat{\bSigma}_{12,ib_m}^{c,m,*} - \bSigma_{12,ib_m}^{m,*}\right| + \left|\hat{\bSigma}_{12,ib_m}^{m,*} - \hat{\bSigma}_{12,ib_m}^{c,m,*}\right| \right) \\
  && \times \left|\hat{\bSigma}_{11,ib_m}^{m,*} - \hat{\bSigma}_{11,ib_m}^{c,m,*}\right| + C  \left|\hat{\bSigma}_{12,ib_m}^{m,*} - \hat{\bSigma}_{12,ib_m}^{c,m,*}\right| ,
\end{eqnarray*}
where the second inequality is due to the triangular inequality and the fact that $\bSigma$ is locally bounded.
By Lemma \ref{lemma:jump-preavg-element} and H\"older's inequality, we have
\begin{eqnarray}\label{D1-discs1}
  \mathbb{E}\left[ \left|\hat{\bSigma}_{11,ib_m}^{m,*} - \hat{\bSigma}_{11,ib_m}^{c,m,*}\right|  \right] &\leq& C \Delta_m ^{([v]-1)(\frac{1}{2} - \varpi_1)} \nonumber \\
  &\leq& C \Delta_m ^{\frac{1}{4} + \varepsilon }
,
\end{eqnarray}
\begin{eqnarray}\label{D1-discs2}
  \mathbb{E}\left[ \left|\hat{\bSigma}_{11,ib_m}^{m,*} - \hat{\bSigma}_{11,ib_m}^{c,m,*}\right|^2  \right] &\leq& C \Delta_m ^{([v]-2)(\frac{1}{2} - \varpi_1)}  \nonumber \\
  &\leq& C \Delta_m ^{\frac{1}{4} + \varepsilon },
\end{eqnarray}
and
\begin{eqnarray}
  && \mathbb{E}\left[ \left|\hat{\bSigma}_{12,ib_m}^{c,m} - \bSigma_{12,ib_m}^{m}\right| \left|\hat{\bSigma}_{11,ib_m}^{m,*} - \hat{\bSigma}_{11,ib_m}^{c,m,*}\right| \right] \nonumber\\
  &\leq&  \mathbb{E}\left[ \left( \hat{\bSigma}_{12,ib_m}^{c,m} - \bSigma_{12,ib_m}^{m} \right)^{3}  \right]^{1/3}  \mathbb{E}\left[ \left(  \hat{\bSigma}_{11,ib_m}^{m,*} - \hat{\bSigma}_{11,ib_m}^{c,m,*}  \right)^{3/2}  \right]^{1/3} \nonumber\\
  &\leq& C (b_m \Delta_m )^{1/3} \Delta_m ^{\frac{2}{3} ([v]-\frac{3}{2} )(\frac{1}{2} - \varpi_{1})}  \nonumber\\
  &\leq& C \Delta_m ^{\frac{1}{2} - \frac{1}{3} \kappa + \frac{1}{12([v]-2)}  }  \nonumber\\
  &\leq& C \Delta_m ^{\frac{1}{4} + \varepsilon }
  ,
\end{eqnarray}
for some $\varepsilon > 0$.
Thus, we have
\begin{eqnarray*}
  \mathbb{E}\left[ \left|\hat{\beta}_{i b_m } - \hat{\beta}_{i b_m }^{c,m}\right|  \right] \leq C \Delta_m ^{1/4 + \varepsilon}
  ,
\end{eqnarray*}
for some $\varepsilon > 0$.
Simple algebra shows that
\begin{eqnarray}\label{B-disc}
  \left|\hat{B}^{m}_{i} - \hat{B}^{c,m}_{i}\right|  &\leq&   C b_m ^{-1} \Delta_m ^{-1/2} \Bigg[ \left|\hat{\bvartheta}_{11,ib_m}^{m}\right|  \left| \frac{\hat{\bSigma}_{12,ib_m}^{m}}{\left( \hat{\bSigma}_{11,ib_m}^{m,*} \right)^2 } - \frac{\hat{\bSigma}_{12,ib_m}^{c,m,*}}{\left( \hat{\bSigma}_{11,ib_m}^{c,m} \right)^2 }  \right| \nonumber\\
  && + \left|\hat{\bvartheta}_{12,ib_m}^{m}\right|  \left| \frac{1}{\left( \hat{\bSigma}_{11,ib_m}^{m,*} \right) } - \frac{1}{\left( \hat{\bSigma}_{11,ib_m}^{c,m,*} \right) }  \right| \nonumber\\
  && + \left( \hat{\bvartheta}_{11,ib_m}^{m} \right)^2  \left| \frac{\hat{\bSigma}_{12,ib_m}^{m}}{\left( \hat{\bSigma}_{11,ib_m}^{m,*} \right)^3 } - \frac{\hat{\bSigma}_{12,ib_m}^{c,m}}{\left( \hat{\bSigma}_{11,ib_m}^{c,m,*} \right)^3 }  \right| \nonumber\\
  && + \left|\hat{\bvartheta}_{11,ib_m}^{m} \hat{\bvartheta}_{12,ib_m}^{m}\right|  \left| \frac{1}{\left( \hat{\bSigma}_{11,ib_m}^{m,*} \right) } - \frac{1}{\left( \hat{\bSigma}_{11,ib_m}^{c,m,*} \right) }  \right| \Bigg] .
\end{eqnarray}
For the third term on the right-hand side of \eqref{B-disc}, we have
\begin{align}\label{fourth-term-Bdiff}
  & \quad b_m ^{-1} \Delta_m ^{-1/2} \mathbb{E}\left[ \left( \hat{\bvartheta}_{11,ib_m}^{m} \right)^2  \left| \frac{\hat{\bSigma}_{12,ib_m}^{m}}{\left( \hat{\bSigma}_{11,ib_m}^{m,*}\right)^3 } - \frac{\hat{\bSigma}_{12,ib_m}^{c,m}}{\left( \hat{\bSigma}_{11,ib_m}^{c,m,*} \right)^3 }  \right|  \right] \nonumber\\
  & \leq C b_m ^{-1} \Delta_m ^{-1/2} \cr
  & \qquad \times \mathbb{E}\bigg[ \left( \hat{\bvartheta}_{11,ib_m}^{m} \right)^2  \Big\{  \left|\hat{\bSigma}_{12,ib_m}^{m} - \hat{\bSigma}_{12,ib_m}^{c,m}\right| + \left( \left|\hat{\bSigma}_{12,ib_m}^{m}\right| + \left|\hat{\bSigma}_{12,ib_m}^{c,m}\right|  \right)  \left|\hat{\bSigma}_{11,ib_m}^{m,*}- \hat{\bSigma}_{11,ib_m}^{c,m,*}\right| \Big\} \bigg] \nonumber\\
  & \leq C b_m ^{-1} \Delta_m ^{-1/2}  \cr
  & \qquad \times \mathbb{E} \bigg[ \left( (\tilde{e}_{11,ib_m}^{m})^2 + (\bvartheta_{11,ib_m}^{m})^2 \right)  \left( |e_{12,ib_m}^{m,d}| + |e_{11,ib_m}^{m,d}| + |e_{12,ib_m}^{m,d}e_{11,ib_m}^{m,d}| + |e_{12,ib_m}^{m}e_{11,ib_m}^{m,d}|  \right)     \bigg], \nonumber\\
\end{align}
where the first and second inequalities are due to Taylor's theorem and triangular inequality, respectively, and for any $x\in \left\lbrace 1,2 \right\rbrace$
\begin{eqnarray}\label{eq:error-expression}
  \tilde{e}_{1x,ib_m}^{m} = \hat{\bvartheta}_{1x,ib_m}^{m} - {\bvartheta}_{1x,ib_m}^{m} \quad \text{and} \quad e_{1x,ib_m}^{m,d} = \hat{\bSigma}_{1x,ib_m}^{m} - \hat{\bSigma}_{1x,ib_m}^{c,m}
  .
\end{eqnarray}
Using H\"older's inequality and Lemmas \ref{lemma:e}, \ref{lemma:jump-preavg-element}, and \ref{spot-noise-estimate}, we have
\begin{eqnarray*}
  && b_m^{-1} \Delta_m^{-1/2} \mathbb{E}\left[ (\tilde{e}_{11,ib_m}^{m})^{2} e_{12,ib_m}^{m,d} \right] \cr
  &\leq& b_m^{-1} \Delta_m^{-1/2}  \mathbb{E}\left[ (\tilde{e}_{11,ib_m}^{m})^{4} \right]^{1/2} \mathbb{E}\left[ (e_{12,ib_m}^{m,d})^{2} \right]^{1/2} \cr
  &\leq& C \Delta_m ^{1/4 + \varepsilon}, \cr
  && b_m^{-1} \Delta_m^{-1/2} \mathbb{E}\left[ (\tilde{e}_{11,ib_m}^{m})^{2} e_{11,ib_m}^{m,d} e_{12,ib_m}^{m,d} \right] \cr
  &\leq& b_m^{-1} \Delta_m^{-1/2} \mathbb{E}\left[ (\tilde{e}_{11,ib_m}^{m})^{4} \right]^{1/2} \mathbb{E}\left[ (e_{11,ib_m}^{m,d})^{4} \right]^{1/4} \mathbb{E}\left[ (e_{12,ib_m}^{m,d})^{4} \right]^{1/4} \cr
  &\leq&  C \Delta_m ^{1/4 + \varepsilon}, \quad \text{and} \quad\cr
  && b_m^{-1} \Delta_m^{-1/2} \mathbb{E}\left[ (\tilde{e}_{11,ib_m}^{m})^{2} e_{11,ib_m}^{m,d} e_{12,ib_m}^{m} \right] \cr
  &\leq& b_m^{-1} \Delta_m^{-1/2} \mathbb{E}\left[ (\tilde{e}_{11,ib_m}^{m})^{4} \right]^{1/2} \mathbb{E}\left[ (e_{11,ib_m}^{m,d})^{4} \right]^{1/4} \mathbb{E}\left[ (e_{12,ib_m}^{m})^{4} \right]^{1/4} \cr
  &\leq&  C \Delta_m ^{1/4 + \varepsilon}
  .
\end{eqnarray*}
Similarly, we can bound all other terms on the right-hand side of \eqref{fourth-term-Bdiff} by $C \Delta_m ^{1/4 + \varepsilon}$ for some $\varepsilon > 0$.
Furthermore, we can bound all rest terms on the right-hand side of \eqref{B-disc}  by $C \Delta_m ^{1/4 + \varepsilon}$ for some $\varepsilon > 0$.
Thus, we have
\begin{eqnarray}\label{D1}
  \mathcal{D}_{m,1} \xrightarrow[]{p} 0
  .
\end{eqnarray}

Consider $\mathcal{D}_{m,2}$.
Simple algebra shows that
\begin{equation*}
  \mathcal{D}_{m,2} =  \bar{\mathcal{D}}_{m,1} + \bar{\mathcal{D}}_{m,2} + \bar{\mathcal{D}}_{m,3} + \bar{\mathcal{D}}_{m,4}  ,
\end{equation*}
where
\begin{eqnarray*}
  \bar{\mathcal{D}}_{m,1} &=& b_m \Delta_m ^{3/4} \sum_{i=0}^{N_m-1} \mathcal{Q}_{1,ib_m}^{m}
  ,\quad \bar{\mathcal{D}}_{m,2} = b_m \Delta_m ^{3/4} \sum_{i=0}^{N_m-1} \mathcal{Q}_{2,ib_m}^{m} - \mathbb{E}\left[ \mathcal{Q}_{2,ib_m}^{m} | \mathcal{K}_{ib_m}^{m}  \right], \\
  \bar{\mathcal{D}}_{m,3} &=& b_m \Delta_m ^{3/4} \sum_{i=0}^{N_m-1} \mathbb{E}\left[ \mathcal{Q}_{2,ib_m}^{m} | \mathcal{K}_{ib_m}^{m}  \right] 
  ,\quad \bar{\mathcal{D}}_{m,4} = b_m \Delta_m ^{3/4} \sum_{i=0}^{N_m-1} B_{ib_m}^{c,m} - \hat{B}_{ib_m}^{c,m}, \\
  \mathcal{Q}_{1,ib_m}^{m} &=&  f(\bSigma_{ib_m}^{m}+e_{ib_m}^{m,*}) - f(\bSigma_{ib_m}^{m}) - \sum_{x=1}^{2}  \partial_{1x}f(\bSigma_{ib_m}^{m}) e_{1x,i b_m}^{m,*}\\
  &&  - \frac{1}{2} \sum_{x,y=1}^{2}  \partial_{1x,1y}^{2}f(\bSigma_{ib_m}^{m}) e_{1x,i b_m}^{m,*} e_{1y,i b_m}^{m,*}, \\
  \mathcal{Q}_{2,ib_m}^{m} &=& \frac{1}{2}  \sum_{x,y=1}^{2}  \partial_{1x,1y}^{2}f(\bSigma_{ib_m}^{m}) \left[ e_{1x,i b_m}^{m,*} e_{1y,i b_m}^{m,*} - \left( 2 b_m \Delta_m ^{1/2} \right)^{-1} \Xi(\bSigma_{ib_m}^{m},\bvartheta_{i,b_m}^{m})_{x,y} \right].
\end{eqnarray*}
By Taylor's theorem, we have
\begin{eqnarray*}
  \mathbb{E}\left[ \left| \mathcal{Q}_{1,ib_m}^{m} \right| \right] &\leq& C \mathbb{E}\left[ \left( \left| \bSigma_{12,ib_m}^{m} \right| + \left| \hat{\bSigma}_{12,ib_m}^{c,m,*} \right|   \right) \left| e_{11,ib_m}^{m,*} \right|^{3} + \left| e_{12,ib_m}^{m,*} \right|^{3}   \right] \\
  &\leq& C \mathbb{E}\left[ \left| e_{11,ib_m}^{m,*} \right|^{3} + \left| e_{12,ib_m}^{m,*} \right|^{3} + \left| e_{11,ib_m}^{m,*} \right|^{3} \left| e_{12,ib_m}^{m,*} \right|  \right] \\
  &\leq& C \mathbb{E}\left[ \left| e_{11,ib_m}^{m,*} \right|^{3} \right] + \mathbb{E}\left[ \left| e_{12,ib_m}^{m,*} \right|^{3}  \right] + E\left| \left| e_{11,ib_m}^{m,*} \right|^{4} \right]^{3/4} \mathbb{E}\left[ \left| e_{12,ib_m}^{m,*} \right|^{4}  \right]^{1/4} \\
  &\leq& C b_m \Delta_m
  .
\end{eqnarray*}
Then, we have $\bar{\mathcal{D}}_{m,1} \xrightarrow[]{p} 0$.
By Burkholder-Davis-Gundy inequality, we have
\begin{eqnarray*}
  \mathbb{E}\left[ \left|\bar{\mathcal{D}}_{m,2} \right|^{2}  \right] &\leq& C b_m^{2} \Delta_m ^{3/2} \sum_{i=0}^{N_m-1}  \mathbb{E}\left[ \left( \mathcal{Q}_{2,ib_m}^{m} \right)^2  \right] - \mathbb{E}\left[ \mathbb{E}\left[ \mathcal{Q}_{2,ib_m}^{m} | \mathcal{K}_{ib_m}^{m}  \right]^2 \right] \\
  &\leq& C b_m^2 \Delta_m ^{3/2} \sum_{i=0}^{N_m-1} \left( \mathbb{E}\left[ \left| e_{11,ib_m}^{m,*} \right|^{4} \right] + \mathbb{E}\left[ \left| e_{11,ib_m}^{m,*} \right|^{2} \left| e_{12,ib_m}^{m,*} \right|^{2} \right] +b_m ^{-2} \Delta_m ^{-1} \right)  \\
  &\leq& C b_m^2 \Delta_m ^{3/2} \sum_{i=0}^{N_m-1} \left( \mathbb{E}\left[ \left| e_{11,ib_m}^{m,*} \right|^{4} \right] + \mathbb{E}\left[ \left| e_{11,ib_m}^{m,*} \right|^{4} \right]^{1/2} \mathbb{E}\left[ \left| e_{12,ib_m}^{m,*} \right|^{4} \right]^{1/2} +b_m ^{-2} \Delta_m ^{-1} \right)  \\
  &\leq& C b_m ^{2} \Delta_m ^{3/2}
  ,
\end{eqnarray*}
where the second inequality is due to the fact that $\bSigma$ and $\bvartheta$ are locally bounded and the third and fourth inequalities are due to H\"older's inequality and Lemma \ref{lemma:e}, respectively.
Thus, we have $\bar{\mathcal{D}}_{m,2} \xrightarrow[]{p} 0$.
By Lemma \ref{lemma:e}, we have
\begin{eqnarray*}
  \mathbb{E}\left[ \left|\bar{\mathcal{D}}_{m,3}\right|  \right] &\leq& C b_m \Delta_m ^{3/4} \sum_{i=0}^{N_m-1} \sum_{x=1}^{2} \mathbb{E}\left[ \left| \mathbb{E}\left[ e_{11,i b_m}^{m,*} e_{1x,i b_m}^{m,*} - \left( 2 b_m \Delta_m ^{1/2} \right)^{-1} \Xi(\bSigma_{ib_m}^{c,m},\bvartheta_{i,b_m}^{m})_{1x} | \mathcal{K}_{ib_m}^{m} \right] \right| \right] \\
  &\leq& C \Delta_m ^{\varepsilon} \quad \text{for some} \quad \varepsilon > 0,
\end{eqnarray*}
where the first inequality is due to the fact that $\bSigma$ is locally bounded.
Therefore, we have $\bar{\mathcal{D}}_{m,3} \xrightarrow[]{p} 0$.
Similar to \eqref{D1}, we can show that $\bar{\mathcal{D}}_{m,4} \xrightarrow[]{p} 0$.
Thus, we have
\begin{equation}\label{D2}
  {\mathcal{D}}_{m,2} \xrightarrow[]{p} 0.
\end{equation}

Consider $\mathcal{D}_{m,3}$.
Since $\beta^c$ is locally bounded, we have
\begin{eqnarray*}
  \mathbb{E}\left[ \left|\Delta_m ^{-1/4} \int_{N_m b_m \Delta_m }^{1}  \beta_{s }^{c}  ds \right|  \right] &\leq& C b_m \Delta_m ^{3/4}
  .
\end{eqnarray*}
Using It\^o's lemma, we have 
\begin{eqnarray*} 
	&&  \Delta_m ^{-1/4} \sum ^{N_m -1}_{i=0}   \int^{(i+1)b_m\Delta_m}_{ib_m\Delta_m} \{ \beta^c_{ib_m\Delta_m} -\beta^c_{s} \} ds \\
	&& = - \Delta_m ^{-1/4}  \sum ^{N_m -1}_{i=0}   \int^{(i+1)b_m\Delta_m}_{ib_m\Delta_m} (t_{(i+1)b_m}-s)d\beta_s^c \\
	&& = - \Delta_m ^{-1/4} \sum ^{N_m -1}_{i=0}   \int^{(i+1)b_m\Delta_m}_{ib_m\Delta_m} (t_{(i+1)b_m}-s) (\mu_{\beta,s}ds+ \sigma_{\beta,s} dW_s).
\end{eqnarray*}
Using It\^o's lemma and It\^o's isometry, we can show
\begin{eqnarray*} 
	&& \mathbb{E}\left[\left( \Delta_m ^{-1/4} \sum ^{N_m -1}_{i=0}  \int^{(i+1)b_m\Delta_m}_{ib_m\Delta_m} (t_{(i+1)b_m}-s) \sigma_{\beta,s} dW_s \right)^2 \right] \\
  && = \mathbb{E}\left[ \Delta_m ^{-1/4} \sum ^{N_m -1}_{i=0}  \int^{(i+1)b_m\Delta_m}_{ib_m\Delta_m} (t_{(i+1)b_m}-s)^2 \sigma_{\beta,s}^2 ds \right] \\
  && \leq C \left( b_m  \Delta_m ^{3/4} \right) ^2
  .
\end{eqnarray*}
Also, we have
\begin{equation*} 
	\mathbb{E}\left[ \Delta_m ^{-1/4} \left|\sum ^{N_m -1}_{i=0}  \int^{(i+1)b_m\Delta_m}_{ib_m\Delta_m} (t_{(i+1)b_m}-s) \mu_{\beta,s}ds\right|  \right] \leq C b_m \Delta_m ^{3/4}
  .
\end{equation*}
Thus, we have
\begin{equation}  \label{D3}
  {\mathcal{D}}_{m,3} \xrightarrow[]{p} 0.
\end{equation}
$\blacksquare$

\textbf{Proof of Lemma \ref{lemma-D4}.}
We have
\begin{equation*}
  \mathcal{D}_{m,4}(p) = \hat{\mathcal{D}}_{m,1}(p) +  \hat{\mathcal{D}}_{m,2}(p)
  ,
\end{equation*}
where
\begin{eqnarray*}
  \hat{\mathcal{D}}_{m,1}(p) &=&  b_m \Delta_m ^{3/4} \sum_{i=0}^{N_m-1} \Biggl( \sum_{x=1}^{2}  \partial_{1x}f(\bSigma_{ib_m}^{m}) \left(  M'(p)_{1x,i b_m }^{m} + \xi_{1x,i b_m }^{m,1} + \xi_{1x,i b_m }^{m,2} \right) \\
  &&- \mathbb{E}\left[ \sum_{x=1}^{2}  \partial_{1x}f(\bSigma_{ib_m}^{m})  \left( M'(p)_{1x,i b_m }^{m} + \xi_{1x,i b_m }^{m,1} + \xi_{1x,i b_m }^{m,2} \right) \Big| \mathcal{K}_{ib_m}^{m}  \right] \Biggl),  \\
  \hat{\mathcal{D}}_{m,2}(p) &=& b_m \Delta_m ^{3/4} \sum_{i=0}^{N_m-1} \mathbb{E}\left[ \sum_{x=1}^{2}  \partial_{1x}f(\bSigma_{ib_m}^{m})  \left( M'(p)_{1x,i b_m }^{m} + \xi_{1x,i b_m }^{m,1} + \xi_{1x,i b_m }^{m,2} \right) \Big| \mathcal{K}_{ib_m}^{m}  \right] 
  .
\end{eqnarray*}
By Burkholder-Davis-Gundy inequality, we have
\begin{eqnarray*}
  \mathbb{E}\left[ \left( \hat{\mathcal{D}}_{m,1}(p) \right)^2  \right] &\leq& C b_m ^2 \Delta_m ^{3/2} \sum_{i=0}^{N_m-1} \sum_{x=1}^{2}  \mathbb{E}\left[ \left(  M'(p)_{1x,i b_m }^{m} \right)^2  + \left( \xi_{1x,i b_m }^{m,1} \right)^2  + \left( \xi_{1x,i b_m }^{m,2} \right)^2  \right] \\
  &\leq& C b_m ^2 \Delta_m ^{3/2} \sum_{i=0}^{N_m-1} \sum_{x=1}^{2} (p^{-1} b_m ^{-1} \Delta_m ^{-1/2} + b_m \Delta_m  + \Delta_m ^{2\tau(v-1)} + \Delta_m ^{\kappa -\varsigma - 2\tau}) \\ 
  &\leq& \frac{C}{p} 
  ,
\end{eqnarray*}
where the second inequality is due to Lemmas \ref{negligible-xi} and \ref{lemma:M}.
By Lemmas \ref{negligible-xi} and \ref{lemma:M}, we have
\begin{eqnarray*}
  \mathbb{E}\left[ \left| \hat{\mathcal{D}}_{m,2}(p) \right|  \right] &\leq& C b_m \Delta_m ^{3/4} \sum_{i=0}^{N_m-1} \sum_{x=1}^{2}  \mathbb{E}\left[ \left| \mathbb{E}\left[  M'(p)_{1x,i b_m }^{m}   +  \xi_{1x,i b_m }^{m,1}  +  \xi_{1x,i b_m }^{m,2} | \mathcal{K}_{ib_m}^{m}  \right] \right| \right] \\
  &\leq& C b_m  \Delta_m ^{3/4} \sum_{i=0}^{N_m-1} \sum_{x=1}^{2} \mathbb{E}\left[ \varPsi_{i,2}^{m} (p^{-1} \Delta_m ^{1/2}  + b_m \Delta_m  + \Delta_m ^{\tau(v-1)} + \Delta_m ^{(v+\frac{1}{2}) \varsigma - \tau} ) \right] \\ 
  &\leq& C ( b_m \Delta_m ^{3/4} + \Delta_m ^{\tau(v-1) - 1/4}) \\
  &\leq& C \Delta_m ^{\varepsilon}
  ,
\end{eqnarray*}
for some positive $\varepsilon$.
Thus, for sufficiently large $m$, we have
\begin{equation*}
  \mathbb{E}\left[ \left| \mathcal{D}_{m,4}(p) \right|  \right] \leq \mathbb{E}\left[ \left| \hat{\mathcal{D}}_{m,1}(p) \right|  \right] + \mathbb{E}\left[ \left| \hat{\mathcal{D}}_{m,2}(p) \right|  \right] \leq  \mathbb{E}\left[ \left( \hat{\mathcal{D}}_{m,1}(p) \right)^2  \right]^{1/2} + \mathbb{E}\left[ \left| \hat{\mathcal{D}}_{m,1}(p) \right|  \right] \leq \frac{C}{\sqrt{p}} ,
\end{equation*}
where the first and second inequality is due to triangular inequality and H\"older's inequality, respectively.
$\blacksquare$

\textbf{Proof of Lemma \ref{lemma-D5}.}
Let
\begin{eqnarray*}
  && \tilde{I}_{p,j}^{m} =  \left[  \frac{j}{L(m,p)} \right]    b_m, \quad \hat{I}_{p,j}^{m} = j-1 -  \left[ \frac{j-1}{L(m,p)} \right]   L(m,p), \quad \bar{I}_{p,j}^{m} = \tilde{I}_{p,j}^{m} + (p+2)k_m \hat{I}_{p,j}^{m} , \\
  && I(m,p) = N_m L(m,p), \quad \tilde{\mathcal{H}}(p)_{j}^{m} = \mathcal{K}_{\bar{I}_{p,j}^{m}}^{m} , \quad \tilde{\eta}(p)_{j}^{m} = \sum_{x=1}^{2}  \partial_{1x}f(\bSigma_{\tilde{I}_{p,j}^{m}}^{c,m}) \hat{\eta}(p)_{1x,\hat{I}_{p,j}^{m}}^{m,\tilde{I}_{p,j}^{m}}
  .
\end{eqnarray*}
Then, $\tilde{\eta}(p)_{j}^{m}$ is a martingale difference sequence with respect to a filtration $\tilde{\mathcal{H}}(p)_{j}^{m}$ and we have
\begin{equation*}
  \mathcal{D}_{m,5}(p) = b_m \Delta_m ^{3/4} \sum_{j=1}^{I(m,p)} \tilde{\eta}(p)_{j}^{m}
  .
\end{equation*}
To prove Lemma \ref{lemma-D5}, it suffices to show the following three convergences:
\begin{eqnarray}
  && b_m ^{2} \Delta_m ^{3/2} \sum_{j=1}^{I(m,p)} \mathbb{E}\left[ \left( \tilde{\eta}(p)_{j}^{m} \right)^2 | \tilde{\mathcal{H}}(p)_{j-1}^{m} \right] \xrightarrow[]{p} \int_{0}^{1} \mathcal{R}(p)_{s}^{2} ds , \label{tri-MDS-2} \\
  && b_m ^{4} \Delta_m ^{3} \sum_{j=1}^{I(m,p)} \mathbb{E}\left[ \left( \tilde{\eta}(p)_{j}^{m} \right)^4 | \tilde{\mathcal{H}}(p)_{j-1}^{m} \right] \xrightarrow[]{p} 0 , \label{tri-MDS-4} \\
  && b_m \Delta_m ^{3/4} \sum_{j=1}^{I(m,p)} \mathbb{E}\left[ \tilde{\eta}(p)_{j}^{m} \Delta(V,p)_{j}^{m}   | \tilde{\mathcal{H}}(p)_{j-1}^{m} \right] \xrightarrow[]{p} 0 \quad \text{for any } V \in \mathcal{M}, \label{tri-MDS-ortho}
\end{eqnarray}
where $\mathcal{M} = \mathcal{M}_1 \cup \left\lbrace (B,W)^{\top} \right\rbrace$ and $\mathcal{M}_1$ is the class of all bounded $(\mathcal{F}_t)$-martingales orthogonal to $(B,W)^{\top}$.
Consider \eqref{tri-MDS-2}.
The left hand side of \eqref{tri-MDS-2} is $H(p)_{1}^{m} + H(p)_{2}^{m} - H(p)_{3}^{m}$, where
\begin{eqnarray*}
  H(p)_{1}^{m} &=&  \frac{b_m ^{2} \Delta_m ^{3/2}}{(b_m  - 2 k_m)^2 \Delta_m ^2 k_m^2 \psi_0^2}  \sum_{j=1}^{I(m,p)} \sum_{x,y=1}^{2} \partial_{1x}f(\bSigma_{\tilde{I}_{p,j}^{m}}^{m}) \partial_{1y}f(\bSigma_{\tilde{I}_{p,j}^{m}}^{m})  \\
  && \times \mathbb{E}\left[ \zeta(p)_{1x,\bar{I}_{p,j}^{m}}^{m} \zeta(p)_{1y,\bar{I}_{p,j}^{m}}^{m} - \varXi(p)_{1x,1y,\bar{I}_{p,j}^{m}}^{m} | \tilde{\mathcal{H}}(p)_{j-1}^{m} \right], \\
  H(p)_{2}^{m} &=&  \frac{b_m ^{2} \Delta_m ^{3/2}}{(b_m  - 2 k_m)^2 \Delta_m ^2 k_m^2 \psi_0^2} \sum_{j=1}^{I(m,p)} \sum_{x,y=1}^{2} \partial_{1x}f(\bSigma_{\tilde{I}_{p,j}^{m}}^{m}) \partial_{1y}f(\bSigma_{\tilde{I}_{p,j}^{m}}^{m})  \varXi(p)_{1x,1y,\bar{I}_{p,j}^{m}}^{m}, \\
  H(p)_{3}^{m} &=&  b_m ^{2} \Delta_m ^{3/2} \sum_{j=1}^{I(m,p)} \sum_{x,y=1}^{2} \partial_{1x}f(\bSigma_{\tilde{I}_{p,j}^{m}}^{m}) \partial_{1y}f(\bSigma_{\tilde{I}_{p,j}^{m}}^{m}) \bar{\eta}(p)_{1x,\hat{I}_{p,j}^{m}}^{m,\tilde{I}_{p,j}^{m}} \bar{\eta}(p)_{1y,\hat{I}_{p,j}^{m}}^{m,\tilde{I}_{p,j}^{m}}
  .
\end{eqnarray*}
By Lemma \ref{lemma:M} and the fact that $\bSigma$ is locally bounded, we have
\begin{eqnarray}\label{H1}
  \mathbb{E}\left[ \left| H(p)_{1}^{m} \right|  \right] &\leq&  C \Delta_m ^{1/2} \sum_{j=1}^{I(m,p)} \sum_{x,y=1}^{2} \mathbb{E}\left[ \left| \partial_{1x}f(\bSigma_{\tilde{I}_{p,j}^{m}}^{m}) \partial_{1y}f(\bSigma_{\tilde{I}_{p,j}^{m}}^{m}) \varPsi_{\bar{I}_{p,j}^{m},1}^{m} \Delta_m ^{1/4}  \right|  \right] \nonumber\\
  &\leq& C \Delta_m ^{1/4}
  .
\end{eqnarray}
Using Lemma \ref{lemma:zeta} and the fact that $\bSigma$ is locally bounded, we have
\begin{equation}\label{H3}
  \mathbb{E}\left[ \left| H(p)_{3}^{m} \right| \right] \leq C  b_m ^{2} \Delta_m ^{3/2}  \sum_{j=1}^{I(m,p)} \sum_{x,y=1}^{2} p^2 b_m ^{-2} \leq C p \Delta_m .
\end{equation}
By Riemann integration, we have
\begin{equation}\label{H2}
  H(p)_2^{m} \rightarrow \int_{0}^{1} \mathcal{R}(p)_{s}^{2} ds
  .
\end{equation}
Then, \eqref{tri-MDS-2} follows from \eqref{H1}, \eqref{H3}, and \eqref{H2}.
Using the same arguments as that of proofs of (A.47) and (A.48) in \citet{jacod2019estimating}, we can show \eqref{tri-MDS-4} and \eqref{tri-MDS-ortho}, respectively.
$\blacksquare$

\subsection{Proof of Proposition \ref{prop-AsympVar}}

\textbf{Proof of Proposition \ref{prop-AsympVar}.}
Simple algebra shows that
\begin{eqnarray*}
  \mathcal{R}_{t_i}^{2} &=& \frac{2C_k}{\psi_{0}^{2}} \Biggl[ \Phi_{00}  \left( \frac{{\bSigma}_{22,i}^{m}}{{\bSigma}_{11,i}^{m}} - \frac{({\bSigma}_{12,i}^{m})^2}{({\bSigma}_{11,i}^{m})^2}  \right)     + \frac{\Phi_{01}}{C_k^{2}}  \left( \frac{{\bvartheta}_{22,i}^{m}}{{\bSigma}_{11,i}^{m}} -  \frac{2{\bSigma}_{12,i}^{m} {\bvartheta}_{12,i}^{m}  }{ ( {\bSigma}_{11,i}^{m}  )^2  }  + \frac{{\bSigma}_{22,i}^{m} {\bvartheta}_{11,i}^{m} }{( {\bSigma}_{11,i}^{m}  )^2 }   \right) \\
  &&  + \frac{\Phi_{11}}{C_k^3}  \left( \frac{2  ({\bSigma}_{12,i}^{m} {\bvartheta}_{11,i}^{m} )^2    }{({\bSigma}_{11,i}^{m} )^4} + \frac{{\bvartheta}_{11,i}^{m} {\bvartheta}_{12,i}^{m} }{({\bSigma}_{11,i}^{m} )^2} - 4 \frac{{\bSigma}_{12,i}^{m} {\bvartheta}_{11,i}^{m} {\bvartheta}_{12,i}^{m}  }{({\bSigma}_{11,i}^{m} )^3}  + \frac{({\bvartheta}_{11,i}^{m} )^2}{({\bSigma}_{11,i}^{m} )^2}  \right)     \Biggl]
  .
\end{eqnarray*}
We have
\begin{align}\label{mathR-decomp}
  &\mathbb{E}\left[ \left|\hat{\mathcal{R}}^{2,m}_{i} - \mathcal{R}_{t_i}^{2} \right| \right] \cr
  &\leq C \Biggl[  \mathbb{E}\left[ \left|\frac{\hat{\bSigma}_{22,i}^{m}}{\hat{\bSigma}_{11,i}^{m,*}} - \frac{{\bSigma}_{22,i}^{m}}{{\bSigma}_{11,i}^{m}}\right|  \right]  - \mathbb{E}\left[ \left|\frac{(\hat{\bSigma}_{12,i}^{m})^2}{(\hat{\bSigma}_{11,i}^{m,*})^2} - \frac{({\bSigma}_{12,i}^{m})^2}{({\bSigma}_{11,i}^{m})^2}\right|  \right] +  \mathbb{E}\left[ \left|\frac{\hat{\bvartheta}_{22,i}^{m}}{\hat{\bSigma}_{11,i}^{m,*}} - \frac{{\bvartheta}_{22,i}^{m}}{{\bSigma}_{11,i}^{m,*}}\right|  \right]  \nonumber\\
  & \quad + \mathbb{E}\left[ \left|\frac{\hat{\bSigma}_{12,i}^{m} \hat{\bvartheta}_{12,i}^{m}  }{ ( \hat{\bSigma}_{11,i}^{m,*}  )^2  } - \frac{{\bSigma}_{12,i}^{m} {\bvartheta}_{12,i}^{m}  }{ ( {\bSigma}_{11,i}^{m,*}  )^2  }\right|  \right] + \mathbb{E}\left[ \left|\frac{\hat{\bSigma}_{22,i}^{m} \hat{\bvartheta}_{11,i}^{m} }{( \hat{\bSigma}_{11,i}^{m,*}  )^2 }  - \frac{{\bSigma}_{22,i}^{m} {\bvartheta}_{11,i}^{m} }{( {\bSigma}_{11,i}^{m,*}  )^2 } \right|  \right] \nonumber\\
  & \quad + \mathbb{E}\left[ \left|\frac{  (\hat{\bSigma}_{12,i}^{m} \hat{\bvartheta}_{11,i}^{m} )^2    }{(\hat{\bSigma}_{11,i}^{m,*} )^4} - \frac{ ({\bSigma}_{12,i}^{m} {\bvartheta}_{11,i}^{m} )^2    }{({\bSigma}_{11,i}^{m,*} )^4}\right|  \right]  + \mathbb{E}\left[ \left|\frac{\hat{\bvartheta}_{11,i}^{m} \hat{\bvartheta}_{12,i}^{m} }{(\hat{\bSigma}_{11,i}^{m,*} )^2} - \frac{{\bvartheta}_{11,i}^{m} {\bvartheta}_{12,i}^{m} }{({\bSigma}_{11,i}^{m,*} )^2}\right|  \right]\nonumber\\
  &\quad  + \mathbb{E}\left[ \left|\frac{\hat{\bSigma}_{12,i}^{m} \hat{\bvartheta}_{11,i}^{m} \hat{\bvartheta}_{12,i}^{m}  }{(\hat{\bSigma}_{11,i}^{m,*} )^3} - \frac{{\bSigma}_{12,i}^{m} {\bvartheta}_{11,i}^{m} {\bvartheta}_{12,i}^{m}  }{({\bSigma}_{11,i}^{m,*} )^3}\right|  \right] + \mathbb{E}\left[ \left|\frac{(\hat{\bvartheta}_{11,i}^{m} )^2}{(\hat{\bSigma}_{11,i}^{m,*} )^2} - \frac{({\bvartheta}_{11,i}^{m} )^2}{({\bSigma}_{11,i}^{m,*} )^2}\right|  \right] \Biggl] 
  .
\end{align}
For the sixth term on the right-hand side of \eqref{mathR-decomp}, by the boundedness of $1/\hat{\bSigma}_{11,i}^{m,*}$, ${\bSigma}_{12,i}^{m}$, and ${\bvartheta}_{11,i}^{m}$ we have
\begin{eqnarray*}
  &&\quad \mathbb{E}\left[ \left|\frac{  (\hat{\bSigma}_{12,i}^{m} \hat{\bvartheta}_{11,i}^{m} )^2    }{(\hat{\bSigma}_{11,i}^{m,*} )^4} - \frac{ ({\bSigma}_{12,i}^{m} {\bvartheta}_{11,i}^{m} )^2    }{({\bSigma}_{11,i}^{m,*} )^4}\right|  \right] \\
  &&\leq \mathbb{E}\left[ \left|\frac{  (\hat{\bSigma}_{12,i}^{m} \hat{\bvartheta}_{11,i}^{m} )^2 - ({\bSigma}_{12,i}^{m} {\bvartheta}_{11,i}^{m} )^2    }{(\hat{\bSigma}_{11,i}^{m,*} )^4} \right| \right] + \mathbb{E}\left[ \left|  \frac{ ({\bSigma}_{12,i}^{m} {\bvartheta}_{11,i}^{m} )^2    }{({\bSigma}_{11,i}^{m,*} \hat{\bSigma}_{11,i}^{m,*}  )^4 } ((\hat{\bSigma}_{11,i}^{m,*} )^4 - ({\bSigma}_{11,i}^{m,*} )^4)\right|  \right] \\
  &&\leq C \mathbb{E}\left[ \left|  (\hat{\bSigma}_{12,i}^{m} \hat{\bvartheta}_{11,i}^{m} )^2 - ({\bSigma}_{12,i}^{m} {\bvartheta}_{11,i}^{m} )^2     \right| \right] + C  \mathbb{E}\left[ \left|  ((\hat{\bSigma}_{11,i}^{m,*} )^4 - ({\bSigma}_{11,i}^{m,*} )^4)\right|  \right] \\
  &&\leq C \mathbb{E}\left[ \left|  (\hat{\bSigma}_{12,i}^{m})^2 ((\hat{\bvartheta}_{11,i}^{m})^2  - ( {\bvartheta}_{11,i}^{m} )^2) \right|\right] + C\mathbb{E}\left[ \left| ({\bvartheta}_{11,i}^{m} )^2 ((\hat{\bSigma}_{12,i}^{m})^2 - ({\bSigma}_{12,i}^{m}  )^2 )    \right| \right] \\
  &&\quad + C  \mathbb{E}\left[ \left|  ((\hat{\bSigma}_{11,i}^{m,*} )^4 - ({\bSigma}_{11,i}^{m,*} )^4)\right|  \right] \\
  &&\leq C \mathbb{E}\left[ \left|  ((\hat{\bSigma}_{12,i}^{m})^2 - ({\bSigma}_{12,i}^{m})^2) ((\hat{\bvartheta}_{11,i}^{m})^2  - ( {\bvartheta}_{11,i}^{m} )^2) \right|\right] + C\mathbb{E}\left[ \left|  (\hat{\bSigma}_{12,i}^{m})^2 - ({\bSigma}_{12,i}^{m}  )^2     \right| \right] \\
  &&\quad + C  \mathbb{E}\left[ \left|  ((\hat{\bSigma}_{11,i}^{m,*} )^4 - ({\bSigma}_{11,i}^{m,*} )^4)\right|  \right] \\
  &&\leq C \sum_{w_1=1}^{2} \sum_{w_2=0}^{2} \mathbb{E}\left[ \left|\hat{\bSigma}_{12,i}^{m} - {\bSigma}_{12,i}^{m}\right|^{w_1} \left|\hat{\bvartheta}_{11,i}^{m} - {\bvartheta}_{11,i}^{m}\right|^{w_2}   \right] + C \sum_{w=1}^{4} \mathbb{E}\left[ \left|\hat{\bSigma}_{11,i}^{m,*} - {\bSigma}_{11,i}^{m,*}\right|^{w_1} \right] \\
  &&\leq C \Delta_m ^{\varepsilon} \quad \text{for some} \quad \varepsilon>0  ,
\end{eqnarray*}
where the sixth inequality is due to Lemmas \ref{lemma:e}, \ref{lemma:jump-preavg-element}, and \ref{spot-noise-estimate}.
Similarly, we can bound all rest terms on the right-hand side of \eqref{mathR-decomp} by $C \Delta_m ^{\varepsilon}$ for some $\varepsilon>0$.
Thus, we have
\begin{equation*}
  b_m \Delta_m  \sum_{i=0}^{\left[ \frac{1}{b_m\Delta_m} \right]  -1 } (\hat{\mathcal{R}}^{2,m}_{i b_m} - \mathcal{R}^2_{t_{ib_m}} ) \xrightarrow[]{p} 0
  .
\end{equation*}
Thus, using Riemann approximation, we can show that $\hat{S}_m \xrightarrow[]{p} \int_{0}^{1} \mathcal{R}^2_{t_{ib_m}} ds $.
$\blacksquare$

\subsection{Proof of Lemmas in Theorem \ref{Theorem-2}}

\textbf{Proof of Lemma \ref{Lemma-6}.}
Since the proof is similar to the case where $p=q=1$, we show the statements for $p=q=1$. 
With Assumption \eqref{Assumption-2}(d) and the iterative relationship in $h_i(\theta_0)$ and $|\alpha^g_0+\gamma^g_0|<1$, we have
\begin{eqnarray} \label{Equation-A.28}
      \mathbb{E}[\left|h_i(\theta_{0})\right|]
     &\leq& |\omega_0|+|\gamma^g_0|\mathbb{E}[\left|D_{i-1}(\theta_{0})\right|]+|\alpha_0^g+\gamma^g_0|(\mathbb{E}[\left|h_{i-1}(\theta_{0})\right|]) \nonumber \\
     &\leq& C  |\omega_0+\gamma^g_0|+|\alpha_0^g+\gamma^g_0|\mathbb{E}[\left|h_{i-1}(\theta_{0})\right|] \nonumber \\
     &\leq& \frac{C\left|\omega_0+\gamma^g_0\right|(1-|\alpha_0^g+\gamma^g_0|^{i-1})}{1-|\alpha_0^g+\gamma^g_0|}+\left|\alpha_0^g+\gamma^g_0 \right|^{i-1}\mathbb{E}[\left|h_1(\theta_{0})\right|] \nonumber \\
     &\leq& 
     \frac{C \left|\omega_0+\gamma^g_0\right|}{1-|\alpha_0^g+\gamma^g_0|}+  \mathbb{E}[\left|h_1(\theta_{0})\right|] < \infty  \quad  \text{a.s.},
\end{eqnarray} 
for any $i$.
Then, \eqref{Equation-A.28} and Assumption \eqref{Assumption-2}(d) derive $ \sup_{i \in \mathbb{N}} \mathbb{E}[\left|I\beta_i\right|] < \infty$. 
Similarly, we can show
\begin{equation*}
    \sup_{i \in \mathbb{N}}\mathbb{E}[ \sup_{\theta \in \Theta} | h_i(\theta) |] < \infty  \quad  \text{a.s.}
\end{equation*}
(b) Consider the first inequality. 
Since $\beta^c_i(\theta)$ is the linear function of $\omega$ and $\alpha^g$, we obtain
\begin{equation*}
 \mathbb{E}\left[ \sup_{\theta \in \Theta} \left| \frac{\partial {{h}_i(\theta)}}{\partial {\theta_{ j}}} \right| \right]  \leq C,
\end{equation*}
for $j=1,3$. 
For $j=2$, we have
\begin{eqnarray*}
    \left| \frac{\partial {{h}_i(\theta)}}{\partial {\theta_{j}}}\right|
    &\leq& \Bigg| \sum_{k=1}^{i-2}[k (\gamma^g)^{k-1}(\omega^g+\alpha^g 
        I\beta_{i-k})]  + (i-1)(\gamma^g)^{i-2} h_1(\theta) + (\gamma^g)^{i-1} \frac{\partial h_1(\theta)}{\partial \gamma^{g}} \Bigg|  \cr
    &\leq&  C \sum_{k=1}^{i-2} k |\gamma^g|^k \{|\omega^g_u|\vee|\omega^g_l|+|\alpha^g_u|\vee|\alpha^g_l|  | I\beta_{i-1-k}|\} + C. 
\end{eqnarray*}
Then,  from Lemma \ref{Lemma-6}(a) and $|\gamma^g|<1$, we obtain 
\begin{equation*}
 \sup_{i \in \mathbb{N}} \mathbb{E}\left[ \sup_{\theta \in \Theta}  \left|  \frac{\partial {{h}_i(\theta)}}{\partial {\theta_{j}}}\right| \right] \leq C  \quad  \text{a.s.}
\end{equation*}
Similarly, we can check the boundedness of the second and third derivatives.
\endpf

\subsection{Proof of Proposition \ref{Proposition-1}}
\textbf{Proof of Proposition \ref{Proposition-1}.}
For $k,n \in \mathbb{N}$, let 
\begin{equation*}
    R(k) \equiv \int ^n_{n-1} \frac{(n-t)^k}{k!}\beta^c_t(\theta)dt.
\end{equation*}
By It\^{o}'s lemma, we have almost surely
\begin{eqnarray*}
     R(k) &=& \frac{\beta_{n-1}^{c}(\theta)}{(k+1)!} + \frac{\omega_{1} + \sum_{i=1}^{q} \gamma_{i} \beta_{n-i}^{c} (\theta) + \sum_{j=2}^{p} \alpha_j \int_{n-j}^{n+1-j} \beta_{t}^{c}(\theta) dt  }{(k+3)! / 2} - \frac{\omega_{2} + \beta_{n-1}^{c}(\theta)}{(k+2)!}  \\
     &&   + \nu \int_{n-1}^{n} \left( \frac{(n-t)^{k+2}}{(k+1)!} - \frac{(n-t)^{k+2}}{(k+2)!}   \right) dZ_t + \alpha_1 R(k+1).
\end{eqnarray*}
Using the iterative relationship and the fact that $\alpha_{1}^{k} R(k) \xrightarrow[]{a.s.} 0$ for any $\alpha_{1} \in \mathbb{R}$, we have
\begin{equation*}
     \int ^n_{n-1} \beta^{c}_{t}(\theta)dt = R(0) = h_n(\theta) + D_n \quad \text{a.s.},
\end{equation*}
where
\begin{eqnarray*}
  h_n(\theta) &=&   \varrho_{1} \beta^{c}_{n-1}(\theta) - \varrho_{2} \left( \omega_{2} + \beta_{n-1}^{c}(\theta) \right)  \\
  && + 2 \varrho_3 \left( \omega_{1} - \beta^{c}_{n-1}(\theta) + \sum_{i=1}^{q} \gamma_{i} \beta^{c}_{n-i}(\theta) + \sum_{j=2}^{p} \alpha_j \int_{n-j}^{n+1-j} \beta_{t}^{c}(\theta) dt \right).
\end{eqnarray*}
By \eqref{EXTENDpq}, we have almost surely
\begin{eqnarray*}
  h_n(\theta) &=& \omega^{(1)} + \sum_{i=1}^{q}  \gamma_{i}^{(1)} \beta^{c}_{n-1-i}(\theta) + \sum_{j=1}^{p} \alpha_{j}^{(1)} \int_{n-j-1}^{n-j} \beta_{t}^{c}(\theta) dt \\
  &=& \omega^{(N)}  + \sum_{i=1}^{q}  \gamma_{i}^{(N)} \beta^{c}_{n-1-i}(\theta) + \sum_{j=1}^{p+N-1} \alpha_{j}^{(N)} \int_{n-j-1}^{n-j} \beta_{t}^{c}(\theta) dt
\end{eqnarray*}
for any integer $N \geq 2$, where $\omega^{(N)}$, $\gamma_{i}^{(N)}$ and $\alpha_{i}^{(N)}$ are recursively defined as follows:
\begin{eqnarray}
  && \gamma_{1}^{(-1)} = 2 \varrho_{3} , \quad \gamma_{1}^{(0)} = \varrho_{1} - \varrho_{2} + 2 \varrho_{3} \gamma_{1} , \nonumber\\
  && \gamma_{i}^{(1)} =  (\varrho_{1} - \varrho_{2} + 2 \varrho_{3} \gamma_{1})  \gamma_{i} + 2 \varrho_{3} \gamma_{i+1}, \quad \gamma_{k} = 0 \quad \text{for} \quad k \geq q+1, \label{gamma-(1)}\\
  && \gamma_{i}^{(N)} = \gamma_{1}^{(N-1)} \gamma_{i} + \gamma_{i+1}^{(N-1)} = \sum_{k=0}^{N \land (q-i)} \gamma_{1}^{(N-1-k)} \gamma_{i+k}, \label{gamma-(N)} \\
  && \omega^{(1)} = \left( \varrho_{1} - \varrho_{2} + 2 \varrho_{3} ( 1+ \gamma_{1})  \right) \omega_{1} - (\varrho_{1} + 2\varrho_{3} \gamma_{1}) \omega_{2}, \quad \omega^{(N)} = \omega^{(1)} + \omega \sum_{k=1}^{N-1} \gamma_1^{(k)}, \nonumber\\
  && \alpha_{j}^{(1)} = (\varrho_{1} - \varrho_{2} + 2 \varrho_{3} \gamma_{1})  \alpha_{j} + 2 \varrho_{3} \alpha_{j+1}, \quad \alpha_{k} = 0 \quad \text{for} \quad k \notin [1,p], \nonumber\\
  && \alpha_{j}^{(N)} = \alpha_{j}^{(N-1)} + \gamma_{1}^{(N-1)} \alpha_{j-N+1} = \sum_{k=0}^{N \land j} \gamma_{1}^{(k-1)} \alpha_{j-k+1} \label{alpha-(N)}.
\end{eqnarray}
Let $\bar{\gamma} = \sum_{i=1}^{q} \left|\gamma_{i}\right| < 1$.
Using the mathematical induction method and \eqref{gamma-(N)} with $i=1$, we can show that there exists $C > 0$ such that
\begin{equation}\label{gamma-1}
  \left|\gamma_{1}^{(N)}\right| \leq C \left( \bar{\gamma} \right)^{\frac{N}{2q} } \quad  \text{for any} \quad N \in \mathbb{N}
  ,
\end{equation}
which implies that there exists $C > 0$ such that
\begin{equation}\label{gamma-i}
  \left|\gamma_{i}^{(N)}\right| \leq C \left( \bar{\gamma} \right)^{\frac{N}{2q} } \quad  \text{for any} \quad i, N \in \mathbb{N}
  .
\end{equation}
Thus, we have
\begin{equation*}
  h_n(\theta) = \omega^{(\infty)} + \sum_{j=1}^{\infty} \alpha_{j}^{(\infty)} \int_{n-j-1}^{n-j} \beta_{t}^{c}(\theta) dt \quad \text{a.s.}
  ,
\end{equation*}
where $\omega^{(\infty)} = \omega^{(1)} + \omega \sum_{k=1}^{\infty} \gamma_{1}^{(k)} $ and $\alpha_{j}^{(\infty)} = \sum_{s=1}^{(j+1) \land p} \gamma_{1}^{(j-s)} \alpha_{s} $.
Simple algebra shows that
\begin{equation*}
  h_n(\theta) - \sum_{i=1}^{q} \gamma_{i} h_{n-i}(\theta) = \bar{\omega}  + \sum_{j=1}^{\infty} \bar{\alpha}_{j} \int_{n-j-1}^{n-j} \beta_{t}^{c}(\theta) dt
  ,
\end{equation*}
where
\begin{equation*}
  \bar{\omega} = \omega^{(\infty)} \left( 1 - \sum_{i=1}^{q} \gamma_{i} \right) 
\end{equation*}
and
\begin{eqnarray*}
  \bar{\alpha}_{j} &=&  \alpha_{j}^{(\infty)} - \sum_{k=1}^{(j-1) \land q} \gamma_{k} \alpha_{j-k}^{(\infty)} \\
  &=& \sum_{s=1}^{(j+1) \land p} \gamma_{1}^{(j-s)} \alpha_{s} - \sum_{k=1}^{(j-1) \land q} \gamma_{k} \sum_{s=1}^{(j-k+1) \land p}  \gamma_{1}^{(j-k-s)}\alpha_{s}
  .
\end{eqnarray*}
For $j \geq (p \lor q) + 1$, we have
\begin{eqnarray*}
  \bar{\alpha}_{j} &=&   \sum_{s=1}^{p} \gamma_{1}^{(j-s)} \alpha_{s} - \sum_{k=1}^{q} \gamma_{k} \sum_{s=1}^{(j-k+1) \land p}  \gamma_{1}^{(j-k-s)}\alpha_{s} \\
  &=& \sum_{s=1}^{p} \left(  \gamma_{1}^{(j-s)} - \sum_{k=1}^{(j-p+1) \land q} \gamma_{k} \gamma_{1}^{(j-s-k)}   \right)  \alpha_{s}  -   \sum_{k=(j-p+1) \land q + 1}^{p} \sum_{s=1}^{(j-k+1) \land p} \gamma_{k} \gamma_{1}^{(j-s-k)} \alpha_{s} \\
  &=& \mathbf{1}_{\left\lbrace j \leq p+q-2 \right\rbrace} \left[ \sum_{s=1}^{p-1} \sum_{k=j-p+2}^{q \land (j-s+1)} \gamma_{k} \gamma_{1}^{(j-s-k)}     \alpha_{s}  -   \sum_{k=j-p+2}^{q} \sum_{s=1}^{j-k+1} \gamma_{k} \gamma_{1}^{(j-s-k)} \alpha_{s} \right]  \\
  &=&  0 
  ,
\end{eqnarray*}
where the third equality is due to \eqref{gamma-(N)}.
Furthermore, after some tedious algebra, we have
\begin{eqnarray*}
  \bar{\alpha}_{j} &=& \mathbf{1}_{\left\lbrace j \leq q \right\rbrace} \gamma_{1}^{(-1)} \gamma_{j} \alpha_{1} + \mathbf{1}_{\left\lbrace j \leq p \right\rbrace} (\varrho_{1} - \varrho_{2}) \alpha_{j} + \mathbf{1}_{\left\lbrace j \leq p-1 \right\rbrace} \gamma_{1}^{(-1)} \alpha_{j+1} \\
  &=& \mathbf{1}_{\left\lbrace j \leq q \right\rbrace} 2 \varrho_{3} \gamma_{j} \alpha_{1} + \mathbf{1}_{\left\lbrace j \leq p \right\rbrace} (\varrho_{1} - \varrho_{2}) \alpha_{j} + \mathbf{1}_{\left\lbrace j \leq p-1 \right\rbrace} 2 \varrho_{3} \alpha_{j+1} \\
  &=& \alpha_{j}^{g}. 
\end{eqnarray*}

On the other hand, by \eqref{gamma-(N)}, we have for any integer $N \geq 2$
\begin{equation*}
  \sum_{i=1}^{q} \gamma_{i}^{(N)} = \gamma_{1}^{(N-1)}\sum_{i=1}^{q} \gamma_{i} + \sum_{i=1}^{q} \gamma_{i}^{(N-1)} - \gamma_{1}^{(N-1)}
  ,
\end{equation*}
which implies that
\begin{equation*}
  \sum_{i=1}^{q} \gamma_{i}^{(1)} - \sum_{i=1}^{q} \gamma_{i}^{(N)} = (1 - \sum_{i=1}^{q} \gamma_{i}) \sum_{k=1}^{N-1} \gamma_{1}^{(k)}
  .
\end{equation*}
By \eqref{gamma-i} and \eqref{gamma-(1)}, we have
\begin{eqnarray*}
  \left(1 - \sum_{i=1}^{q} \gamma_{i}\right) \sum_{k=1}^{\infty} \gamma_{1}^{(k)} &=&  \sum_{i=1}^{q} \gamma_{i}^{(1)} \\
  &=& \left( \varrho_{1} - \varrho_{2} + 2 \varrho_{3} ( \gamma_{1} + 1) \right) \sum_{i=1}^{q} \gamma_{i} - 2 \varrho_{3}\gamma_{1}
  .
\end{eqnarray*}
Thus, we have
\begin{eqnarray*}
  \bar{\omega} &=&  \left( \omega^{(1)} + \omega \sum_{k=1}^{\infty} \gamma_{1}^{(k)}  \right) \left( 1 - \sum_{i=1}^{q} \gamma_{i} \right)  \\
  &=&  \left( \varrho_{1} - \varrho_{2} + 2 \varrho_{3} ( 1+ \gamma_{1})  \right) \omega  \left( 1 - \sum_{i=1}^{q} \gamma_{i} \right) + (2 \varrho_{3} - \varrho_{2}) \omega_{2}  \left( 1 - \sum_{i=1}^{q} \gamma_{i} \right) \\
  && + \omega \sum_{k=1}^{\infty} \gamma_{1}^{(k)}   \left( 1 - \sum_{i=1}^{q} \gamma_{i} \right)  \\
  &=&  \left( \varrho_{1} - \varrho_{2} + 2 \varrho_{3} ( 1+ \gamma_{1})  \right) \omega  \left( 1 - \sum_{i=1}^{q} \gamma_{i} \right) + (2 \varrho_{3} - \varrho_{2}) \omega_{2}  \left( 1 - \sum_{i=1}^{q} \gamma_{i} \right) \\
  && + \omega \left( \left( \varrho_{1} - \varrho_{2} + 2 \varrho_{3} ( \gamma_{1} + 1) \right) \sum_{i=1}^{q} \gamma_{i} - 2 \varrho_{3}\gamma_{1} \right)   \\
  &=& (\varrho_{1} - \varrho_{2} + 2 \varrho_{3}) \omega + (2\varrho_{3} - \varrho_{2}) \left( 1 - \sum_{i=1}^{q} \gamma_{i} \right)  \omega_2 \\
  &=& \omega^g 
  ,
\end{eqnarray*}
and the proof of Proposition \ref{Proposition-1}(a) is complete.

Similar to the proof of  Proposition 2.3 \citep{francq2013garch}, Proposition \ref{Proposition-1}(b) can be shown with the result of Proposition \ref{Proposition-1}(a).

Using It\^o's isometry and It\^o's lemma, we have
\begin{eqnarray} \label{Equation-A.27}
    \mathbb{E}[D_i^2|\mathcal{F}_{i-1}] 
    &=& \mathbb{E}[4\nu_0^2\alpha_0^{-4} \int^i_{i-1}\{\alpha_0(i-t-\alpha_0^{-1})e^{\alpha_0(i-t)}+1\}^2 Z_t^2dt|\mathcal{F}_{i-1}] \cr
    &=& 4\nu_0^2\alpha_0^{-4} \int^i_{i-1}\{\alpha_0(i-t-\alpha_0^{-1})e^{\alpha_0(i-t)}+1\}^2 (t-i+1)dt \cr
    &\leq&  C \quad  \text{a.s.},
\end{eqnarray}
which completes the proof of Proposition \ref{Proposition-1}(c).

\endpf

\textbf{Proof of Lemma \ref{Lemma-7}.}
Consider \eqref{Equation-A.29}. We have
\begin{eqnarray*}
    \left| \hat{L}_{n,m} (\theta) - \hat{L}_{n} (\theta) \right| 
    &\leq& \frac{1}{n} \sum^n_{i=1} \left|RIB^2_i- I\beta^2_i \right| + \frac{2}{n} \sum^n_{i=1} \left| RIB_i\{h_i(\theta)-\hat{h}_i(\theta)\} \right| \cr
    & &  +\frac{2}{n} \sum^n_{i=1}  \left|  h_i(\theta)( I\beta_i-RIB_i)\right|+ \frac{1}{n} \sum^n_{i=1} | \hat{h}^2_i(\theta)-h^2_i(\theta)|  \cr
    &=& \left( I\right) +\left( II\right) +\left( III\right) + \left( IV\right).
\end{eqnarray*}
For $\left( I\right)$, we have 
\begin{eqnarray*}
    \frac{1}{n} \sum^n_{i=1} \left| RIB^2_i - I\beta^2_i\right|
    &=& \frac{1}{n} \sum^{n}_{i=1} \Big|  \{RIB_i -I\beta_i \}    \{RIB_i +I\beta_i \} \Big| \\
    &=& O_p(m^{-1/4} ),
\end{eqnarray*}
where the last inequality is due to Assumption \ref{Assumption-2}(f).
Also, we obtain
\begin{eqnarray} \label{Equation-A.32}
    |\sup_{\theta \in \Theta} \big( \hat{h}_i(\theta)-h_i(\theta) \big)  | 
    &\leq& C \sum^{i-2}_{k=0} |\gamma_u|^k \vee |\gamma_l|^k 
    |  RIB_{i-1-k}-I\beta_{i-1-k} |\cr
    &=& O_p( m^{-1/4} )  \quad  \text{a.s.}
\end{eqnarray}
For  $\left( II\right)$, by \eqref{Equation-A.32}, we have 
\begin{eqnarray*} 
     \sup_{\theta \in \Theta} \frac{2}{n} \sum^n_{i=1} \left| RIB_i\{h_i(\theta)-\hat{h}_i(\theta)\} \right| 
     &=& O_p(m^{-1/4}  ).
\end{eqnarray*}
For $(III)$, by Lemma \ref{Lemma-6}(a) and Assumption \ref{Assumption-2}(f), we have
\begin{eqnarray*}
    \sup_{\theta \in \Theta} \frac{2}{n} \sum^n_{i=1}  \left|  h_i(\theta)( I\beta_i-RIB_i)\right| 
    &=& O_p(m^{-1/4} ).
\end{eqnarray*}
For $(IV)$, by \eqref{Equation-A.32} and Lemma \ref{Lemma-6}(a), we have
\begin{eqnarray*}
    \sup_{\theta \in \Theta} \sum^n_{i=1} \left| \hat{h}^2_i(\theta)-h^2_i(\theta) \right|
    &=& \frac{1}{n} \sum^{n}_{i=1}\sup_{\theta \in \Theta} \Big|  \{\hat{h}_i(\theta)-h_i(\theta) \} \{\hat{h}_i(\theta) + h_i(\theta) \} \Big|  \cr
    &=& O_p(m^{-1/4}     ).
\end{eqnarray*}
Hence, we have 
\begin{equation*}
    \sup_{\theta \in \Theta} \left| \hat{L}_{n,m} (\theta) - \hat{L}_{n} (\theta) \right|= O_{p}\left( m^{-1/4}  \right).
\end{equation*}

Consider \eqref{Equation-A.30}. 
We have 
\begin{equation*}
    \hat{L} _{n}(\theta)-L _{n}(\theta)=-\frac{2}{n} \sum_{i=1}^{n} D_i\{h_i(\theta_0)-h_i(\theta) \}.
\end{equation*}
Since $h_i(\theta)$ is adapted to $\mathcal{F}_{i-1}$, $D_i\{h_i(\theta_0)-h_i(\theta) \}$ is also a martingale difference. 
Also, $D_i\{h_i(\theta_0)-h_i(\theta) \}$ is uniform integrable.
Then,  by application of Theorem 2.22 in \citet{hall2014martingale}, we can show 
\begin{equation*}
    \left|\hat{L} _{n}(\theta)-L _{n}(\theta)\right| \rightarrow 0 \quad \text{in probability}.
\end{equation*}
Define 
\begin{equation*}
    G_n(\theta)=\hat{L} _{n}(\theta)-L _{n}(\theta).
\end{equation*}
From Theorem 3 in \citet{andrews1992generic}, the stochastic equicontinuity of $G_n(\theta)$ implies that $G_n(\theta)$ uniformly converges to 0.
Thus, it is enough to show that $G_n(\theta)$ is stochastic equicontinuous.
By the mean value theorem and Taylor expansion, there exists $\theta^{\ast }$ between $\theta$ and $\theta'$ such that 
\begin{eqnarray*}
    \left| G_n(\theta) -G_n(\theta') \right|
    &=&  \frac{2}{n}\sum_{i=1}^{n} \left| \dfrac{\partial h_i(\theta^{\ast })}{\partial \theta} D_i(\theta-\theta') \right| \cr
    &\leq&  \frac{C}{n}\sum_{i=1}^{n} \left \|  \dfrac{\partial h_i(\theta^{\ast })}{\partial \theta} D_i \right\|_{\max} \|\theta-\theta'\|_{\max} .
\end{eqnarray*}
Similar to  the proofs of Lemma \ref{Lemma-6}(b) and Assumption \ref{Assumption-2}(d), we can show that 
\begin{equation*} 
    \mathbb{E}\left[ \left\|  \sup_{\theta^{\ast } \in \Theta} \left| \frac{\partial {h_i(\theta^{\ast })}}{\partial {\theta_{ j}}} D_i \right| \right\|_{\max} \right] \leq C \quad \text{a.s.},
\end{equation*}
which implies that $G_n(\theta)$ is stochastic equicontinuous.

Finally, the triangular inequality concludes \eqref{Equation-A.31}.
\endpf

\textbf{Proof of Proposition \ref{Proposition-2}.}
First, we show that there is a unique maximizer of $L_n(\theta)$. 
$L_n(\theta)$ is concave and the solution $\theta$ of $ \partial L_n (\theta) / \partial \theta = 0$ should satisfy $h_i(\theta)=h_i(\theta_0)$ for all $i=1, \ldots , n$. 
Thus, the maximizer $\theta^{\ast}$ should satisfy $h_i(\theta^{\ast })=h_i(\theta_0)$ for all $i=1, \ldots , n$. 
By Proposition \ref{Proposition-1}(a), we have
\begin{eqnarray*}
	h_i (\theta)= \frac{\omega^g}{\varphi_{\theta} (1) } + \frac{\Upsilon_{\theta} (B) }{\varphi_{\theta} (B) } D_{i} \text{ a.s.},
\end{eqnarray*}
where $B$ is the back operator. 
Since $\varphi_{\theta_0} (B)$ and $\Upsilon_{ \theta_0} (B)$  do not have the common root and $D_i$'s are nondegenerating,  to satisfy $h_i(\theta^{\ast })=h_i(\theta_0)$ for all $i=1, \ldots , n$, we should have $\theta^{\ast } = \theta_0$.
 Thus, there is a unique maximizer, $\theta_0$. 
Then, the statement can be shown by Theorem 1 in \citet{xiu2010quasi} with the result of Lemma \ref{Lemma-7}.
\endpf
\\

\subsection{Proof of Theorem \ref{thm:BoundedMoment}}

\textbf{Proof of Lemma \ref{lemma:BoundE2BoundCE}.}
By Jensen's inequality and tower property, we have
\begin{eqnarray*}
  \mathbb{E}\left[ | \mathbb{E}\left[ |P_{t+u}|^{w} | \mathcal{F}_{t} \right] - \mathbb{E}\left[ |P_{t+u}|^{w} \right] |^{k} \right] &\leq& C_{k} \left( \mathbb{E}\left[ \mathbb{E}\left[ |P_{t+u}|^{kw} | \mathcal{F}_{t} \right] \right]  + \mathbb{E}\left[ |P_{t+u}|^{kw} \right]  \right) \cr
  &\leq& C_{k}
  .
\end{eqnarray*}
Thus, we have
\begin{equation*}
  \mathbb{E}\left[ |P_{t+u}|^{w} | \mathcal{F}_{t} \right] \leq C' \varPhi_{i}^{m,k} 
  ,
\end{equation*}
for some $C' >0$ and $\mathcal{F}_{t}$-measurable random variable $\varPhi_{i}^{m,k} $.
$\blacksquare$

\textbf{Proof of Lemma \ref{lemma:moment-itoproperty}.}
Consider the first part of \eqref{eq:moment-itoproperty}.
Using Jensen's inequality and Burkholder-Davis-Gundy inequality, we have
\begin{eqnarray*}
  \mathbb{E}\left[ \sup_{u \in [0,s]} |X_{t+u} - X_{t}|^{w} \Big| \mathcal{F}_{t} \right] &\leq& C \bigg( s^{w-1} \int_{t}^{t+s} \mathbb{E}\left[ |\mu_{u}|^{w} | \mathcal{F}_{t}  \right]  du + s^{w/2-1} \int_{t}^{t+s} \mathbb{E}\left[ |\sigma_{u}|^{w} | \mathcal{F}_{t}  \right]  du \cr
  && + \int_{t}^{t+s} du \int_{E} \mathbb{E}\left[ |\mathfrak{d}(u,z)|^{w} | \mathcal{F}_{t} \right] \lambda(dz)   \bigg) \cr
  &\leq& C \varPhi_{t}^{k} s
  ,
\end{eqnarray*}
where the last inequality is due to Lemma \ref{lemma:BoundE2BoundCE}.
Similarly, we can show the second part of \eqref{eq:moment-itoproperty}.
$\blacksquare$

\subsubsection{A key decomposition}\label{key-decomp-longspan}
Similar to \eqref{key-decomp-thm1}, we have
\begin{equation}\label{key-decomp-BoundedMoment}
  \Delta_m ^{-1/4} \left( RIB_{1} - I\beta_{1} \right) = \sum_{z=1}^{8} \mathcal{Z}_{m,z}.
  ,
\end{equation}
where 
\begin{align*}
  &\mathcal{Z}_{m,1} = b_m \Delta_m ^{3/4} \sum_{i=0}^{N_m-1} \left\lbrace \left( {B}^{c,m}_{ib_m} - \hat{B}^{m}_{ib_m}  \right) - \mathbb{E}\left[ {B}^{c,m}_{ib_m} - \hat{B}^{m}_{ib_m}   | \mathcal{K}_{ib_m}^{m} \right] \right\rbrace ,\cr
  &\mathcal{Z}_{m,2} = b_m \Delta_m ^{3/4} \sum_{i=0}^{N_m-1} \left\lbrace \left( \hat{\beta}_{i b_m } - \hat{\beta}_{i b_m }^{c,m}   \right) - \mathbb{E}\left[ \hat{\beta}_{i b_m } - \hat{\beta}_{i b_m }^{c,m}    | \mathcal{K}_{ib_m}^{m} \right]  \right\rbrace , \cr
  &\mathcal{Z}_{m,3} = b_m \Delta_m ^{3/4} \sum_{i=0}^{N_m-1} \mathbb{E}\left[ \hat{\beta}_{i b_m } - \hat{\beta}_{i b_m }^{c,m}  + {B}^{c,m}_{ib_m} - \hat{B}^{m}_{ib_m} | \mathcal{K}_{i}^{m}  \right] ,\cr
  &\mathcal{Z}_{m,4} = b_m \Delta_m ^{3/4} \sum_{i=0}^{N_m-1} \Bigg\lbrace  \hat{\beta}_{i b_m }^{c,m} - \beta_{i b_m }^{c,m} - \sum_{x=1}^{2}  \partial_{1x}f(\bSigma_{ib_m}^{c,m}) e_{1x,i b_m}^{m,*} - \frac{1}{2}  \sum_{x,y=1}^{2} \partial^{2}_{1x,1y}f(\bSigma_{ib_m}^{c,m}) e_{1x,i b_m}^{m,*} e_{1y,i b_m}^{m,*}   \cr
  &\qquad\quad - \mathbb{E}\left[ \hat{\beta}_{i b_m }^{c,m} - \beta_{i b_m }^{c,m} - \sum_{x=1}^{2}  \partial_{1x}f(\bSigma_{ib_m}^{c,m}) e_{1x,i b_m}^{m,*} - \frac{1}{2}  \sum_{x,y=1}^{2} \partial^{2}_{1x,1y}f(\bSigma_{ib_m}^{c,m}) e_{1x,i b_m}^{m,*} e_{1y,i b_m}^{m,*} \bigg| \mathcal{K}_{ib_m}^{m} \right] ,\cr
  &\mathcal{Z}_{m,5} = b_m \Delta_m ^{3/4} \sum_{i=0}^{N_m-1} \mathbb{E}\Bigg[ \hat{\beta}_{i b_m }^{c,m} - \beta_{i b_m }^{c,m} - \sum_{x=1}^{2}  \partial_{1x}f(\bSigma_{ib_m}^{c,m}) e_{1x,i b_m}^{m,*} \cr
  &\qquad\qquad\qquad\qquad\qquad - \frac{1}{2}  \sum_{x,y=1}^{2} \partial^{2}_{1x,1y}f(\bSigma_{ib_m}^{c,m}) e_{1x,i b_m}^{m,*} e_{1y,i b_m}^{m,*}  \bigg| \mathcal{K}_{ib_m}^{m}  \Bigg] ,\cr
  &\mathcal{Z}_{m,6} = b_m \Delta_m ^{3/4} \sum_{i=0}^{N_m-1} \sum_{x,y=1}^{2} \frac{1}{2}  \partial^{2}_{1x,1y}f(\bSigma_{ib_m}^{c,m}) \left( e_{1x,i b_m}^{m,*} e_{1y,i b_m}^{m,*} - b_m ^{-1} \Delta_m ^{-1/2} \Xi (\bSigma_{ib_m}^{c,m}, \bvartheta_{ib_m}^{m}) \right) ,\cr
  &\mathcal{Z}_{m,7} = b_m \Delta_m ^{3/4} \sum_{i=0}^{N_m-1} \sum_{x=1}^{2}  \partial_{1x}f(\bSigma_{ib_m}^{c,m}) e_{1x,ib_m}^{m,*}, \cr
  &\mathcal{Z}_{m,8} = \Delta_m ^{-1/4} \left[ \sum_{i=0}^{N_m - 1} \int_{i b_m \Delta_m }^{(i+1) b_m \Delta_m } \beta_{i b_m }^{c,m} - \beta_{s }^{c} ds + \int_{N_m b_m \Delta_m }^{1}  \beta_{s }^{c}  ds \right]
  .
\end{align*}
Using Burkholder-Davis-Gundy inequality, we have
\begin{eqnarray}\label{mathZ1-burk}
  \mathbb{E}\left[ \left|\mathcal{Z}_{m,1}\right|^{2}  \right] &\leq&  C b_m^{2} \Delta_m ^{3/2} \sum_{i=0}^{N_m - 1} \mathbb{E}\left[ \left\lbrace \left( {B}^{c,m}_{ib_m} - \hat{B}^{m}_{ib_m}  \right) - \mathbb{E}\left[ {B}^{c,m}_{ib_m} - \hat{B}^{m}_{ib_m}   | \mathcal{K}_{ib_m}^{m} \right] \right\rbrace^{2} \right] \cr
  &\leq& C \Delta_m ^{1/2} \sum_{i=0}^{N_m - 1} \mathbb{E}\left[ b_m ^{2} \Delta_m   | \hat{B}^{m}_{ib_m} - {B}^{c,m}_{ib_m} |^{2} \right]
  .
\end{eqnarray}
By Talyor's theorem, we have
\begin{eqnarray}\label{mathZ1-taylor}
  && \mathbb{E}\left[ b_m ^{2} \Delta_m \left|\hat{B}^{m}_{ib_m} - {B}^{c,m}_{ib_m}\right|^{2}  \right] \cr
  &\leq& C \mathbb{E} \Bigg[  |\hat{\bSigma}_{11,ib_m}^{m,*}-{\bSigma}_{11,ib_m}^{c,m}|^2 \left( \left| \frac{{\tilde{\bvartheta}_{11}}^{2} {\tilde{\bSigma}_{12}}^{2}}{{\tilde{\bSigma}_{11}}^6}  \right| + \left|\frac{{\tilde{\bvartheta}_{11}}^4 {\tilde{\bSigma}_{12}}^2}{{\tilde{\bSigma}_{11}}^8} \right| + \left|\frac{{\tilde{\bvartheta}_{12}}^2}{{\tilde{\bSigma}_{11}}^4} \right| + \left|\frac{{\tilde{\bvartheta}_{11}}^2 {\tilde{\bvartheta}_{12}}^2}{{\tilde{\bSigma}_{11}}^6} \right|       \right) \cr
  &&+ |\hat{\bSigma}_{12,ib_m}^{m}-{\bSigma}_{12,ib_m}^{c,m}|^{2} \left( \left|\frac{{\tilde{\bvartheta}_{11}}^2}{{\tilde{\bSigma}_{11}}^4}  \right| + \left|\frac{{\tilde{\bvartheta}_{11}}^4}{{\tilde{\bSigma}_{11}}^6} \right|  \right) + |\hat{\bvartheta}_{12,ib_m}^{m}-{\bvartheta}_{12,ib_m}^{m}|^2 \left( \left|\frac{1}{{\tilde{\bSigma}_{11}}^2} + \frac{1}{{\tilde{\bSigma}_{11}}^4} \right|  \right) \cr
  &&+ |\hat{\bvartheta}_{11,ib_m}^{m}-{\bvartheta}_{11,ib_m}^{m}|^{2} \left( \left|\frac{{\tilde{\bSigma}_{12}}^2}{{\tilde{\bSigma}_{11}}^4} \right| + \left|\frac{{\tilde{\bvartheta}_{11}}^2 {\tilde{\bSigma}_{12}}^2}{{\tilde{\bSigma}_{11}}^6} \right| + \left|\frac{{\tilde{\bvartheta}_{12}}^2}{{\tilde{\bSigma}_{11}}^4} \right|    \right) \Bigg]
  ,
\end{eqnarray}
where
\begin{eqnarray*}
  && \tilde{\bSigma}_{11} = a \hat{\bSigma}_{11,ib_m}^{m,*} + (1-a) {\bSigma}_{11,ib_m}^{c,m}, \quad \tilde{\bSigma}_{12} = a \hat{\bSigma}_{12,ib_m}^{m,*} + (1-a) {\bSigma}_{12,ib_m}^{c,m}, \cr
  && \tilde{\bvartheta}_{11} = a \hat{\bvartheta}_{11,ib_m}^{m} + (1-a) {\bvartheta}_{11,ib_m}^{m} ,\quad \tilde{\bvartheta}_{12} = a \hat{\bvartheta}_{12,ib_m}^{m} + (1-a) {\bvartheta}_{11,ib_m}^{m}
  ,
\end{eqnarray*}
for some $a \in (0,1)$.
Then, we can bound the right-hand side of \eqref{mathZ1-taylor} by $C b_m \Delta_m ^{1/2 }$.
For example,
\begin{eqnarray*}
  && \mathbb{E}\left[ |\hat{\bSigma}_{11,ib_m}^{m,*}-{\bSigma}_{11,ib_m}^{c,m}|^2 \left|\frac{{\tilde{\bvartheta}_{11}}^4 {\tilde{\bSigma}_{12}}^2}{{\tilde{\bSigma}_{11}}^8} \right|  \right] \cr
  &\leq& \mathbb{E}\left[ |\hat{\bSigma}_{11,ib_m}^{m,*}-{\bSigma}_{11,ib_m}^{c,m}|^2 \left|\frac{{\tilde{\bvartheta}_{11}}^4 {\tilde{\bSigma}_{12}}^2}{({{\bSigma}_{11,ib_m}^{c,m}})^8} \right|  \right] + \delta_m^{-8} \mathbb{E}\left[ |\hat{\bSigma}_{11,ib_m}^{m,*}-{\bSigma}_{11,ib_m}^{c,m}|^2 \left|{\tilde{\bvartheta}_{11}}^4 {\tilde{\bSigma}_{12}}^2 \right|  \right] \cr
  &\leq& C \mathbb{E}\left[ ((e_{11,ib_m}^{m,*})^{2} + (e_{11,ib_m}^{d,m,*})^{2}) (\bvartheta_{11}^{4} + (\tilde{e}_{11,ib_m}^{m})^{4}) ((e_{12,ib_m}^{m})^{2} + (e_{12,ib_m}^{d,m})^{2} + (\bSigma_{12,ib_m}^{c,m})^{2}) (\bSigma_{11,ib_m}^{c,m})^{-8} \right] \cr
  &&+ C \delta_{m}^{-8} \mathbb{E}\left[ ((e_{11,ib_m}^{m,*})^{2} + (e_{11,ib_m}^{d,m,*})^{2}) (\bvartheta_{11}^{4} + (\tilde{e}_{11,ib_m}^{m})^{4}) ((e_{12,ib_m}^{m})^{2} + (e_{12,ib_m}^{d,m})^{2}  + (\bSigma_{12,ib_m}^{c,m})^{2}) \right] \cr
  &\leq& C \bigg( \mathbb{E}\left[ (e_{11,ib_m}^{m,*})^{2} (\tilde{e}_{11,ib_m}^{m})^{4} (e_{12,ib_m}^{m})^{2} (\bSigma_{11,ib_m}^{c,m})^{-8} \right] + \mathbb{E}\left[ (e_{11,ib_m}^{d,m,*})^{2} (\tilde{e}_{11,ib_m}^{m})^{4}(e_{12,ib_m}^{d,m})^{2} (\bSigma_{11,ib_m}^{c,m})^{-8} \right] \bigg) \cr
  &&+ C \delta_{m}^{-8} \bigg( \mathbb{E}\left[ (e_{11,ib_m}^{m,*})^{2} (\tilde{e}_{11,ib_m}^{m})^{4} (e_{12,ib_m}^{m})^{2} \right] + \mathbb{E}\left[ (e_{11,ib_m}^{d,m,*})^{2} (\tilde{e}_{11,ib_m}^{m})^{4}(e_{12,ib_m}^{d,m})^{2}  \right] \bigg) \cr
  &=& C \bigg( \mathbb{E}\left[ \mathbb{E}\left[ (e_{11,ib_m}^{m,*})^{2} (\tilde{e}_{11,ib_m}^{m})^{4} (e_{12,ib_m}^{m})^{2} | \mathcal{K}_{ib_m}^{m} \right] (\bSigma_{11,ib_m}^{c,m})^{-8} \right] \cr
  &&+ \mathbb{E}\left[ \mathbb{E}\left[ (e_{11,ib_m}^{d,m,*})^{2} (\tilde{e}_{11,ib_m}^{m})^{4}(e_{12,ib_m}^{d,m})^{2}  | \mathcal{K}_{ib_m}^{m} \right](\bSigma_{11,ib_m}^{c,m})^{-8} \right] \bigg) \cr
  &&+ C \delta_{m}^{-8} \bigg( \mathbb{E}\left[ (e_{11,ib_m}^{m,*})^{2} (\tilde{e}_{11,ib_m}^{m})^{4} (e_{12,ib_m}^{m})^{2} \right] + \mathbb{E}\left[ (e_{11,ib_m}^{d,m,*})^{2} (\tilde{e}_{11,ib_m}^{m})^{4}(e_{12,ib_m}^{d,m})^{2}  \right] \bigg) \cr
  &\leq& C \bigg( \mathbb{E}\left[ \mathbb{E}\left[ (e_{11,ib_m}^{m,*})^{8} | \mathcal{K}_{ib_m}^{m} \right]^{1/4} \mathbb{E}\left[ (\tilde{e}_{11,ib_m}^{m})^{8} | \mathcal{K}_{ib_m}^{m} \right]^{1/2} \mathbb{E}\left[ (e_{12,ib_m}^{m})^{8} | \mathcal{K}_{ib_m}^{m} \right]^{1/4} (\bSigma_{11,ib_m}^{c,m})^{-8} \right] \cr
  &&+ \mathbb{E}\left[ \mathbb{E}\left[ (e_{11,ib_m}^{d,m,*})^{8} | \mathcal{K}_{ib_m}^{m} \right]^{1/4} \mathbb{E}\left[ (\tilde{e}_{11,ib_m}^{m})^{8} | \mathcal{K}_{ib_m}^{m} \right]^{1/2} \mathbb{E}\left[  (e_{12,ib_m}^{d,m})^{8}  | \mathcal{K}_{ib_m}^{m} \right]^{1/4} (\bSigma_{11,ib_m}^{c,m})^{-8} \right] \bigg) \cr
  &&+ C \delta_{m}^{-8} \bigg( \mathbb{E}\left[ (e_{11,ib_m}^{m,*})^{8} \right]^{1/4} \mathbb{E}\left[ (\tilde{e}_{11,ib_m}^{m})^{8} \right]^{1/2} \mathbb{E}\left[ (e_{12,ib_m}^{m})^{8} \right]^{1/4} \cr
  &&+ \mathbb{E}\left[ (e_{11,ib_m}^{d,m,*})^{8} \right]^{1/4} \mathbb{E}\left[ (\tilde{e}_{11,ib_m}^{m})^{8} \right]^{1/2} \mathbb{E}\left[ (e_{12,ib_m}^{d,m})^{8}  \right]^{1/4} \bigg) \cr
  &\leq& C b_m ^{1/2}  \Delta_m  ^{1/4} %
  ,
\end{eqnarray*}
where $e_{12,i}^{d,m} = \hat{\bSigma}_{12,i}^{m} - \hat{\bSigma}_{12,i}^{m,c}$, $e_{11,i}^{d,m,*} = \hat{\bSigma}_{11,i}^{m,*} - \hat{\bSigma}_{11,i}^{m,c}$, $\tilde{e}_{1x,ib_m}^{m}$ is defined in \eqref{eq:error-expression}, the second, fourth, and fifth inequalities are due to Jensen's inequality, H\"older's inequality, and Lemmas \ref{lemma:e}, \ref{lemma:jump-preavg-element}, and \ref{spot-noise-estimate}, respectively, and the equality is due to tower property
Thus, by \eqref{mathZ1-burk} and \eqref{mathZ1-taylor}, we have
\begin{eqnarray*}
  \mathbb{E}\left[ |\mathcal{Z}_{m,1}|^{2} \right] &\leq&  C \Delta_m ^{1/2} \sum_{i=0}^{N_m - 1} \mathbb{E}\left[ b_m^{2} \Delta_m  \left|\hat{B}_{ib_m}^{m} - {B}_{ib_m}^{c,m} \right|^{2}  \right] \cr
  &\leq& C
  .
\end{eqnarray*}
Similarly, we can bound the second moment of $\mathcal{Z}_{m,2}$ and $\mathcal{Z}_{m,4}$ by some constant $C$.
Further, similar to proof of \eqref{D1}, we can show that $| \mathcal{Z}_{m,3} | \leq C \varPsi_{i}^{m,2}$, which implies that the second moment of $\mathcal{Z}_{m,3}$ is bounded by some constant $C$.

For $\mathcal{Z}_{m,5}$, using Taylor's theorem, we have
\begin{eqnarray}\label{mathZ5-taylor}
  && \left| \hat{\beta}_{i b_m }^{c,m} - \beta_{i b_m }^{c,m} - \sum_{x=1}^{2}  \partial_{1x}f(\bSigma_{ib_m}^{c,m}) e_{1x,i b_m}^{m,*} - \frac{1}{2}  \sum_{x,y=1}^{2} \partial^{2}_{1x,1y}f(\bSigma_{ib_m}^{c,m}) e_{1x,i b_m}^{m,*} e_{1y,i b_m}^{m,*} \right| \cr
  &=& \left|\frac{1}{6} \sum_{x,y,z=1}^{2} \partial^{2}_{1x,1y,1z}f(\tilde{\bSigma}_{ib_m}) e_{1x,i b_m}^{m,*} e_{1y,i b_m}^{m,*} e_{1z,i b_m}^{m,*} \right| \cr
  &\leq& C \sum_{x,y,z=1}^{2}  \frac{|\bSigma_{12,ib_m}^{c,m}| + |\bSigma_{11,ib_m}^{c,m}| + |e_{12,i b_m}^{m,*}|}{(\bSigma_{11,ib_m}^{c,m})^{4}} \left|e_{1x,i b_m}^{m,*} e_{1y,i b_m}^{m,*} e_{1z,i b_m}^{m,*}\right| \cr
  &&+ C \sum_{x,y,z=1}^{2}  \frac{|\bSigma_{12,ib_m}^{c,m}| + |\bSigma_{11,ib_m}^{c,m}| + |e_{12,i b_m}^{m,*}|}{(\delta_m)^{4}} \left|e_{1x,i b_m}^{m,*} e_{1y,i b_m}^{m,*} e_{1z,i b_m}^{m,*}\right|
  ,
\end{eqnarray}
where $\tilde{\bSigma}_{ib_m} =  a \hat{\bSigma}_{ib_m}^{c,m} + (1-a) \bSigma_{ib_m}^{c,m} $ for some $a \in [0,1]$ and the inequality is due to triangular inequality.
Using Jensen's inequality, tower property, and H\"older's inequality, sequentially, we can show that the second moment of $\mathcal{Z}_{m,5}$ is bounded by some constant $C$.
For example, we have
\begin{eqnarray*}
  \mathbb{E}\left[ \delta_m^{-4} |\bSigma_{12,ib_m}^{c,m}| \left|e_{12,i b_m}^{m,*}\right|^{3} \big| \mathcal{K}_{i}^{m} \right] &=&  \delta_m^{-4} |\bSigma_{12,ib_m}^{c,m}| \mathbb{E}\left[ \left|e_{12,i b_m}^{m,*}\right|^{3} \big| \mathcal{K}_{i}^{m} \right] \cr
  &\leq& C \delta_m^{-4} |\bSigma_{12,ib_m}^{c,m}| \varPsi_{i}^{m,4}  (b_m \Delta_m) \quad \text{a.s.} \quad \text{and} \quad
\end{eqnarray*}
\begin{eqnarray*}
  \mathbb{E}\left[ \mathbb{E}\left[ \delta_m^{-4} |\bSigma_{12,ib_m}^{c,m}| \left|e_{12,i b_m}^{m,*}\right|^{3} \big| \mathcal{K}_{i}^{m} \right]^{2} \right] &\leq& C \delta_m^{-8} (b_m \Delta_m )^{2} \mathbb{E}\left[  |\bSigma_{12,ib_m}^{c,m}|^{2} (\varPsi_{i}^{m,4})^{2}   \right] \cr
  &\leq& C \delta_m^{-8} (b_m \Delta_m )^{2} \mathbb{E}\left[  |\bSigma_{12,ib_m}^{c,m}|^{2} \right] \mathbb{E}\left[(\varPsi_{i}^{m,4})^{2}\right] \cr
  &\leq& C \delta_m^{-8} (b_m \Delta_m )^{2}
  .
\end{eqnarray*}
Similarly, using Burkholder-Davis-Gundy inequality and Lemma \ref{lemma:e}, we can bound the second moment of $\mathcal{Z}_{m,z}$ by some constant $C$ for $z\in \left\lbrace 6,7 \right\rbrace$.
Further, similar to the proof of \eqref{D3}, we can bound the second moment of $\mathcal{Z}_{m,8}$.
$\blacksquare$

\end{spacing}

\end{document}